\documentclass[
12pt, english,
onehalfspacing, 
nolistspacing, 
liststotoc, 
headsepline, 
consistentlayout, 
]{Thesis} 

\usepackage[utf8]{inputenc} 
\usepackage[T1]{fontenc} 
\usepackage{libertine} 
\usepackage[autostyle=true]{csquotes}
\usepackage{abstract}   
\usepackage{graphicx}   
\usepackage{psfrag}    
\usepackage{subfigure} 
\usepackage{url}       
\usepackage{stfloats} 
\usepackage{amsmath}
\usepackage{amsthm}
\usepackage{amssymb}
\usepackage{braket}
\usepackage{array}
\usepackage{lettrine}
\usepackage{siunitx}
\usepackage{physics}
\usepackage{dsfont}
\usepackage{pdfpages}
\usepackage{lipsum}
\usepackage{verbatim}
\usepackage{cancel}
\usepackage{mathrsfs}
\usepackage[shortlabels]{enumitem}
\usepackage{ragged2e}
\usepackage{pdfpages}

\usepackage[backend=biber, biblabel=brackets, maxbibnames=6, style=phys,sorting=none]{biblatex}
\AtEveryBibitem{%
  \clearfield{doi}%
  \clearfield{url}%
  \clearfield{isbn}%
  \clearfield{issn}%
}
\usepackage[autostyle=true]{csquotes} 

\thesistitle{Decoherence and entropy production due to quantum fluctuations of spacetime} 

\supervisor{Prof. Dr. Lucas Chibebe \textsc{Céleri}} 

\examiner{Prof. Dr. Daniel Augusto Turolla Vanzella \\ Prof. Dr. Fernando da Rocha Vaz Bandeira de Melo \\ Prof. Dr. Nelson de Oliveira Yokomizo \\ Prof. Dr. Rômulo César Rougemont Pereira} 

\degree{PhD in Physics} 

\author{Thiago Henrique \textsc{Moreira}} 

\addresses{Goiânia, 2026} 

\subject{Quantum Stuff} 


\university{\href{http://www.ufg.br}{Federal University of Goiás}} 

\department{\href{http://if.ufg.br}{Institute of Physics}} 

\group{\href{http://qpequi.com/}{Quantum Pequi Group}} 

\AtBeginDocument{
\hypersetup{pdftitle=\ttitle} 
\hypersetup{pdfauthor=\authorname} 
\hypersetup{pdfkeywords=\keywordnames} 
}

\numberwithin{equation}{section}

\begin{document}

\frontmatter 
\pagestyle{plain} 

\thispagestyle{empty}
\begin{figure}[t!]
\centering
\includegraphics[scale=0.15]{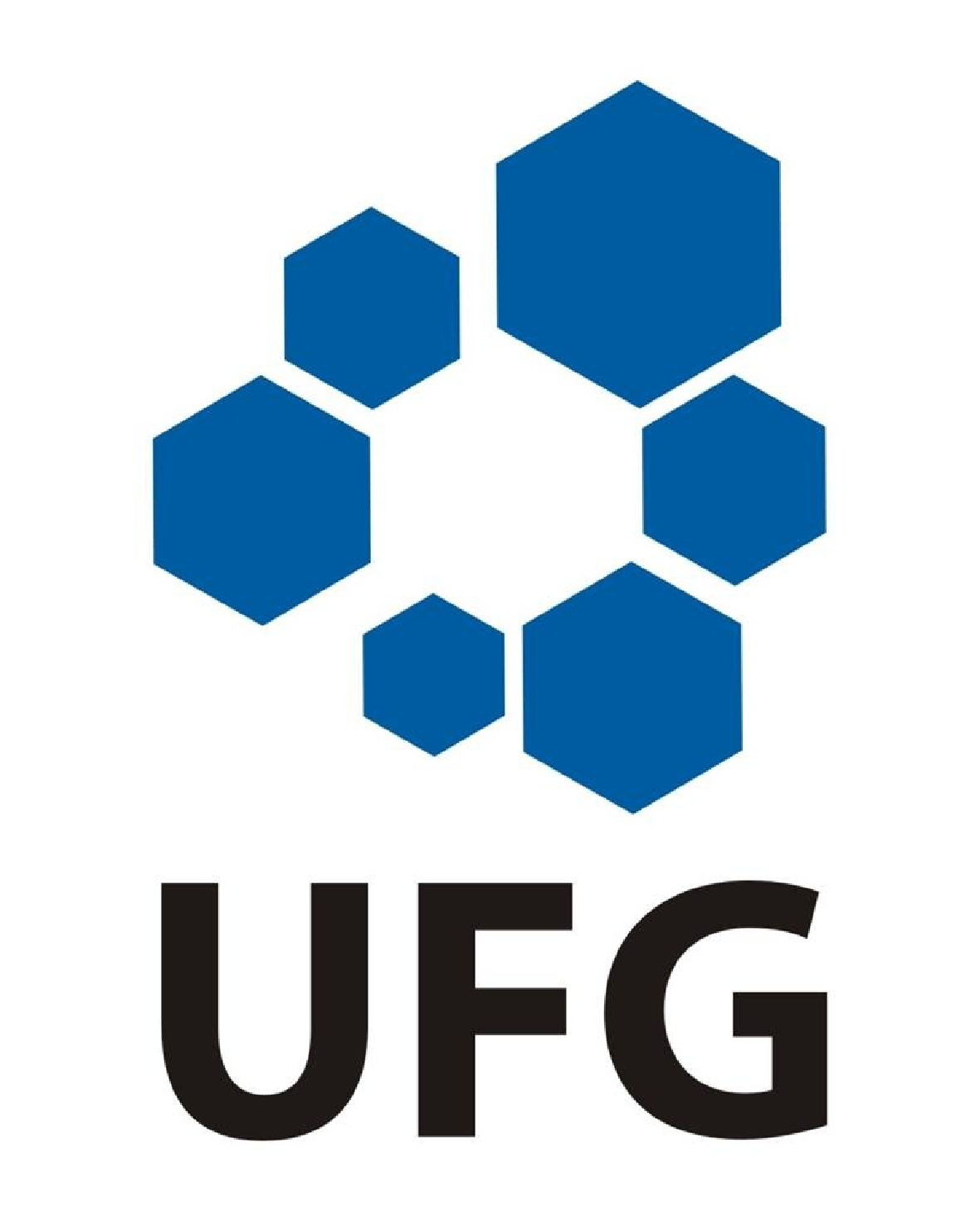} \hspace{5cm} \includegraphics[scale=0.05]{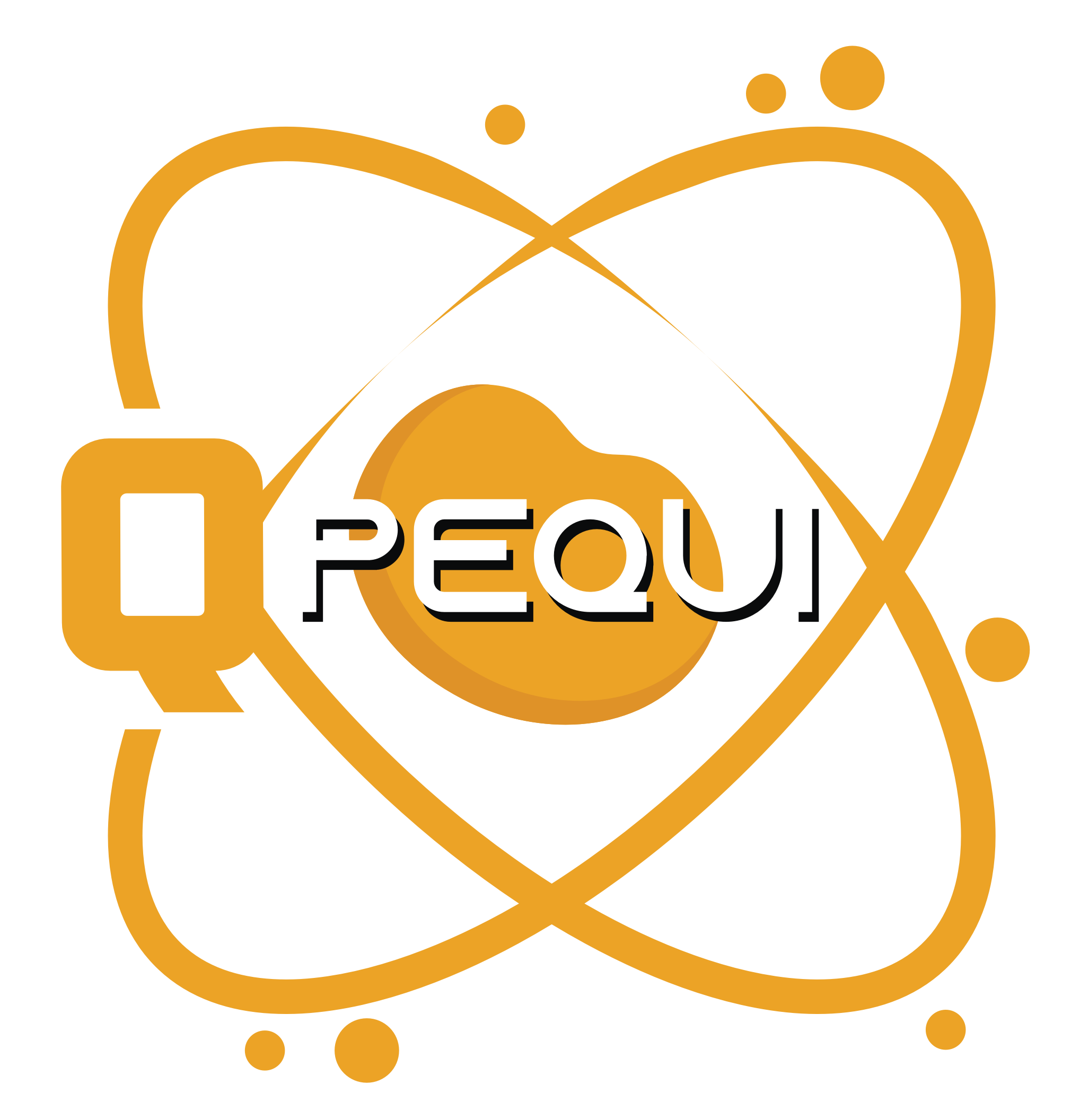}
\end{figure}

\vspace{2cm}

\begin{center}
{\Large \bf \univname}\\
\vspace{0.5cm}
{\large \deptname}\\
\vspace{0.5cm}
{\large \groupname}

\vspace{2cm}
\noindent\makebox[\linewidth]{\rule{0.8\paperwidth}{1pt}}
\vspace{1cm}

\parbox{14cm}{\centering \bf \Large \ttitle} \\

\vspace{1cm}
\noindent\makebox[\linewidth]{\rule{0.8\paperwidth}{1pt}}
\vspace{2cm}

{\large \degreename}\\
\vspace{1.5cm}
 
{\bf \Large \authorname}\\ 
\vspace{1.5cm}

{\bf \large Advisor: \supname}\\
\vspace{0.5cm}

{\large \addressname}\\
\vspace{0.2cm}
{\large Brazil}
\end{center}

\newpage
\null
\thispagestyle{empty}
\newpage

\includepdf[pages={-}, pagecommand={\thispagestyle{empty}}]{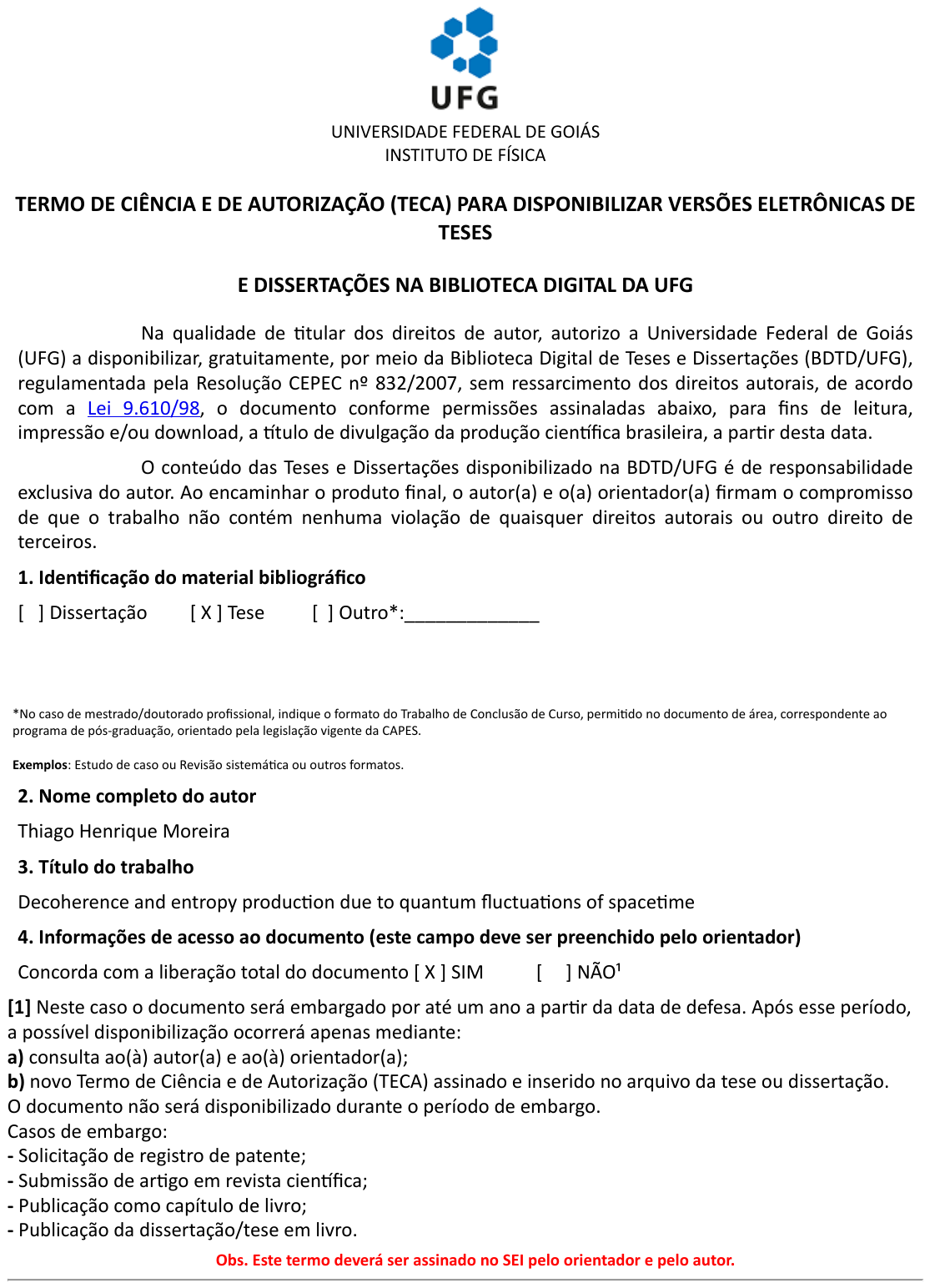}

\newpage
\thispagestyle{empty} 
\begin{center}
    {\large THIAGO HENRIQUE MOREIRA}
    
    \vspace{6cm}
    
    {\Large \textbf{\ttitle}}
    
    \vspace{3cm}
    
    \hspace{.45\textwidth}
    \begin{minipage}{.5\textwidth}
        Tese apresentada ao Programa de Pós-Graduação em Física, do Instituto de Física (IF), da Universidade Federal de Goiás (UFG), como requisito para obtenção do título de Doutor em Física.
        \vspace{0.5cm}
        
        \textbf{Área de concentração:} Física.
        \vspace{0.5cm}
        
        \textbf{Linha de pesquisa:} Óptica Quântica e Informação Quântica.
        \vspace{0.5cm}
        
        \textbf{Orientador:} Professor Doutor Lucas Chibebe Céleri.
    \end{minipage}
    
    \vfill
    
    {\large GOIÂNIA \\ 2026}
\end{center}

\includepdf[pages={-}, pagecommand={\thispagestyle{empty}}]{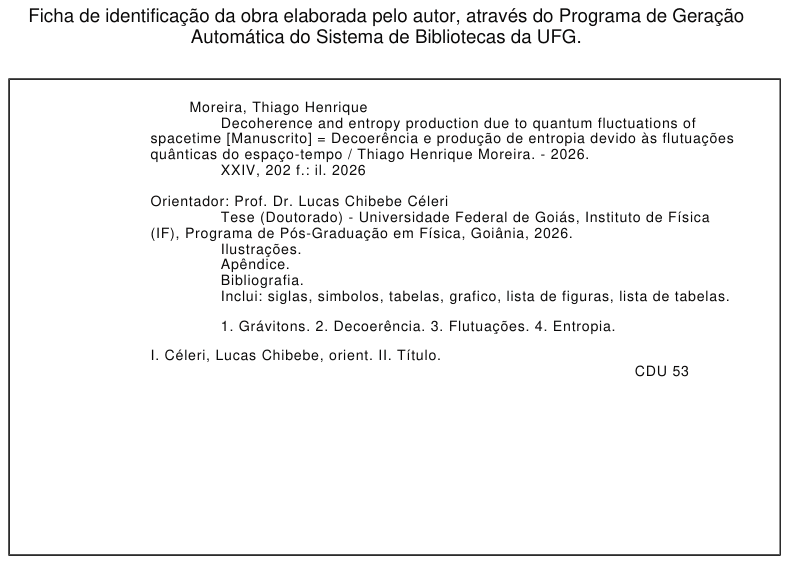}

\includepdf[pages={-}, pagecommand={\thispagestyle{empty}}]{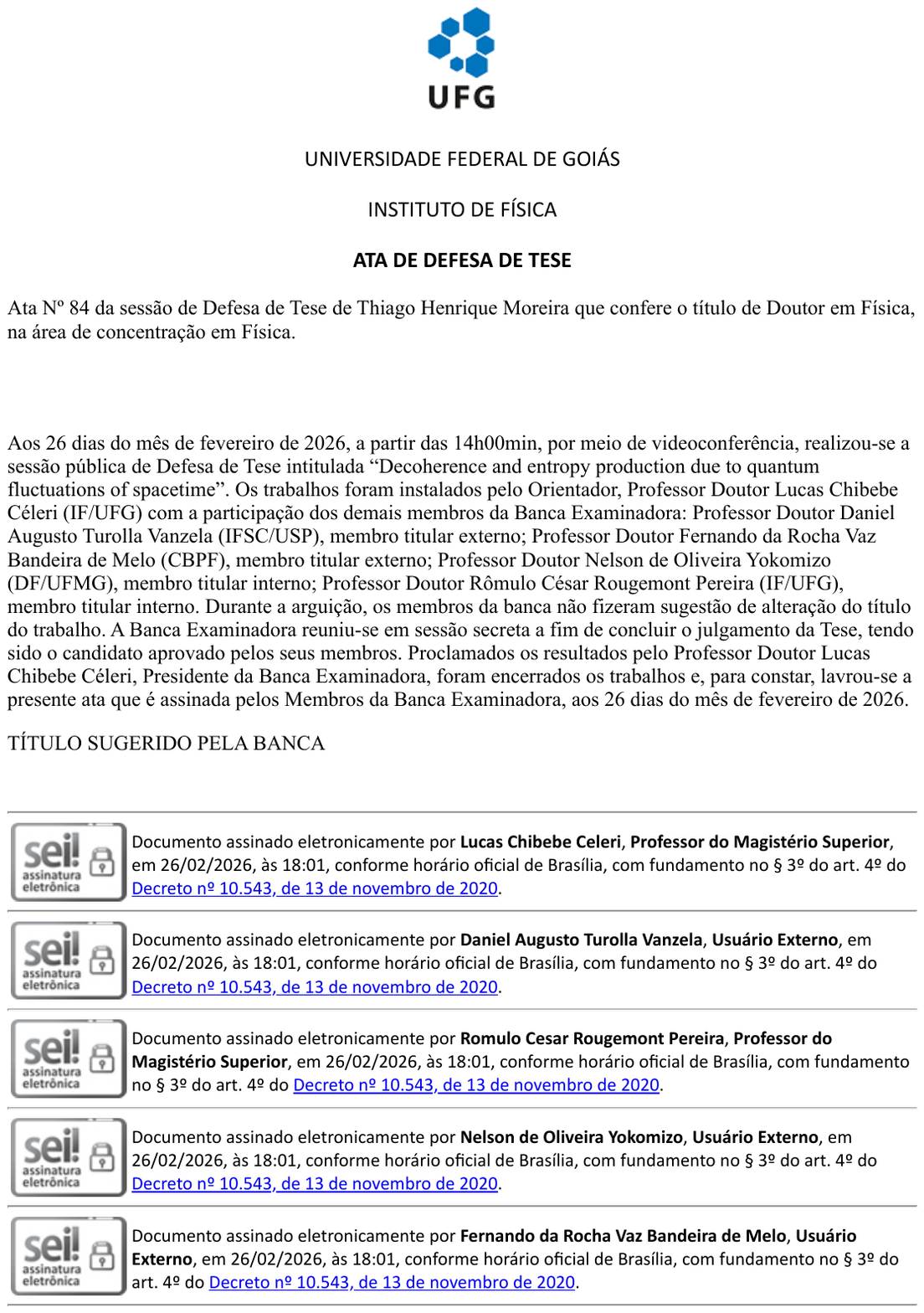}

\begin{center}
{\bf \Large Acknowledgments}\footnote{This work was supported by the Coordination of Superior Level Staff Improvement (CAPES) - Finance Code 001.}
\end{center}

I would like to begin by acknowledging that I owe a great deal to the Institute of Physics of the Federal University of Goiás, which has been my second home for the past ten years, where I went from an undergraduate student, to a master's student, to a substitute professor and finally to a PhD candidate.

I thank all the professors who contributed significantly to my physics education in the past years. Special mention must be made to Professor Renato Pontes, who was my supervisor during my internship in teaching quantum mechanics. Not only did he teach me a lot on the subject, he also gave me some great career advice. I would also like to thank Professors Hermann Freire and Rômulo Rougemont for their useful comments on this work during my qualifying exam.

I thank Konrad Schlichtholz, Tomasz Linowski, and Professor Łukasz Rudnicki, from the International Centre for Theory of Quantum Technologies of the University of Gdańsk, for their hospitality during the time I spent as a visiting PhD student in Poland.

Next, I can hardly express how grateful I am to my advisor, Professor Lucas Céleri. During these last four years, he taught me a lot of physics and has been guiding me in the critical thinking process that is expected of a scientist. From theme suggestions, physics discussions, career and life advice, to the support and understanding during my times of need and the independence he allowed me to pursue, his guidance went far beyond what is strictly required of an advisor.

Now, to the people who were there for me during the last few years, either accompanying me to parties or simply watching movies and TV shows, drinking in bars or sharing a great meal, laughing at stupid things or just endlessly talking about physics, books, movies, TV shows, music, and video games, I cannot express my gratitude enough. This is the family that I have chosen and, although I certainly will not be able to mention everyone, there are some whom I simply cannot fail to acknowledge. To Ary, Igor, Pedro Augusto, Gabi, Babi, Maria, Felipe, Henrique, Sarah, Lucão, Gustavo, Frank, Dib, Lago, Mateus, Edson, Mazetti, Gratão, Arthur, Maykon, Murilo, and Larissa, my most sincere thanks for being part of my life. And, finally, a very special mention to my boyfriend Adal, whom I thank for all of his love and support during this past year.

Naturally, I owe pretty much everything to my family: my mother Nádia, my father Jeneir, my sister Fernanda and her husband Lucas, and, of course, my four-legged daughter Fiona. I do not know who I would be without them.

Lastly, I thank myself for not giving up.

\newpage
\newpage


\vspace*{0.2\textheight}
\thispagestyle{empty}

{\large \itshape Learning all that really matters

Is a slow and painful lesson

}\bigbreak

\hfill {\large Twenty One Pilots, \emph{RAWFEAR}}

\newpage
\newpage


\begin{center}
{\bf \Large Abstract}
\end{center}

\vspace{2cm}

The intersection between quantum mechanics and gravitational physics has been providing challenging puzzles for decades. In this context, we study the dynamics of an open quantum system coupled with a bath of gravitons, the quanta of the gravitational field in the linear limit of general relativity. We focus on two main aspects. First, we analyze the decoherence induced by gravitons when we consider the open system to be described by both external and internal degrees of freedom. Since gravity is universal, the internal variables also interact with the gravitons, and here we show that this interaction leads to the decoherence of spatial superpositions of microscopic systems in the long-time regime, even when the graviton bath alone does not. We then proceed to the second main aspect, which is the entropy production that arises when an external agent drives a quantum system through the graviton bath. This irreversibility comes from quantum fluctuations of spacetime itself and, as such, has a fundamentally universal aspect.
    
\vspace{1cm}
    
   
\newpage
\newpage

\begin{center}
{\bf \Large Resumo}
\end{center}

\vspace{2cm}

A interseção entre a mecânica quântica e a física gravitacional tem proporcionado enigmas desafiadores há décadas. Neste contexto, estudamos a dinâmica de um sistema quântico aberto acoplado a um banho de grávitons, os \emph{quanta} do campo gravitacional no limite linearizado da relatividade geral. Focamos em dois aspectos principais. Primeiro, analisamos a decoerência induzida por grávitons quando consideramos que o sistema aberto é descrito por graus de liberdade tanto externos quanto internos. Como a gravidade é universal, as variáveis internas também interagem com os grávitons, e aqui mostramos que essa interação leva à decoerência de superposições espaciais de sistemas microscópicos no regime de tempos longos, mesmo quando o banho de grávitons isoladamente não o faz. Procedemos então para o segundo aspecto principal, que é a produção de entropia que surge quando um agente externo conduz um sistema quântico através do banho de grávitons. Essa irreversibilidade provém das flutuações quânticas do próprio espaço-tempo e, como tal, possui um aspecto universal fundamental.

\vspace{1cm}

\textbf{Palavras-chave:} Grávitons; Decoerência; Flutuações; Entropia.

\newpage
\newpage


\tableofcontents 

\listoffigures 

\listoftables 

\cleardoublepage
\phantomsection
\addcontentsline{toc}{chapter}{List of Publications}
\chapter*{List of Publications}

\paragraph{Related to this thesis}

\begin{itemize}
    \item MOREIRA, T. H.; CÉLERI, L. C. \textbf{Decoherence of a composite particle induced by a weak quantized gravitational field}. \textit{Classical and Quantum Gravity}, v. 41, p. 015006, 2023.
    \item MOREIRA, T. H.; CÉLERI, L. C. \textbf{Entropy production due to spacetime fluctuations}. \textit{Classical and Quantum Gravity}, v. 42, p. 025022, 2024.
    \item MOREIRA, T. H.; CÉLERI, L. C. \textbf{Graviton induced decoherence of a composite particle}. In: Alexandre Dodonov; Lucas Chibebe Céleri. (Ed.). \emph{Proceedings QNS III International Workshop on Quantum Nonstationary Systems}. 1ed.: LF Editorial, 2025, p. 53-70.
    \item MOREIRA, T. H.; CÉLERI, L. C. \textbf{Gravitational decoherence and recoherence of a composite particle: the interplay between gravitons and a classical Newtonian potential}, arXiv:2602.22517. 
\end{itemize}

\paragraph{Other publications}

\begin{itemize}
    \item MOREIRA, T. H.; BRAGHIN, F. L. \textbf{Magnetic field induced corrections to the NJL model coupling constant from vacuum polarization}. \textit{Physical Review D}, v. 105, p. 114009, 2022.
    \item DE OLIVEIRA, G.; MOREIRA, T. H.; CÉLERI, L. C. \textbf{Dynamical Casimir Effect Under the Action of Gravitational Waves}. \textit{Entropy}, v. 28, p. 177, 2026.
\end{itemize}

\newpage
\newpage

\cleardoublepage
\phantomsection
\addcontentsline{toc}{chapter}{Notations and Conventions}
\chapter*{Notations and Conventions}

\begin{itemize}
    \item \textbf{Index notation.} Four-vectors are denoted by $x^\mu=\qty(x^0,\vb{x})$, where $\vb{x}$ refers to its spatial components and $x^0$ to the temporal one. Partial derivatives are usually denoted by $\partial_\mu\equiv\pdv*{x^\mu}$.
    When working with tensor fields on an $n-$dimensional spacetime, Greek indices run from $0$ to $n-1$ while Latin indices run from $1$ to $n-1$. Repeated indices are summed over (Einstein summation convention) unless stated otherwise.
    The box symbol is reserved for the flat d'Alembertian, $\Box\equiv\eta^{\mu\nu}\partial_\mu\partial_\nu$, where $\eta_{\mu\nu}$ denotes the Minkowski spacetime metric tensor.
    \item \textbf{Symmetrization and antisymmetrization.} For an arbitrary rank tensor ${T_{{\mu_1}\dots{\mu_n}\rho}}^\sigma$ we define
    \begin{equation*}
        {T_{({\mu_1}\dots{\mu_n})\rho}}^\sigma=\frac{1}{n!}\qty({T_{{\mu_1}\dots{\mu_n}\rho}}^\sigma+\textrm{sum over permutations of indices $\mu_1\dots\mu_n$}),
    \end{equation*}
    and
    \begin{equation*}
    \begin{split}
        {T_{[{\mu_1}\dots{\mu_n}]\rho}}^\sigma=\frac{1}{n!}( {T_{{\mu_1}\dots{\mu_n}\rho}}^\sigma+\,&\textrm{alternating sum over} \\
        &\textrm{permutations of indices $\mu_1\dots\mu_n$}) .
    \end{split}
    \end{equation*}
    \item \textbf{Metric signature.} Throughout this work we use the mostly-plus metric convention,
    \begin{equation*}
        \eta_{\mu\nu}=\textrm{diag}\qty(-1,+1,+1,\dots).
    \end{equation*}
    \item \textbf{Curvature.} Useful tensors and symbols constructed from the spacetime metric tensor $g_{\mu\nu}$ and its inverse $g^{\mu\nu}$:
    \begin{table}[!ht]
        \centering 
        \begin{tabular}{ll}
            \toprule 

            \multicolumn{2}{c}{\textbf{Curvature symbols and tensors}} \\
            \cmidrule(lr){1-2} 

            Christoffel symbols   & $\Gamma_{\mu\nu}^\sigma=\frac{1}{2}g^{\sigma\rho}\qty(\partial_\mu g_{\nu\rho}+\partial_\nu g_{\rho\mu}-\partial_\rho g_{\mu\nu})$ \\
            Riemann tensor        & ${R^\rho}_{\sigma\mu\nu}=\partial_\mu\Gamma_{\nu\sigma}^\rho-\partial_\nu\Gamma_{\mu\sigma}^\rho+\Gamma_{\mu\lambda}^\rho\Gamma_{\nu\sigma}^\lambda-\Gamma_{\nu\lambda}^\rho\Gamma_{\mu\sigma}^\lambda$ \\
            Ricci tensor          & $R_{\mu\nu}={R^\lambda}_{\mu\lambda\nu}$ \\
            Ricci scalar          & $R=g^{\mu\nu}R_{\mu\nu}$ \\
            \bottomrule 
        \end{tabular}
    \end{table}
    \item \textbf{Time derivatives.}  Dots over functions of a time variable denote derivatives with respect to such time. For example, $\dot{f}(t)\equiv\dv*{f(t)}{t}$, $\Ddot{h}(t,\vb{x})\equiv\pdv*[2]{h(t,\vb{x})}{t}$, and so on.
    \item \textbf{Fourier transforms.} The $n-$dimensional Fourier transform of a spacetime function $F(x)$, denoted $\Tilde{F}(k)$, is defined such that
    \begin{equation*}
    \begin{split}
        F(x)&=\int\frac{\dd^nk}{(2\pi)^n}\Tilde{F}(k)e^{ikx}, \\
        \Tilde{F}(k)&=\int\dd^nx\,F(x)e^{-ikx},
    \end{split}
    \end{equation*}
    with $kx\equiv\eta_{\mu\nu}k^\mu x^\nu$ and $k^\mu=\qty(\omega,\vb{k})$. The $n-$dimensional delta function satisfies
    \begin{equation*}
        \delta^n(k)=\frac{1}{(2\pi)^n}\int\dd^nx\,e^{ikx}.
    \end{equation*}
    \item \textbf{Units.} Unless explicitly stated otherwise, we work in units such that
    \begin{equation*}
        \hbar=c=k_B=G=1,
    \end{equation*} 
    where $\hbar$ is the reduced Planck constant, $c$ is the speed of light in vacuum, $k_B$ is the Boltzmann constant and $G$ is the gravitational constant. Sometimes it will be useful to restore these universal constants, which can appear in terms of the so-called Planck quantities:
    \begin{table}[!ht]
        \centering 
        \begin{tabular}{llc}
            \toprule 

            \multicolumn{3}{c}{\textbf{Planck quantities}} \\
            \cmidrule(lr){1-3} 

            \textbf{Quantity} & \textbf{Definition} & \textbf{Approximate value in SI units} \\
            \midrule 

            Length      & $L_{\rm P}=\sqrt{\hbar G/c^3}$      & $1.6\times10^{-35}\,\textrm{m}$ \\
            Time        & $t_{\rm P}=\sqrt{\hbar G/c^5}$      & $5.4\times10^{-44}\,\textrm{s}$ \\
            Mass        & $M_{\rm P}=\sqrt{\hbar c/G}$        & $2.2\times10^{-8}\,\textrm{kg}$ \\
            Energy      & $E_{\rm P}=\sqrt{\hbar c^5/G}$      & $2.0\times10^9\,\textrm{J}$ \\
            Temperature & $T_{\rm P}=\sqrt{\hbar c^5/Gk_B^2}$ & $1.4\times10^{32}\,\textrm{K}$ \\
            \bottomrule 
        \end{tabular}
    \end{table}
\end{itemize}

\newpage
\newpage


\mainmatter 

\pagestyle{thesis} 


\cleardoublepage
\phantomsection
\addcontentsline{toc}{chapter}{Introduction}
\chapter*{Introduction}
\markboth{Introduction}{Introduction}

Topics that fall within the intersection of general relativity and quantum mechanics, arguably the two main pillars of modern physics, have been puzzling a fair number of scientists for the last few decades. Perhaps the most famous one refers to the pursuit of a quantum theory of gravity, which remains one of the greatest current open problems in physics. Despite theoretical efforts~\cite{Basile2025,Buoninfante2025,Eppley1977,Albers2008,Kiefer2012,Penrose2014,Armas2021,Oppenheim2023,Salvio_2018,Donoghue2022,Percacci2017,Reuter2019,Ambjoern2012,Loll_2019,Rovelli2007,Ashtekar2021,Polchinski2005,Polchinski2005_2}, an experimental confirmation of the quantum nature of the gravitational interaction is still lacking, most likely due to its weakness relative to the other three known fundamental interactions, all of which are satisfactorily described by quantum field theory. Without guiding experimental results, there are many candidates for a quantum theory of gravity with significant differences among them, and there is no general consensus on how to properly quantize gravity, or even if it requires such a quantum treatment.

Whatever theory of quantum gravity (if any) is ever proved to be the most suitable, it is natural to expect that the usual quantum field theoretical treatment will hold in the limit of weak gravitational fields. In this limit, the total spacetime metric is described by an expansion around some known solution to Einstein's equation, in which the perturbation represents the propagating degrees of freedom. This is the classical description of the so-called gravitational waves, which were detected by LIGO in 2015~\cite{Abbott_2016}. At the quantum level, these waves give rise to a spin 2 excitation called the \emph{graviton}, in the same way that the quantization of electromagnetic waves introduces the concept of the photon. This is the framework of \emph{perturbative quantum gravity}~\cite{Basile2025}, a formalism that faces problems concerning perturbative non-renormalizability; nonetheless, it stands as a genuine predictive effective field theory~\cite{Donoghue1994}. However, the weakness of gravity poses a significant obstacle to verifying such predictions with current technology.

Even the detection of gravitational waves by LIGO was only possible after considerable efforts by hundreds of scientists over many years, so one can wonder how far we are from actually detecting single gravitons (if such a detection is even possible~\cite{Dyson_2013}). This scenario motivates different proposals for detecting quantum aspects of gravity in more indirect ways that do not require the direct detection of gravitons, such as gravity-induced entanglement~\cite{Bose2017,Marletto2017}, graviton noise affecting the geodesic deviation of test particles~\cite{Parikh2020,Parikh_2021,Parikh2021,Cho2022,Cho_2023,Chawla2023,Kanno2021}, and gravitational decoherence~\cite{Bassi_2017,Hsiang_2024,Blencowe_2013,Anastopoulos_2013,Kanno2021}.

\emph{Decoherence} refers to the phenomenon of irreversible loss of quantum coherence due to the entanglement between a system of interest and an environment, which is usually a quantum system whose degrees of freedom are either not of interest or intractable from a practical point of view. When this phenomenon is somewhat tied to gravitational interactions, we call it \emph{gravitational decoherence}. For example, Kanno et al.~\cite{Kanno2021} analyzed the decoherence of a spatial superposition of a quantum point particle induced by interaction with a graviton bath. However, the term "gravitational decoherence" is not always tied to cases where the environment is described by quantum gravitational degrees of freedom. Particularly, Pikovski et al.~\cite{Pikovski2015,Pikovski2017} studied the decoherence of a composite particle due to the interaction with its own internal structure, which is mediated by the time dilation induced by a classical static gravitational potential. Following these two distinct mechanisms of gravitational decoherence, we pose the question of whether a graviton bath can also induce a coupling between the internal and external degrees of freedom of a composite system while acting as an environment itself, and what the consequences are for the decoherence of quantum superpositions of the center-of-mass coordinate of such a composite particle.

In this work, we investigate the decoherence in the external degrees of freedom of a quantum system due to its coupling with its own internal structure and with a bath of gravitons. We make use of the Feynman-Vernon influence functional approach~\cite{Feynman1963,Feynman2010,Calzetta2008} (which generates the non-equilibrium effective action for the system of interest within the Schwinger-Keldysh closed time path framework~\cite{Schwinger_1961,Keldysh2023}) to obtain the reduced density matrix of the relevant variables. Such a scenario involves dealing with an open quantum system interacting with two environments. However, due to the universal aspect of gravity, the field also couples with the internal degrees of freedom, meaning that the two environments also interact with each other and the solution to the problem is not easily obtained by simply adding the individual influences of each of them. We show how to obtain the decoherence function, as well as the decoherence time, for the loss of quantum coherence of the system initially in a superposition of spatially separated spacetime events by considering four possible initial graviton states.

Another question we pose is regarding the entropy production due to such spacetime quantum fluctuations. Particularly, the irreversibility of gravitational time dilation was analyzed in Ref.~\cite{Basso2023}, which was later generalized to classical curved spacetimes in Ref.~\cite{Basso2025} and applied to quantum fields in Ref.~\cite{Costa2025}. Here, by considering the open quantum system acted upon by an external agent during a finite time interval, we obtain an expression for the work dissipated when the system is bound to move through a bath of gravitons. This is accomplished by analyzing how the fluctuation theorem~\cite{Jarzynski1997,Crooks1999,Horowitz2007,Jarzynski2007,Jarzynski2008,Campisi2011} applies to the problem at hand.

This thesis is divided into three main parts, which we organize as follows:
\begin{itemize}
    \item Part~\ref{Part:Theory} presents the theoretical background, the tools and concepts that will be used in the remainder of this work.
    \begin{itemize}
        \item Chapter~\ref{chap:quantum-grav} is a literature review of classical linearized gravity and its quantum treatment, the framework of perturbative quantum gravity, which introduces the concept of the graviton.
        \item Chapter~\ref{chap:dec-and-FV} reviews the topic of open quantum systems with focus on the phenomenon of decoherence, particularly in the context of the quantum Brownian motion. Chapters~\ref{chap:quantum-grav} and~\ref{chap:dec-and-FV} are completely unrelated and do not need to be read in any particular order.
        \item Chapter~\ref{chap:gravdec} puts together the concepts introduced in the previous ones and presents a brief and selected literature review of the topic of gravitational decoherence.
    \end{itemize}
    \item Part~\ref{Part:Dec-and-ent} is the core of this work, where we analyze the graviton-induced decoherence of a composite system, as well as the graviton-induced entropy production.
    \begin{itemize}
        \item In Chapter~\ref{chap:quantum-system} we consider a quantum composite particle, described by external and internal degrees of freedom, coupled with a weak quantized gravitational field in a Newtonian background. By integrating out all but the external system's variables, using the Feynman-Vernon influence functional, we find the reduced system density matrix in which all environment influences are encoded in noise kernels.
        \item In Chapter~\ref{chap:decoherence} we analyze the decoherence function for a spatial superposition of the system's center-of-mass variable by considering specific configurations of the superposition state, as well as different initial graviton states. Both Chapters~\ref{chap:quantum-system} and~\ref{chap:decoherence} are adapted from two original papers~\cite{Moreira_2023,Moreira2026}.
        \item In Chapter~\ref{chap:entropy} we analyze the work dissipated when an external agent moves a system through the graviton bath and quantify the entropy production. This chapter is adapted from another original paper~\cite{Moreira2024}.
        \item We close Part~\ref{Part:Dec-and-ent} with some concluding remarks as well as with some perspectives for future work.
    \end{itemize}
    \item Part~\ref{Part:App} contains the appendices.
    \begin{itemize}
        \item Appendix~\ref{app:Differential} is a review of useful concepts in differential geometry, the mathematics behind general relativity.
        \item Appendix~\ref{app:Scattering} explores the interaction between the quantum gravitational radiation degrees of freedom and the Newtonian background. Here we explicitly compute the differential cross section for graviton scattering, which we briefly discuss in Chapters~\ref{chap:quantum-grav} and~\ref{chap:quantum-system}.
        \item In Appendix~\ref{app:noise-kernels} we explicitly compute the noise kernels that encode the environmental influences on the quantum system of interest. We compute the gravitational noise kernel for vacuum, thermal, coherent and squeezed states, and also the internal degrees of freedom noise kernel by considering those to be in a thermal state.
        \item Finally, Appendix~\ref{app:fluctuation-theorem} briefly reviews the Crooks (classical and quantum) fluctuation relation and Jarzynski's equality.
        \item The Bibliography is listed at the end of the text.
    \end{itemize}
\end{itemize}
\let\oldthispagestyle\thispagestyle
\renewcommand{\thispagestyle}[1]{\oldthispagestyle{empty}}
\part{Theoretical background} \label{Part:Theory}
\let\thispagestyle\oldthispagestyle
\chapter{Perturbative quantum gravity}
\label{chap:quantum-grav}

Developed by Albert Einstein in the 20th century, the general theory of relativity quickly became one of the pillars of modern physics, alongside quantum mechanics. By providing a successful and modern description of the gravitational interaction, general relativity has led to some remarkable predictions that have been experimentally verified, such as the precession of Mercury's perihelion, the bending of light by gravitational fields, and the existence of gravitational waves~\cite{Carroll,Wald1984,Weinberg2013,Misner,Hartle2003,Zee2013,Schutz2022,Stewart1993,Maggiore2007,Maggiore2018}. Being the weakest of the four known fundamental interactions, gravity is the only one among them that lacks a satisfactory and experimentally confirmed quantum description. While the electromagnetic, weak, and strong forces are successfully described by the Standard Model of particle physics~\cite{Griffiths2008,Schwartz2013,Navas_2024}, the question of whether gravity even admits a quantum description or retains its classical status is still under debate~\cite{Eppley1977,Albers2008,Kiefer2012,Penrose2014,Armas2021,Oppenheim2023}. Naturally, this debate can only be resolved by experiments. Although standard dimensional analysis seems to indicate that quantum gravity is expected to become relevant only at the Planck scale~\cite{Wald1984}, which is significantly beyond the reach of current technology, recent years have witnessed increasing attention to the idea of testing quantum features of gravity in table-top experiments, motivated by quantum information concepts~\cite{Pikovski2012,Bose2017,Marletto2017,Carney2019,Anastopoulos2020,Chevalier2020,Carney2021,Pedernales2022,Danielson_2022,Christodoulou2023,Christodoulou2023b,Kaku2023,Bose2025,Beyer2025,Aziz2025,Marletto2025}. However, conclusive experimental evidence remains elusive due to the difficulties in probing the weak gravitational interaction in the quantum regime.

From the theoretical side, there are also some conceptual difficulties. Gravity differs from the other interactions since the latter are described by fields on a fixed spacetime background, while general relativity is a theory of spacetime itself, and it is not clear what it means to quantize it. There are many candidates for a full theory of quantum gravity, such as quadratic gravity~\cite{Salvio_2018,Donoghue2022}, asymptotically safe quantum gravity~\cite{Percacci2017,Reuter2019}, causal dynamical triangulations~\cite{Ambjoern2012,Loll_2019}, loop quantum gravity~\cite{Rovelli2007,Ashtekar2021}, string theory~\cite{Polchinski2005,Polchinski2005_2}, among others~\cite{Basile2025,Buoninfante2025}. Each one faces its own conceptual issues, and all of them lack experimental corroboration. 

Nevertheless, whatever full description we may have for quantum gravity at all energy scales (if needed), one may expect to recover, in the weak-field limit, the usual quantum field theory description on a fixed background. To be more precise, the weak-field limit involves a metric expansion around some classical solution to Einstein's field equations, and only the dynamics of the small perturbation is considered. Classically, this leads to the description of gravitational waves. Since this is mathematically equivalent to describing a field over a fixed background, the quantization procedure follows straightforwardly in complete analogy with the quantization of the electromagnetic field, for instance. This is the framework of \textbf{perturbative quantum gravity}, and the quantum excitations of the metric perturbation field are what we call the \textbf{gravitons}.

This quantum field theoretical approach to general relativity is to be viewed as an effective field theory since it suffers from perturbative non-renormalizability; hence, it cannot be taken seriously up to any energy scale. This means that, at the Planck scale, a UV completion is necessary, i.e., a full theory of quantum gravity. Nevertheless, perturbative quantum gravity makes testable predictions. For instance, this framework was used in Ref.~\cite{Donoghue1994} to explicitly compute radiative corrections to the Newtonian potential (see also Ref.~\cite{Schwartz2013}). For a potential of the form\footnote{Keeping the universal constants explicit.} $\phi(r)=-GM/r$ this correction is of the order of $\hbar G/c^3r^2$, which is too small to be measured any time soon, but it stands as a genuine prediction of perturbative quantum gravity nonetheless.

The present chapter is devoted to a review of this framework, which serves as the basis for the subsequent analysis. We begin by quickly introducing the general theory of relativity formulated from the action principle, and then proceed to consider the case of \emph{linearized gravity}, in which we study small perturbations to Minkowski spacetime. These small perturbations are the classically known gravitational waves, so we discuss how these ripples in spacetime propagate, how they interact with test particles, how they are produced, and how they are detected. Next, we discuss their quantization, which gives rise to the gravitons, and we also outline the limitations of the theory. Finally, we generalize the framework of linearized gravity for a general curved background. Then, we take the Newtonian limit and study how gravitons interact and are scattered by a classical Newtonian potential. This discussion will be essential for when we consider the problem of graviton-induced decoherence near some Newtonian source (like Earth) in Part~\ref{Part:Dec-and-ent}.

\section{The Einstein-Hilbert action}

One of the main ideas behind Einstein's theory of gravitation is the \textbf{equivalence principle}. It dates back to Newtonian mechanics by stating that the inertial mass, which measures how much any given force can influence the motion of a system, is the same as the gravitational mass, which is related to the property that any massive system has to interact with and produce a gravitational field in all space (in a way analogous to the electric charge in Coulomb's law). The equivalence principle provides a special status to the gravitational interaction, for which a more modern statement reads~\cite{Carroll}
\begin{center}
    \emph{In small enough regions of spacetime, the laws of physics reduce to those of special relativity; it is impossible to detect the existence of a gravitational field by means of local experiments.}
\end{center}

The equivalence principle implies that gravity is universal, affecting all forms of energy and momentum. If that were not the case, one could perform a local experiment involving two systems (one affected by gravity and the other immune to it), observe their different behaviors, and infer the presence of the gravitational field, thereby violating the principle. Under this paradigm, it becomes natural to think that such a universal interaction must therefore be linked to the structure of spacetime on which matter and radiation propagate.

The connection with Riemannian geometry arises due to the similarity between the equivalence principle and Gauss's guiding principle of non-Euclidean geometry: at any point on a curved space, we may construct a locally Cartesian coordinate system in which distances obey the law of Pythagoras~\cite{Weinberg2013}. If we postulate that spacetime events are points in a curved differentiable manifold with a pseudo-Riemannian metric, the equivalence principle arises naturally from the construction of locally inertial coordinates once we identify gravity with the curvature of spacetime (see Appendix~\ref{app:Differential} for a quick review on differential geometry). The spacetime curvature is affected by the distribution of energy and momentum in a way that is described by Einstein's equation.

The postulates of general relativity can be listed as follows:
\begin{enumerate}
    \item Spacetime is a four-dimensional differentiable manifold $M$ endowed with a Lorentzian metric $g_{\mu\nu}$.
    \item The curvature of $(M,g_{\mu\nu})$ is related to the matter distribution in spacetime by Einstein's equation, which reads\footnote{In this work, we only consider solutions with null cosmological constant.}
    \begin{equation} \label{Einstein-equation}
        R_{\mu\nu}-\frac{1}{2}Rg_{\mu\nu}=8\pi T_{\mu\nu}.
    \end{equation}
\end{enumerate}
In Eq.~\eqref{Einstein-equation}, $R_{\mu\nu}$ is the Ricci tensor and $R$ is the Ricci scalar. The $(0,2)$ symmetric tensor $T_{\mu\nu}$ is called the \textbf{energy-momentum tensor}. The Bianchi identity, together with Einstein's equation, implies energy-momentum conservation,
\begin{equation}
    \nabla_\mu T^{\mu\nu}=0,
\end{equation}
with $\nabla_\mu$ denoting the (metric-compatible) covariant derivative.

Einstein's equation states that the metric structure of spacetime is governed by the energy density distribution encoded in the energy-momentum tensor; however, this tensor also depends on the metric itself. For a perfect fluid, for instance, it reads
\begin{equation}
    T^{\mu\nu}=(\rho+P)U^\mu U^\nu+g^{\mu\nu}P,
\end{equation}
being described by its rest-frame energy density $\rho$ and an isotropic rest-frame pressure $P$, as well as its four-velocity vector field $U^\mu(x)$. Due to the explicit dependence on the metric, and also to the highly non-linear nature of the Einstein tensor, finding general solutions to Einstein's equation is a very difficult task, in general.

General relativity is an example of a \emph{classical field theory} in which the dynamical field is the spacetime metric. As is usual in field theories, one can derive the equations of motion from the principle of least action. For general relativity, we write the total action as
\begin{equation}
    S=S_{\rm EH}+S_{\rm matter},
\end{equation}
where the first term on the right-hand side describes the metric field, while the second term describes the matter that propagates in spacetime. The action describing the gravitational field is called the Einstein-Hilbert action, and it is given by
\begin{equation} \label{Einstein-Hilbert-action}
    S_{\rm EH}=\frac{1}{16\pi}\int\dd^4x\,\sqrt{-g}R,
\end{equation}
where the integral extends throughout all spacetime and $g$ is the metric determinant.

According to the principle of least action, the equations of motion arise for field configurations that satisfy $\delta S=0$, where the $\delta$ notation denotes the variation of the action functional with respect to the metric field. For the Einstein-Hilbert action, we have
\begin{equation} \label{deltaSEH}
    \delta S_{\rm EH}=\frac{1}{16\pi}\int\dd^4x\,\qty[\qty(\delta\sqrt{-g})R+\sqrt{-g}\qty(\delta g^{\mu\nu})R_{\mu\nu}+\sqrt{-g}g^{\mu\nu}\qty(\delta R_{\mu\nu})].
\end{equation}
For the first term on the right-hand side, we take the variation on both sides of the identity $\ln g={\rm Tr}\,\qty(\ln g_{\mu\nu})$ to obtain $\delta g=g\qty(g^{\mu\nu}\delta g_{\mu\nu})$. Now use the fact that $\delta(g^{\mu\lambda}g_{\lambda\nu})=0$, which then leads to $\delta g_{\mu\nu}=-g_{\mu\rho}g_{\nu\lambda}\delta g^{\rho\lambda}$ and, finally, to $g^{\mu\nu}\delta g_{\mu\nu}=-g_{\mu\nu}\delta g^{\mu\nu}$. Then,
\begin{equation}
    \delta \sqrt{-g}=-\frac{1}{2\sqrt{-g}}\delta g=-\frac{1}{2}\sqrt{-g}g_{\mu\nu}\delta g^{\mu\nu}.
\end{equation}
Now, for the third term on the right-hand side of Eq.~\eqref{deltaSEH}, we need to compute $\delta R_{\mu\nu}$. First, note that a linear metric variation, $g_{\mu\nu}\to g_{\mu\nu}+\delta g_{\mu\nu}$, leads to a linear variation of the Christoffel symbols, $\Gamma_{\mu\nu}^\rho\to\Gamma_{\mu\nu}^\rho+\delta\Gamma_{\mu\nu}^\rho$. Being the difference between two connections, the variation $\delta\Gamma_{\mu\nu}^\rho$ is a tensor. Then, the corresponding variation of the Riemann tensor reads
\begin{align}
    \delta {R^\rho}_{\mu\lambda\nu}&=\partial_\lambda\qty(\delta\Gamma_{\nu\mu}^\rho)+\Gamma_{\lambda\sigma}^\rho\qty(\delta\Gamma_{\nu\mu}^\sigma)+\Gamma_{\nu\mu}^\sigma\qty(\delta\Gamma_{\lambda\sigma}^\rho)-\qty(\lambda\leftrightarrow\nu) \nonumber \\
    &=\nabla_\lambda\qty(\delta\Gamma_{\nu\mu}^\rho)-\nabla_\nu\qty(\delta\Gamma_{\lambda\mu}^\rho),
\end{align}
where the covariant derivative is compatible with $g_{\mu\nu}$, not with $g_{\mu\nu}+\delta g_{\mu\nu}$.

Putting everything back together into Eq.~\eqref{deltaSEH} leads to
\begin{align}
    \delta S_{\rm EH}&=\frac{1}{16\pi}\int\dd^4x\,\sqrt{-g}\qty(R_{\mu\nu}-\frac{1}{2}g_{\mu\nu}R)\delta g^{\mu\nu} \nonumber \\
    &+\frac{1}{16\pi}\int\dd^4x\,\sqrt{-g}\nabla_\rho\qty[g^{\mu\nu}\qty(\delta\Gamma_{\nu\mu}^\rho)-g^{\mu\rho}\qty(\delta\Gamma_{\lambda\mu}^\lambda)].
\end{align}
Note that, by Stokes's theorem, the term on the last line is a boundary contribution. However, it does not vanish for general variations with $g_{\mu\nu}$ held fixed because it also depends on the first derivatives of the metric. For a spacetime with a non-null boundary, for which the first derivatives of $g_{\mu\nu}$ are not held fixed on it, one needs to include a boundary term in the Einstein-Hilbert action in order to cancel this contribution and obtain Einstein's equation. This occurs because the Ricci scalar, the Lagrangian density of the Einstein-Hilbert action, depends on the second derivatives of the metric field, while the Lagrangian densities of other field theories usually depend only on the field and its first derivatives (see Refs.~\cite{Wald1984,Poisson2004}). Having mentioned this subtlety, let us simply set the boundary term to zero, which will be enough for our purposes.

The variation of the total action reads
\begin{equation}
    \delta S=\frac{1}{16\pi}\int\dd^4x\,\sqrt{-g}\qty(R_{\mu\nu}-\frac{1}{2}Rg_{\mu\nu})\delta g^{\mu\nu}+\delta S_{\rm matter}.
\end{equation}
The energy-momentum tensor $T_{\mu\nu}$ can then be defined from the variation of the matter action $S_{\rm matter}$ under a change of the metric according to
\begin{equation} \label{Definition-energy-momentum-tensor}
    \delta S_{\rm matter}=-\frac{1}{2}\int\dd^4x\,\sqrt{-g}\,T_{\mu\nu}\delta g^{\mu\nu}.
\end{equation}
Then, setting $\delta S=0$ immediately yields Einstein's equation~\eqref{Einstein-equation}.

\section{Linearized gravity in flat spacetime} \label{Sec:LG-flat-spacetime}

Having established the foundations of general relativity, we now address the dynamical degrees of freedom of the metric field in the weak-field limit, which leads to the phenomenon of \emph{gravitational radiation}. The study of gravitational radiation faces an immediate difficulty due to the non-linearity of Einstein's equation, which is not encountered when studying electromagnetic radiation, for instance. The main difference between the former and the latter is that, while electromagnetic waves carry no electric charge, a gravitational wave is itself a distribution of energy and momentum, which then affects its own field. We shall then be interested in studying the weak-field solutions to Einstein's equation, representing gravitational waves whose energy and momentum are not strong enough to significantly alter their field. In this section, we study the metric expansion around \emph{flat spacetime}. The general case of expansion around a curved metric will be treated in Section~\ref{Sec:LG-curved-spacetime}.

\subsection{Weak-field limit}

The linearized version of general relativity describes a theory of a symmetric tensor field $h_{\mu\nu}(x)$ propagating on a fixed background spacetime, which we choose to be flat spacetime for now. We begin by expanding the metric as
\begin{equation} \label{g=n+h}
    g_{\mu\nu}=\eta_{\mu\nu}+h_{\mu\nu},
\end{equation}
where we assume that $\abs{h_{\mu\nu}}\ll1$ such that we can neglect higher order terms in the equations of motion. The inverse metric reads
\begin{equation}
    g^{\mu\nu}=\eta^{\mu\nu}-h^{\mu\nu},
\end{equation}
where $h^{\mu\nu}=\eta^{\mu\rho}\eta^{\nu\sigma}h_{\rho\sigma}$. To linear order in $h_{\mu\nu}$, the Christoffel symbols read
\begin{equation} \label{Christoffel-linear-in-h}
    \Gamma_{\mu\nu}^\rho=\frac{1}{2}\eta^{\rho\sigma}\qty(\partial_\mu h_{\nu\sigma}+\partial_\nu h_{\sigma\mu}-\partial_\sigma h_{\mu\nu}),
\end{equation}
and the Riemann tensor becomes
\begin{equation} \label{Riemann-linear-in-h}
    R_{\rho\sigma\mu\nu}=\frac{1}{2}\qty(\partial_\mu\partial_\sigma h_{\rho\nu}+\partial_\nu\partial_\rho h_{\mu\sigma}-\partial_\mu\partial_\rho h_{\nu\sigma}-\partial_\nu\partial_\sigma h_{\rho\mu}).
\end{equation}

In order to obtain an expression for the action describing the field $h_{\mu\nu}(x)$, we need to expand the Einstein-Hilbert action up to second order in $h_{\mu\nu}$, since the equations of motion are expected to be first order in the metric perturbation. After a long but straightforward calculation, we find
\begin{align}
    R&=\partial_\mu\partial_\nu h^{\mu\nu}-\Box h-\frac{1}{2}h^{\mu\nu}\partial_\mu\partial^\rho h_{\rho\nu}+\frac{1}{4}h\Box h+\frac{1}{4}h^{\mu\nu}\Box h_{\mu\nu} \nonumber \\
    &+\textrm{total derivative of second order terms},
\end{align}
where $h=\eta^{\mu\nu}h_{\mu\nu}$ and we are omitting total derivatives of terms of order $O(h^2)$, since these will either lead to boundary terms or terms of order $O(h^3)$ in the Lagrangian. Furthermore, we can use the identity\footnote{This follows from the identity $\det \mathbb{G}=\exp[\textrm{Tr}(\ln\mathbb{G})]$ for a matrix $\mathbb{G}$.}
\begin{equation}
    \sqrt{-g}=1+\frac{1}{2}h+O(h^2),
\end{equation}
to obtain
\begin{equation} \label{Einstein-Hilbert-linearized}
    S_{\rm EH}=\frac{1}{64\pi}\int \dd^4x\,\qty(h_{\mu\nu}\Box h^{\mu\nu}+2h^{\mu\nu}\partial_\mu\partial_\nu h-h\Box h-2h_{\mu\nu}\partial_\rho\partial^\mu h^{\nu\rho}),
\end{equation}
where we dropped boundary terms as well as higher order ones. Eq.~\eqref{Einstein-Hilbert-linearized} is the Einstein-Hilbert action linearized around flat spacetime, and it describes the free dynamics of the metric perturbation field $h_{\mu\nu}$. The equation of motion \emph{in vacuum} ($T_{\mu\nu}=0$) obtained from this action, for instance, takes the form
\begin{equation} \label{Linearized-Mink-vacuum-equation}
    \Box h_{\mu\nu}=\partial^\sigma\partial_\nu h_{\sigma\mu}+\partial^\sigma\partial_\mu h_{\sigma\nu}-\partial_\mu\partial_\nu h.
\end{equation}

Next, let us see how diffeomorphism invariance of general relativity can actually simplify the Einstein-Hilbert action even further.

\subsection{Gauge invariance}

Recall that Einstein's equation can be written as $G_{\mu\nu}=8\pi T_{\mu\nu}$, where $G_{\mu\nu}$ is the Einstein tensor [see Eq.~\eqref{Einstein-tensor}], which is a symmetric tensor with ten algebraically independent components. These do not suffice to determine $g_{\mu\nu}$ uniquely because these components are related by four differential identities, namely the Bianchi identities, $\nabla^\mu G_{\mu\nu}=0$. This means that the solution of Einstein's equation leaves us with four degrees of freedom in $g_{\mu\nu}$ that are not uniquely determined. This arises from diffeomorphism invariance, which physically corresponds to the fact that if $g_{\mu\nu}$ solves Einstein's equation, then so will $g_{\mu\nu}'$, which is derived from $g_{\mu\nu}$ by a general coordinate transformation.

To see the consequences of this observation for linearized gravity, let us write the metric expansion~\eqref{g=n+h} as
\begin{equation} \label{Metric-expansion-parameterized-by-e}
    g_{\mu\nu}(\epsilon)=\eta_{\mu\nu}+\epsilon h_{\mu\nu},
\end{equation}
such that
\begin{equation}
    h_{\mu\nu}=\eval{\dv{g_{\mu\nu}(\epsilon)}{\epsilon}}_{\epsilon=0}.
\end{equation}
One can then think of the parameter $\epsilon$ as generating a family of spacetime metrics.

Let $\phi:M\to M$ be a diffeomorphism. Due to diffeomorphism invariance, $(M,g_{\mu\nu})$ and $(M,(\phi^*g)_{\mu\nu})$ represent the same physical spacetime, with $(\phi^*g)_{\mu\nu}$ denoting the pullback of the metric tensor\footnote{For notation simplicity, we leave implicit the fact that if $g_{\mu\nu}$ is evaluated at the spacetime point $p$, then $(\phi^*g)$ is evaluated at $\phi(p)$.} (see Appendix~\ref{app:Differential}). Now consider a one-parameter family of spacetimes, $(M,g_{\mu\nu}(\epsilon))$. Then $(M,(\phi_\epsilon^*g(\epsilon))_{\mu\nu})$ represents the same spacetime family, with $\phi_\epsilon$ denoting the one-parameter group of diffeomorphisms generated by an arbitrary vector field $\varepsilon^\mu(x)$.

Diffeomorphism invariance then implies that $h_{\mu\nu}'=\eval{\dv*{(\phi_\epsilon^*g(\epsilon))_{\mu\nu}}{\epsilon}}_{\epsilon=0}$ and $h_{\mu\nu}$ represent the same physical metric perturbation. We can obtain a relation between them by computing
\begin{align}
    h_{\mu\nu}'-h_{\mu\nu}&=\eval{\dv{\epsilon}\qty[(\phi_\epsilon^*g(\epsilon))_{\mu\nu}-g_{\mu\nu}(\epsilon)]}_{\epsilon=0} \nonumber \\
    &=\lim_{\Delta\epsilon\to0}\qty[\frac{(\phi_{\Delta\epsilon}^*g(\Delta\epsilon))_{\mu\nu}-g_{\mu\nu}(\Delta\epsilon)}{\Delta\epsilon}] \nonumber \\
    &=\mathcal{L}_\varepsilon g_{\mu\nu}(0),
\end{align}
where $\mathcal{L}_\varepsilon$ is the Lie derivative along the vector field $\varepsilon^\mu(x)$. Here, we used the fact that $\phi_0$ is the identity map. We can use Eq.~\eqref{Lie-derivative-of-the-metric} to write
\begin{equation} \label{General-gauge-transformation}
    h_{\mu\nu}'=h_{\mu\nu}+\nabla_\mu\varepsilon_\nu+\nabla_\nu\varepsilon_\mu,
\end{equation}
with $\nabla_\mu$ denoting the covariant derivative compatible with $g_{\mu\nu}(0)$. In our case, this is simply the Minkowski metric, $g_{\mu\nu}(0)=\eta_{\mu\nu}$, and we are left with\footnote{Although the vector field $\varepsilon^\mu(x)$ is arbitrary, we restrict ourselves to those for which $\partial_\nu\varepsilon^\mu\sim O(h)$ in order to keep the condition $\abs{h_{\mu\nu}}\ll1$ satisfied.}
\begin{equation} \label{Gauge-transformation}
    h_{\mu\nu}'=h_{\mu\nu}+\partial_\mu\varepsilon_\nu+\partial_\nu\varepsilon_\mu.
\end{equation}
This holds a close resemblance to the gauge freedom of electromagnetism, where Maxwell's equations are left invariant under the gauge transformation $A_\mu\to A_\mu+\partial_\mu\Lambda$, with $A_\mu$ being the electromagnetic potential field and $\Lambda$ an arbitrary scalar function. We shall then refer to Eq.~\eqref{Gauge-transformation} as a \textbf{gauge transformation}, and the fact that it leaves the equations of motion invariant is called \textbf{gauge invariance}.

\subsubsection{Lorenz gauge}

Gauge freedom allows us to simplify the equations of motion. For instance, one can choose the \textbf{Lorenz gauge} where
\begin{equation} \label{Lorenz-gauge}
    \partial_\mu h^{\mu\nu}-\frac{1}{2}\partial^\nu h=0.
\end{equation}
To see how this is possible, suppose we start with a given metric perturbation $h_{\mu\nu}$ that does not satisfy the condition~\eqref{Lorenz-gauge}. Then one can perform a gauge transformation such that the new field satisfies
\begin{equation}
    \partial_{\mu}{h'}^{\mu\nu}-\frac{1}{2}\partial^{\nu}h'=\partial_\mu h^{\mu\nu}-\frac{1}{2}\partial^\nu h+\Box\varepsilon^\nu.
\end{equation}
We can then choose the vector $\varepsilon^\mu(x)$ such that
\begin{equation}
    \Box\varepsilon^\nu=-\qty(\partial_\mu h^{\mu\nu}-\frac{1}{2}\partial^\nu h),
\end{equation}
and the gauge-transformed metric perturbation will satisfy Eq.~\eqref{Lorenz-gauge} automatically.

With this gauge choice, the Einstein-Hilbert action~\eqref{Einstein-Hilbert-linearized} takes the form
\begin{equation}
    S_{\rm EH}=\frac{1}{64\pi}\int\dd^4x\,h_{\rm L}^{\mu\nu}\Box\qty(h^{\rm L}_{\mu\nu}-\frac{1}{2}\eta_{\mu\nu}h^{\rm L}),
\end{equation}
where $h^{\rm L}_{\mu\nu}$ denotes the metric perturbation in Lorenz gauge. In the presence of a matter source described by the energy-momentum tensor $T_{\mu\nu}$, we have, from Eq.~\eqref{Definition-energy-momentum-tensor},
\begin{equation}
    \delta S_{\rm matter}=\frac{1}{2}\int\dd^4x\,T_{\mu\nu}\delta h^{\mu\nu},
\end{equation}
and the principle of least action leads to
\begin{equation} \label{Grav-wave-equation-with-source}
    \Box\Bar{h}^{\rm L}_{\mu\nu}=-16\pi T_{\mu\nu},
\end{equation}
where we defined
\begin{equation} \label{H_munu(h_munu)}
    \Bar{h}_{\mu\nu}\equiv h_{\mu\nu}-\frac{1}{2}\eta_{\mu\nu}h
\end{equation}
and the Lorenz gauge condition can be written as $\partial_\mu\Bar{h}_{\rm L}^{\mu\nu}=0$. Note that Eq.~\eqref{Grav-wave-equation-with-source} is a wave equation sourced by the energy-momentum tensor.

\subsubsection{Transverse-traceless gauge}

When interested in regions outside the gravitational source, $T_{\mu\nu}=0$, one is left with
\begin{equation}
    \Box\Bar{h}^{\rm L}_{\mu\nu}=0.
\end{equation}
Because the flat d'Alembertian commutes with partial derivatives, this equation naturally implies $\partial_\mu\Bar{h}_{\rm L}^{\mu\nu}=0$ for suitable boundary conditions, and the Lorenz condition does not completely fix the gauge anymore (we say that there is a \emph{residual} gauge freedom). We may then perform a further gauge transformation $h^{\rm L}_{\mu\nu}\to h^{\rm L}_{\mu\nu}+\partial_\mu\xi_\nu+\partial_\nu\xi_\mu$, as long as $\Box\xi^\mu=0$, and use the four functions $\xi^\mu(x)$ to impose four more conditions on the metric perturbation, reducing from six to two degrees of freedom. For instance, we can choose $\xi^0$ such that the new perturbation satisfies $h=0$ (implying $\Bar{h}_{\mu\nu}=h_{\mu\nu}$) and $\xi^i$ such that $h_{0i}=0$. The $\nu=0$ component of the Lorenz condition, $\partial_\mu h^{\mu\nu}=0$, now reads $\partial_0 h^{00}=0$, meaning that $h_{00}$ is constant in time. In the context of gravitational wave propagation, the physical meaning behind the constant $h_{00}$ is related to the static part of the gravitational interaction, namely the Newtonian potential of the source that generated the waves~\cite{Maggiore2007,Carroll}. For a region with no sources, this automatically reads $h_{00}=0$, which, together with $h_{0i}=0$, can be written as $h_{0\mu}=0$. The Lorenz condition, automatically enforced by the equations of motion, becomes $\partial_ih^{ij}=0$.

In summary, our choice of gauge is defined by
\begin{subequations} \label{TT-gauge}
\begin{align}
    h^{\rm TT}_{0\mu}=0, \\
    \eta^{\mu\nu}h^{\rm TT}_{\mu\nu}=0, \\
    \partial^i h^{\rm TT}_{ij}=0.
\end{align}
\end{subequations}
This choice is called the \textbf{transverse-traceless gauge}, or simply the TT gauge, and the metric perturbation satisfying such requirements is denoted by $h^{\rm TT}_{\mu\nu}$. As we will see, this gauge choice greatly simplifies our study of gravitational waves.

\subsection{Plane wave solutions}

For the dynamics in regions outside of any source, the equations of motion take the form
\begin{equation} \label{Wave-equation}
    \Box h^{\rm TT}_{\mu\nu}=0,
\end{equation}
namely a wave equation, whose solutions are of the form\footnote{Naturally, any superposition of such plane wave solutions also solves the wave equation.}
\begin{equation} \label{Gravitational-waves}
    h^{\rm TT}_{\mu\nu}(x)=C_{\mu\nu}e^{ik_\rho x^\rho},
\end{equation}
which we call \textbf{gravitational waves}. Here $C_{\mu\nu}$ is a constant, symmetric, $(0,2)$ tensor, and the constant vector $k_\rho$ is called the wave vector. We can insert the solution~\eqref{Gravitational-waves} into the wave equation~\eqref{Wave-equation} to obtain $k_\rho k^\rho h^{\rm TT}_{\mu\nu}=0$. By requiring non-trivial solutions for $h^{\rm TT}_{\mu\nu}$, this translates to
\begin{equation} \label{Null-wave-vector}
    k_\rho k^\rho=0,
\end{equation}
that is, the wave vector is null, implying that gravitational waves propagate at the speed of light in vacuum. The timelike component of the wave vector is called the \textbf{frequency} of the wave, and we write $k^\rho=\qty(\omega,k^1,k^2,k^3)$, such that
\begin{equation}
    \omega^2=\delta_{ij}k^ik^j.
\end{equation}

As we discussed before, our gauge choices reduce the degrees of freedom of a gravitational wave from ten to two. In terms of the plane wave solutions~\eqref{Gravitational-waves}, this means that the constant tensor $C_{\mu\nu}$ has only two independent components, which we shall call the two polarizations of the gravitational wave. We can then write $C_{\mu\nu}=\sum_sq_s\epsilon_{\mu\nu}^s$, with the index $s$ labeling the polarizations, $q_s$ being the wave amplitude, and the tensor $\epsilon^s_{\mu\nu}$ being called the \textbf{polarization tensor}. It is traceless and purely spatial, while also satisfying the transversality condition,
\begin{subequations} \label{Polarization-tensor-conditions}
\begin{align}
    \epsilon_{0\nu}=0, \\
    \eta^{\mu\nu}\epsilon_{\mu\nu}=0, \\
    k^i\epsilon_{ij}=0.
\end{align}
\end{subequations}
These follow immediately from Eqs.~\eqref{TT-gauge} and~\eqref{Gravitational-waves}. Note that the transversality condition necessarily yields the polarization tensor dependent on the spatial part of the wave vector, $\epsilon_{\mu\nu}=\epsilon_{\mu\nu}(\vb{k})$.

In order to better understand the propagation of gravitational waves, let us choose spatial coordinates such that the wave is traveling in the $x^3$ direction. Thus, the wave vector is of the form $k^\rho=\qty(\omega,0,0,\omega)$. In that case, the transversality condition of the polarization tensor requires $\epsilon_{3j}=0$. Also, since the polarization is purely spatial, symmetric, and traceless, we conclude that $C_{\mu\nu}$ is of the form
\begin{equation} \label{C_munu}
    C_{\mu\nu}=\mqty(0 & 0 & 0 & 0 \\ 0 & q_+ & q_\times & 0 \\ 0 & q_\times & -q_+ & 0 \\ 0 & 0 & 0 & 0),
\end{equation}
where $q_+$ and $q_\times$ are constants (the reason behind the choice of notation will be clear soon). Together with the wave frequency $\omega$, these completely characterize the wave. It is common to write the two polarizations separately~\cite{Isi_2023},
\begin{equation} \label{Polarization-tensors}
\begin{split}
    \epsilon_{ij}^+=\hat{\epsilon}_i^{(1)}\otimes\hat{\epsilon}_j^{(1)}-\hat{\epsilon}_i^{(2)}\otimes\hat{\epsilon}_j^{(2)}, \\
    \epsilon_{ij}^\times=\hat{\epsilon}_i^{(1)}\otimes\hat{\epsilon}_j^{(2)}+\hat{\epsilon}_i^{(2)}\otimes\hat{\epsilon}_j^{(1)},
\end{split}
\end{equation}
where the spatial polarization unit vectors $\hat{\epsilon}^{(1)}$ and $\hat{\epsilon}^{(2)}$ are orthogonal to the direction of propagation $\hat{k}=\vb{k}/\abs{\vb{k}}$ (for $\hat{k}=\vu{e}_3$, the polarization unit vectors are $\vu{e}_1$ and $\vu{e}_2$). The plane wave solutions can then be written in the form
\begin{equation} \label{Gravitational-wave-sum-over-polarizations}
    h^{\rm TT}_{\mu\nu}(x)=\sum_{s=+,\times}q_{s}\epsilon_{\mu\nu}^s(\vb{k})e^{ik_\rho x^\rho},
\end{equation}
and the polarization tensors are conventionally normalized according to
\begin{equation} \label{Polarization-tensor-normalization}
    \epsilon_{ij}^s\epsilon^{ij}_{s'}=2\delta^s_{s'}.
\end{equation}
These "plus" and "cross" polarizations are jointly called the linear polarization basis\footnote{We can also consider right- and left-handed circularly polarized modes by defining $q_R=\frac{1}{\sqrt{2}}\qty(q_++iq_\times)$ and $q_L=\frac{1}{\sqrt{2}}\qty(q_+-iq_\times)$.}.

\subsection{Interaction of gravitational waves with test masses}

We now  turn our attention to the effects of gravitational waves on the motion of test particles. We begin by taking a single test mass, described by coordinates $x^\mu$ in a TT frame (a frame in which the metric perturbation satisfies the conditions~\eqref{TT-gauge}), which is at rest at $\tau=0$, where $\tau$ denotes its proper time. The geodesic equation implies
\begin{equation}
    \eval{\dv[2]{x^i}{\tau}}_{\tau=0}=-\eval{\qty[\Gamma^i_{\mu\nu}\dv{x^\mu}{\tau}\dv{x^\nu}{\tau}]}_{\tau=0}=-\eval{\qty[\Gamma^i_{00}\qty(\dv{x^0}{\tau})^2]}_{\tau=0},
\end{equation}
where we used the fact that, by assumption, $\dv*{x^i}{\tau}=0$ at $\tau=0$. From the expansion  of the Christoffel symbols in first order in $h_{\mu\nu}$, Eq.~\eqref{Christoffel-linear-in-h}, we find
\begin{equation}
    \Gamma_{00}^i=\frac{1}{2}\delta^{ij}\qty(2\partial_0 h_{0j}-\partial_j h_{00}).
\end{equation}
By imposing the TT gauge conditions~\eqref{TT-gauge}, we are left with $\Gamma_{00}^i=0$ and, consequently,
\begin{equation}
    \eval{\dv[2]{x^i}{\tau}}_{\tau=0}=0.
\end{equation}
Therefore, if at time $\tau=0$ the particle is at rest, it will remain at rest at all times in the TT frame. If we want to observe the physical effect of gravitational waves on test masses, a single one does not suffice.

Let us then consider the relative motion of nearby particles as described by the geodesic deviation equation\footnote{See Ref.~\cite{Flanagan2005} for a slightly different (but equivalent) approach.}. Take two test masses, each of which is traveling along its geodesic parametrized by $t$, with four-velocities described by a single vector field $U^\mu(x)$. Denoting the separation vector by $S^\mu$, the geodesic deviation equation reads (Appendix~\ref{app:Differential})
\begin{equation}
    \frac{\textrm{D}^2}{\textrm{d}t^2}S^\mu={R^\mu}_{\nu\rho\sigma}U^\nu U^\rho S^\sigma,
\end{equation}
where $\textrm{D}/\textrm{d}t$ denotes the directional covariant derivative along the tangent vector to the geodesics. Next, we consider the test masses to be moving slowly such that $U^\mu(x)=\delta^\mu_0+O(h)$. Since the Riemann tensor is already $O(h)$, the geodesic deviation equation reduces to
\begin{equation}
    \pdv[2]{t}S_\mu=R_{\mu00\sigma}S^\sigma,
\end{equation}
where we dropped higher order terms on the left-hand side as well. Up to first order in $h_{\mu\nu}$, the Riemann tensor is given by Eq.~\eqref{Riemann-linear-in-h}. In the TT gauge,
\begin{equation} \label{Riemann-tensor-TT-gauge}
    R^{\rm TT}_{\mu00\sigma}=\frac{1}{2}\partial_0^2h^{\rm TT}_{\mu\sigma},
\end{equation}
and the geodesic deviation equation becomes
\begin{equation} \label{Geodesic-deviation-GW}
    \pdv[2]{t}S_\mu=\frac{1}{2}S^\sigma\pdv[2]{t}h^{\rm TT}_{\mu\sigma}.
\end{equation}
It is then clear from Eqs.~\eqref{Gravitational-waves} and~\eqref{C_munu} that, for a wave traveling in the $x^3$ direction, only $S^1$ and $S^2$ will be affected; the test masses are only disturbed in directions orthogonal to the wave vector.

\begin{figure}[!ht]
    \centering
    \includegraphics[width=0.7\linewidth]{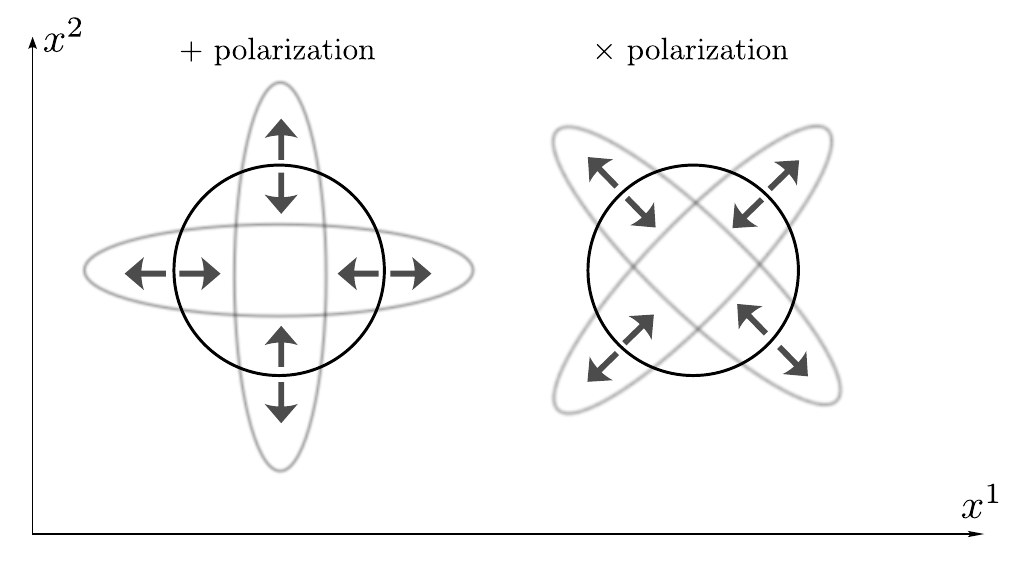}
    \caption{Effect of a gravitational wave on a ring of stationary particles in the $x^1$-$x^2$ plane, depending on the polarization of the wave.}
    \label{polarizations}
\end{figure}

We can now plug in the expression for $h^{\rm TT}_{\mu\nu}$, Eq.~\eqref{Gravitational-wave-sum-over-polarizations}. It will be instructive to consider the effect of each polarization separately. For instance, by taking $q_\times=0$ and keeping $q_+\neq0$, the solution of Eq.~\eqref{Geodesic-deviation-GW} to the lowest order in $h_{\mu\nu}$ is of the form
\begin{subequations} \label{S_+}
\begin{equation}
    S_+^1(t)=S^1(0)+\frac{1}{2}S^1(0)q_+e^{ik_\rho x^\rho},
\end{equation}
\begin{equation}
    S_+^2(t)=S^2(0)-\frac{1}{2}S^2(0)q_+e^{ik_\rho x^\rho}.
\end{equation}
\end{subequations}
Similarly, by taking $q_+=0$ and keeping $q_\times\neq0$, the solution of Eq.~\eqref{Geodesic-deviation-GW} to the lowest order in $h_{\mu\nu}$ is of the form
\begin{subequations} \label{S_X}
\begin{equation}
    S_\times^1(t)=S^1(0)+\frac{1}{2}S^2(0)q_\times e^{ik_\rho x^\rho},
\end{equation}
\begin{equation}
    S_\times^2(t)=S^2(0)+\frac{1}{2}S^1(0)q_\times e^{ik_\rho x^\rho}.
\end{equation}
\end{subequations}
We can get an image of the solutions above by picturing a ring of stationary particles and considering the effect of a passing gravitational wave. For the plus polarization, Eqs.~\eqref{S_+} show that the particles will bounce back and forth in the shape of a '$+$', since geodesics initially separated in the $x^1$ ($x^2$) direction will oscillate in the $x^1$ ($x^2$) direction. For the cross polarization, Eqs.~\eqref{S_X} show that the particles will bounce back and forth in the shape of a '$\times$', since geodesics initially separated in the $x^1$ ($x^2$) direction will oscillate in the $x^2$ ($x^1$) direction. These effects are illustrated in Figure~\ref{polarizations} and the notation $q_+$ and $q_\times$ should now be clear.

\subsection{Production of gravitational waves}

Having discussed the propagation of gravitational waves and their interaction with test masses, let us now understand how they are produced. Recall that, in the presence of sources, we cannot impose the TT gauge anymore. Instead, we shall use the Lorenz gauge, for which the equations of motion read
\begin{equation}
    \Box\Bar{h}^{\rm L}_{\mu\nu}=-16\pi T_{\mu\nu},
\end{equation}
where $\Bar{h}_{\mu\nu}$ is defined by Eq.~\eqref{H_munu(h_munu)} and the Lorenz condition is $\partial_\mu\Bar{h}_{\rm L}^{\mu\nu}=0$. The solution of the inhomogeneous wave equation is of the form
\begin{equation} \label{Solution-of-the-inhomogeneous-wave-equation}
    \Bar{h}^{\rm L}_{\mu\nu}(x)=-16\pi\int\dd^4y\,G(x-y)T_{\mu\nu}(y),
\end{equation}
where $G(x-y)$ is the Green function for the d'Alembertian satisfying
\begin{equation}
    \Box_xG(x-y)=\delta^4(x-y),
\end{equation}
with $\Box_x$ denoting the d'Alembertian with respect to the $x$ coordinates. Since we are interested in the effect of signals from the past of the point under consideration, we consider only the retarded Green function, given by~\cite{Carroll,Jackson1998}
\begin{equation} \label{Green-function-d'Alembertian}
    G(x-y)=-\frac{1}{4\pi\abs{\vb{x}-\vb{y}}}\delta\qty[y^0-\qty(x^0-\abs{\vb{x}-\vb{y}})]\theta(x^0-y^0),
\end{equation}
where $\theta(x)$ denotes the Heaviside step function. Plugging Eq.~\eqref{Green-function-d'Alembertian} into Eq.~\eqref{Solution-of-the-inhomogeneous-wave-equation} leads to
\begin{equation} \label{Solution-of-the-inhomogeneous-wave-equation-2}
    \Bar{h}^{\rm L}_{\mu\nu}(t,\vb{x})=4\int\dd^3y\,\frac{1}{\abs{\vb{x}-\vb{y}}}T_{\mu\nu}(t_r,\vb{y}),
\end{equation}
where $t_r=t-\abs{\vb{x}-\vb{y}}$ ($t=x^0$) is called the "retarded time". Its appearance in the general solution~\eqref{Solution-of-the-inhomogeneous-wave-equation-2} serves to ensure that the disturbance in the metric perturbation field at $(t,\vb{x})$ is only influenced by sources of energy and momentum at the point $(t_r,\abs{\vb{x}-\vb{y}})$ on the past light cone in order to maintain causality.

The analysis of the general solution~\eqref{Solution-of-the-inhomogeneous-wave-equation-2} is better carried out in frequency space, where we introduce the temporal Fourier transform
\begin{equation}
   \Tilde{\Bar{h}}^{\rm L}_{\mu\nu}(\omega,\vb{x})=\int_{-\infty}^\infty\dd t\,e^{i\omega t}\Bar{h}^{\rm L}_{\mu\nu}(t,\vb{x}).
\end{equation}
Using Eq.~\eqref{Solution-of-the-inhomogeneous-wave-equation-2} and changing the integration variable, we find
\begin{align}
    \Tilde{\Bar{h}}^{\rm L}_{\mu\nu}(\omega,\vb{x})&=4\int_{-\infty}^\infty\dd t\int\dd^3y\,e^{i\omega t}\frac{1}{\abs{\vb{x}-\vb{y}}}T_{\mu\nu}(t_r,\vb{y}) \nonumber \\
    &=4\int\dd^3y\,\frac{e^{i\omega\abs{\vb{x}-\vb{y}}}}{\abs{\vb{x}-\vb{y}}}\int_{-\infty}^\infty\dd t_r\,e^{i\omega t_r}T_{\mu\nu}(t_r,\vb{y}) \nonumber \\
    &=4\int\dd^3y\,\frac{e^{i\omega\abs{\vb{x}-\vb{y}}}}{\abs{\vb{x}-\vb{y}}}\Tilde{T}_{\mu\nu}(\omega,\vb{y}),
\end{align}
where we introduced the temporal Fourier transform of the energy-momentum tensor, $\Tilde{T}_{\mu\nu}(\omega,\vb{y})$.

\begin{figure}[!ht]
    \centering
    \includegraphics[width=0.7\linewidth]{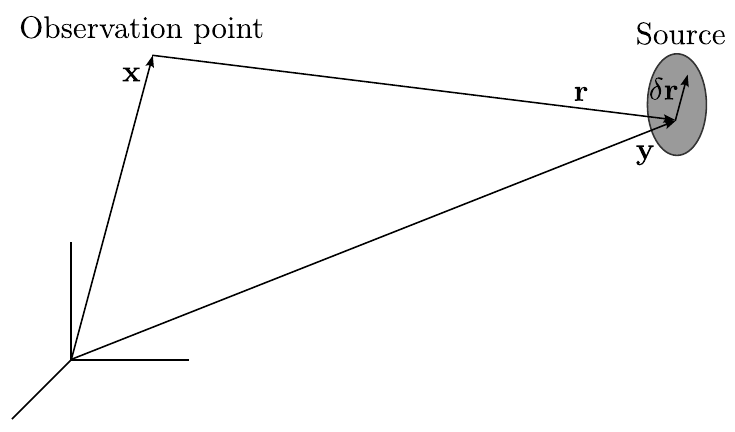}
    \caption{The size of the source is roughly $\delta r\ll r$, where $r$ is the distance to the observation point.}
    \label{isolated-source}
\end{figure}

We can now simplify our analysis by making a few reasonable approximations. Consider an isolated, slowly moving source, far away from the observation point, centered at a point $\vb{r}$ with its internal points at $\vb{r}+\delta\vb{r}$ such that $\delta r\ll r$ (Figure~\ref{isolated-source}). Then the distance $r=\abs{\vb{x}-\vb{y}}$ varies very little when integrating over the source, and we may write
\begin{equation}
    \Tilde{\Bar{h}}^{\rm L}_{\mu\nu}(\omega,\vb{x})\simeq4\frac{e^{i\omega r}}{r}\int \dd^3y\,\Tilde{T}_{\mu\nu}(\omega,\vb{y}).
\end{equation}

We can take advantage of the fact that we only need to compute the spatial components of the metric perturbation field since the remaining ones can be obtained from those by using the Lorenz condition. Now, in frequency space, the conservation of energy-momentum, $\partial^\mu T_{\mu\nu}=0$, takes the form $\partial^j\Tilde{T}_{j\nu}=-i\omega\Tilde{T}_{0\nu}$, so that
\allowdisplaybreaks
\begin{align}
    \Tilde{\Bar{h}}^{\rm L}_{ij}(\omega,\vb{x})&=4\frac{e^{i\omega r}}{r}\int\dd^3y\,\Tilde{T}_{ij}(\omega,\vb{y})=4\frac{e^{i\omega r}}{r}\int\dd^3y\qty[\partial^k\qty(y_i\Tilde{T}_{kj})-y_i\partial^k\Tilde{T}_{kj}] \nonumber \\
    &=4i\omega\frac{e^{i\omega r}}{r}\int\dd^3y\,y_i\Tilde{T}_{0j}=2i\omega\frac{e^{i\omega r}}{r}\int\dd^3y\,\qty(y_i\Tilde{T}_{0j}+y_j\Tilde{T}_{0i}) \nonumber \\
    &=2i\omega\frac{e^{i\omega r}}{r}\int\dd^3y\,\qty[\partial^k\qty(y_iy_j\Tilde{T}_{0k})-y_iy_j\partial^k\Tilde{T}_{0k}] \nonumber \\
    &=-2\omega^2\frac{e^{i\omega r}}{r}\int\dd^3y\,y_iy_j\Tilde{T}_{00}(\omega,\vb{y}),
\end{align}
where we dropped boundary terms, as usual. We may now compute the inverse Fourier transform and define the \textbf{quadrupole moment tensor} of the energy density of the source as
\begin{equation}
    Q_{ij}(t)=\int \dd^3y\,y_iy_jT_{00}(t,\vb{y}).
\end{equation}
This allows us to write
\begin{equation} \label{Quadrupole-formula}
    \Bar{h}^{\rm L}_{ij}(t,\vb{x})=\frac{2}{r}\ddot{Q}_{ij}(t_r).
\end{equation}

Eq.~\eqref{Quadrupole-formula} shows that the gravitational radiation emitted by an isolated, slowly moving source is proportional to the second time derivative of the quadrupole moment tensor of the energy density $T_{00}$. Compare this result with the analogous case of electromagnetic radiation, for which the leading order contribution comes from the dipole moment of the electric charge density. Typically, the quadrupole moment is smaller than the dipole moment, which, combined with the weak coupling of gravity to matter, explains why gravitational radiation is much weaker than electromagnetic radiation, making its detection a highly non-trivial task.

\begin{figure}[!ht]
    \centering
    \includegraphics[width=0.5\linewidth]{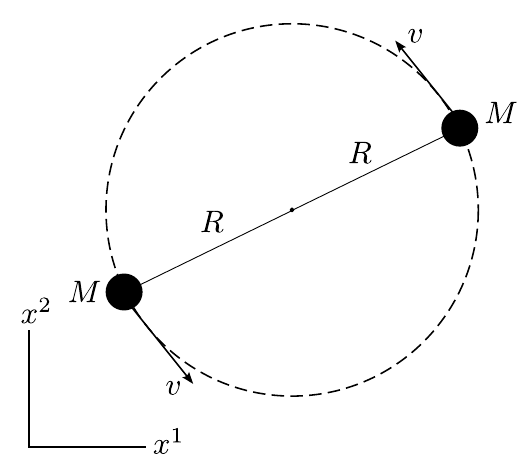}
    \caption{A binary star system.}
    \label{binary-star}
\end{figure}

A typical example is the emission of gravitational radiation by a system of two stars in orbit around each other (a binary star). Take, for instance, two stars of mass $M$ in a circular orbit in the $x^1$-$x^2$ plane, at a distance $R$ from their center-of-mass and with a velocity $v$ (Figure~\ref{binary-star}). Their motion can be treated in the Newtonian approximation, and the metric perturbation can be obtained by using Eq.~\eqref{Quadrupole-formula}. The result is~\cite{Carroll}
\begin{equation} \label{H-binary-star}
    \Bar{h}^{\rm L}_{ij}(t,\vb{x})=\frac{8M}{r}\Omega^2R^2\mqty(-\cos2\Omega t_r & -\sin2\Omega t_r & 0 \\ -\sin2\Omega t_r & \cos2\Omega t_r & 0 \\ 0 & 0 & 0),
\end{equation}
where $\Omega=v/R$ is the angular frequency of the orbit.

\subsection{Detection of gravitational waves}

In recent years, gravitational waves have become a new source of astrophysical information concerning the universe outside the solar system, in addition to electromagnetic radiation, neutrinos, and cosmic rays. Due to their importance, it is worth mentioning a few words about how these are detected in modern experiments. We will simply outline the basic principles, since understanding the actual mechanism behind the workings of a real-life gravitational wave detector is outside our scope.

First, it will be useful to estimate how sensitive a gravitational wave detector must be. It comes as no surprise that, due to the weakness of gravitational radiation, building a detector is not an easy task. In order to provide numerical estimations, we take the binary star system from the previous section (Figure~\ref{binary-star}) as the source of the gravitational wave to be measured. According to Eq.~\eqref{H-binary-star}, the frequency of the wave is $f=\Omega/2\pi$, where $\Omega$ is the angular frequency of the orbit of the sources. Since our goal is only to provide numerical estimates, we can use Newton's law of motion to write\footnote{In this subsection, we are restoring factors of $G$ and $c$ in order to obtain numerical estimates in SI units.}
\begin{equation}
    \frac{GM^2}{(2R)^2}=\frac{Mv^2}{R},
\end{equation}
where the left-hand side is simply Newton's law of gravitation, and the right-hand side is the expression for the centripetal force. From this, we find
\begin{equation}
    \Omega=c\sqrt{\frac{R_S}{8R^3}},
\end{equation}
where $R_S=2MG/c^2$ denotes the Schwarzschild radius\footnote{The Schwarzschild radius is a parameter in the spherically symmetric vacuum solution to Einstein's equation that defines the event horizon of a Schwarzschild black hole.} of the star with mass $M$. Then the wave frequency is estimated to be
\begin{equation}
    f=\frac{cR_S^{1/2}}{2\pi\sqrt{8}R^{3/2}}\sim\frac{cR_S^{1/2}}{10R^{3/2}}.
\end{equation}
Now, for the wave amplitude, we have, from Eq.~\eqref{H-binary-star},
\begin{equation}
    q=\frac{8GM}{c^4r}\Omega^2R^2\sim\frac{R_S^2}{rR}.
\end{equation}

As an example of the gravitational wave source, consider the coalescence of a black hole binary\footnote{The coalescence of a black hole binary occurs when two black holes rotate around each other due to mutual gravitational attraction until they merge into a single black hole.} for which typical parameters are $R_S\sim10^4\,\rm m$, $R\sim 10^5\,\rm m$, and $r\sim10^{24}\,\rm m$~\cite{Carroll}. Then,
\begin{equation}
    f\sim10^2\,\textrm{s}^{-1},\hspace{1cm}q\sim10^{-21}.
\end{equation}
Therefore, a detector capable of measuring the gravitational radiation emitted by a black hole coalescence with the parameters specified above must be sensitive to frequencies around $100\,\textrm{Hz}$ and wave amplitudes around $10^{-21}$.

\begin{figure}[!ht]
    \centering
    \includegraphics[width=0.7\linewidth]{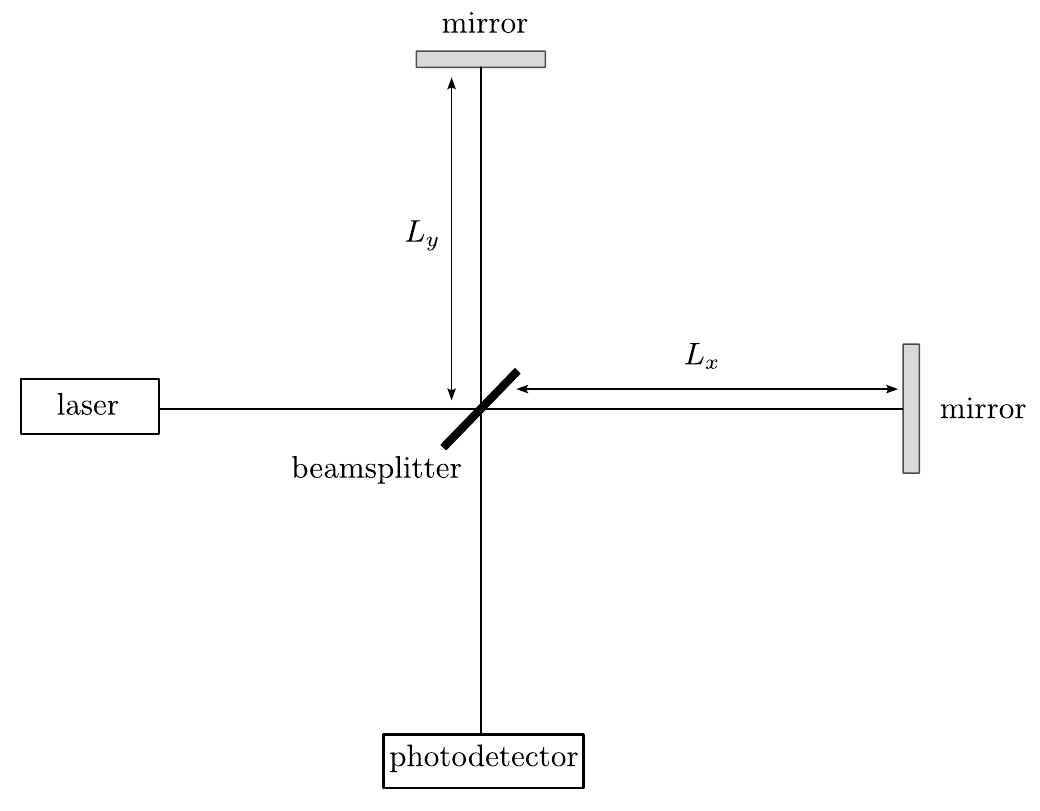}
    \caption{Michelson interferometer.}
    \label{michelson}
\end{figure}

The prospect of observing gravitational waves began when Joseph Weber developed the concept of resonant-mass detectors in the 1960s, but soon it became clear that these would be sensitive only to rare, strong neighboring sources. On the other hand, modern gravitational wave detectors are usually based on a laser interferometer, such as the Michelson interferometer schematized in Figure~\ref{michelson}. Basically, a laser is pointed at a beamsplitter, which then sends photons on two different paths of sizes $L_x$ and $L_y$. These are reflected at the mirrors at the end of each path and recombined back in the beamsplitter, where they destructively interfere while the signal is sent to a photodetector. The effect of a passing gravitational wave in such an apparatus is to modify the lengths of the orthogonal paths traveled by light, disturbing the interference and yielding a phase shift
\begin{equation} \label{Phase-shift}
    \delta\phi\sim\qty(\frac{2\pi}{\lambda})\delta L,
\end{equation}
where $\lambda$ is the wavelength of the laser and $\delta L=\delta L_x-\delta L_y$, with $\delta L_x$ ($\delta L_y$) denoting the difference in $L_x$ ($L_y$) due to the passing gravitational wave.

In order to get an estimate for $\delta L$, let us consider a plus-polarized gravitational wave
\begin{equation}
    h_{ij}(t,x,y,z)=\mqty(q & 0 & 0 \\ 0 & -q & 0 \\ 0 & 0 & 0)\cos\qty(\omega t-kz),
\end{equation}
where we are denoting $x^1=x$, $x^2=y$, and $x^3=z$. This solution was obtained in the TT gauge, which we are allowed to use since we are supposedly very far from the source. For this metric, the spacetime interval between any two points is
\begin{align}
    \dd s^2&=g_{\mu\nu}\dd x^\mu\dd x^\nu=\qty(\eta_{\mu\nu}+h_{\mu\nu})\dd x^\mu\dd x^\nu \nonumber \\
    &=-c^2\dd t^2+\qty(1+q)\dd x^2+\qty(1-q)\dd y^2+\dd z^2.
\end{align}

A photon traveling along the $x-$direction from the beamsplitter to the mirror in a proper distance $L_x$ will take a time $\Delta t_x$ such that
\begin{equation*}
    c\int_0^{\Delta t_x}\dd t=\int_0^{L_x}\sqrt{1+q}\,\dd x=\int_0^{L_x}\qty(1+\frac{1}{2}q)\,\dd x+O(q^2),
\end{equation*}
\begin{equation}
    c\Delta t_x=\qty(1+\frac{1}{2}q)L_x,
\end{equation}
where we used the fact that, for a light beam, $\dd s^2=0$. This means that the difference in the path traveled by light on the $x-$axis is $\delta L_x=qL_x/2$. A completely analogous analysis leads to the difference in the path traveled by light in the $y$-axis being given by $\delta L_y=-qL_y/2$. If we now assume\footnote{This guarantees that, in the absence of any disturbance of the paths, the recombined beam will undergo completely destructive interference.} $L_x=L_y\equiv L$, then $\delta L=\delta L_x-\delta L_y=qL$, or
\begin{equation}
    \frac{\delta L}{L}=q.
\end{equation}
For a detector with an arm length of the order of a few kilometers, we have $\delta L\sim10^{-18}\,\rm m$ for $q\sim10^{-21}$. If we use a laser with a typical wavelength $\lambda\sim10^{-6}\,\rm m$ and we allow the photons to travel the arm lengths about $100$ times before returning to the beamsplitter\footnote{This is usually accomplished by the addition of a partially reflective mirror on each path.}, we estimate from Eq.~\eqref{Phase-shift} that the interferometer must be sensitive to a phase shift of the order
\begin{equation}
    \delta\phi\sim10^{-9},
\end{equation}
which can be measured with current technology for a sufficiently large number of photons~\cite{Carroll}.

Now, real-life gravitational wave detectors have a more complicated structure than a simple Michelson interferometer, which is usually combined with Fabry-Pérot interferometers~\cite{Cahillane2022}. As it happens in almost every experiment, most detector work is dedicated to eliminating noise, which is essentially any laser power fluctuation that is not due to the gravitational waves. Noise sources include Poisson fluctuations in the arrival time of discrete objects (this is called shot noise), displacement noise arising from quantum fluctuations of the electric field in the arms, thermal noise from the atoms making up the mirrors, seismic noise due to the motion of the Earth, and control noise in general\footnote{The different types of noise are distinguished by their power spectrum.}. We refer the reader to Ref.~\cite{Maggiore2007} for details concerning data analysis in gravitational waves experiments.

The first experimental evidence of the existence of gravitational waves came in 1974 from the Hulse-Taylor binary pulsar (highly magnetized and rapidly rotating neutron stars), whose orbital decay matched the predictions of energy loss due to the emission of gravitational waves~\cite{Hulse1975}. Russell A. Hulse and Joseph H. Taylor Jr. were awarded the Nobel Prize in Physics in 1993 for this discovery. The first \emph{direct} detection of a gravitational wave from a binary black hole merger was made in 2015 by the Advanced Laser Interferometer Gravitational-Wave Observatory (LIGO)~\cite{Abbott_2016}, a long-baseline laser interferometer with two 4 km long orthogonal arms. For decisive contributions to the LIGO detector and the direct observation of gravitational waves, the 2017 Nobel Prize in Physics was awarded to Rainer Weiss, Barry C. Barish, and Kip C. Thorne\footnote{We refer the reader to Ref.~\cite{Levin2016} for the fascinating history of the development of the ideas that led to the first direct observation of gravitational waves by the scientists of the LIGO collaboration.}. The field of experimental gravitational wave physics is experiencing significant growth, bolstered by the anticipated sensitivity of future observations. For instance, while LIGO has a peak sensitivity to frequencies of about $100$ Hz, the space-based Laser Interferometer Space Antenna (LISA) is expected to probe lower frequency (mHz) gravitational waves when it launches in the late 2030s~\cite{Bailes2021}.

\section{Quantum gravitational radiation} \label{Sec:Quantum-gravitational-radiation}

The description of gravitational waves we have discussed so far can be seen as a classical field theory of a symmetric $(0,2)$ tensor propagating on a flat background metric. This allows us to provide a quantum description of the metric perturbation field using the machinery of modern quantum field theory. The particle arising from the quantization of gravitational radiation is called the \textbf{graviton}. It is important to emphasize that there is a significant difference between graviton physics (perturbative quantum gravity) and quantum gravity \emph{per se}. While graviton physics refers to the quantization of the metric perturbation in the weak-field limit, a full quantum theory of gravity refers to the quantization of the entire metric field. In this sense, gravitons are not necessarily to be thought of as the basic constituents of spacetime. Nevertheless, one may say that any quantum gravity theory should recover graviton physics when taking some kind of low energy limit.

\subsection{Helicity and spin}

In the canonical formalism of quantum field theory, the classical fields are promoted to operators on a Hilbert space and are required to satisfy the canonical equal-time commutation relations. The mode excitations of a given field are what we understand as particles. The kind of particle a classical field will represent upon quantization depends on its behavior under Poincaré transformations, which include general Lorentz transformations and spacetime translations. We say that different kinds of particles transform under different irreducible unitary representations of the Poincaré group\footnote{To be more precise, the representation actually refers to the set of operators on the Hilbert space of particle states that perform the Poincaré transformations on those states. For a review of concepts in basic representation theory, we refer the reader to Refs.~\cite{Jeevanjee2015,Costa2012}.}~\cite{Schwartz2013}. For instance, since the generator of spatial rotations is the angular momentum operator, the spin of the particle obtained upon the quantization of gravitational radiation can be inferred by checking how the waves change under such Lorentz transformations.

Consider a gravitational wave propagating in the $x^3-$direction, Eqs.~\eqref{Gravitational-waves} and~\eqref{C_munu}. Under a Lorentz transformation $\Lambda$, the metric perturbation field changes according to $h_{\mu'\nu'}={\Lambda_{\mu'}}^\mu{\Lambda_{\nu'}}^\nu h_{\mu\nu}$. Let $\Lambda$ represent a spatial rotation by an angle $\theta$ around the $x^3-$axis (naturally, the temporal components are left unchanged), so that $\Lambda_1^1=\Lambda_2^2=\cos\theta$, $\Lambda_1^2=-\Lambda_2^1=-\sin\theta$, $\Lambda_3^3=1$, and $\Lambda_i^j=0$ otherwise. This transformation leads to
\begin{equation}
    C_{i'j'}=\mqty(q_+' & q_\times' & 0 \\ q_\times' & -q_+' & 0 \\ 0 & 0 & 0),
\end{equation}
with
\begin{equation}
\begin{split}
    q_+'&=q_+\cos2\theta+q_\times\sin2\theta, \\
    q_\times'&=q_\times\cos2\theta-q_+\sin2\theta.
\end{split}
\end{equation}
In terms of the linear combinations $C_\pm\equiv q_+\mp iq_\times$, this can be written as
\begin{equation}
    C_\pm'=\exp(\pm2i\theta)C_\pm.
\end{equation}

In general, any plane wave $\psi$ that is transformed by a rotation of an angle $\theta$ around the direction of its propagation into
\begin{equation}
    \psi'=e^{ih\theta}\psi
\end{equation}
is said to have \textbf{helicity}\footnote{Not to be confused with the trace of the metric perturbation.} $h$~\cite{Weinberg2013}. Upon quantization, the helicity represents the spin of the particle projected along the direction of motion. We have thus shown that gravitational waves have helicity $h=2$, and therefore, the graviton is a spin-2 particle.

It is interesting to mention that we could have taken an inverse approach. If we had asked instead what kind of field represents a massless spin-2 particle, we would have been led to the theory of a symmetric $(0,2)$ tensor $h_{\mu\nu}$ for which the unique Lagrangian is given in Eq.~\eqref{Einstein-Hilbert-linearized}. Here, we will simply outline the procedure. For more details, we refer to Refs.~\cite{Schwartz2013,Fierz1939}.

According to Wigner's theorem~\cite{Schwartz2013,Weinberg12005,Wigner1939}, the unitary irreducible representations of the Poincaré group are uniquely classified by two parameters $m$ and $J$, where $m$ is a non-negative real number, while $J$ is a non-negative integer or half-integer, which we call mass and spin, respectively. These representations depend on the momentum $p$ of the particle states, and thus, they are infinite dimensional. If $J>0$, for each value of the momentum with $p^2=-m^2$, there are $2J+1$ independent states in the representation if $m>0$ and exactly $2$ states for $m=0$. If $J=0$, there is only one independent state for any $m$. This classification tells us what kind of particles we can describe in a Lorentz invariant theory. The next step is to build Lagrangians by embedding these representations into fields.

For $J=0$, we only have 1 degree of freedom, which is naturally embedded into a scalar field. For $J=\frac{1}{2}$ (2 degrees of freedom), the embedding is done into spinor fields. For $J=1$ and $m\neq0$, we have 3 degrees of freedom. These are embedded into a vector field, which has four components. The Lagrangian for a massive spin 1 particle must then be built in a way to enforce the removal of one degree of freedom in the equations of motion. The Proca Lagrangian does precisely that~\cite{Schwartz2013}. By taking the limit $m\to0$, we are left with the theory of a massless spin 1 particle, such as the photon. According to Wigner's theorem, a photon is described by two degrees of freedom, which we call its two polarizations. This requires the removal of an additional degree of freedom in the vector field representation. That is where gauge invariance plays its important role. It is necessary to remove the unwanted degrees of freedom so that a vector field can, in fact, represent a massless spin 1 particle. The case $J=3/2$ describes spin $3/2$ particles that are embedded in Rarita–Schwinger fields, and examples include Delta and Sigma baryons, as well as the gravitino, a hypothesized gauge fermion partner of the graviton in supersymmetric theories of gravity~\cite{Weinberg2013a}.

For $J=2$ and $m\neq0$, there are 5 degrees of freedom, which clearly does not fit into a vector representation. Let us then take a symmetric $(0,2)$ tensor field, which has 10 independent components. The Lagrangian can be built in a way to remove 5 degrees of freedom in the equations of motion~\cite{Schwartz2013,Fierz1939}. But according to Eq.~\eqref{Null-wave-vector}, gravitational waves travel at the speed of light in vacuum, and so the graviton must be a massless particle. In the massless limit, the equations of motion are still able to remove 4 independent components\footnote{As in the spin 1 case, the mass term is responsible for the removal of one independent component. That component is no longer removed when taking $m=0$.}, leaving us with 6 of them. Since $J>0$ and $m=0$, we must have only 2, and the removal of the extra 4 degrees of freedom is once again accomplished by the introduction of gauge invariance. Then it can be shown that the unique Lagrangian describing a massless spin 2 particle is the one given in Eq.~\eqref{Einstein-Hilbert-linearized}. Finally, by adopting a geometrical point of view, this gauge invariance can be understood in terms of diffeomorphism invariance, as we have discussed in the previous section.

\subsection{Free gravitons}

The (canonical) quantization of gravitational radiation is achieved by promoting the metric perturbation field to a quantum operator in a plane-wave expansion~\cite{Hsiang_2024}
\begin{equation} \label{h-operator-plane-wave-expansion}
    \Hat{h}_{\mu\nu}(x)=\int\frac{\dd^3p}{(2\pi)^3}\frac{1}{\sqrt{2\omega_p}}\sum_s\qty[\epsilon_{\mu\nu}^s(p)\Hat{a}_{p,s}e^{ipx}+{\epsilon_{\mu\nu}^s}^*(p)\Hat{a}^\dagger_{p,s}e^{-ipx}],
\end{equation}
where $p^\mu=\qty(\omega_p,\vb{p})$, $px\equiv\eta_{\mu\nu}p^\mu x^\nu$, and the operator coefficients $\hat{a}_{p,s}$ and $\hat{a}_{p,s}^\dagger$ are the annihilation and creation operators satisfying
\begin{equation}
    \comm{\hat{a}_{p,s}}{\hat{a}_{p',s'}^\dagger}=(2\pi)^3\delta_{ss'}\delta^3(\vb{p}-\vb{p}').
\end{equation}
From the vacuum state $\ket{0}$, which is defined as the state in Fock's space that is annihilated by all $\hat{a}'$s, we obtain the one-particle state
\begin{equation}
    \ket{p,\epsilon^s}=\sqrt{2\omega_p}\hat{a}_{p,s}^\dagger\ket{0},
\end{equation}
which represents a graviton of momentum $p$ and polarization $s$.

As far as the free theory goes, this is pretty much it. One can use the plane wave expansion~\eqref{h-operator-plane-wave-expansion} to calculate any relevant two-point function, such as the Feynman propagator~\cite{Basile2025} or the Wightman function~\cite{Hsiang_2024}, depending on the problem at hand. However, physical predictions come from interacting theories, and in quantum field theory, these are usually given a perturbative treatment, which involves Feynman diagrams and Feynman rules for computing probability amplitudes for scattering and decay processes~\cite{Schwartz2013,Weinberg12005,Srednicki2007,Peskin2007}. These rules lead to integrals over large virtual momenta, which are usually divergent. In some cases, these divergences can be absorbed into coefficients of terms in the Lagrangian. For instance, in quantum electrodynamics (QED), all divergences can be dealt with by the renormalization of mass, charge, and wave function. If this is possible for all orders in perturbation theory\footnote{The expansion is on powers of the coupling constant.}, the theory is said to be \textbf{perturbatively renormalizable}. If further divergences arise, one can add new terms into the Lagrangian, called counterterms, in order to absorb them. A theory that requires an infinite number of counterterms is \textbf{perturbatively non-renormalizable}~\cite{Srednicki2007}.

\subsection{Interacting gravitons}

Let us now consider gravitational interactions. In this subsection, it will be interesting to keep factors of $G$ while retaining $\hbar=c=1$ (the so-called natural units usually used in quantum field theory texts). Also, let us rescale the metric perturbation field in a way that the expansion~\eqref{g=n+h} becomes
\begin{equation} \label{Metric-expansion-with-kappa}
    g_{\mu\nu}=\eta_{\mu\nu}+\kappa_{\rm g} h_{\mu\nu},
\end{equation}
where $\kappa_{\rm g}=\sqrt{32\pi G}$. Note that now the field $h_{\mu\nu}$ has mass dimension $+1$ since $\kappa_{\rm g}$ has mass dimension $-1$. The Einstein-Hilbert action,
\begin{equation}
    S_{\rm EH}=\frac{2}{\kappa_{\rm g}^2}\int\dd^4x\,\sqrt{-g}R,
\end{equation}
when linearized according to Eq.~\eqref{Metric-expansion-with-kappa}, becomes
\begin{equation} \label{Linearized-EH-action-canonical-normalization}
    S_{\rm EH}=\frac{1}{2}\int \dd^4x\,\qty(-\partial_\rho h_{\mu\nu}\partial^\rho h^{\mu\nu}+2h^{\mu\nu}\partial_\mu\partial_\nu h-h\Box h-2h_{\mu\nu}\partial_\rho\partial^\mu h^{\nu\rho}).
\end{equation}
Note that the rescaled fields are now canonically normalized.

We can obtain gravity-gravity interactions by keeping higher order terms in the Einstein-Hilbert action. An $n$th order interaction term ($n\geq3$) will then be proportional to $\kappa_{\rm g}^{n-2}h^n$, so the coupling constant always has a negative mass dimension. This is a strong indication that the theory is perturbatively non-renormalizable, for the following reason.

Higher-order Feynman diagrams in perturbative quantum field theory contain \emph{loop integrals}, which are four-dimensional integrals over internal momenta. It is common for these integrals to be formally infinite, which requires some regularization procedure, such as cutting off the internal momenta at a single scale $\Lambda$, which we would desire to set to infinity at the end of the calculation. A loop diagram, which contributes to some desired scattering amplitude, scales as $\Lambda^D$ as we take $\Lambda\to\infty$ if $D\neq0$ and as $\ln\Lambda$ if $D=0$. The factor $D$ is called a \emph{superficial degree of divergence}. For pure gravity, a diagram with $E$ external graviton legs and with $V_n$ insertions of the vertices with dimension $\Delta_n$ has~\cite{Schwartz2013,Basile2025}
\begin{equation}
    D=4-E-\sum_nV_n\Delta_n.
\end{equation}
Since $\Delta_n=-(n-2)$, which is the mass dimension of the coupling constant of the $n$th order interaction, and $n\geq3$, we always have $\Delta_n<0$. Then there can be an infinite number of values of $E$ and $V_n$ for which $D>0$, meaning that there are an infinite number of Green functions with $D>0$ that contribute to scattering amplitudes, thus requiring the addition of an infinite number of counterterms to cancel all infinities.

It turns out that pure gravity (gravity with no matter coupling) can actually be renormalized at one loop by the cancellation of infinities through a field redefinition; however, such cancellations do not occur at two loops. Now, if we include coupling with matter, for which the first-order metric expansion of the matter actions reads
\begin{equation}
    S_{\rm matter}=\frac{\kappa_{\rm g}}{2}\int\dd^4x\,T_{\mu\nu}h^{\mu\nu},
\end{equation}
it turns out that renormalization is not possible even at the level of one loop diagrams~\cite{Basile2025}.

Perturbative non-renormalizability does not render the theory useless and certainly does not imply that there is some kind of inconsistency between general relativity and quantum mechanics. A non-renormalizable theory is still able to make useful predictions at energies below some ultraviolet (UV) cutoff~\cite{Srednicki2007}. For instance, the Fermi theory of weak interactions, originally developed to describe beta decay while taking the existence of neutrinos into account, is a perturbatively non-renormalizable theory that makes good predictions below an energy scale of $E\sim300\,\rm GeV$. The electroweak theory of Weinberg, Salam, and Glashow is what we call the ultraviolet (UV) completion of Fermi theory, which is renormalizable once we include a Higgs boson~\cite{Schwartz2013}. Similarly, graviton physics makes genuine predictions below some UV cutoff ($E\sim M_{\rm P}$, where $M_{\rm P}$ denotes the Planck mass~\cite{Schwartz2013}), such as the radiative corrections to the Newtonian potential~\cite{Donoghue1994}. In this sense, perturbative quantum gravity is a consistent effective field theory that is valid below the UV cutoff.

While general relativity is perturbatively non-renormalizable, there is a metric compatible, torsion free, diffeomorphism invariant quantum field theory of gravity that is perturbatively renormalizable in four spacetime dimensions, which is \emph{quadratic gravity}, although it has some puzzles concerning unitarity~\cite{Salvio_2018,Donoghue2022}. On the other hand, a quantum field theoretical treatment of general relativity could still be non-perturbatively renormalizable if it has a non trivial UV fixed point\footnote{A fixed point is a point in the space of parameters of a given theory in which the renormalization group flow ceases, which means that, at that point, the theory is scale invariant. When this occurs for parameters set to zero, we call it a trivial fixed point.}. In that case, a non-perturbative framework would have to be adopted, which is the approach of asymptotically safe quantum gravity~\cite{Percacci2017,Reuter2019}. Lastly, it could be that the UV behavior of quantum gravity cannot be described by the tools of quantum field theory at all, and a completely distinct approach would be required, such as string theory, for instance~\cite{Polchinski2005,Polchinski2005_2}.

In this work, we shall be satisfied with perturbative quantum gravity as an effective field theory while its UV completion remains outside our scope.

\subsection{The PWZ approach}

We now briefly turn to an important question: whether it is possible to detect single gravitons. An argument by Freeman Dyson~\cite{Dyson_2013} seems to lead to the conclusion that the detection of an individual graviton is bound to be impossible. For instance, a gravitational wave detector like LIGO would require the separation between the two mirrors to be less than their Schwarzschild radii in order to detect a graviton, which would lead them to collapse into black holes before the measurement could be completed. Although the matter of single graviton detection is still under debate~\cite{Carney2024,Tobar2024}, another approach was recently taken by Parikh, Wilczek, and Zahariade (PWZ)~\cite{Parikh2020,Parikh_2021,Parikh2021}.

The basic idea behind the PWZ approach is to investigate the possible quantum nature of the gravitational field by detecting the effect of the quantum noise induced by the gravitons on classical particles, in the same spirit as quantum Brownian motion. In this way, the quantization of gravity in the weak-field limit is manifest in the equations of motion followed by the classical system with which it interacts. By using a formalism due to Feynman and Vernon~\cite{Feynman1963,Feynman2010}, PWZ derived a Langevin-like stochastic equation characterizing the geodesic deviation between two test particles. A similar analysis was conducted in Ref.~\cite{Kanno2021} through an alternative approach and subsequently in Ref.~\cite{Cho2022}, where all graviton modes and polarizations were taken into account.

In this work, we will follow the PWZ approach in order to investigate the decoherence of a composite particle induced by a weak quantized gravitational field, as well as the entropy production arising in such a system. This will be the subject of Part~\ref{Part:Dec-and-ent} of this thesis.

\section{Linearized gravity in curved spacetime} \label{Sec:LG-curved-spacetime}

Now that we have discussed linearized gravity in flat spacetime, let us next consider a general background. We can proceed in the same way we did in Section~\ref{Sec:LG-flat-spacetime} by expanding the Einstein-Hilbert action around a background metric $\gamma_{\mu\nu}$. However, such expansion at the level of the action needs to be up to second order in the metric perturbation, and for a general background, this can become quite involved. We will take a different approach in this section to avoid an excess of mathematical details. The metric expansion will be made at the level of the equations of motion (Einstein's equation), meaning that we will keep terms only up to the first order in the perturbation. Then, after a convenient choice of gauge, we can write an action whose variation leads to such a linearized equation. For the full action expansion up to second order around a general background, we refer the reader to Ref.~\cite{Basile2025}.

The final goal of this section is to describe the interaction of gravitons with a classical static Newtonian potential. This means that, although we will begin by considering a general background $\gamma_{\mu\nu}$, we will eventually take the Newtonian limit, where $\gamma_{\mu\nu}$ is described by another small (now fixed) perturbation to Minkowski spacetime. Thus, the quantization procedure of the gravitational radiation discussed in Section~\ref{Sec:Quantum-gravitational-radiation} will remain valid for our purposes. In this sense, we will not discuss the subtleties associated with quantum fields in general curved spacetimes~\cite{Wald1994,Birrell1984}.

Finally, a word on notation. In this section, we will need to distinguish between curvature symbols and tensors that are associated with $\gamma_{\mu\nu}$, the background metric, and those associated with $g_{\mu\nu}$, the \emph{full} spacetime metric that includes the propagating gravitational radiation. For the latter, we will simply continue to use the same notation as we have been so far, but for the Christoffel symbols associated with the background metric, for instance, we will denote $\Gamma_{\mu\nu}^\rho[\gamma]$, ${R^\rho}_{\sigma\mu\nu}[\gamma]$ for the background Riemann tensor, and so on. The covariant derivative compatible with $\gamma_{\mu\nu}$ will be denoted by $D_\mu$.

\subsection{Linearized Einstein's equation}

Suppose that $\gamma_{\mu\nu}$ is a known solution to Einstein's equation (in vacuum, for simplicity) and we wish to study small metric perturbations around it. In that case, the full metric tensor takes the form $g_{\mu\nu}=\gamma_{\mu\nu}+h_{\mu\nu}$, where $h_{\mu\nu}$ denotes the perturbation we are interested in, with $\abs{h_{\mu\nu}}\ll\abs{\gamma_{\mu\nu}}$. In the same spirit as Eq.~\eqref{Metric-expansion-parameterized-by-e}, we can consider a one-parameter family of solutions,
\begin{equation} \label{g=gamma+h}
    g_{\mu\nu}(\epsilon)=\gamma_{\mu\nu}+\epsilon h_{\mu\nu},
\end{equation}
such that, once again,
\begin{equation} \label{hmunu-definition}
    h_{\mu\nu}=\eval{\dv{g_{\mu\nu}(\epsilon)}{\epsilon}}_{\epsilon=0}.
\end{equation}

The one-parameter family of solutions satisfies Einstein's equation, which is a non-linear operation of the form
\begin{equation}
    \mathcal{E}[g(\epsilon)]=0.
\end{equation}
Differentiating this equation with respect to $\epsilon$ and setting $\epsilon=0$ leads to a linear equation for $h_{\mu\nu}$, which is the desired linearized Einstein's equation.

Einstein's equation in vacuum reads
\begin{equation}
    R_{\sigma\nu}=0.
\end{equation}
The Ricci tensor depends on $\epsilon$ through $g_{\mu\nu}(\epsilon)$. A linearized equation is then obtained by
\begin{equation} \label{Linearized-Einstein-equation-curved-0}
    \eval{\dv{R_{\sigma\nu}}{\epsilon}}_{\epsilon=0}=0.
\end{equation}

Now, recall that the Riemann tensor is defined such that
\begin{equation}
    \nabla_\mu\nabla_\nu V^\rho-\nabla_\nu\nabla_\mu V^\rho={R^\rho}_{\sigma\mu\nu}V^\sigma,
\end{equation}
for some arbitrary vector $V^\mu$. The covariant derivative $\nabla_\mu$, associated with $g_{\mu\nu}(\epsilon)$, can be related to the covariant derivative $D_\mu$, associated with $\gamma_{\mu\nu}$, through the tensor field (see Appendix~\ref{app:Differential})
\begin{equation} \label{Tensor-C}
    {C^\rho}_{\mu\nu}=\frac{1}{2}g^{\rho\sigma}(\epsilon)\qty[D_\mu g_{\nu\sigma}(\epsilon)+D_\nu g_{\sigma\mu}(\epsilon)-D_\sigma g_{\mu\nu}(\epsilon)].
\end{equation}
Explicitly, we find
\begin{equation}
    \nabla_\mu\nabla_\nu V^\rho-\nabla_\nu\nabla_\mu V^\rho=D_\mu D_\nu V^\rho-D_\nu D_\mu V^\rho+\qty(2D_{[\mu}{C^\rho}_{\nu]\sigma}+2{C^\rho}_{\delta[\mu}{C^\delta}_{\nu]\sigma})V^\sigma,
\end{equation}
and thus
\begin{equation} \label{Riemann-in-terms-of-background-Riemann}
    {R^\rho}_{\sigma\mu\nu}={R^\rho}_{\sigma\mu\nu}[\gamma]+2D_{[\mu}{C^\rho}_{\nu]\sigma}+2{C^\rho}_{\delta[\mu}{C^\delta}_{\nu]\sigma}.
\end{equation}
The Ricci tensor associated with $g_{\mu\nu}(\epsilon)$ is given by
\begin{align}
    R_{\sigma\nu}=R_{\sigma\nu}[\gamma]+2D_{[\mu}{C^\mu}_{\nu]\sigma}+2{C^\mu}_{\delta[\mu}{C^\delta}_{\nu]\sigma}.
\end{align}
Since $\gamma_{\mu\nu}$ is assumed to be an exact solution of the vacuum Einstein equation, we must have $R_{\sigma\nu}[\gamma]=0$ and
\begin{equation}
    R_{\sigma\nu}=-2D_{[\nu}{C^\mu}_{\mu]\sigma}+2{C^\mu}_{\delta[\mu}{C^\delta}_{\nu]\sigma}.
\end{equation}

Plugging this result back into Eq.~\eqref{Linearized-Einstein-equation-curved-0} and using Eqs.~\eqref{hmunu-definition} and~\eqref{Tensor-C} leads to the linear equation for $h_{\mu\nu}$,
\begin{equation}
    -\frac{1}{2}D_\nu D_\sigma h-\frac{1}{2}D^\mu D_\mu h_{\sigma\nu}+D^\mu D_{(\nu}h_{\sigma)\mu}=0,
\end{equation}
where $h=\gamma^{\mu\nu}h_{\mu\nu}$. Note that in the limit of flat spacetime, where $\gamma_{\mu\nu}\to\eta_{\mu\nu}$ and $D_\mu\to\partial_\mu$, we recover Eq.~\eqref{Linearized-Mink-vacuum-equation}.

We can rewrite
\begin{align}
    D^\mu D_{(\nu}h_{\sigma)\mu}&=\frac{1}{2} D_\nu D^\mu h_{\sigma\mu}+\frac{1}{2}\qty( D^\mu D_\nu- D_\nu D^\mu)h_{\sigma\mu} \nonumber \\
    &\hspace{0.5cm}+\frac{1}{2} D_\sigma D^\mu h_{\nu\mu}+\frac{1}{2}\qty( D^\mu D_\sigma- D_\sigma D^\mu)h_{\nu\mu} \nonumber \\
    &=D_{(\nu} D^\mu h_{\sigma)\mu}-R_{\rho\sigma\mu\nu}[\gamma]h^{\rho\mu}.
\end{align}
Also, since $h$ is a scalar, $ D_\nu D_\sigma h= D_\sigma D_\nu h$, and thus
\begin{align}
    D_{(\nu} D^\mu h_{\sigma)\mu}-\frac{1}{2} D_\nu D_\sigma h&=\frac{1}{2} D_\nu\qty( D^\mu h_{\sigma\mu}-\frac{1}{2} D_\sigma h)+\frac{1}{2} D_\sigma\qty( D^\mu h_{\nu\mu}-\frac{1}{2} D_\nu h) \nonumber \\
    &=D_{(\nu} D^\mu\Bar{h}_{\sigma)\mu},
\end{align}
where
\begin{equation}
    \Bar{h}_{\mu\nu}=h_{\mu\nu}-\frac{1}{2}\gamma_{\mu\nu}h.
\end{equation}
Finally, one is left with
\begin{equation} \label{Linearized-Einstein-equation}
    D_{(\nu} D^\mu\Bar{h}_{\sigma)\mu}-\frac{1}{2} D^\mu D_\mu h_{\sigma\nu}-R_{\rho\sigma\mu\nu}[\gamma]h^{\rho\mu}=0.
\end{equation}
This is the linearized Einstein's equation in curved spacetime.

\subsection{TT gauge}

Let us now discuss gauge invariance in this more general case. It turns out that the equations of motion are still left invariant by the gauge transformation~\eqref{General-gauge-transformation}. The difference now is that $g_{\mu\nu}(0)=\gamma_{\mu\nu}$, and so the gauge transformation takes the form
\begin{equation}
    h_{\mu\nu}'=h_{\mu\nu}+D_\mu\varepsilon_\nu+D_\nu\varepsilon_\mu,
\end{equation}
for some arbitrary vector $\varepsilon_\mu$. Once again, gauge freedom allows one to simplify the equations of motion for $h_{\mu\nu}$. For instance, starting with any given $h_{\mu\nu}$ such that $D^\nu\Bar{h}_{\mu\nu}\neq0$ in general, one can perform a gauge transformation for which the new perturbation will satisfy
\begin{equation}
    D^\nu\Bar{h}'_{\mu\nu}=D^\nu\Bar{h}_{\mu\nu}+D^\nu D_\nu\varepsilon_\mu,
\end{equation}
where we used the fact that $R_{\mu\nu}[\gamma]=0$.
We may then choose $\varepsilon^\mu$ to satisfy
\begin{equation}
    D^\nu D_\nu\varepsilon_\mu=-D^\nu\Bar{h}_{\mu\nu},
\end{equation}
such that, for the perturbation in the new gauge, we now have $D^\nu\Bar{h}^{\rm L}_{\mu\nu}=0$, which is the generalization of the Lorenz condition to curved spacetime.

In this gauge, the linearized Einstein equation~\eqref{Linearized-Einstein-equation} becomes
\begin{equation} \label{Gauged-linearized-Einstein-equation}
    D^\mu D_\mu h^{\rm L}_{\sigma\nu}+2R_{\rho\sigma\mu\nu}[\gamma]h_{\rm L}^{\rho\mu}=0,
\end{equation}
with the metric perturbation satisfying the gauge condition
\begin{equation} \label{Gauge-condition}
    D^\nu h^{\rm L}_{\mu\nu}-\frac{1}{2}D_\mu h^{\rm L}=0.
\end{equation}

In the limit of flat spacetime, Eq.~\eqref{Gauged-linearized-Einstein-equation} reduces to $\Box h^{\rm L}_{\sigma\nu}=0$. \emph{In that limit}, as we saw in Section~\ref{Sec:LG-flat-spacetime}, there is a residual gauge freedom; i.e., one can perform a further gauge transformation parametrized by a vector $\xi^\nu$ as long as $\Box\xi_\nu=0$, such that both the gauge condition and the equation of motion remain satisfied. Then, one uses this freedom to choose $h=0$ and $h_{0i}=0$, which, together with the equations of motion, implies $h_{00}=0$.

In the case of a curved background, however, this is not generally possible. We can try to perform such additional gauge transformation parametrized by a vector $\xi^\nu$ satisfying $D^\mu D_\mu\xi_\nu=0$ (which is necessary for keeping Eq.~\eqref{Gauge-condition} satisfied), but now the left-hand side of the equation of motion~\eqref{Gauged-linearized-Einstein-equation} transforms to
\begin{align} \label{GaugedGauged}
    D^\mu D_\mu h'_{\sigma\nu}+2R_{\rho\sigma\mu\nu}[\gamma]h'^{\rho\mu}&=D^\mu D_\mu h^{\rm L}_{\sigma\nu}+2R_{\rho\sigma\mu\nu}[\gamma]h_{\rm L}^{\rho\mu} \nonumber \\
    &-\qty{D^\mu\qty(R_{\mu\sigma\rho\nu}[\gamma]+R_{\rho\sigma\mu\nu}[\gamma])}\xi^\rho.
\end{align}
We can see that the term on the last line of Eq.~\eqref{GaugedGauged} prevents us from making such a gauge transformation while simultaneously keeping the equations of motion invariant.

Let us see what happens to this term when the background itself can be viewed as a small perturbation of Minkowski spacetime,
\begin{equation} \label{Tilde-g}
    \gamma_{\mu\nu}=\eta_{\mu\nu}+h^{(B)}_{\mu\nu},
\end{equation}
with $|h^{(B)}_{\mu\nu}|\ll1$ as usual. Then a direct calculation yields
\begin{align} \label{Annoying-term}
    D^\mu\qty{R_{\rho\sigma\mu\nu}[\gamma]+R_{\mu\sigma\rho\nu}[\gamma]}&=\frac{1}{2}\partial_\sigma\Box h^{(B)}_{\rho\nu}+\frac{1}{2}\partial_\nu\partial_\rho\partial^\mu h^{(B)}_{\mu\sigma}+\frac{1}{2}\partial_\rho\partial_\sigma\partial^\mu h^{(B)}_{\mu\nu} \nonumber \\
    &+\frac{1}{2}\partial_\nu\Box h^{(B)}_{\rho\sigma}-\partial_\rho\Box h^{(B)}_{\nu\sigma}-\partial_\nu\partial_\sigma\partial^\mu h^{(B)}_{\rho\mu} \nonumber \\
    &=0,
\end{align}
where we used the fact that, since $\gamma_{\mu\nu}$ is a solution to Einstein's equation in vacuum, the background perturbation $h^{(B)}_{\mu\nu}$ must satisfy the linearized equation~\eqref{Linearized-Mink-vacuum-equation}.

In other words, \emph{when the background metric can be treated as a small perturbation of Minkowski spacetime}, the term preventing us from performing the residual gauge transformation on $h_{\mu\nu}$ vanishes up to first order in $h^{(B)}_{\mu\nu}$. One is then allowed to choose $\xi^\mu$ satisfying $D^\mu D_\mu\xi_\nu=0$, such that $h=0$ and $h_{0i}=0$. The gauge condition then becomes
\begin{equation}
    D_\mu h^{\mu\nu}=0.
\end{equation}
Particularly, the $\nu=0$ component of this equation reads
\begin{equation}
    \partial_0h^{00}+\qty{\Gamma_{\mu0}^\mu[\gamma]+\Gamma_{00}^0[\gamma]}h^{00}+\Gamma_{ij}^0[\gamma]h^{ij}=0.
\end{equation}
Explicitly, one has
\begin{subequations}
\begin{align}
    \Gamma_{\mu\nu}^0[\gamma]&=\frac{1}{2}\qty(\partial_\mu h^{(B)}_{\nu0}+\partial_\nu h^{(B)}_{0\mu}-\partial_0h^{(B)}_{\mu\nu}), \\
    \Gamma_{\mu\nu}^\mu[\gamma]&=\frac{1}{2}\partial_\nu h^{(B)}.
\end{align}
\end{subequations}
\emph{If the metric perturbation is static}, one obtains $\Gamma_{\mu0}^\mu[\gamma]=\Gamma_{00}^0[\gamma]=0$. Additionally, \emph{if $h^{(B)}_{0i}=0$}, we also have $\Gamma_{ij}^0[\gamma]=0$, and thus
\begin{equation}
    \partial_0h^{00}=0.
\end{equation}
Then, since $h_{00}$ is just a constant, one can choose suitable boundary conditions such that $h_{00}=0$.

In summary, \emph{for a background metric which is a small static perturbation to flat spacetime with $h^{(B)}_{0i}=0$}, the Einstein equation for $h_{\mu\nu}$ reads
\begin{equation} \label{Linearized-Einstein-equation-in-TT-gauge}
    D^\mu D_\mu h^{\rm TT}_{ij}+2R_{ikjl}[\gamma]h_{\rm TT}^{kl}=0,
\end{equation}
with the TT gauge conditions
\begin{subequations} \label{TT-gauge-curved}
\begin{align}
    D_\mu h_{\rm TT}^{\mu\nu}=0, \\
    \gamma^{\mu\nu}h^{\rm TT}_{\mu\nu}=0, \\
    h^{\rm TT}_{0\mu}=0.
\end{align}
\end{subequations}
Particularly, this shows that we can impose the TT gauge conditions when the background is described in the Newtonian limit, for instance.

Finally, note that the linearized vacuum equation in TT gauge can be obtained from the extremization of the action\footnote{The normalization factor was chosen to match the linearized Einstein-Hilbert action in the limit of flat spacetime. Furthermore, we note that this action matches the one presented in Ref.~\cite{Basile2025} (which was obtained from the second-order expansion at the level of the action from the beginning) if we consider the TT gauge and match the normalization conventions.}
\begin{equation}
    S_{\rm EH}=\frac{1}{32\pi}\int\dd^4x\,\qty(\frac{1}{2}h^{\rm TT}_{ij}D^\mu D_\mu h_{\rm TT}^{ij}+h^{\rm TT}_{ij}R^{ikjl}[\gamma]h^{\rm TT}_{kl}).
\end{equation}
Setting $\delta S_{\rm EH}=0$ immediately leads to Eq.~\eqref{Linearized-Einstein-equation-in-TT-gauge}.

\subsection{Newtonian limit}

The Newtonian limit of general relativity is described by Eq.~\eqref{Tilde-g}, for which case we have
\begin{equation}
    h^{(B)}_{\mu\nu}=-2\phi\delta_{\mu\nu},
\end{equation}
with $\phi=\phi(\vb{x})$ being the time-independent Newtonian potential. Then we can show that
\begin{equation}
    h^{\rm TT}_{ij}R^{ikjl}[\gamma]h^{\rm TT}_{kl}=-2\phi(\partial_kh^{\rm TT}_{ij})(\partial^ih_{\rm TT}^{jk})+\textrm{boundary terms},
\end{equation}
where we used the gauge conditions
\begin{subequations}
\begin{equation}
    0=\gamma_{\mu\nu}h_{\rm TT}^{\mu\nu}=(1-2\phi)\delta_{ij}h_{\rm TT}^{ij}\implies\delta_{ij}h_{\rm TT}^{ij}=0,
\end{equation}
and
\begin{equation} \label{phi*dihij}
    0=\phi\,D_\mu h_{\rm TT}^{\mu\nu}=\phi\,\partial_\mu h^{\mu\nu}+O(\phi^2)\implies\phi\,\partial_ih^{ij}=0.
\end{equation}
\end{subequations}
Additionally, for the background metric given by Eq.~\eqref{Tilde-g}, we have, to first order in $h^{(B)}_{\mu\nu}$,
\begin{align}
    h^{\rm TT}_{ij}D_\mu D^\mu h_{\rm TT}^{ij}&=\gamma^{\mu\nu}h^{\rm TT}_{ij}\partial_\mu\partial_\nu h_{\rm TT}^{ij}+2h^{\rm TT}_{ij}(\partial^\mu\Gamma_{\mu k}^i[\gamma])h_{\rm TT}^{kj} \nonumber \\
    &-h^{\rm TT}_{ij}\eta^{\mu\nu}\Gamma_{\mu\nu}^\lambda[\gamma]\partial_\lambda h_{\rm TT}^{ij}+4h^{\rm TT}_{ij}\Gamma_{\mu k}^i[\gamma]\partial^\mu h_{\rm TT}^{kj}.
\end{align}
In the Newtonian limit, the only non-vanishing Christoffel symbols are
\begin{subequations}
\begin{align}
    \Gamma_{00}^i[\gamma]&=\partial^i\phi, \\
    \Gamma_{0i}^0[\gamma]=\Gamma_{i0}^0[\gamma]&=\partial_i\phi, \\
    \Gamma_{ij}^k[\gamma]&=\delta_{ij}\partial^k\phi-\delta_j^k\partial_i\phi-\delta_i^k\partial_j\phi.
\end{align}
\end{subequations}
Then, after some manipulations involving integration by parts and the gauge conditions, while keeping only terms up to the first order in $\phi$, we find
\begin{align} \label{Graviton-action-position-space}
    S_{\rm EH}&=\frac{1}{64\pi}\int\dd^4x\,\gamma^{\mu\nu}h^{\rm TT}_{ij}\partial_\mu\partial_\nu h_{\rm TT}^{ij} \nonumber \\
    &=\frac{1}{64\pi}\int\dd^4x\,\qty(h^{\rm TT}_{ij}\Box h_{\rm TT}^{ij}+2\phi\,h^{\rm TT}_{ij}\delta_{\mu\nu}\partial^\mu\partial^\nu h_{\rm TT}^{ij}).
\end{align}

\begin{figure}[!ht]
    \centering
    \includegraphics[width=0.8\linewidth]{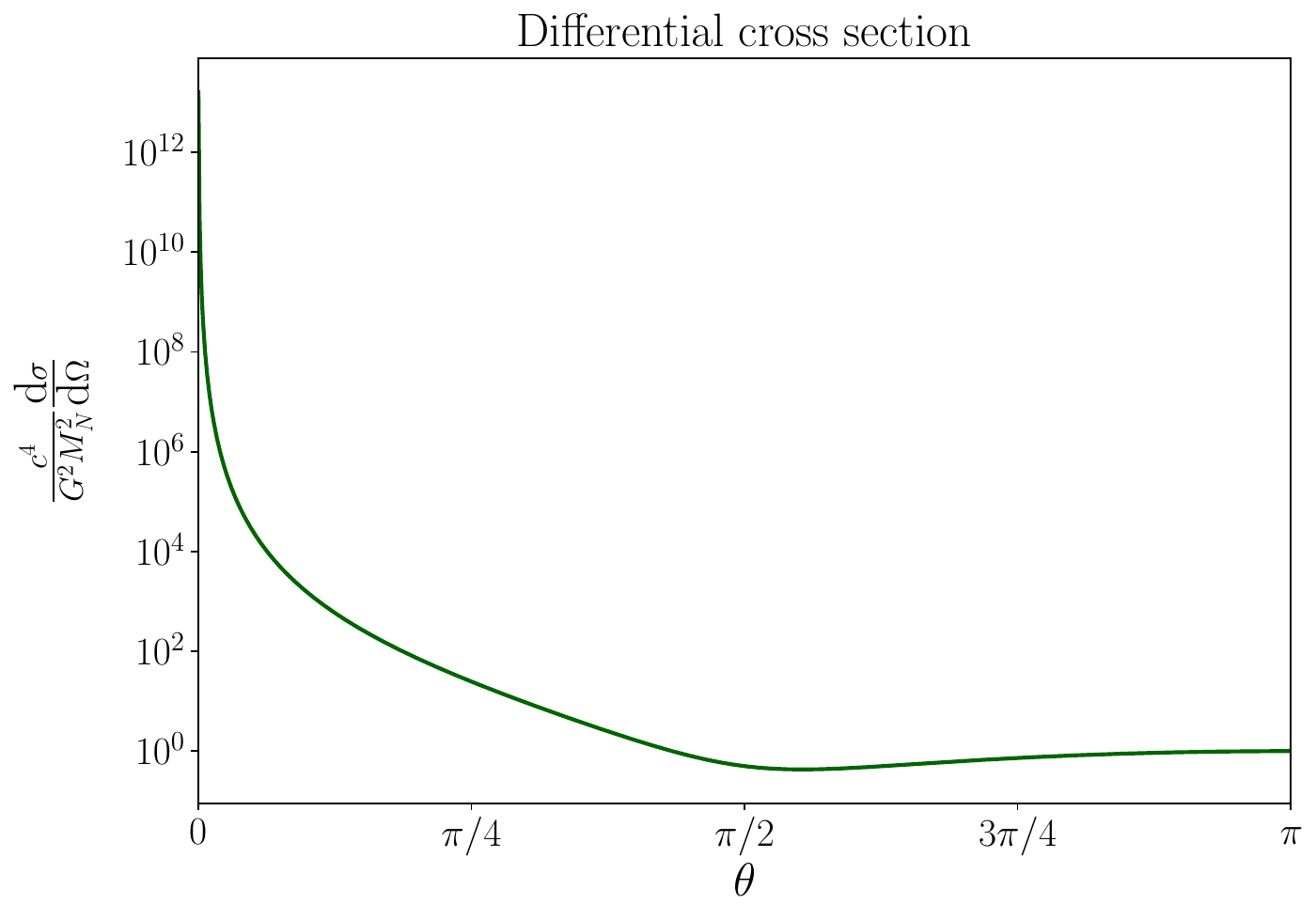}
    \caption{Differential cross section as a function of the scattering angle.}
    \label{Fig:Diff-cross-sec}
\end{figure}

The action Eq.~\eqref{Graviton-action-position-space} can be thought of as describing gravitational radiation in flat spacetime interacting with a static Newtonian potential. At the quantum level, the quantization procedure follows as described in Section~\ref{Sec:Quantum-gravitational-radiation}. This action then describes graviton scattering by a Newtonian potential. In Appendix~\ref{app:Scattering} we compute the differential cross section for a graviton scattered by a Newtonian source behaving like $\phi(r)=-M_N/r$, with $r^2=\delta_{ij}x^ix^j$ and $M_N$ being the mass of the source. The result is\footnote{Restoring the universal constants once again.}~\cite{Westervelt_1971,Ragusa_2003}
\begin{equation}
    \dv{\sigma}{\Omega}=\frac{G^2M_N^2}{c^4\sin^4\frac{\theta}{2}}\qty(\cos^8\frac{\theta}{2}+\sin^8\frac{\theta}{2}),
\end{equation}
with $\theta$ being the scattering angle. The behavior of the differential cross section can be seen from Figure~\ref{Fig:Diff-cross-sec} to be strongly dominated by regions in which $\theta\ll1$, going as
\begin{equation}
    \dv{\sigma}{\Omega}\sim\frac{G^2M_N^2}{c^4}\frac{1}{\theta^4}.
\end{equation}
Such small probability for higher values of the scattering angle reflects the weakness of the gravitational interaction.

\chapter{Environment-induced decoherence}
\label{chap:dec-and-FV}

The concept of a closed system is a useful but sometimes oversimplified idealization in physics, since it is based on the idea that one can isolate an object of interest from any interaction with its surroundings. In practice, one tries to minimize these unwanted influences in order to study such \emph{approximately} closed systems. Now, while classical systems are subject to momentum and energy exchanges with the environment, which can be neglected in most cases (as, for instance, in the collisions between dust particles and a tennis ball), quantum systems, in addition to these kinds of influences, can also become entangled with the environment.

Entanglement is a strong correlation that has no classical analogue. If two quantum systems are found in an entangled state, then measurements performed on one of the subsystems will reveal some amount of information about the other, even if they are spatially separated over large distances. Now, consider a quantum system that is initially prepared in a superposition of mutually orthogonal states. Assuming that the quantum description of nature is universal, the system of interest will, in general, become entangled after interacting with its (quantum) environment. The environmental degrees of freedom (DoFs) can then be thought of as probing the quantum system while acquiring some measure of "which-path information" (or "which-state information", more generally). There is no fundamental principle that prevents us from measuring these environmental DoFs and acquiring such information, even though one typically does not do so, either by choice or due to some practical limitation (in fact, this is often the criterion used to determine which system is the system of interest and which one is the environment). As a consequence, measurements performed on the system will be less and less able to detect interference terms between the components of the superposition (the coherences) as the environment acquires more and more which-state information. In the limiting case, all coherences are somewhat "destroyed"\footnote{To be more precise, the interaction with the environment does not destroy the initial quantum superposition but rather enlarges it to include the environment itself. Decoherence happens when such superposition becomes unobservable at the level of the system of interest alone, giving the impression that it has been destroyed (although it is still there).} and the system can only be found in definite "semi-classical" states. This process is called \textbf{environment-induced decoherence}, or simply \textbf{decoherence}~\cite{K_bler_1973,Zurek_1981,Zurek_1982,Joos_1985,Zurek_1991,Zurek_2003,Schlosshauer2008,Breuer2002}.

Environment-induced decoherence refers to the phenomenon of the irreversible loss of quantum coherence from a system that is coupled to an environment, as well as the dynamical selection of the observable properties of the system. It is a purely quantum phenomenon, and it can occur alongside some "classical" aspects of environmental interactions, such as dissipation. Now, since a quantum state can be expressed as different superpositions of different bases of the Hilbert space, decoherence is a basis-dependent phenomenon. In general, it is the way the system interacts with a given environment that dictates which basis will be more susceptible to the loss of quantum coherence. For instance, if the interaction Hamiltonian couples environment operators with the system position operator, then we can think of the environment as probing the position of the system at each time instant, and decoherence will eventually occur in the position basis, thereby localizing the system. This selection of observable properties is the aspect of decoherence that is most related to the quantum-to-classical transition, the dynamical selection of a few "classical" observable quantities like energy, position, and momentum. For discussions concerning this connection, and further connections with different interpretations of quantum mechanics, see Ref.~\cite{Schlosshauer_2005} and references therein.

In this chapter, we present a somewhat general overview of the decoherence program in order to establish the physics behind the phenomenon before we specialize in the case of a gravitational environment for the remainder of this work. We begin by exploring the basic ideas behind decoherence in Section~\ref{Sec:Basics-of-decoherence}, making our discussion more precise and explicitly showing the consequences of the system-environment entanglement. Then, in Section~\ref{Sec:The-FV-IF}, we introduce the influence functional formalism to open quantum systems. Since more common techniques involve the so-called master equations, we show the equivalence of the two approaches in Section~\ref{Sec:Master-equation}, where we also explore the physical interpretation of each term encompassing the influence of the environment on the open quantum system. We do this by considering the canonical model in which the system of interest is described by continuous phase-space variables, while the environment is described by a set of non-interacting harmonic oscillators. This model is usually referred to as \textbf{quantum Brownian motion}, and it bears many resemblances to the problem of a quantum particle in a bath of gravitons that we wish to consider in this work. In Section~\ref{Sec:Langevin-equation} we explore how the localization of the system due to decoherence can lead to the system being described by trajectories that are obtained as the solution of the Langevin equation, a Newtonian-like equation of motion governed by a stochastic force variable entering as noise from the environment. Finally, the physical reality of such trajectories is discussed in Section~\ref{Sec:Decoherent-histories} in light of the decoherent histories approach to quantum mechanics. This will be crucial for when we define work for open quantum systems in Chapter~\ref{chap:entropy}.


\section{Basics of decoherence} \label{Sec:Basics-of-decoherence}

Let us consider a quantum system $\mathcal{S}$ coupled with an environment $\mathcal{E}$. The Hilbert space of the total system $\mathcal{S}+\mathcal{E}$ is given by the tensor product $\mathcal{H}=\mathcal{H}_{\mathcal{S}}\otimes\mathcal{H}_{\mathcal{E}}$, where $\mathcal{H}_{\mathcal{S}}$ ($\mathcal{H}_{\mathcal{E}}$) denotes the Hilbert space of the system (environment) alone. This separation between the system and the environment requires us to distinguish between the variables that describe the system of interest and the other (usually infinitely many) variables that can affect the system, but whose detailed dynamics are of somewhat less relevance or may even be intractable from a practical point of view. From now on, let us assume that this choice has been made in a reasonable manner. Moreover, let us assume that one can also introduce some coarse-graining measures that characterize the environment (as, for instance, when we describe a bath by its thermodynamic variables).

Suppose that the total Hamiltonian can be written as
\begin{equation}
    \hat{H}=\hat{H}_{\mathcal{S}}\otimes \hat{I}_{\mathcal{E}}+\hat{I}_{\mathcal{S}}\otimes \hat{H}_{\mathcal{E}}+\hat{H}_{\mathcal{I}},
\end{equation}
where $\hat{H}_{\mathcal{S}}$ ($\hat{H}_{\mathcal{E}}$) is the free Hamiltonian of the system (environment), while $\hat{I}_{\mathcal{S}}$ ($\hat{I}_{\mathcal{E}}$) denotes the identity on the system (environment) subspace. The interaction between them is described by $\hat{H}_{\mathcal{I}}$. All the operators are written in the Schrödinger picture, and we have assumed that the Hamiltonian contains no explicit time dependence. The interaction Hamiltonian can be written as
\begin{equation}
    \hat{H}_{\mathcal{I}}=\sum_\alpha\hat{S}_\alpha\otimes\hat{E}_\alpha,
\end{equation}
which is the most general form, namely a diagonal decomposition of the system and environment Hermitian operators $\hat{S}_\alpha$ and $\hat{E}_\alpha$. This interaction Hamiltonian singles out a specific set of orthonormal basis vectors of the system, $\qty{\ket{s_n}}$, which satisfy
\begin{equation}
    \hat{S}_\alpha\ket{s_n}=s_n^{(\alpha)}\ket{s_n}\hspace{0,5cm}\textrm{for all $n$ and $\alpha$}.
\end{equation}

For simplicity, let us assume that $[\Hat{H}_{\mathcal{S}},\Hat{S}_\alpha]=0$. In the interaction picture, we have
\begin{equation}
    \hat{H}_{\mathcal{I}}(t)=e^{i\hat{H}_0t}\hat{H}_{\mathcal{I}}e^{-i\hat{H}_0t}=\sum_\alpha\hat{S}_\alpha\otimes\hat{E}_\alpha(t),
\end{equation}
where $\hat{H}_0=\hat{H}_{\mathcal{S}}\otimes \hat{I}_{\mathcal{E}}+\hat{I}_{\mathcal{S}}\otimes \hat{H}_{\mathcal{E}}$ and $\hat{E}_\alpha(t)=e^{i\hat{H}_{\mathcal{E}}t}\hat{E}_\alpha e^{-i\hat{H}_{\mathcal{E}}t}$. The interaction picture time-evolution operator for the total system is then
\begin{equation} \label{Interaction-picture-time-evolution-operator}
    \hat{U}(t)=\textrm{T}\exp\qty[-i\int_0^t\dd t'\,\sum_\alpha\hat{S}_\alpha\otimes\hat{E}_\alpha(t')],
\end{equation}
with T denoting the time ordering operation, which takes any product of time dependent operators and changes the order such that later times are on the left.

Now consider an initial state $\ket{\psi(0)}=\ket{s_n}\otimes\ket{E(0)}$, where $\ket{E(0)}$ stands for the environment initial state. Since this is a product state, it means that initially there is no entanglement between the system and the environment. According to Eq.~\eqref{Interaction-picture-time-evolution-operator}, this state evolves to
\begin{align}
    \ket{\psi(t)}&=\hat{U}(t)\ket{\psi(0)} \nonumber \\
    &=\textrm{T}\exp\qty[-i\int_0^t\dd t'\,\sum_\alpha\hat{S}_\alpha\otimes\hat{E}_\alpha(t')]\ket{s_n}\otimes\ket{E(0)} \nonumber \\
    &=\ket{s_n}\otimes\qty{\textrm{T}\exp\qty[-i\int_0^t\dd t'\,\sum_\alpha s_n^{(\alpha)}\hat{E}_\alpha(t')]\ket{E(0)}}.
\end{align}
Since the final state is still described by a tensor product, this means that a system initially in the state $\ket{s_n}$ does not become entangled with the environment. We say it represents an environment-superselected preferred state~\cite{Schlosshauer2008}.

However, let us now consider the initial state
\begin{equation}
    \ket{\Psi(0)}=\qty(\sum_nc_n\ket{s_n})\otimes\ket{E(0)},
\end{equation}
which is still a product state and therefore contains no entanglement. The difference here is that the initial state of the system alone is described by a superposition of the state vectors $\ket{s_n}$. It is not hard to see that the total state evolves to
\begin{equation} \label{Entangled-state}
    \ket{\Psi(t)}=\sum_nc_n\ket{s_n}\otimes\ket{E_n(t)},
\end{equation}
where
\begin{equation}
    \ket{E_n(t)}=\textrm{T}\exp\qty[-i\int_0^t\dd t'\,\sum_\alpha s_n^{(\alpha)}\hat{E}_\alpha(t')]\ket{E(0)}.
\end{equation}

The state~\eqref{Entangled-state} cannot be written as a tensor product since it is a superposition of the states $\ket{s_n}\otimes\ket{E_n(t)}$, and thus it represents an entangled system-environment state. We see that the initial superposition of system states has been enlarged to include the environment, and one can no longer attribute a specific state to the system alone. However, if one wishes to perform measurements only in the system of interest, then one needs to compute expectation values of system observables by using the reduced density matrix, which is obtained from the total density matrix $\hat{\rho}(t)=\ket{\Psi(t)}\bra{\Psi(t)}$ by performing a partial trace with respect to the environment variables~\cite{Cohen1},
\begin{align} \label{Reduced-density-matrix}
    \hat{\rho}_{\mathcal{S}}(t)&=\textrm{Tr}_{\mathcal{E}}\qty{\ket{\Psi(t)}\bra{\Psi(t)}} \nonumber \\
    &=\sum_i\bra{i}\qty(\sum_nc_n\ket{s_n}\otimes\ket{E_n(t)})\qty(\sum_mc_m^*\bra{s_m}\otimes\bra{E_m(t)})\ket{i} \nonumber \\
    &=\sum_n\sum_m\sum_ic_nc_m^*\ket{s_n}\bra{s_m}\braket{i}{E_n(t)}\braket{E_m(t)}{i} \nonumber \\
    &=\sum_{n,m}c_nc_m^*\ket{s_n}\bra{s_m}\braket{E_m(t)}{E_n(t)},
\end{align}
where $\qty{\ket{i}}$ denotes a complete set of environment orthonormal basis vectors. Since $\braket{E_n(t)}{E_n(t)}=1$, the diagonal elements of $\hat{\rho}_{\mathcal{S}}(t)$ are constant over time. However, the off-diagonal elements, which are called \emph{coherences}, do change over time. The time dependence of the matrix element $\mel{s_n}{\hat{\rho}_{\mathcal{S}}(t)}{s_m}$ is given by the overlap of the corresponding environment states $\ket{E_n(t)}$ and $\ket{E_m(t)}$, which is usually written as
\begin{equation}
    \abs{\braket{E_n(t)}{E_m(t)}}=\exp[-\Gamma_{nm}(t)].
\end{equation}
When the quantity $\Gamma_{nm}(t)$ satisfies $\Gamma_{nm}(t)\geq0$, we call it the \textbf{decoherence function}~\cite{Breuer2002}. The matrix element $\rho_{\mathcal{S}}^{nm}=\mel{s_n}{\hat{\rho}_{\mathcal{S}}}{s_m}$ can then be written as
\begin{equation} \label{Coherences-decay}
    \rho_{\mathcal{S}}^{nm}(t)=\rho_{\mathcal{S}}^{nm}(0)e^{-\Gamma_{nm}(t)},
\end{equation}
where $\rho_{\mathcal{S}}^{nm}(0)=c_nc_m^*$.

The time dependence of the decoherence function depends on many variables, such as the system-environment coupling and the total system initial state. For many physical systems of interest, the system-environment interaction leads to a rapid decrease in the overlap between the environment states $\ket{E_n(t)}$ and $\ket{E_m(t)}$, $m\neq n$, which can be found to vanish after times that are long compared to a typical timescale $\tau_{\rm dec}$,
\begin{equation}
    \braket{E_n(t)}{E_m(t)}\to\delta_{nm}\hspace{0,5cm}\textrm{for $t\gg\tau_{\rm dec}$}.
\end{equation}
The timescale $\tau_{\rm dec}$ is called the \textbf{decoherence time}, usually defined by the condition $\Gamma_{nm}(\tau_{\rm dec})=1$. For times much larger than $\tau_{\rm dec}$, the reduced system density matrix approaches
\begin{equation} \label{Decoherence}
    \hat{\rho}_{\mathcal{S}}(t)\to\sum_n\abs{c_n}^2\ket{s_n}\bra{s_n}\hspace{0,5cm}\textrm{for $t\gg\tau_{\rm dec}$}.
\end{equation}
Physically, this means that superpositions of the states $\ket{s_n}$, which were initially present, are effectively "destroyed" in the sense that they are no longer observable for any measurement performed on the system $\mathcal{S}$. The coherent superposition has transitioned to an incoherent statistical mixture, and the reduced density matrix has become diagonal in the particular set of basis states $\ket{s_n}$. This environment induced transition is the phenomenon known as \emph{decoherence}.

In the following sections, we will discuss a specific canonical model that exhibits and illustrates environment-induced decoherence in a quantum system. To do so, we will first need to establish a formalism for describing open quantum systems.

\section{The Feynman-Vernon influence functional} \label{Sec:The-FV-IF}

Having introduced the ideas behind the phenomenon of decoherence, let us next present the formalism of the Feynman–Vernon influence functional in order to study the general dynamics of an open quantum system~\cite{Feynman1963,Feynman2010,Calzetta2008}. We begin with a brief review of the path integral formulation of quantum mechanics, followed by the study of open quantum systems using the influence functional approach. Lastly, we present an example of an open system linearly coupled with an environment, which will be very relevant to this work.

\subsection{Path integrals}

In the Schrödinger picture of quantum mechanics, the states evolve in time according to
\begin{equation} \label{Time-evolution-of-states}
    \ket{\psi(t)}=\hat{U}(t,t_0)\ket{\psi(t_0)},
\end{equation}
where the unitary time evolution operator from initial time $t_0$ to time $t$ reads
\begin{equation} \label{Time-evolution-operator}
    \hat{U}(t,t_0)=\textrm{T}\qty[\exp\qty(-i\int_{t_0}^t\dd t'\,\hat{H}(t'))]
\end{equation}
with $\hat{H}(t)$ being the Hamiltonian operator and T standing for time ordering.

For simplicity, let us consider a system described by a single degree of freedom $x$, which denotes the spectral family of the position operator $\hat{X}$, $\hat{X}\ket{x}=x\ket{x}$. The inner product $\psi(x)=\braket{x}{\psi}$ is what we call the wavefunction in the position representation. In terms of the wavefunction, the time evolution~\eqref{Time-evolution-of-states} reads
\begin{equation}
    \psi(x,t)=\braket{x}{\psi(t)}=\int_{-\infty}^\infty \dd x_0\,\mel{x}{\hat{U}(t,t_0)}{x_0}\braket{x_0}{\psi(t_0)},
\end{equation}
where we introduced a completeness relation. We can rewrite the last equation as
\begin{equation} \label{Wavefunction}
    \psi(x,t)=\int_{-\infty}^\infty \dd x_0\,K(x,t;x_0,t_0)\psi(x_0,t_0).
\end{equation}
The function $K(x,t;x_0,t_0)=\mel{x}{\hat{U}(t,t_0)}{x_0}$ is called the \textbf{propagator}, which is an amplitude to move from point $x_0$ at $t_0$ to point $x$ at $t$.

In the path integral formulation of quantum mechanics, the propagator is obtained from
\begin{equation} \label{Propagator-path-integral}
    K(x_f,t_f;x_0,t_0)=\int\displaylimits_{\substack{x(0)\,=\,x_0}}^{\substack{x(t_f)\,=\,x_f}}\mathcal{D}x\,e^{iS[x(t)]},
\end{equation}
where the integral over the measure $\mathcal{D}x$ denotes a sum over all paths $x(t)$ that lead from $x_0$ to $x_f$, and $S[x(t)]$ is the classical action (the time integral from $t_0$ to $t_f$ of the system's Lagrangian) computed for each path. One can arrive at~\eqref{Propagator-path-integral} by dividing the time interval $[t_0,t_f]$ into $N$ infinitesimal intervals of length $\delta t$, inserting a completeness relation for each intermediate point $x_i$, $i\in[0,N]$, into $\mel{x}{\hat{U}(t,t_0)}{x_0}$, computing all matrix elements by using the expression for the time evolution operator \eqref{Time-evolution-operator}, and finally taking the limit $N\to\infty$ ($\delta t\to0$) \cite{Schwartz2013,Srednicki2007,Peskin2007}. Alternatively, one can take Eq.~\eqref{Propagator-path-integral} as a time evolution postulate of quantum mechanics and show that the wavefunction~\eqref{Wavefunction} must obey Schrödinger's equation~\cite{Feynman2010}. Both formulations of quantum mechanics are equivalent.

The Feynman path integral formalism, as described by Eq.~\eqref{Propagator-path-integral}, remarkably tells us that the time evolution of the quantum state $\ket{\psi}$ is described by an integral over all paths from the beginning to the endpoint, weighted by the exponential of the \emph{classical} action. However, our description of the quantum system may not be given by a single ket in Hilbert space, but rather by a given class of states $\ket{\psi_i}$ for which our knowledge does not allow us to go beyond assigning a probability of occurrence $p_i$ to each member of this class. We then describe the system using a density matrix,
\begin{equation} \label{Density-matrix}
    \hat{\rho}=\sum_ip_i\ket{\psi_i}\bra{\psi_i}.
\end{equation}
We always have $\textrm{Tr}\,\hat{\rho}=1$, while in general $\textrm{Tr}\,\hat{\rho}^2\leq1$. Kets in the Hilbert space are represented by particular cases of density matrices with $\textrm{Tr}\,\hat{\rho}^2=1$ (also called \emph{pure states}, while states with $\textrm{Tr}\,\hat{\rho}^2<1$ are called \emph{mixed states}). It follows from Eqs.~\eqref{Time-evolution-of-states} and~\eqref{Density-matrix} that the time evolution of the density matrix is given by
\begin{equation}
    \hat{\rho}(t)=\hat{U}(t)\hat{\rho}(0)\hat{U}^\dagger(t),
\end{equation}
where we are taking the initial time instant as $t_0=0$ and denoting $\hat{U}(t)\equiv\hat{U}(t,0)$.

We can obtain a path integral representation for the time evolution of the density matrix by considering the matrix element $\rho(x,x',t)\equiv\mel{x}{\hat{\rho}(t)}{x'}$. By inserting completeness relations and using the path integral representation for the propagator~\eqref{Propagator-path-integral}, we arrive at\footnote{A word on notation, as it can become quite confusing from this point on: when we have variables like $x$ and $q$ in a path integral measure, like $\mathcal{D}x$ and $\mathcal{D}q$, or inside square brackets like in $S[x,q]$, we are referring to the functions $x(t)$, $q(t)$, and so on. Everywhere else, we are referring to the spatial points $x$ or $q$, which are often the endpoints of paths $x(t)$ and $q(t)$.}
\begin{align} \label{Path-integral-evolution-of-density-matrices}
    \rho(x,x',t)&=\mel{x}{\hat{U}(t)\hat{\rho}(0)\hat{U}^\dagger(t)}{x'} \nonumber \\
    &=\mel{x}{\hat{U}(t)\qty(\int_{-\infty}^\infty \dd x_0\ket{x_0}\bra{x_0})\hat{\rho}(0)\qty(\int_{-\infty}^\infty \dd x_0'\ket{x_0'}\bra{x_0'}) \hat{U}^\dagger(t)}{x'} \nonumber \\
    &=\int_{-\infty}^\infty \dd x_0\dd x_0'\,\mel{x}{\hat{U}(t)}{x_0}\mel{x_0'}{\hat{U}^\dagger(t)}{x'}\mel{x_0}{\hat{\rho}(0)}{x_0'} \nonumber \\
    &=\int_{-\infty}^\infty \dd x(0)\dd x'(0)\,\int\displaylimits_{\substack{x(0)\,=\,x_0 \\ x'(0)\,=\,x_0'}}^{\substack{x(t)\,=\,x \\ x'(t)\,=\,x'}}\mathcal{D}x\mathcal{D}x'\,e^{i\qty(S[x]-S[x'])}\rho(x(0),x'(0),0).
\end{align}
As we can see, the time evolution of the density matrix involves two histories of the system, rather than a single one, as in the time evolution of the wave function~\eqref{Wavefunction}.

\subsection{The influence functional}

Let us again consider an open quantum system $\mathcal{S}$ coupled with a quantum environment $\mathcal{E}$. While the dynamics of the combined (closed) total system $\mathcal{S}+\mathcal{E}$ are unitary, the state of the system $\mathcal{S}$, a subsystem of the combined one, will evolve according to its internal dynamics as well as to its interactions with the surroundings. These interactions lead to system-environment correlations such that the dynamics of the system $\mathcal{S}$ will not be unitary in general. The dynamics of the system $\mathcal{S}$ are called the \emph{reduced system dynamics}, and the system $\mathcal{S}$ itself is referred to as the \emph{reduced system}~\cite{Breuer2002}.

The action of the total system is usually a sum of three terms: one describing the system alone, another for the dynamics of the environment, and the interaction between them. Let us then consider a system $\mathcal{S}$ described by a set of variables $x=\qty{x_n}$ that interacts with an environment $\mathcal{E}$ described by variables $q=\qty{q_n}$. The classical action takes the form
\begin{equation}
    S[x,q]=S_{\mathcal{S}}[x]+S_{\mathcal{E}}[q]+S_{\mathcal{I}}[x,q],
\end{equation}
where the action $S_{\mathcal{I}}[x,q]$ describes the interaction between the system and the environment. The quantum state of the total system is described by the density matrix $\rho(xq,x'q',t)$ depending on both system and environment variables. It evolves unitarily under the total Hamiltonian, according to~\cite{Calzetta2008}
\begin{equation}
    \rho(xq,x'q',t)=\int_{-\infty}^\infty \dd x_0\dd q_0\int_{-\infty}^\infty \dd x_0'\dd q_0'\,\mathcal{J}(xq,x'q',t|x_0q_0,x_0'q_0',0)\,\rho(x_0q_0,x_0'q_0',0),
\end{equation}
where $\rho(xq,x'q',t)\equiv\mel{xq}{\hat{\rho}(t)}{x'q'}$, and $\mathcal{J}$ is seen as a time evolution kernel for the total combined system, given by
\begin{equation}
    \mathcal{J}(xq,x'q',t|x_0q_0,x_0'q_0',0)=\int\displaylimits_{\substack{x(0)\,=\,x_0 \\ x'(0)\,=\,x_0'}}^{\substack{x(t)\,=\,x \\ x'(t)\,=\,x'}}\mathcal{D}x\mathcal{D}x'\int\displaylimits_{\substack{q(0)\,=\,q_0 \\ q'(0)\,=\,q_0'}}^{\substack{q(t)\,=\,q \\ q'(t)\,=\,q'}}\mathcal{D}q\mathcal{D}q'\,e^{i\qty(S[x,q]-S[x',q'])}.
\end{equation}

However, when dealing with open quantum system dynamics, one is usually interested in computing expectation values of system observables while taking no interest in the environment dynamics. The expectation value of such observables is computed with the reduced density matrix $\hat{\rho}_{\mathcal{S}}$, which is obtained from the total density matrix by a partial trace over the environment variables,
\begin{equation}
    \rho_{\mathcal{S}}(x,x',t)=\int_{-\infty}^\infty \dd q\,\rho(xq,x'q,t).
\end{equation}

Assuming that at $t=0$ the system and environment are uncorrelated,
\begin{equation}
    \rho(x_0q_0,x_0'q_0',0)=\rho_{\mathcal{S}}(x_0,x_0',0)\rho_{\mathcal{E}}(q_0,q_0',0),
\end{equation}
we can rearrange the order of integration to write the reduced density matrix as
\begin{equation} \label{Evolution-of-reduced-density-matrix}
    \rho_{\mathcal{S}}(x,x',t)=\int_{-\infty}^\infty \dd x_0\dd x_0'\,\mathcal{J}_{\mathcal{S}}(x,x',t|x_0,x_0',0)\rho_{\mathcal{S}}(x_0,x_0',0),
\end{equation}
where the time evolution kernel for the reduced system reads
\begin{equation}
    \mathcal{J}_{\mathcal{S}}(x,x',t|x_0,x_0',0)\equiv\int\displaylimits_{\substack{x(0)\,=\,x_0 \\ x'(0)\,=\,x_0'}}^{\substack{x(t)\,=\,x \\ x'(t)\,=\,x'}}\mathcal{D}x\mathcal{D}x'\,e^{i\qty(S_{\mathcal{S}}[x]-S_{\mathcal{S}}[x'])}\mathcal{F}[x,x'].
\end{equation}
The functional $\mathcal{F}[x,x']$ is called the \textbf{Feynman-Vernon influence functional}, and it is given by
\begin{align} \label{Influence-functional-1}
    \mathcal{F}[x,x']&=e^{iS_{\rm IF}[x,x',t]} \nonumber \\
    &=\int_{-\infty}^\infty \dd q\dd q_0\dd q_0'\,\rho_{\mathcal{E}}(q_0,q_0',0) \nonumber \\
    &\hspace{1cm}\times\int\displaylimits_{\substack{q(0)\,=\,q_0 \\ q'(0)\,=\,q_0'}}^{q(t)\,=\,q'(t)\,=\,q}\mathcal{D}q\mathcal{D}q'\,e^{i\qty(S_{\mathcal{E}}[q]+S_{\mathcal{I}}[x,q]-S_{\mathcal{E}}[q']-S_{\mathcal{I}}[x',q'])}.
\end{align}
Here, $S_{\rm IF}$ is called the \textbf{influence action}, and it encodes all influence from the environment on the system. Its presence in the time evolution~\eqref{Evolution-of-reduced-density-matrix} induces an interaction between the two histories $x$ and $x'$, being responsible for the non-unitary evolution.

We can list some general properties of the influence functional~\cite{Feynman1963,Feynman2010,Calzetta2008}:
\begin{enumerate}
    \item In terms of the time evolution operators $\hat{\mathcal{U}}(t)$ and $\hat{\mathcal{U}}'(t)$ for $S_{\mathcal{E}}[q]+S_{\mathcal{I}}[x,q]$ and $S_{\mathcal{E}}[q]+S_{\mathcal{I}}[x',q]$, respectively, the influence functional can be expressed as
    \begin{align} \label{Basis-independent-influence-functional}
        \mathcal{F}[x,x']&=\int_{-\infty}^\infty \dd q\dd q_0\dd q_0'\,\rho_{\mathcal{E}}(q_0,q_0',0)\,\mel{q}{\hat{\mathcal{U}}(t)}{q_0}\mel{q_0'}{\hat{\mathcal{U}}'^\dagger(t)}{q} \nonumber \\
        &=\int_{-\infty}^\infty \dd q\dd q_0\dd q_0'\,\mel{q_0'}{\hat{\mathcal{U}}'^\dagger(t)}{q}\mel{q}{\hat{\mathcal{U}}(t)}{q_0}\mel{q_0}{\hat{\rho}_{\mathcal{E}}(0)}{q_0'} \nonumber \\
        &=\textrm{Tr}_{\mathcal{E}}\qty{\hat{\mathcal{U}}(t)\hat{\rho}_{\mathcal{E}}(0)\hat{\mathcal{U}}'^\dagger(t)}=\expval{\hat{\mathcal{U}}'^\dagger(t)\hat{\mathcal{U}}(t)}_\mathcal{E},
    \end{align}
    where $\expval{\cdot}_\mathcal{E}$ denotes the average with respect to the environment initial state $\hat{\rho}_{\mathcal{E}}(0)$.
    \item The influence functional satisfies
    \begin{equation}
        \mathcal{F}[x,x']=\mathcal{F}^*[x',x],
    \end{equation}
    which follows immediately from Eq.~\eqref{Basis-independent-influence-functional}.
    \item The influence action satisfies
    \begin{equation}
        S_{\rm IF}[x,x,t]=0,
    \end{equation}
    which can be easily seen from Eq.~\eqref{Basis-independent-influence-functional} by recalling that $\textrm{Tr}_{\mathcal{E}}\,\hat{\rho}_{\mathcal{E}}(0)=1$.
    \item If a number of statistically and dynamically independent environments act on the system at the same time, and if $\mathcal{F}_k$ denotes the influence functional of the $k$th environment alone, the total influence is given by the product of the individual $\mathcal{F}_k$,
    \begin{equation} \label{Independent-environments}
        \mathcal{F}=\prod_k\mathcal{F}_k.
    \end{equation}
\end{enumerate}

The path integral representation~\eqref{Influence-functional-1} involves two histories $q$ and $q'$ that may be described as an integral over a single history defined on a closed time path (CTP) due to the boundary condition $q(t)=q'(t)$. Throughout the remainder of this section, we shall follow Ref.~\cite{Calzetta2008} and denote $x(t)=x^1(t)$, $x'(t)=x^2(t)$ such that $x^a$, $a=1,2$, can be thought of as a single doublet field defined on a single time path. We also define a metric tensor $c_{ab}=\textrm{diag}\qty(1,-1)$, which, together with its inverse $c^{ab}=\textrm{diag}\qty(1,-1)$, may be used to raise and/or lower indices, as in $x_1=c_{1a}x^a=c_{11}x^1=x^1=x$ and $x_2=c_{2a}x^a=c_{22}x^2=-x^2=-x'$. In this CTP notation, the kinetic terms in the system Lagrangian, for instance, will be written as $c_{ab}\dot{x}^a\dot{x}^b=\dot{x}_a\dot{x}^a=\dot{x}^2-\dot{x}'^2$. We refer to the \emph{CTP action} $S[x^a]\equiv S[x]-S[x']$ when considering the contributions from both branches of the closed time path. In CTP notation, the influence functional~\eqref{Influence-functional-1} is written as
\begin{equation} \label{Influence-functional-2}
\begin{split}
    \mathcal{F}[x^a]&=\int_{-\infty}^\infty \dd q^1(t)\dd q^1(0)\dd q^2(0)\,\rho_{\mathcal{E}}(q^1(0),q^2(0),0) \\
    &\hspace{1cm}\times\int\displaylimits_{q^1(t)\,=\,q^2(t)}\mathcal{D}q^a\,e^{i\qty(S_{\mathcal{E}}[q^a]+S_{\mathcal{I}}[x^a,q^a])}.
\end{split}
\end{equation}

Having discussed the general properties of the influence functional, let us now consider a specific example of an environment that is linearly coupled to the system.

\subsection{Linear coupling model} \label{Subsec:linear-coupling-model}

The linear coupling model is described by the assumptions that the environmental action is quadratic in the $q$ variables, the initial environmental density matrix is Gaussian, and the interaction term is bilinear (linear in the system and environmental variables). This is the case for the model of quantum Brownian motion, for instance, in which the environmental action describes harmonic oscillators, and it will also be the case for the problem of a quantum particle in a bath of gravitons, as we will see in Chapter~\ref{chap:quantum-system}.

Let us then work under all these assumptions and take an interaction of the form\footnote{A more precise notation would be $S_{\mathcal{I}}[x,q,t]=\int_0^t \dd t'\,x^a(t')Q_a[q(t')]$.}
\begin{equation} \label{Linear-bath-action}
    S_{\mathcal{I}}[x,q]=\int \dd t\,x^a(t)Q_a[q(t)],
\end{equation}
where the $Q$'s are linear combinations of the $q$'s. In that case, the influence functional~\eqref{Influence-functional-1} becomes a functional Fourier transform of a Gaussian functional of histories $Q(t)$ and $Q'(t)$. Since the Fourier transform of a Gaussian is another Gaussian, we conclude that, under all these assumptions, the influence action must also be quadratic in $x$ and $x'$~\cite{Calzetta2008}. Therefore, we write
\begin{equation} \label{Linear-coupling-influence-action-1}
    S_{\textrm{IF}}[x^a]=\frac{1}{2}\int \dd t\dd t'\,x^a(t)G_{ab}(t,t')x^b(t'),
\end{equation}
where
\begin{equation}
    G_{ab}(t,t')=-i\frac{\delta^2}{\delta x^a(t)\delta x^b(t')}\eval{e^{iS_{\rm IF}[x^a]}}_{x_a\,=\,0},
\end{equation}
with the $\delta$ notation denoting functional derivatives. A direct variation from Eq.~\eqref{Influence-functional-2} with $S_{\mathcal{I}}$ given by Eq.~\eqref{Linear-bath-action} yields
\begin{align}
    &\frac{\delta^2}{\delta x^a(t)\delta x^b(t')}\eval{e^{iS_{\rm IF}[x^a]}}_{x_a\,=\,0}=-\int_{-\infty}^\infty \dd q^1(t_f)\dd q^1(0)\dd q^2(0) \nonumber \\
    &\hspace{2cm}\times\int\displaylimits_{q^1(t_f)\,=\,q^2(t_f)}\mathcal{D}q^a\,e^{iS_{\mathcal{E}}[q^a]}Q_a(t)Q_b(t')\,\rho_{\mathcal{E}}(q^1(0),q^2(0),0).
\end{align}
Explicitly,
\begin{subequations} \label{CTP-Green-functions}
\begin{equation}
\begin{split}
    G_{11}(\tau,\tau')&=i\int_{-\infty}^\infty \dd q\dd q(0)\dd q'(0)\,\rho_{\mathcal{E}}(q(0),q'(0),0) \\
    &\hspace{1cm}\times\int\displaylimits_{q(t)\,=\,q'(t)\,=\,q}\mathcal{D}q\mathcal{D}q'\,e^{i\qty(S_{\mathcal{E}}[q]-S_{\mathcal{E}}[q'])}Q(\tau)Q(\tau'),
\end{split}
\end{equation}
\begin{equation}
\begin{split}
    G_{12}(\tau,\tau')&=-i\int_{-\infty}^\infty \dd q\dd q(0)\dd q'(0)\,\rho_{\mathcal{E}}(q(0),q'(0),0) \\
    &\hspace{1cm}\times\int\displaylimits_{q(t)\,=\,q'(t)\,=\,q}\mathcal{D}q\mathcal{D}q'\,e^{i\qty(S_{\mathcal{E}}[q]-S_{\mathcal{E}}[q'])}Q(\tau)Q'(\tau'),
\end{split}
\end{equation}
\begin{equation}
\begin{split}
    G_{21}(\tau,\tau')&=-i\int_{-\infty}^\infty \dd q\dd q(0)\dd q'(0)\,\rho_{\mathcal{E}}(q(0),q'(0),0) \\
    &\hspace{1cm}\times\int\displaylimits_{q(t)\,=\,q'(t)\,=\,q}\mathcal{D}q\mathcal{D}q'\,e^{i\qty(S_{\mathcal{E}}[q]-S_{\mathcal{E}}[q'])}Q'(\tau)Q(\tau'),
\end{split}
\end{equation}
\begin{equation}
\begin{split}
    G_{22}(\tau,\tau')&=i\int_{-\infty}^\infty \dd q\dd q(0)\dd q'(0)\,\rho_{\mathcal{E}}(q(0),q'(0),0) \\
    &\hspace{1cm}\times\int\displaylimits_{q(t)\,=\,q'(t)\,=\,q}\mathcal{D}q\mathcal{D}q'\,e^{i\qty(S_{\mathcal{E}}[q]-S_{\mathcal{E}}[q'])}Q'(\tau)Q'(\tau').
\end{split}
\end{equation}
\end{subequations}

In order to compute these in canonical form, let us use the fact that the propagator~\eqref{Propagator-path-integral} satisfies~\cite{Feynman2010}
\begin{equation}
    K(x_b,t_b;x_a,t_a)=\int_{-\infty}^\infty \dd x_c\,K(x_b,t_b;x_c,t_c)K(x_c,t_c;x_a,t_a),
\end{equation}
where $x_c=x(t_c)$ and $t_a<t_c<t_b$. Then, by assuming $\tau>\tau'$, we may write
\begin{align}
    &\int\displaylimits_{q(t)\,=\,q'(t)}\mathcal{D}q\mathcal{D}q'\,e^{i\qty(S_{\mathcal{E}}[q]-S_{\mathcal{E}}[q'])}Q(\tau)Q(\tau') \nonumber \\
    &\hspace{0,5cm}=\int_{-\infty}^\infty \dd q(\tau')\dd q(\tau)\,\int\displaylimits_{0\,\leq\,t'\,\leq\,\tau'}\mathcal{D}q(t')\,e^{iS_{\mathcal{E}}[q(t')]}Q(\tau')\int\displaylimits_{\tau'\,\leq\,t'\,\leq\,\tau}\mathcal{D}q(t')\,e^{iS_{\mathcal{E}}[q(t')]}Q(\tau) \nonumber \\
    &\hspace{1cm}\times\int\displaylimits_{\tau\,\leq\,t'\,\leq\,t}\mathcal{D}q(t')\,e^{iS_{\mathcal{E}}[q(t')]}\int\displaylimits_{0\,\leq\,t'\,\leq\,t}\mathcal{D}q'(t')e^{-iS_{\mathcal{E}}[q'(t')]},
\end{align}
where $q(t')$ and $q'(t')$ denote the integration (path) variables, and the subscripts on the right-hand side of the equation indicate the time interval in which the path integral is being computed. We may now identify each path integral as a matrix element for the environment free time evolution operator $\hat{U}_{\mathcal{E}}(t)=e^{-i\hat{H}_{\mathcal{E}}t}$ and write
\allowdisplaybreaks
\begin{align}
    -iG_{11}(\tau,\tau')&=\int_{-\infty}^\infty \dd q(t)\dd q(0)\dd q'(0)\dd q(\tau')\dd q(\tau) \nonumber \\
    &\hspace{1cm}\times\mel{q(t)}{\hat{U}_{\mathcal{E}}(t,\tau)}{q(\tau)}\,Q(\tau)\,\mel{q(\tau)}{\hat{U}_{\mathcal{E}}(\tau,\tau')}{q(\tau')}\,Q(\tau') \nonumber \\
    &\hspace{1cm}\times\mel{q(\tau')}{\hat{U}_{\mathcal{E}}(\tau',0)}{q(0)}\mel{q(0)}{\rho_{\mathcal{E}}(0)}{q'(0)}\mel{q'(0)}{\hat{U}_{\mathcal{E}}(0,t)}{q(t)} \nonumber \\
    &=\int_{-\infty}^\infty \dd q(t)\dd q(\tau')\dd q(\tau)\,\mel{q(t)}{\hat{U}_{\mathcal{E}}(t,\tau)\hat{Q}}{q(\tau)} \nonumber \\
    &\hspace{1cm}\times\mel{q(\tau)}{\hat{U}_{\mathcal{E}}(\tau,\tau')\hat{Q}}{q(\tau')}\mel{q(\tau')}{\hat{U}_{\mathcal{E}}(\tau',0)\rho_{\mathcal{E}}(0)\hat{U}_{\mathcal{E}}(0,t)}{q(t)} \nonumber \\
    &=\int_{-\infty}^\infty \dd q(t)\,\mel{q(t)}{\hat{U}_{\mathcal{E}}(t,\tau)\hat{Q}\hat{U}_{\mathcal{E}}(\tau,\tau')\hat{Q}\hat{U}_{\mathcal{E}}(\tau',0)\rho_{\mathcal{E}}(0)\hat{U}_{\mathcal{E}}(0,t)}{q(t)} \nonumber \\
    &=\textrm{Tr}_{\mathcal{E}}\qty{\hat{U}_{\mathcal{E}}(t,\tau)\hat{Q}\hat{U}_{\mathcal{E}}(\tau,\tau')\hat{Q}\hat{U}_{\mathcal{E}}(\tau',0)\rho_{\mathcal{E}}(0)\hat{U}_{\mathcal{E}}(0,t)} \nonumber \\
    &=\textrm{Tr}_{\mathcal{E}}\qty{\hat{U}_{\mathcal{E}}(0,t)\hat{U}_{\mathcal{E}}(t,\tau)\hat{Q}\hat{U}_{\mathcal{E}}(\tau,\tau')\hat{Q}\hat{U}_{\mathcal{E}}(\tau',0)\rho_{\mathcal{E}}(0)} \nonumber \\
    &=\textrm{Tr}_{\mathcal{E}}\qty{\hat{U}_{\mathcal{E}}(0,\tau)\hat{Q}\hat{U}_{\mathcal{E}}(\tau,0)\hat{U}_{\mathcal{E}}(0,\tau')\hat{Q}\hat{U}_{\mathcal{E}}(\tau',0)\rho_{\mathcal{E}}(0)} \nonumber \\
    &=\textrm{Tr}_{\mathcal{E}}\qty{\hat{U}_{\mathcal{E}}^\dagger(\tau)\hat{Q}\hat{U}_{\mathcal{E}}(\tau)\hat{U}_{\mathcal{E}}^\dagger(\tau')\hat{Q}\hat{U}_{\mathcal{E}}(\tau')\rho_{\mathcal{E}}(0)} \nonumber \\
    &=\expval{\hat{Q}(\tau)\hat{Q}(\tau')}_{\mathcal{E}}, \hspace{1cm}(\tau>\tau')
\end{align}
where $\hat{Q}(\tau)$ is an operator in the interaction picture, $\hat{Q}(\tau)=\hat{U}_{\mathcal{E}}^\dagger(\tau)\hat{Q}\hat{U}_{\mathcal{E}}(\tau)$. In the expressions above, we used the composition rule $\hat{U}_{\mathcal{E}}(t,t')\hat{U}_{\mathcal{E}}(t',t_0)=\hat{U}_{\mathcal{E}}(t,t_0)$. It is interesting to note that if we had not specified the temporal relation between $\tau$ and $\tau'$, the path integral would have automatically set the latest time to the left. This means that we can write the general result
\begin{equation}
    G_{11}(\tau,\tau')=i\expval{\textrm{T}\qty[\hat{Q}(\tau)\hat{Q}(\tau')]}_{\mathcal{E}}.
\end{equation}

By completely analogous procedures, we can write the other components~\eqref{CTP-Green-functions} in canonical form and obtain
\begin{equation}
    \mathbb{G}(t,t')=i\mqty(\expval{\textrm{T}\qty[\hat{Q}(t)\hat{Q}(t')]}_{\mathcal{E}} & -\expval{\hat{Q}(t')\hat{Q}(t)}_{\mathcal{E}} \\ -\expval{\hat{Q}(t)\hat{Q}(t')}_{\mathcal{E}} & \expval{\Tilde{\textrm{T}}\qty[\hat{Q}(t)\hat{Q}(t')]}_{\mathcal{E}}),
\end{equation}
where $\Tilde{\textrm{T}}$ denotes \emph{anti-time} ordering, which takes any product of time dependent operators and changes the order such that later times are on the \emph{right}. Also $\mathbb{G}$ denotes the matrix whose elements are the $G_{ab}$'s.

The influence action~\eqref{Linear-coupling-influence-action-1} can then be written as
\begin{align}
    S_{\rm IF}[x,x']&=\frac{i}{2}\int \dd t\dd t'\,\left\{ \expval{\textrm{T}\qty[\hat{Q}(t)\hat{Q}(t')]}_{\mathcal{E}}x(t)x(t')-\expval{\hat{Q}(t')\hat{Q}(t)}_{\mathcal{E}}x(t)x'(t')\right. \nonumber \\
    &\hspace{0.5cm}\left.-\expval{\hat{Q}(t)\hat{Q}(t')}_{\mathcal{E}}x'(t)x(t')+\expval{\Tilde{\textrm{T}}\qty[\hat{Q}(t)\hat{Q}(t')]}_{\mathcal{E}}x'(t)x'(t')\right\} .
\end{align}
We can write it in a more compact form by introducing the variables $u=x-x'$ and $X=(x+x')/2$. Additionally, we use the property $\theta(-x)=1-\theta(x)$ of the step function to write
\begin{subequations}
\begin{equation}
    \textrm{T}\qty[\hat{Q}(t)\hat{Q}(t')]=\theta(t-t')\comm{\hat{Q}(t)}{\hat{Q}(t')}+\hat{Q}(t')\hat{Q}(t),
\end{equation}
and
\begin{equation}
    \Tilde{\textrm{T}}\qty[\hat{Q}(t)\hat{Q}(t')]=-\theta(t-t')\comm{\hat{Q}(t)}{\hat{Q}(t')}+\hat{Q}(t)\hat{Q}(t'),
\end{equation}
\end{subequations}
with $\comm{\cdot}{\cdot}$ denoting the commutator between operators. Then, after a fair amount of algebraic manipulations, we arrive at
\begin{equation} \label{Linear-coupling-influence-action-2}
    S_{\rm IF}[x,x']=\int \dd t\dd t'\,\qty[u(t)D(t,t')X(t')+\frac{i}{2}u(t)N(t,t')u(t')],
\end{equation}
where we have defined the \textbf{dissipation} and \textbf{noise kernels}
\begin{subequations} \label{Dissipation-and-noise-kernels}
    \begin{equation} \label{Dissipation-kernel}
        D(t,t')=i\expval{\comm{\hat{Q}(t)}{\hat{Q}(t')}}_{\mathcal{E}}\theta(t-t'),
    \end{equation}
    \begin{equation} \label{Noise-kernel}
        N(t,t')=\frac{1}{2}\expval{\acomm{\hat{Q}(t)}{\hat{Q}(t')}}_{\mathcal{E}},
    \end{equation}
\end{subequations}
respectively. Here $\acomm{\cdot}{\cdot}$ denotes the anti-commutator between operators.

The physical interpretation of these kernels, as well as the reason for their nomenclature, will become more evident once we derive a master equation for the reduced density matrix in the next section.

\section{The master equation} \label{Sec:Master-equation}

For closed quantum systems, the time evolution can be described entirely in terms of the density matrix by the \emph{Liouville-von Neumann equation},
\begin{equation}
    \dv{t}\hat{\rho}(t)=-i\comm{\hat{H}}{\hat{\rho}(t)},
\end{equation}
where $\hat{H}$ is the total Hamiltonian of the system. This time evolution is unitary, and the same cannot be said when the system is open. In that case, the Liouville-von Neumann equation is replaced by the so-called master equations. For most cases of interest, the time evolution of the reduced density matrix for a system coupled with an environment takes the form~\cite{Schlosshauer2008,Breuer2002}
\begin{equation} \label{General-master-equation}
    \dv{t}\hat{\rho}_{\mathcal{S}}(t)=-i\comm{\hat{H}'_{\mathcal{S}}}{\hat{\rho}_{\mathcal{S}}(t)}+\hat{\mathcal{D}}\qty[\hat{\rho}_{\mathcal{S}}(t)].
\end{equation}
Here $\hat{H}'_{\mathcal{S}}$ is the part of the Hamiltonian that acts only on the system of interest. In general, $\hat{H}'_{\mathcal{S}}\neq\hat{H}_{\mathcal{S}}$, since the interaction with the environment can lead to a renormalization of the energy levels of the system, an effect usually referred to as the \emph{Lamb-shift} contribution~\cite{Schlosshauer2008}. Additionally, the presence of the environment introduces the second term on the right-hand side of Eq.~\eqref{General-master-equation}, which is responsible for the non-unitary time evolution of the reduced system. The super-operator\footnote{A super-operator refers to an operator that acts on another operator.} $\hat{\mathcal{D}}$ is sometimes referred to as the dissipator, and it is responsible for the effects of dissipation and decoherence.

A master equation like Eq.~\eqref{General-master-equation} is usually obtained from the Liouville-von Neumann equation for the total combined system by taking the partial trace over the environment variables, assuming an initial product state (as we have done in the previous section), and further imposing some other approximations, like the Born and Markov approximations\footnote{The Born approximation assumes that the interaction between the system and the environment is sufficiently weak, and the environment is sufficiently large in comparison with the size of the system, such that the total state remains approximately a product state throughout the time evolution. The Markov approximation assumes that the environmental correlation functions drop to zero much faster than the characteristic timescale over which the reduced density matrix of the system changes appreciably. It basically neglects the memory effects of the environment and usually transforms the integro-differential master equation into a much simpler differential one.} (see Refs.~\cite{Schlosshauer2008,Breuer2002} for the details). Here we will simply show how we can obtain a master equation for the reduced density matrix from the influence functional approach.

\subsection{Deriving the master equation}

We begin by writing an expression for $\rho_{\mathcal{S}}(x_f,x'_f,t+\Delta t)$ in terms of $\rho_{\mathcal{S}}(x,x',t)$,
\begin{equation} \label{DeltarhoS}
    \rho_{\mathcal{S}}(x_f,x'_f,t+\Delta t)=\int_{-\infty}^\infty \dd x\dd x'\,\mathcal{J}_{\mathcal{S}}(x_f,x'_f,t+\Delta t|x,x',t)\rho_{\mathcal{S}}(x,x',t),
\end{equation}
where
\begin{equation}
    \mathcal{J}_{\mathcal{S}}(x_f,x'_f,t+\Delta t|x,x',t)=\int\displaylimits_{\substack{x(t)\,=\,x \\ x'(t)\,=\,x'}}^{\substack{x\qty(t+\Delta t)\,=\,x_f \\ x'(t+\Delta t)\,=\,x_f'}}\mathcal{D}x\mathcal{D}x'\,e^{i\qty(\Delta S_{\mathcal{S}}[x]-\Delta S_{\mathcal{S}}[x'])}e^{i\Delta S_{\rm IF}[x,x']},
\end{equation}
with $\Delta S_{\mathcal{S}}[x]\equiv S_{\mathcal{S}}[x,t+\Delta t]-S_{\mathcal{S}}[x,t]$ and the same holds for $\Delta S_{\rm IF}[x,x']$.

Since our goal is to obtain a first order differential equation for the time evolution of the reduced density matrix, we want to keep terms up to first order in $\Delta t$. Now, $\Delta S_{\rm IF}[x,x']\propto\Delta t$ and thus
\begin{align} \label{DeltaJ}
    \mathcal{J}_{\mathcal{S}}(x_f,x'_f,t+\Delta t|x,x',t)&\simeq K_{\mathcal{S}}(x_f,t+\Delta t|x,t)K_{\mathcal{S}}^*(x_f',t+\Delta t|x',t) \nonumber \\
    &+i\int\mathcal{D}x\mathcal{D}x'\,e^{i\qty(\Delta S_{\mathcal{S}}[x]-\Delta S_{\mathcal{S}}[x'])}\Delta S_{\rm IF}[x,x'],
\end{align}
where
\begin{align} \label{K_S}
    K_{\mathcal{S}}(x_f,t+\Delta t|x,t)&=\int\displaylimits_{x(t)\,=\,x}^{x\qty(t+\Delta t)\,=\,x_f}\mathcal{D}x\,e^{i\Delta S_{\mathcal{S}}[x]}=\mel{x_f}{\hat{U}_{\mathcal{S}}(t+\Delta t,t)}{x} \nonumber \\
    &\simeq\delta(x_f-x)-i\Delta t\mel{x_f}{\hat{H}_{\mathcal{S}}}{x}.
\end{align}

If we work with the linear coupling model, the influence action is given by Eq.~\eqref{Linear-coupling-influence-action-2}, and we find
\begin{equation} \label{DeltaS_IF}
    \Delta S_{\rm IF}[x,x']=\Delta t\,u(t)\int_0^t\dd t'\,\qty[D(t,t')X(t')+iN(t,t')u(t')],
\end{equation}
where we used the fact that $D(t,t')\propto\theta(t-t')$, i.e., $D(t,t')$ is explicitly causal.

By plugging Eqs.~\eqref{K_S} and~\eqref{DeltaS_IF} into Eq.~\eqref{DeltarhoS}, using Eq.~\eqref{Evolution-of-reduced-density-matrix} in order to write the time evolution in terms of the initial system density matrix, and then taking the limit where $\Delta t\to0$ while letting $x_f\to x$ and $x_f'\to x'$, we finally arrive at
\begin{align}
    &\pdv{t}\rho_{\mathcal{S}}(x,x',t)=-i\mel{x}{\comm{\hat{H}_{\mathcal{S}}}{\hat{\rho}_{\mathcal{S}}(t)}}{x'} \nonumber \\
    &-(x-x')\int_0^t\dd t'\,\qty[N(t,t')\qty(\mathbf{X}-\mathbf{X}')(x,x',t')-\frac{i}{2}D(t,t')\qty(\mathbf{X}+\mathbf{X}')(x,x',t')],
\end{align}
with
\begin{equation}
    \mathbf{X}(x,x',t')=\int_{-\infty}^\infty\dd x_0\dd x_0'\,\int\mathcal{D}x\mathcal{D}x'\,e^{i\qty(S_{\mathcal{S}}[x]-S_{\mathcal{S}}[x']+S_{\rm IF}[x,x'])}\rho_{\mathcal{S}}(x_0,x_0',0)x(t').
\end{equation}

Note that the evolution of $\mathbf{X}(x,x',t')$ also depends on the interaction with the environment through the influence action. However, for sufficiently weak interactions, we can take advantage of the fact that both the noise and dissipation kernels are, in general, of second order in the operator $\hat{Q}$ that couples with the system. Therefore, we can neglect the influence action in the exponential since this would lead to higher order contributions and write~\cite{Calzetta2008}
\begin{align} \label{X-weak-coupling}
    \mathbf{X}(x,x',t')&\simeq\int_{-\infty}^\infty\dd x_0\dd x_0'\,\int\mathcal{D}x\mathcal{D}x'\,e^{i\qty(S_{\mathcal{S}}[x]-S_{\mathcal{S}}[x'])}\rho_{\mathcal{S}}(x_0,x_0',0)x(t') \nonumber \\
    &=\int_{-\infty}^\infty\dd x_0\dd x_0'\dd x(t')\mel{x(t')}{\hat{U}_{\mathcal{S}}(t',0)}{x_0}\mel{x}{\hat{U}_{\mathcal{S}}(t,t')\hat{X}}{x(t')} \nonumber \\
    &\hspace{1cm}\times\mel{x_0}{\hat{U}_{\mathcal{S}}(0,t)}{x'}\mel{x_0}{\hat{\rho}_{\mathcal{S}}(0)}{x_0'} \nonumber \\
    &=\mel{x}{e^{-i\hat{H}_{\mathcal{S}}(t-t')}\hat{X}e^{-i\hat{H}_{\mathcal{S}}t'}\hat{\rho}_{\mathcal{S}}(0)e^{i\hat{H}_{\mathcal{S}}t}}{x'} \nonumber \\
    &=\mel{x}{\hat{X}(-\tau)\hat{\rho}_{\mathcal{S}}(t)}{x'},
\end{align}
where $\tau=t-t'$ and $\hat{X}(t)=e^{i\hat{H}_{\mathcal{S}}t}\hat{X}e^{-i\hat{H}_{\mathcal{S}}t}$ is the interaction picture position operator. Similarly, one can show that $\mathbf{X}'(x,x',t')$ is the $(x,x')$ matrix element of the operator $\hat{\rho}_{\mathcal{S}}(t)\hat{X}(-\tau)$.

At last, putting everything together leads to the master equation
\begin{align} \label{Master-equation}
    \dv{t}\hat{\rho}_{\mathcal{S}}(t)=-i\comm{\hat{H}_{\mathcal{S}}}{\hat{\rho}_{\mathcal{S}}(t)}&-\int_0^t\dd\tau\,\left\{ N(t,t-\tau)\comm{\hat{X}}{\comm{\hat{X}(-\tau)}{\hat{\rho}_{\mathcal{S}}(t)}}\right. \nonumber \\
    &\left. -\frac{i}{2}D(t,t-\tau)\comm{\hat{X}}{\acomm{\hat{X}(-\tau)}{\hat{\rho}_{\mathcal{S}}(t)}}\right\} .
\end{align}
This is of the form presented in Eq.~\eqref{General-master-equation}, with the dissipation and noise kernels encoding the non-unitary time evolution. To further explore the physical meaning behind these terms, let us apply the master equation for the case of a well-known model of the system of interest, namely a harmonic oscillator.

\subsection{Dissipation, diffusion and decoherence}

Suppose now that our system of interest is a harmonic oscillator with mass $M$ and frequency $\Omega$, such that its free Hamiltonian takes the form
\begin{equation}
    \hat{H}_{\mathcal{S}}=\frac{\hat{P}^2}{2M}+\frac{1}{2}M\Omega^2\hat{X}^2,
\end{equation}
with $\hat{X}$ and $\hat{P}$ being the (Schrödinger picture) position and momentum operators of the system satisfying the canonical commutation relation $[\hat{X},\hat{P}]=i$. The interaction picture position operator (which is the same as the Heisenberg picture operator when the system evolves only according to its self-Hamiltonian $\hat{H}_{\mathcal{S}}$) reads~\cite{Sakurai2020}
\begin{equation}
    \hat{X}(t)=\hat{X}\cos(\Omega t)+\frac{1}{M\Omega}\hat{P}\sin(\Omega t).
\end{equation}

Using this result in the master equation~\eqref{Master-equation} yields
\begin{align} \label{Master-equation-harmonic-oscillator}
    \dv{t}\hat{\rho}_{\mathcal{S}}(t)&=-i\comm{\hat{H}_{\mathcal{S}}+\frac{1}{2}M\delta\Omega^2(t)\hat{X}^2}{\hat{\rho}_{\mathcal{S}}(t)}-i\gamma(t)\comm{\hat{X}}{\acomm{\hat{P}}{\hat{\rho}_{\mathcal{S}}(t)}} \nonumber \\
    &-\sigma^2(t)\comm{\hat{X}}{\comm{\hat{X}}{\hat{\rho}_{\mathcal{S}}(t)}}-\Sigma^2(t)\comm{\hat{X}}{\comm{\hat{P}}{\hat{\rho}_{\mathcal{S}}(t)}},
\end{align}
where
\begin{subequations} \label{Coefficients-master-equation}
\begin{align}
    \delta\Omega^2(t)&=-\frac{1}{M}\int_0^t\dd\tau\,D(t,t-\tau)\cos(\Omega\tau), \label{Renormalization-coefficient} \\
    \gamma(t)&=\frac{1}{2M\Omega}\int_0^t\dd\tau\,D(t,t-\tau)\sin(\Omega\tau), \label{Dissipation-coefficient} \\
    \sigma^2(t)&=\int_0^t\dd\tau\,N(t,t-\tau)\cos(\Omega\tau), \label{Normal-diffusion-coefficient} \\
    \Sigma^2(t)&=-\frac{1}{M\Omega}\int_0^t\dd\tau\,N(t,t-\tau)\sin(\Omega\tau). \label{Anomalous-diffusion-coefficient}
\end{align}
\end{subequations}
In the position representation, the master equation~\eqref{Master-equation-harmonic-oscillator} reads
\begin{align} \label{Master-equation-position-representation}
    \pdv{t}\rho_{\mathcal{S}}(x,x',t)=&\left[ \frac{i}{2M}\qty(\pdv[2]{x}-\pdv[2]{{x'}})-\frac{i}{2}M\qty(\Omega^2+\delta\Omega^2(t))\qty(x^2-x'^2)\right. \nonumber \\
    &-\gamma(t)\qty(x-x')\qty(\pdv{x}-\pdv{{x'}})-\sigma^2(t)\qty(x-x')^2 \nonumber \\
    &\left. +i\Sigma^2(t)\qty(x-x')\qty(\pdv{x}+\pdv{{x'}})\right] \rho_{\mathcal{S}}(x,x',t).
\end{align}

It is worth mentioning that an exact master equation for quantum Brownian motion can be obtained without imposing the weak coupling approximation, as we have done in Eq.~\eqref{X-weak-coupling}. In that case, the expressions for the coefficients~\eqref{Coefficients-master-equation} become much more complicated, although the master equation maintains the same form~\eqref{Master-equation-harmonic-oscillator} (see Refs.~\cite{Hu_1992,Calzetta2008}). Remarkably, the exact master equation is found to be local in time even though it exhibits non-Markovian effects.

Let us now discuss the physical interpretation of each of the coefficients in Eq.~\eqref{Coefficients-master-equation}. First, let us note that the coefficient $\delta\Omega^2(t)$, given by Eq.~\eqref{Renormalization-coefficient}, amounts to a renormalization of the natural frequency of the system (the Lamb-shift contribution). The unitary part of the time evolution is then that of a quantum harmonic oscillator with physical frequency $\qty(\Omega^2+\delta\Omega^2(t))^{1/2}$.

The coefficient $\gamma(t)$, given by Eq.~\eqref{Dissipation-coefficient}, describes dissipation through momentum damping. To see this, let us consider the time evolution of the expectation value $\langle\hat{P}\rangle(t)=\textrm{Tr}_{\mathcal{S}}[\hat{P}\hat{\rho}_{\mathcal{S}}(t)]$. Using the master equation~\eqref{Master-equation-harmonic-oscillator} and working out the details using the canonical commutation relation, one is led to
\begin{align}
    \dv{t}\langle\hat{P}\rangle(t)&=\textrm{Tr}_{\mathcal{S}}\qty[\hat{P}\dv{t}\hat{\rho}_{\mathcal{S}}(t)] \nonumber \\
    &=-M\qty(\Omega^2+\delta\Omega^2(t))\langle\hat{X}\rangle(t)-2\gamma(t)\langle\hat{P}\rangle(t).
\end{align}
While the first term on the right-hand side describes the usual unitary evolution of the oscillator, the second term describes momentum damping. This is more evident when we take the full Markovian approximation and extend the integral upper limit to infinity, rendering the coefficient~\eqref{Dissipation-coefficient} constant~\cite{Schlosshauer2008}. In that case, we have $\langle\hat{P}\rangle(t)\propto e^{-2\gamma t}\langle\hat{P}\rangle(0)$. Note that $\gamma(t)$ is completely determined by the kernel $D(t,t')$, which is why we call it the \emph{dissipation} kernel.

The coefficient $\sigma^2(t)$ describes decoherence in the position basis, as can be seen from the position representation of the master equation~\eqref{Master-equation-position-representation}. Not only does it amount to environmental monitoring of the system position operator, but it also, as a consequence, describes diffusion\footnote{Since normal diffusion is observed in classical Brownian motion~\cite{Kubo1995}, it is this similarity that rendered this canonical model the nomenclature of "quantum Brownian motion"~\cite{Schlosshauer2008}.} in momentum. This can be seen from the time evolution of $\langle\hat{P}^2\rangle(t)$~\cite{Schlosshauer2008},
\begin{equation}
    \dv{t}\langle\hat{P}^2\rangle(t)=-M\qty(\Omega^2+\delta\Omega^2(t))\langle\hat{X}\hat{P}+\hat{P}\hat{X}\rangle(t)-4\gamma(t)\langle\hat{P}^2\rangle(t)+2\sigma^2(t),
\end{equation}
and, therefore, $\langle\hat{P}^2\rangle(t)\propto\sigma^2t$ in the Markovian limit, when this coefficient becomes constant over time. For this reason, $\sigma^2(t)$ is sometimes called the (normal) diffusion coefficient. It is also possible to show that the Wigner function for such a system obeys a diffusion equation, with the diffusion coefficient given by Eq.~\eqref{Normal-diffusion-coefficient}~\cite{Schlosshauer2008}.

Lastly, the coefficient $\Sigma^2(t)$ is also related to decoherence and diffusion. However, in the Wigner representation of the master equation, this coefficient is tied to a double derivative, each with respect to a different variable (position and momentum) instead of a single one. For this reason, it is called the anomalous diffusion coefficient, and it leads, in many physical situations of interest, to a negligible influence compared to the one induced by the normal diffusion coefficient~\cite{Schlosshauer2008}.

Now, if we wish to gain a more insightful physical interpretation of the kernels $D(t,t')$ and $N(t,t')$, we will need to specify the model for the environment. In quantum Brownian motion, the environment is described by a set of harmonic oscillators in thermal equilibrium. This model will allow us to derive an important theorem relating the dissipation and noise kernels, namely the fluctuation-dissipation theorem.

\subsection{The fluctuation-dissipation theorem}

Our derivation of the influence action~\eqref{Linear-coupling-influence-action-2} only assumes that the free Hamiltonian of the environment is quadratic, the initial state is Gaussian, and the coupling is linear, with no further specification. Now let us choose this environment to represent a set of harmonic oscillators with masses $m_i$ and frequencies $\omega_i$ such that the environment Hamiltonian reads
\begin{equation}
    \hat{H}_{\mathcal{E}}=\sum_i\qty(\frac{\hat{p}_i^2}{2m_i}+\frac{1}{2}m_i\omega_i^2\hat{q}_i^2),
\end{equation}
with $\hat{q}_i$ ($\hat{p}_i$) being the position (momentum) operator of the $i$th oscillator. Suppose now that the $i$th mode couples with the system with a coupling constant $c_i$, so that the environment operator $\hat{Q}$ is given by $\hat{Q}=\sum_ic_i\hat{q}_i$. For the initial environment state, let us assume that the modes are in thermal equilibrium at a temperature $\beta^{-1}$ such that $\hat{\rho}_{\mathcal{E}}(0)=e^{-\beta\hat{H}_{\mathcal{E}}}/Z$, where $Z=\textrm{Tr}_{\mathcal{E}}(e^{-\beta\hat{H}_{\mathcal{E}}})$ is the canonical partition function.

Now, since the modes do not directly interact with each other, we have, for $i\neq j$,
\begin{equation}
    \expval{q_i(t)q_j(t')}_{\mathcal{E}}=\expval{q_i(t)}_{\mathcal{E}}\expval{q_j(t')}_{\mathcal{E}}=0,
\end{equation}
and the dissipation and noise kernel can be written as
\begin{subequations}
\begin{equation}
    D(t,t')=i\sum_ic_i^2\expval{\comm{\hat{q}_i(t)}{\hat{q}_i(t')}}_{\mathcal{E}}\theta(t-t'),
\end{equation}
\begin{equation}
    N(t,t')=\frac{1}{2}\sum_ic_i^2\expval{\acomm{\hat{q}_i(t)}{\hat{q}_i(t')}}_{\mathcal{E}}.
\end{equation}
\end{subequations}

The interaction picture position operators are given by~\cite{Sakurai2020}
\begin{equation}
    \hat{q}_i(t)=\sqrt{\frac{1}{2m_i\omega_i}}\qty(\hat{a}_ie^{-i\omega_it}+\hat{a}_i^\dagger e^{i\omega_it}),
\end{equation}
where the $\hat{a}$'s ($\hat{a}^\dagger$'s) are annihilation (creation) operators satisfying the commutation relations
\begin{equation}
\begin{split}
    \comm{\hat{a}_i}{\hat{a}_j}&=\comm{\hat{a}_i^\dagger}{\hat{a}_j^\dagger}=0, \\
    \comm{\hat{a}_i}{\hat{a}_j^\dagger}&=\delta_{ij}.
\end{split}
\end{equation}
A direct calculation then yields
\begin{subequations}
\begin{equation}
    \expval{\comm{\hat{q}_i(t)}{\hat{q}_i(t')}}_{\mathcal{E}}=-\frac{i}{m_i\omega_i}\sin[\omega_i(t-t')],
\end{equation}
\begin{align}
    \expval{\acomm{\hat{q}_i(t)}{\hat{q}_i(t')}}_{\mathcal{E}}&=\frac{2}{m_i\omega_i^2}\langle \hat{H}_i\rangle_{\mathcal{E}}\,\cos[\omega_i(t-t')] \nonumber \\
    &+\frac{1}{m_i\omega_i}\qty[\langle \hat{a}_i^2\rangle_{\mathcal{E}}\,e^{-i\omega_i(t+t')}+\langle(\hat{a}^\dagger_i)^2\rangle_{\mathcal{E}}\,e^{i\omega_i(t+t')}],
\end{align}
\end{subequations}
where $\hat{H}_i=\omega_i\qty(\hat{a}_i^\dagger \hat{a}_i+\frac{1}{2})=\frac{\omega_i}{2}\{\hat{a}_i,\hat{a}_i^\dagger\}$
is the free Hamiltonian operator of the harmonic oscillator with frequency $\omega_i$. Note that, in this model, the dissipation kernel is independent of the initial environmental state.

Since the environment is initially in a thermal state, we have $\langle \hat{a}_i^2\rangle_{\mathcal{E}}=\langle(\hat{a}^\dagger_i)^2\rangle_{\mathcal{E}}=0$. Furthermore, the partition function for the $i$th mode reads
\begin{equation}
    Z_i=\sum_{n_i=0}^\infty\exp\qty[-\omega_i\beta\qty(n_i+\frac{1}{2})]=\frac{e^{-\omega_i\beta/2}}{1-e^{-\omega_i\beta}},
\end{equation}
from which we find~\cite{Pathria2021}
\begin{equation}
    \langle\hat{H}_i\rangle_{\mathcal{E}}=-\pdv{\beta}\ln Z_i=\frac{\omega_i}{2}\coth\qty(\frac{\omega_i\beta}{2}).
\end{equation}

Now, putting everything together into the dissipation and noise kernels leads to
\begin{subequations}
\begin{equation}
    D(\tau)=\int_0^\infty\dd\omega\,J(\omega)\sin(\omega\tau)\,\theta(\tau),
\end{equation}
\begin{equation} \label{Thermal-noise-kernel}
    N(\tau)=\frac{1}{2}\int_0^\infty\dd\omega\,J(\omega)\coth\qty(\frac{\omega\beta}{2})\cos(\omega\tau),
\end{equation}
\end{subequations}
where $\tau=t-t'$ and we have introduced the \textbf{spectral density}~\cite{Schlosshauer2008,Breuer2002}
\begin{equation} \label{Spectral-density}
    J(\omega)\equiv\sum_i\frac{c_i^2}{m_i\omega_i}\delta(\omega-\omega_i).
\end{equation}

Note that the effects of thermal fluctuations are completely encoded in the kernel $N(\tau)$, which is why we called it the \emph{noise} kernel. However, although the dissipation kernel is completely determined by the spectral density and does not contain any explicit dependence on the environmental temperature, these kernels are connected through an important relation. To see this, first note that $J(\omega)$ is simply the sine Fourier transform of the dissipation kernel,
\begin{subequations}
\begin{equation}
    J(\omega)=\Tilde{D}_S(\omega)=\frac{2}{\pi}\int_0^\infty\dd\tau\,D(\tau)\sin(\omega\tau).
\end{equation}
Similarly,
\begin{equation}
    \frac{1}{2}J(\omega)\coth\qty(\frac{\omega\beta}{2})=\Tilde{N}_C(\omega)=\frac{2}{\pi}\int_0^\infty\dd\tau\,N(\tau)\cos(\omega\tau),
\end{equation}
\end{subequations}
with $\Tilde{N}_C(\omega)$ being the cosine Fourier transform of the noise kernel. Both equations together imply that
\begin{equation}
    \Tilde{N}_C(\omega)=\frac{1}{2}\Tilde{D}_S(\omega)\coth\qty(\frac{\omega\beta}{2}).
\end{equation}
This expression is known as the \textbf{fluctuation-dissipation theorem}~\cite{Kubo1995,LandauStat1,Caldeira_1983}, as it essentially relates the thermal fluctuations of a system, encoded in the noise kernel, to dissipation.

The fluctuation-dissipation theorem arises from the fact that both effects depend on the spectral density $J(\omega)$. Typically, instead of working with the definition given by Eq.~\eqref{Spectral-density}, one usually replaces the discrete sum with a continuous function of the environmental frequencies, often following some phenomenological motivation~\cite{Schlosshauer2008,Calzetta2008}. Let us see an example before closing this section.

\subsection{The Caldeira-Leggett master equation}

Usually, one takes a power-law expression for the spectral density, $J(\omega)\propto\omega^{\alpha}$, for some constant $\alpha$. The case $\alpha=1$ is referred to as an \emph{Ohmic spectral density}. For instance, we can take
\begin{equation} \label{Ohmic-spectral-density}
    J(\omega)=\frac{4M\gamma_0}{\pi}\omega\frac{\Lambda^2}{\Lambda^2+\omega^2},
\end{equation}
where $\gamma_0$ is an effective coupling constant measuring the interaction between the system and the environment. The factor $\Lambda^2/(\Lambda^2+\omega^2)$ was added to avoid the spectral density from growing without bound, since this would be non-physical. One then introduces a frequency cutoff $\Lambda$, usually much greater than the natural frequencies of the system, $\Lambda\gg\Omega$. There are multiple ways to do that, and the prescription in Eq.~\eqref{Ohmic-spectral-density} is called the Lorentz-Drude form~\cite{Schlosshauer2008}.

For the Ohmic spectral density~\eqref{Ohmic-spectral-density}, we find
\begin{subequations}
\begin{equation}
    D(\tau)=2M\gamma_0\Lambda^2e^{-\Lambda\tau},
\end{equation}
\begin{equation}
    N(\tau)\simeq2M\gamma_0\frac{\Lambda}{\beta}e^{-\Lambda\tau},
\end{equation}
\end{subequations}
where we took the high temperature limit for simplicity, $\beta\omega\ll1$. As a further simplification, let us take the Markovian approximation versions of the coefficients \eqref{Coefficients-master-equation},
\begin{subequations}
\begin{align}
    \delta\Omega^2&=-\frac{1}{M}\int_0^\infty\dd\tau\,D(\tau)\cos(\Omega\tau)=-2\gamma_0\frac{\Lambda^3}{\Lambda^2+\Omega^2}, \\
    \gamma&=\frac{1}{2M\Omega}\int_0^\infty\dd\tau\,D(\tau)\sin(\Omega\tau)=\gamma_0\frac{\Lambda^2}{\Lambda^2+\Omega^2}, \\
    \sigma^2&=\int_0^\infty\dd\tau\,N(\tau)\cos(\Omega\tau)=\frac{2M\gamma_0}{\beta}\frac{\Lambda^2}{\Lambda^2+\Omega^2}, \\
    \Sigma^2&=-\frac{1}{M\Omega}\int_0^\infty\dd\tau\,N(\tau)\sin(\Omega\tau)=-\frac{2\gamma_0}{\beta}\frac{\Lambda}{\Lambda^2+\Omega^2}.
\end{align}
\end{subequations}
We can obtain even simpler expressions by recalling that $\Lambda\gg\Omega$. Particularly, we find that the anomalous diffusion coefficient behaves as $\Sigma^2=-\sigma^2/M\Lambda$, leading to an effect that is much smaller than normal diffusion.

Finally, the master equation~\eqref{Master-equation-harmonic-oscillator} can be written as
\begin{equation} \label{Caldeira-Leggett master equation}
    \dv{t}\hat{\rho}_{\mathcal{S}}(t)=-i\comm{\hat{H}'_{\mathcal{S}}}{\hat{\rho}_{\mathcal{S}}(t)}-i\gamma_0\comm{\hat{X}}{\acomm{\hat{P}}{\hat{\rho}_{\mathcal{S}}(t)}}-\frac{2M\gamma_0}{\beta}\comm{\hat{X}}{\comm{\hat{X}}{\hat{\rho}_{\mathcal{S}}(t)}},
\end{equation}
where $\hat{H}'_{\mathcal{S}}$ is the harmonic oscillator Hamiltonian with its frequency shifted by an amount $-2\gamma_0\Lambda$. Eq.~\eqref{Caldeira-Leggett master equation} is known as the \textbf{Caldeira-Leggett master equation}~\cite{Caldeira_1983}, and it describes the interaction of a system (here represented by a harmonic oscillator) with a Markovian Ohmic bath of non-interacting modes in thermal equilibrium at high temperature $\beta^{-1}$.

Note that our choice of spectral density implies that the diffusion coefficient is given by $\sigma^2=\gamma_0/\lambda_{\rm B}^2$, with $\lambda_{\rm B}=1/\sqrt{2M\beta^{-1}}$ being the thermal de Broglie wavelength. Since the decoherence rate is proportional to $\sigma^2$, and macroscopic systems have an extremely small thermal de Broglie wavelength, we have shown a way to \emph{describe} how decoherence occurs very quickly for such a system, which is compatible with the observation that one does not usually detect macroscopic objects in quantum superpositions~\cite{Schlosshauer2008}.

\section{The Langevin equation} \label{Sec:Langevin-equation}

The localization of a quantum system due to decoherence raises interesting questions: if we no longer observe superpositions in the position basis, can we then make sense of the concept of "trajectory" for the quantum system? If yes, how can one obtain these trajectories? We will address the former question in Section~\ref{Sec:Decoherent-histories}. In this section, we show how to obtain such trajectories from the so-called Langevin equation.

Regardless of the possibility of making sense of quantum trajectories, it is possible to take the classical limit of the system under consideration\footnote{Although mathematically the solutions to the Langevin equation fall under both cases of a "quantum trajectory" (to be discussed in Section~\ref{Sec:Decoherent-histories}) and the trajectory in the classical limit, the physical meaning is distinct for each of them. While in the latter case one imposes the classical limit by hand, in the former case this notion arises only due to the decoherence mechanism, without any further assumption other than quantum theory itself~\cite{Hu2012}.}. In the path integral formulation, this is done by the method of stationary phase, which is based on the fact that the classical paths are the ones that render the total action stationary~\cite{Feynman2010}. Let us now see how to obtain the equation of motion for the classical limit of a reduced quantum system coupled with an environment.

Recall from Eq.~\eqref{Evolution-of-reduced-density-matrix} that the reduced density matrix evolves in time according to
\begin{equation}
    \rho_{\mathcal{S}}(x,x',t)=\int_{-\infty}^\infty \dd x_0\dd x_0'\,\int\mathcal{D}x\mathcal{D}x'\,e^{i\qty(S_{\mathcal{S}}[x]-S_{\mathcal{S}}[x']+S_{\rm IF}[x,x'])}\rho_{\mathcal{S}}(x_0,x_0',0),
\end{equation}
where all environmental influence is encoded in the influence action $S_{\rm IF}[x,x']$. Particularly, if we consider the linear coupling model, the influence action is given by Eq.~\eqref{Linear-coupling-influence-action-2} in terms of the dissipation and noise kernels. Assuming that this is the case, let us note that we can rewrite the term containing the noise kernel as~\cite{Calzetta2008}
\begin{equation}
    e^{-\frac{1}{2}\int\dd t\dd t'\,u(t)N(t,t')u(t')}=\mathcal{C}\int\mathcal{D}\mathcal{N}e^{-\frac{1}{2}\int\dd t\dd t'\,\mathcal{N}(t)N^{-1}(t,t')\mathcal{N}(t')}e^{i\int\dd t\mathcal{N}(t)u(t)},
\end{equation}
where $\mathcal{C}$ is a normalization constant and $\mathcal{D}\mathcal{N}$ denotes the path integral measure for the stochastic variable $\mathcal{N}(t)$. This simple functional identity allows us to rewrite the noise kernel influence as a stochastic average with Gaussian probability density given by
\begin{equation}
    \mathscr{P}[\mathcal{N}]=\mathcal{C}\,e^{-\frac{1}{2}\int \dd t\dd t'\,\mathcal{N}(t)N^{-1}(t,t')\mathcal{N}(t')}.
\end{equation}
For example, the one point and two point correlation functions are
\begin{subequations}
\begin{equation}
    \expval{\mathcal{N}(t)}_{\rm sto}=\int\mathcal{D}\mathcal{N}\,\mathscr{P}[\mathcal{N}]\mathcal{N}(t)=0,
\end{equation}
\begin{equation}
    \expval{\mathcal{N}(t)\mathcal{N}(t')}_{\rm sto}=\int\mathcal{D}\mathcal{N}\,\mathscr{P}[\mathcal{N}]\mathcal{N}(t)\mathcal{N}(t')=N(t,t').
\end{equation}
\end{subequations}

The time evolution of the reduced density matrix can then be written as
\begin{equation}
    \rho_{\mathcal{S}}(x,x',t)=\int_{-\infty}^\infty \dd x_0\dd x_0'\,\int\mathcal{D}x\mathcal{D}x'\mathcal{D}\mathcal{N}\,\rho_{\mathcal{S}}(x_0,x_0',0)e^{iS_{\rm SEA}[x,x',\mathcal{N}]},
\end{equation}
with the stochastic effective action
\begin{align}
    S_{\rm SEA}[x,x',\mathcal{N}]&=S_{\mathcal{S}}[x]-S_{\mathcal{S}}[x']+\int\dd t\,\mathcal{N}(t)\qty[x(t)-x'(t)] \nonumber \\
    &+\frac{1}{2}\int\dd t\dd t'\,\qty[x(t)-x'(t)]D(t,t')\qty[x(t')+x'(t')].
\end{align}
The equation of motion is obtained by setting $\delta S_{\rm SEA}=0$ as usual, resulting in
\begin{equation} \label{Langevin-equation}
    \dv{t}\qty(\pdv{L_{\mathcal{S}}}{\dot{x}})-\pdv{L_{\mathcal{S}}}{x}-\frac{1}{2}\int_0^t\dd t'\,D(t,t')\qty[x(t')+x'(t')]=\mathcal{N}(t),
\end{equation}
where $L_{\mathcal{S}}$ is the system's Lagrangian. Eq.~\eqref{Langevin-equation} is the \textbf{Langevin equation}. If the system of interest is a harmonic oscillator, for instance, we have
\begin{equation}
    M\Ddot{x}(t)+M\Omega^2x(t)-\frac{1}{2}\int_0^t\dd t'\,D(t,t')\qty[x(t')+x'(t')]=\mathcal{N}(t).
\end{equation}
Note that the $D$ term generates non-local, memory-dependent dissipation, while the $N$ term acts as a stochastic force on the system. Once again, the name \emph{noise kernel} for the two-point function $N(t,t')$ proves to be quite fitting.

Finally, before we close this chapter, let us address the question of the observability of individual solutions to the Langevin equation for \emph{quantum systems}, i.e., when the classical limit is not imposed by hand, yet the trajectory description is still fitting. To do so, we shall turn to the \textbf{decoherent} (or \textbf{consistent}) \textbf{histories} approach to quantum mechanics.

\section{Decoherent histories} \label{Sec:Decoherent-histories}

Let us now present the decoherent histories formalism, which was developed primarily by Griffiths, Omnès, Gell-Mann, and Hartle~\cite{Griffiths1984,Omnes1990,Omnes1992,Gell-Mann-Hartle1990} (see also Ref.~\cite{Dowker1992} for a review). We begin by introducing the concept of history in quantum mechanics and then define the so-called decoherence functional while addressing its role in the formalism. Finally, we discuss the physical conditions for making sense of trajectories of quantum systems.

\subsection{Histories in quantum mechanics}

A quantum mechanical history is a sequence of quantum mechanical events at successive moments in time, thus characterized by a sequence of successive projection operators. Consider a closed quantum system, for instance, which at an initial time $t_0$ is in the state described by the density matrix $\hat{\rho}_0$. At time $t_1>t_0$, the system will be described by
\begin{equation}
    \hat{\rho}(t_1)=\hat{U}(t_1,t_0)\hat{\rho}_0\hat{U}^\dagger(t_1,t_0),
\end{equation}
where $\hat{U}(t_i,t_j)$ is the unitary time-evolution operator from time $t_j$ to time $t_i$. We may now wonder whether the event corresponding to a set of mutually orthogonal projection operators $\hat{\Pi}_{\alpha_1}$ occurs at $t_1$. The probability of such an occurrence is given by
\begin{align} \label{p(alpha1,t1)}
    p(\alpha_1t_1)&=\textrm{Tr}\qty[\hat{\Pi}_{\alpha_1}\hat{\rho}(t_1)]=\textrm{Tr}\qty[\hat{\Pi}_{\alpha_1}\hat{U}(t_1,t_0)\hat{\rho}_0\hat{U}^\dagger(t_1,t_0)] \nonumber \\
    &=\textrm{Tr}\qty[\hat{U}^\dagger(t_1,t_0)\hat{\Pi}_{\alpha_1}\hat{U}(t_1,t_0)\hat{\rho}_0] \nonumber \\
    &=\textrm{Tr}\qty[\hat{\Pi}_{\alpha_1}(t_1)\hat{\rho}_0],
\end{align}
where we have defined $\hat{\Pi}_{\alpha_1}(t_1)\equiv\hat{U}^\dagger(t_1,t_0)\hat{\Pi}_{\alpha_1}\hat{U}(t_1,t_0)$ and the trace is taken over an orthonormal basis of the system's Hilbert space. Once the projective measurement has been made, the system undergoes the usual non-unitary evolution,
\begin{equation}
    \hat{\rho}(t_1)\to\hat{\rho}_{\alpha_1}(t_1)=\frac{\hat{\Pi}_{\alpha_1}\hat{\rho}(t_1)\hat{\Pi}_{\alpha_1}}{\textrm{Tr}\qty[\hat{\Pi}_{\alpha_1}\hat{\rho}(t_1)]}.
\end{equation}

Now suppose we let the system evolve further to time $t_2>t_1$, where we ask about the event corresponding to projectors $\hat{\Pi}_{\alpha_2}$. The probability of this additional occurrence is
\begin{align}
    p(\alpha_2t_2|\alpha_1t_1)&=\textrm{Tr}\qty[\hat{\Pi}_{\alpha_2}\hat{U}(t_2,t_1)\hat{\rho}_{\alpha_1}(t_1)\hat{U}^\dagger(t_2,t_1)] \nonumber \\
    &=\frac{1}{p(\alpha_1t_1)}\textrm{Tr}[\hat{\Pi}_{\alpha_2}\hat{U}(t_2,t_1)\hat{\Pi}_{\alpha_1}\hat{U}(t_1,t_0)\hat{\rho}_0\hat{U}^\dagger(t_1,t_0)\hat{\Pi}_{\alpha_1}\hat{U}^\dagger(t_2,t_1)] \nonumber \\
    &=\frac{1}{p(\alpha_1t_1)}\textrm{Tr}\left[ \hat{\Pi}_{\alpha_2}\hat{U}(t_2,t_1)\hat{U}(t_1,t_0)\hat{U}^\dagger(t_1,t_0)\hat{\Pi}_{\alpha_1}\hat{U}(t_1,t_0)\hat{\rho}_0\hat{U}^\dagger(t_1,t_0)\right. \nonumber \\
    &\hspace{2cm}\times \left. \hat{\Pi}_{\alpha_1}\hat{U}(t_1,t_0)\hat{U}^\dagger(t_1,t_0)\hat{U}^\dagger(t_2,t_1)\right] \nonumber \\
    &=\frac{1}{p(\alpha_1t_1)}\textrm{Tr}\left[ \hat{U}^\dagger(t_2,t_0)\hat{\Pi}_{\alpha_2}\hat{U}(t_2,t_0)\hat{U}^\dagger(t_1,t_0)\hat{\Pi}_{\alpha_1}\hat{U}(t_1,t_0)\hat{\rho}_0] \right. \nonumber \\
    &\hspace{2cm}\times \left. \hat{U}^\dagger(t_1,t_0)\hat{\Pi}_{\alpha_1}\hat{U}(t_1,t_0)\right] \nonumber \\
    &=\frac{1}{p(\alpha_1t_1)}\textrm{Tr}\qty[\hat{\Pi}_{\alpha_2}(t_2)\hat{\Pi}_{\alpha_1}(t_1)\hat{\rho}_0\hat{\Pi}_{\alpha_1}(t_1)] \nonumber \\
    &=\frac{1}{p(\alpha_1t_1)}\textrm{Tr}\qty[\hat{\Pi}_{\alpha_2}(t_2)\hat{\Pi}_{\alpha_1}(t_1)\hat{\rho}_0\hat{\Pi}_{\alpha_1}(t_1)\hat{\Pi}_{\alpha_2}(t_2)],
\end{align}
where we used the unitarity of the time-evolution operators, as well as their composition rule, the cyclic property of the trace, and the idempotence of the projection operators.

The probability
\begin{align} \label{p(alpha2t2,alpha1t1)}
    p(\alpha_2t_2,\alpha_1t_1)&=p(\alpha_2t_2|\alpha_1t_1)p(\alpha_1t_1) \nonumber \\
    &=\textrm{Tr}\qty[\hat{\Pi}_{\alpha_2}(t_2)\hat{\Pi}_{\alpha_1}(t_1)\hat{\rho}_0\hat{\Pi}_{\alpha_1}(t_1)\hat{\Pi}_{\alpha_2}(t_2)]
\end{align}
is what we call the probability of the history $\qty(\rho_0,t_0)\to\qty(\alpha_1,t_1)\to\qty(\alpha_2,t_2)$. Strictly speaking, this is an abuse of nomenclature since the quantity in Eq.~\eqref{p(alpha2t2,alpha1t1)} does not satisfy Kolmogorov's third axiom of probability theory, the probability sum rule.  For instance, let us consider another history in which no projection is made at time $t_1$, that is, the history $\qty(\rho_0,t_0)\to\qty(\alpha_2,t_2)$. In complete analogy with Eq.~\eqref{p(alpha1,t1)}, the probability of $\alpha_2$ occurrence at $t_2$ is now given by
\begin{equation}
    p(\alpha_2t_2)=\textrm{Tr}\qty[\hat{\Pi}_{\alpha_2}(t_2)\hat{\rho}_0].
\end{equation}
We can use the identity $\sum_{\alpha_1}\hat{\Pi}_{\alpha_1}(t_1)=1$ to write this as
\begin{align}
    p(\alpha_2t_2)&=\textrm{Tr}\qty[\hat{\Pi}_{\alpha_2}(t_2)\hat{\rho}_0\hat{\Pi}_{\alpha_2}(t_2)] \nonumber \\
    &=\textrm{Tr}\qty[\hat{\Pi}_{\alpha_2}(t_2)\sum_{\alpha_1}\hat{\Pi}_{\alpha_1}(t_1)\hat{\rho}_0\sum_{\alpha_1'}\hat{\Pi}_{\alpha_1'}(t_1)\hat{\Pi}_{\alpha_2}(t_2)] \nonumber \\
    &=\sum_{\alpha_1}\textrm{Tr}\qty[\hat{\Pi}_{\alpha_2}(t_2)\hat{\Pi}_{\alpha_1}(t_1)\hat{\rho}_0\hat{\Pi}_{\alpha_1}(t_1)\hat{\Pi}_{\alpha_2}(t_2)] \nonumber \\
    &\hspace{1cm}+\sum\displaylimits_{\substack{\alpha_1,\alpha_1' \\ \alpha_1\neq\alpha_1'}}\textrm{Tr}\qty[\hat{\Pi}_{\alpha_2}(t_2)\hat{\Pi}_{\alpha_1}(t_1)\hat{\rho}_0\hat{\Pi}_{\alpha_1'}(t_1)\hat{\Pi}_{\alpha_2}(t_2)].
\end{align}
By comparing this result with Eq.~\eqref{p(alpha2t2,alpha1t1)}, we conclude that, in general,
\begin{equation}
    p(\alpha_2t_2)\neq\sum_{\alpha_1}p(\alpha_2t_2,\alpha_1t_1)
\end{equation}
due to the term
\begin{equation} \label{Pre-decoherence-functional}
    \textrm{Tr}\qty[\hat{\Pi}_{\alpha_2}(t_2)\hat{\Pi}_{\alpha_1}(t_1)\hat{\rho}_0\hat{\Pi}_{\alpha_1'}(t_1)\hat{\Pi}_{\alpha_2}(t_2)],
\end{equation}
which is generally non-zero and represents interference between different quantum-mechanical histories. This is a key feature of quantum mechanics, namely the presence of interference terms that prevent probabilities from being assigned to quantum histories. For some applications, however, we may attempt to identify sets of histories that suffer negligible interference with each other and, therefore, to which probabilities can be assigned. These may be found by studying the quantity in Eq.~\eqref{Pre-decoherence-functional}.


\subsection{Decoherence functional}

For a pair of histories $[\alpha]$, $[\beta]$, where $[\alpha]$ denotes the sequence of events $\alpha_1$, $\alpha_2$, $\dots$, and $\alpha_n$ at times $t_1<t_2<\dots<t_n$, and analogously for $[\beta]$, we define the \textbf{decoherence functional} as
\begin{equation} \label{Decoherence-functional-1}
    \mathcal{D}\qty[\alpha,\beta]=\textrm{Tr}\qty[\hat{\Pi}_{\alpha_n}(t_n)\dots\hat{\Pi}_{\alpha_1}(t_1)\hat{\rho}_0\hat{\Pi}_{\beta_1}(t_1)\dots \hat{\Pi}_{\beta_n}(t_n)],
\end{equation}
where
\begin{equation}
    \hat{\Pi}_{\alpha_k}(t_k)=\hat{U}^\dagger(t_k,t_0)\hat{\Pi}_{\alpha_k}\hat{U}(t_k,t_0).
\end{equation}
The decoherence functional may also be written in a more compact form as~\cite{Calzetta2008}
\begin{equation} \label{Decoherence-functional-2}
    \mathcal{D}\qty[\alpha,\beta]=\textrm{Tr}\qty{\Tilde{\textrm{T}}\qty[\prod_{j=1}^n\hat{\Pi}_{\beta_j}(t_j)]\textrm{T}\qty[\prod_{i=1}^n\hat{\Pi}_{\alpha_i}(t_i)]\hat{\rho}_0}.
\end{equation}
It is then straightforward to show that this functional satisfies the following properties:
\begin{enumerate}[i)]
    \item $\mathcal{D}\qty[\alpha,\beta]=\mathcal{D}^*\qty[\beta,\alpha]$,
    \item $\sum_{[\alpha],[\beta]}\mathcal{D}\qty[\alpha,\beta]=1$,
    \item $\mathcal{D}\qty[\alpha,\alpha]\geq0$,
    \item $\sum_{[\alpha]}\mathcal{D}\qty[\alpha,\alpha]=1$,
\end{enumerate}
where we are denoting $\sum_{[\alpha]}\equiv \sum_{\alpha_1}\sum_{\alpha_2}\dots\sum_{\alpha_n}$.

The last two properties suggest that we identify the diagonal elements $\mathcal{D}[\alpha,\alpha]$ as the probability for the history $(\rho_0,t_0)\to\qty(\alpha_1,t_1)\to\dots\to\qty(\alpha_n,t_n)$, namely $p(\alpha)=\mathcal{D}[\alpha,\alpha]$. This does not satisfy the probability sum rule, however, and, in general, we have~\cite{Dowker1992}
\begin{align}
    p(\alpha\vee\beta)&=\mathcal{D}[\alpha,\alpha]+\mathcal{D}[\beta,\beta]+2\textrm{Re}\mathcal{D}[\alpha,\beta] \nonumber \\
    &=p(\alpha)+p(\beta)+2\textrm{Re}\mathcal{D}[\alpha,\beta].
\end{align}
As we discussed before, this violation prevents us from making sense of the notion of trajectory in quantum mechanics. On the other hand, this notion acquires meaning when there is strong decoherence, $\mathcal{D}[\alpha,\beta]=0$. One may then write the fundamental formula for the quantum mechanics of history as
\begin{equation}
    \textrm{Re}\qty{\mathcal{D}[\alpha,\beta]}=p(\alpha)\delta_{\alpha_1,\beta_1}\dots\delta_{\alpha_n,\beta_n}.
\end{equation}
This equation expresses the necessary and sufficient condition for us to assign probabilities to individual histories; at the same time, it tells us what those probabilities are. For most applications of the formalism, one may be satisfied to claim that a pair of mutually exclusive histories is consistent when
\begin{equation}
    \textrm{Re}\qty{\mathcal{D}[\alpha,\beta]}\ll\mathcal{D}[\alpha,\alpha],\,\mathcal{D}[\beta,\beta]\hspace{0,5cm}\textrm{for}\hspace{0,5cm}\alpha\neq\beta.
\end{equation}

From now on, let us consider projections in the position basis, which is a kind of history implemented naturally in the path integral formalism. The projectors are represented by window functions $\hat{\Pi}_{\alpha_k}(t_k)\to w_\alpha[x(t_k)]$, which, for example, can be unity if the configuration at $t_k$ satisfies the requirement of history $\alpha$ and zero otherwise~\cite{Hu2012,Calzetta2008}. Eq.~\eqref{Decoherence-functional-2} can then be written as
\begin{align}
    \mathcal{D}[\alpha,\beta]&=\int \dd y(0)\dd y'(0) \nonumber \\
    &\times\int\mathcal{D}y\mathcal{D}y'\,e^{i\qty(S[y]-S[y'])}\rho\qty(y(0),y'(0),0)\qty{\prod_iw_\alpha\qty[y(t_i)]}\qty{\prod_jw_\beta\qty[y'(t_j)]}.
\end{align}
Here $S$ is the action for the total system described by the generic coordinate $y$. Let us now study the decoherence functional for an open quantum system in order to establish under what conditions one can actually observe the quasi-classical trajectories arising from the Langevin equation.

\subsection{Consistent histories for open quantum systems}

Let us consider the open quantum system in the linear coupling model that we treated in previous sections. Let us also assume that the system of interest is a harmonic oscillator for illustration purposes. In that case, we are interested in histories where the system variable $x(t)$ follows a trajectory $\chi(t)$, which is obtained by solving the Langevin equation with a given accuracy $\sigma_\chi(t)$. Now take the window functions to be Gaussians so that we can make the replacement~\cite{Calzetta2008,Hu2012}
\begin{align}
    \rho_{\mathcal{S}}\qty(x(0),x'(0),0)&\qty{\prod_iw_\alpha\qty[x(t_i)]}\qty{\prod_jw_\beta\qty[x'(t_j)]}\to \nonumber \\
    &\to\exp\qty{-\int\dd t\frac{1}{2\sigma^2_\chi(t)}\qty[(x-\chi)^2+(x'-\chi')^2]}
\end{align}
in the decoherence functional of two histories represented by trajectories $\chi(t)$ and $\chi'(t)$. The integration over the environmental variables is done as outlined in Section~\ref{Sec:The-FV-IF}, leading to
\begin{align}
    \mathcal{D}[\chi,\chi']&=\int\mathcal{D}x\mathcal{D}x'\,\exp\left\{ \int\dd t\dd t'\qty[u(t)iL(t,t')X(t')-\frac{1}{2}u(t)N(t,t')u(t')]\right. \nonumber \\
    &\left. -\int\dd t\frac{1}{2\sigma^2_\chi(t)}\qty[(x-\chi)^2+(x'-\chi')^2]\right\} ,
\end{align}
where we defined~\cite{Hu2012}
\begin{equation}
    L(t,t')\equiv-M\qty(\dv[2]{t}+\Omega^2)\delta(t-t')+D(t,t').
\end{equation}
Also, recall that $X(t)=\qty[x(t)+x'(t)]/2$ and $u(t)=x(t)-x'(t)$.

Now define $Y(t)=\qty[\chi(t)+\chi'(t)]/2$ and $v(t)=\chi(t)-\chi'(t)$ and change the path integral variables $(x,x')\to(X,u)$ so that the decoherence functional can be written as
\begin{align} \label{Dec-functional-Gaussian-integral}
    \mathcal{D}[\chi,\chi']&=\exp{-\int\dd t\frac{1}{2\sigma^2_\chi(t)}\qty[2Y^2(t)+\frac{1}{2}v^2(t)]} \nonumber \\
    &\times\int\mathcal{D}X\mathcal{D}u\,\exp\qty[-\frac{1}{2}\int\dd t\dd t'\,\mqty(X(t) & u(t))\mathbb{M}(t,t')\mqty(X(t') \\ u(t'))] \nonumber \\
    &\times\exp\qty{\int\dd t\,\qty[\mathbb{J}(t)]^T\mqty(X(t) \\ u(t))},
\end{align}
where we defined
\begin{subequations}
\begin{equation}
    \mathbb{M}(t,t')=\mqty(\frac{2}{\sigma^2_\chi(t)}\delta(t-t') & -iL(t,t') \\ -iL(t,t') & \Tilde{N}(t,t')),
\end{equation}
with $\Tilde{N}(t,t')=N(t,t')+[2\sigma_\chi^2(t)]^{-1}\delta(t-t')$, and also
\begin{equation}
    \mathbb{J}(t)=\frac{1}{2\sigma_{\chi}^2(t)}\mqty(4Y(t) \\ v(t)).
\end{equation}
\end{subequations}

The Gaussian path integral in Eq.~\eqref{Dec-functional-Gaussian-integral} is a traditional one in field theory with the well known result~\cite{Schwartz2013}
\begin{align}
    \mathcal{D}[\chi,\chi']&\propto\exp{-\int\dd t\frac{1}{2\sigma^2_\chi(t)}\qty[2Y^2(t)+\frac{1}{2}v^2(t)]} \nonumber \\
    &\times\exp\qty[\frac{1}{2}\int\dd t\dd t'\,\qty[\mathbb{J}(t)]^T\mathbb{M}^{-1}(t,t')\mathbb{J}(t)].
\end{align}
Explicitly, we have
\begin{equation}
    \mathbb{M}^{-1}(t,t')=\qty[\det\mathbb{M}(t,t')]^{-1}\mqty(\Tilde{N}(t,t') & iL(t,t') \\ iL(t,t') & \frac{2}{\sigma^2_\chi(t)}\delta(t-t'))
\end{equation}
with $\det\mathbb{M}(t,t')=\frac{2}{\sigma^2_\chi(t)}\delta(t-t')\Tilde{N}(t,t')+L^2(t,t')$. Then we can write
\begin{align}
    \mathcal{D}[\chi,\chi']&\propto\exp\left\{ -\int\dd t\dd t'\,\frac{Y(t)}{\sigma_\chi^2(t)}\qty[\delta(t-t')-\frac{2}{\sigma_\chi^2(t')}\qty[\det\mathbb{M}(t,t')]^{-1}\Tilde{N}(t,t')]Y(t')\right. \nonumber \\
    &+\int\dd t\dd t'\frac{Y(t)}{\sigma_\chi^2(t)\sigma_\chi^2(t')}\qty[\det\mathbb{M}(t,t')]^{-1}iL(t,t')v(t') \nonumber \\
    &\left. -\int\dd t\frac{v(t)}{4\sigma_\chi^2(t)}\qty[1-\frac{1}{\sigma_\chi^4(t)}\qty[\det\mathbb{M}(t,t)]^{-1}]v(t)\right\} .
\end{align}
For simplicity, let us assume that the (dissipative) dynamics is much smaller than the noise, $L(t,t')\ll\Tilde{N}(t,t')$~\cite{Calzetta2008}, such that
\begin{align} \label{D[chi,chi']}
    \mathcal{D}[\chi,\chi']&\sim\exp\left\{ -\frac{1}{2}\int\dd t\,\qty[N(t)+\qty(2\sigma_\chi^2(t))^{-1}]^{-1}\qty[L(t)Y(t)]^2\right. \nonumber \\
    &+\int\dd t\,\frac{i}{2\sigma_\chi^2(t)}Y(t)\qty[N(t)+\qty(2\sigma_\chi^2(t))^{-1}]^{-1}L(t)v(t) \nonumber \\
    &\left. -\frac{1}{2}\int\dd t\,N(t)\qty[2\sigma_\chi^2(t)N(t)+1]^{-1}v^2(t)\right\} .
\end{align}

The probability of history $\chi$ is obtained by setting $v=0$ in Eq.~\eqref{D[chi,chi']},
\begin{equation}
    \mathcal{D}[\chi,\chi]\sim\exp\qty{-\frac{1}{2}\int\dd t\,\qty[N(t)+\qty(2\sigma_\chi^2(t))^{-1}]^{-1}\qty[L(t)\chi(t)]^2}.
\end{equation}
Thus, the highest probabilities are associated with those trajectories that satisfy the equation of motion (without the noise), $L\chi=0$, while the deviations from determinism are measured by the factor $N+\qty(2\sigma_\chi^2)^{-1}$.

Consistency between the two histories $\chi$ and $\chi'$ follows from the behavior of Eq.~\eqref{D[chi,chi']} for increasing values of $v$. Particularly, two histories are approximately consistent when $v^2\geq2\sigma_\chi^2+N^{-1}$. Now, if the noise is weak, $N^{-1}\gg\sigma_\chi^2$, then $v\gtrsim N^{-1}\gg\sigma_\chi^2$, which means that histories of accuracy $\sim\sigma_\chi$ do not decohere.  On the other hand, if we have strong noise, $N^{-1}\ll\sigma_\chi^2$, then $v\gtrsim \sigma_\chi$, which means that any history whose accuracy can be probed is consistent and, therefore, the solutions to the Langevin equation represent trajectories that actually describe physical reality. Of course, strong noise leads to less predictability, and one can find balance by saying that this "physical reality aspect" of the Langevin trajectories arises when we choose to follow these paths with accuracy given by $\sigma_\chi^2\sim N^{-1}$~\cite{Calzetta2008,Hu2012}. 
\chapter{Gravitational decoherence}
\label{chap:gravdec}

We saw in Chapter~\ref{chap:dec-and-FV} how decoherence is linked to the quantum-to-classical transition, which can explain why one does not usually detect macroscopic systems in quantum superpositions. Decoherence occurs due to interactions with an environment, such as photons, dust particles, and sometimes even the system's own internal degrees of freedom. If this process is related to gravitational interactions, either directly or indirectly, we call it \textbf{gravitational decoherence}.

As we discussed in Chapter~\ref{chap:quantum-grav}, gravity is the weakest of the four known interactions, and, usually, other decoherence sources are much stronger than gravitational ones. This raises an immediate challenge for observing the effects of gravity in quantum superpositions: all other competing decoherence processes must be suppressed (see Ref.~\cite{Pfister2016} for a proposal of a universal test of gravitational decoherence). From the theoretical side, the interest comes from the fact that gravity is universal and, therefore, cannot be shielded.

Gravitational decoherence is a broad term that encompasses any loss of coherence in a quantum system, either directly caused by the coupling with a (classical or quantum) gravitational field or related to gravitational effects in some way (see Refs.~\cite{Bassi_2017,Hsiang_2024} for a review). It can refer, for instance, to collapse models in which the wavefunction collapse is understood as a physical process that occurs due to gravity. Quantum mechanics, as opposed to classical physics, seems to assign a special role to measurement processes, which are not described by the dynamical Schrödinger equation. Collapse models are modifications of quantum mechanics that typically introduce stochastic terms in the Schrödinger equation to account for the measurement postulate, or what is sometimes referred to as the wavefunction collapse. This is the viewpoint that the collapse is indeed a physical process. Since such processes are expected to be universal and more significant for macroscopic systems, some authors have proposed that gravity is behind the wavefunction collapse since it is also universal and scales with the mass of the quantum system~\cite{Karolyhazy1966,Diosi1984,Diosi1989,Frenkel1990,Penrose1996,Diosi2007,Diosi2014,Adler2014}. Recently, a different viewpoint was proposed in which quantum and classical mechanics belong to different regimes, both emerging from some yet unknown physics at the Planck scale (as opposed to the viewpoint that classical physics emerges from its quantum counterpart), with a gravitational self-decoherence model that describes decoherence of quantum systems as they approach a Heisenberg cut $M_{\rm C}\sim M_{\rm P}$ while maintaining the coherence of microscopic systems for which $m\ll M_{\rm C}$~\cite{Aguiar2025}.

In this work, we consider a different type of gravitational decoherence. We take the viewpoint that, to the best of our knowledge, the quantum description is, in principle, valid for all systems in nature, including the gravitational dynamical degrees of freedom. We work within the limits of weak gravitational fields, and so its quantum description falls within the framework of perturbative quantum gravity as described in Chapter~\ref{chap:quantum-grav}. Our interest will be in investigating the decoherence of quantum and non-relativistic systems due to quantum fluctuations of spacetime (of course, one can also study decoherence induced by stochastic \emph{classical} spacetime fluctuations; see Refs.~\cite{Linet1976,Stodolsky1979,Cai1989,Power2000,Reynaud2004,Lamine2006,Breuer2009}).

This chapter is not supposed to be an extensive review of the literature on gravitational decoherence, nor a complete description of selected works in the field. In Section~\ref{Sec:Master-eq-grav-dec} we outline the findings of Blencowe~\cite{Blencowe_2013}, and Anastopoulos and Hu~\cite{Anastopoulos_2013}, in which the authors obtained a Markovian master equation describing the decoherence of a non-relativistic system induced by a weak quantum gravitational field. In Section~\ref{Sec:Kanno} we introduce another approach taken by Kanno et al.~\cite{Kanno2021} that describes spatial localization induced by gravitons using the Feynman-Vernon influence functional. Finally, in Section~\ref{Sec:Dec-grav-time-dilation} we describe a different kind of gravitational decoherence mechanism proposed by Pikovski et al.~\cite{Pikovski2015,Pikovski2017}. Here, decoherence occurs due to the coupling of the center-of-mass coordinate of a quantum system to its internal degrees of freedom induced by time dilation caused by a classical static gravitational potential. This is different from decoherence due to spacetime fluctuations, but it will serve as motivation for the problem of gravitational decoherence of a composite particle that we will study in Part~\ref{Part:Dec-and-ent} of this thesis.

\section{Master equation for gravitational decoherence} \label{Sec:Master-eq-grav-dec}

We introduce the topic of gravitational decoherence by outlining the approach of Blencowe~\cite{Blencowe_2013}, and also mentioning the equivalent findings of Anastopoulos and Hu~\cite{Anastopoulos_2013}.

Blencowe~\cite{Blencowe_2013} considers quantum matter as described by a massive scalar field $\varphi(x)$, with a mass parameter $m$. The metric field is expanded as in Eq.~\eqref{Metric-expansion-with-kappa}, i.e., the author considers the weak field limit where the dynamical degrees of freedom are a perturbation of Minkowski spacetime, $g_{\mu\nu}=\eta_{\mu\nu}+\kappa_{\rm g} h_{\mu\nu}$. The total action is then given by
\begin{equation}
    S[\varphi,h_{\mu\nu}]=S_{\mathcal{S}}[\varphi]+S_{\mathcal{E}}[h_{\mu\nu}]+S_{\mathcal{I}}[\varphi,h_{\mu\nu}],
\end{equation}
where $S_{\mathcal{S}}[\varphi]$ is the free Klein-Gordon action,
\begin{equation}
    S_{\mathcal{S}}[\varphi]=-\frac{1}{2}\int\dd^4x\qty(\partial_\mu\varphi\partial^\mu\varphi+m^2\varphi^2),
\end{equation}
$S_{\mathcal{E}}[h_{\mu\nu}]$ is the linearized Einstein-Hilbert action, Eq.~\eqref{Linearized-EH-action-canonical-normalization}, and the interaction between the scalar and the tensor fields is described by\footnote{Note that the author also works with second-order terms in the metric expansion in the interaction term.}
\begin{equation}
    S_{\mathcal{I}}[\varphi,h_{\mu\nu}]=\int\dd^4x\qty[\frac{\kappa_{\rm g}}{2}T^{\mu\nu}(\varphi)h_{\mu\nu}+\frac{\kappa_{\rm g}^2}{4}U^{\mu\nu\rho\sigma}(\varphi)h_{\mu\nu}h_{\rho\sigma}],
\end{equation}
with
\begin{subequations}
\begin{equation}
    T_{\mu\nu}(\varphi)=\partial_\mu\varphi\partial_\nu\varphi-\frac{1}{2}\eta_{\mu\nu}\partial_\rho\varphi\partial^\rho\varphi-\frac{1}{2}\eta_{\mu\nu}m^2\varphi^2
\end{equation}
being the scalar field energy-momentum tensor and~\cite{Arteaga2004}
\begin{align}
    U^{\rho\sigma\mu\nu}(\varphi)&=-2\eta^{\sigma\mu}\partial^\rho\varphi\partial^\nu\varphi+\eta^{\rho\sigma}\partial^\mu\varphi\partial^\nu\varphi \nonumber \\
    &+\qty(\frac{1}{2}\eta^{\rho\mu}\eta^{\sigma\nu}-\frac{1}{4}\eta^{\rho\sigma}\eta^{\mu\nu})\qty(\partial^\lambda\varphi\partial_\lambda\varphi+m^2\varphi^2).
\end{align}
\end{subequations}

The author then proceeds to quantize both the scalar field and the metric perturbation using the closed time path integral approach while integrating over the gravitational variables, thereby treating the gravitons as the environment. As usual, one considers the initial state to be separable, $\hat{\rho}(0)=\hat{\rho}_{\mathcal{S}}(0)\otimes\hat{\rho}_{\mathcal{E}}(0)$, while the gravitons are considered to be initially in a thermal state with temperature $T_{\rm g}$ (we will discuss the interpretation of this parameter near the end of this section). The result is the time evolution of the reduced density matrix $\rho_{\mathcal{S}}[\varphi,\varphi',t]$ in terms of the influence functional, in complete analogy with\footnote{See Ref.~\cite{Calzetta2008} for the influence functional approach applied to quantum fields.} Eq.~\eqref{Evolution-of-reduced-density-matrix}. Then, by evaluating the influence action to lowest order in $\kappa_{\rm g}$ and introducing a Lorenz gauge fixing term in the linearized Einstein-Hilbert action, Blencowe obtains a Born-approximated master equation for the scalar field given by
\allowdisplaybreaks
\begin{align}
    \dv{t}\hat{\rho}_{\mathcal{S}}(t)&=-i\comm{\hat{H}_{\mathcal{S}}}{\hat{\rho}_{\mathcal{S}}(t)} \nonumber \\
    &-\int_0^t\dd\tau\int\dd^3r\dd^3r'\left\{ N(\tau,\vb{r}-\vb{r}')\left( 2\comm{\hat{T}_{\mu\nu}(\vb{r})}{\comm{\hat{T}^{\mu\nu}(-\tau,\vb{r}')}{\hat{\rho}_{\mathcal{S}}(t)}}\right. \right. \nonumber \\
    &\left. -\comm{{\hat{T}_\mu}\,^\mu(\vb{r})}{\comm{{\hat{T}_\nu}\,^\nu(-\tau,\vb{r}')}{\hat{\rho}_{\mathcal{S}}(t)}}\right) \nonumber \\
    &-iD(\tau,\vb{r}-\vb{r}')\left( 2\comm{\hat{T}_{\mu\nu}(\vb{r})}{\acomm{\hat{T}^{\mu\nu}(-\tau,\vb{r}')}{\hat{\rho}_{\mathcal{S}}(t)}}\right. \nonumber \\
    &\left. \left. -\comm{{\hat{T}_\mu}\,^\mu(\vb{r})}{\acomm{{\hat{T}_\nu}\,^\nu(-\tau,\vb{r}')}{\hat{\rho}_{\mathcal{S}}(t)}}\right) \right\} .
\end{align}
where $\hat{H}_{\mathcal{S}}$ is the free scalar field Hamiltonian and
\begin{subequations}
\begin{equation}
    D(t,\vb{r})=\qty(\frac{\kappa_{\rm g}}{4})^2\int\frac{\dd^3k}{(2\pi)^3}\sin(\omega_kt)\frac{e^{i\vb{k}\vdot\vb{r}}}{\omega_k},
\end{equation}
\begin{equation}
    N(t,\vb{r})=\qty(\frac{\kappa_{\rm g}}{4})^2\int\frac{\dd^3k}{(2\pi)^3}\cos(\omega_kt)\qty[1+2n(\omega_k)]\frac{e^{i\vb{k}\vdot\vb{r}}}{\omega_k},
\end{equation}
\end{subequations}
are the dissipation and noise kernels, with $\omega_k=\abs{\vb{k}}$ and $n(\omega_k)$ being the Bose-Einstein distribution at temperature $T_{\rm g}$.

For the system, Blencowe proceeds to consider a specific class of coherent states that model stationary material objects as Gaussian matter "balls"~\cite{Blencowe_2013}. Then, in the non-relativistic limit in which the dominant contribution comes from $T_{00}\simeq m^2\varphi^2/2$, and within the Markovian approximation and high temperature limit, the author finds
\begin{equation}
    \pdv{t}\rho_{\mathcal{S}}[\varphi,\varphi',t]=-\frac{T_{\rm g}}{2\pi}\qty(\frac{\kappa_{\rm g}}{4})^2\qty{\int\dd^3r\qty[\frac{1}{2}m^2\varphi^2(\vb{r})-\frac{1}{2}m^2{\varphi'}^2(\vb{r})]}^2\rho_{\mathcal{S}}[\varphi,\varphi',t]+\dots,
\end{equation}
where we only wrote explicitly the terms that are relevant to decoherence. For two matter "ball" states with energies $E_1$ and $E_2$, this Markovian master equation describes the decay of the off-diagonal density matrix elements with decoherence time
\begin{equation} \label{Dec-time-Blencowe}
    t_{\rm dec}=\frac{\hbar}{k_BT_{\rm g}}\qty(\frac{E_{\rm P}}{\Delta E})^2,
\end{equation}
with $E_{\rm P}$ being the Planck energy, $\Delta E=E_1-E_2$ and we restored the universal constants.

Blencowe~\cite{Blencowe_2013} argues that this result is "sufficiently basic" such that one can expect it to hold for more general cases than the scalar field model. In fact, this result was used to constrain quantum spacetime induced oscillation damping in (ultra-relativistic) neutrinos experiments in Refs.~\cite{D_Esposito_2024,Domi2024}, for instance.

Note from Eq.~\eqref{Dec-time-Blencowe} that decoherence does not occur for superposition states with $\Delta E\ll E_{\rm P}$. Additionally, \emph{gravitational decoherence occurs in the energy basis}, meaning that a spatial superposition of states $\ket{\vb{r}_1}$ and $\ket{\vb{r}_2}$ will not decohere \emph{if they have the same energy}, for example.

By taking a slightly different path, Anastopoulos and Hu~\cite{Anastopoulos_2013} arrive at the same conclusions. The authors also model the system as a massive scalar field, which interacts with the graviton thermal bath through a $3+1$ decomposition of the total action. After obtaining the master equation, the authors restrict the analysis to the one-particle Hilbert subspace and take the non-relativistic limit, which leads to (within the Born-Markov approximation and in the high temperature limit)
\begin{equation}
    \dv{t}\hat{\rho}_{\mathcal{S}}(t)=-\frac{i}{2m_R}\comm{\hat{P}^2}{\hat{\rho}_{\mathcal{S}}(t)}-\frac{4\pi T_{\rm g}}{9m_R^2}\comm{\hat{P}^2}{\comm{\hat{P}^2}{\hat{\rho}_{\mathcal{S}}(t)}},
\end{equation}
for motion in one spatial dimension. Here $m_R$ is the renormalized mass of the system and $\hat{P}$ is the one-particle momentum operator.

If the initial state is a quantum superposition of two states that are localized in $p_1$ and $p_2$, or equivalently with velocities $v_1$ and $v_2$, decoherence will take place after a time
\begin{equation}
    t_{\rm dec}=\frac{\hbar^2c^5}{Gk_BT_{\rm g}m_R^2V^2(\Delta v)^2},
\end{equation}
where $V=(v_1+v_2)/2$ and $\Delta v=v_1-v_2$. By using the fact that $m_R^2V^2(\Delta v)^2=(\Delta E)^2$, this decoherence time can be easily seen to be equivalent to the one obtained by Blencowe~\cite{Blencowe_2013}, Eq.~\eqref{Dec-time-Blencowe}.

Now, let us turn to the interpretation of the parameter $T_{\rm g}$, which we loosely referred to as the temperature of the graviton bath. As Anastopoulos and Hu~\cite{Anastopoulos_2013} point out, the thermalization of the graviton environment over typical timescales cannot be assumed, since they interact very weakly. Instead, the parameter $T_{\rm g}$ is to be thought of as a phenomenological one that simply characterizes the power spectral density of the gravitational noise, sometimes called \emph{noise temperature}. It serves the same role as the temperature of usual thermal baths, but without necessarily being associated with any precise thermodynamic definition. Nevertheless, the observation of gravitational decoherence, as characterized by the parameter $T_{\rm g}$, would offer valuable information concerning the initial graviton state, which Anastopoulos and Hu~\cite{Anastopoulos_2013} call the "textures of spacetime".

Naturally, the literature on master equations for gravitational decoherence is not limited to the works we discussed here. The ones we outlined in this section were obtained within the Born-Markov approximations in the non-relativistic limit, but they can also be applied to relativistic systems, such as photons, for instance~\cite{Lagouvardos2021}. Additionally, one can also obtain master equations without restricting to scalar fields as the matter system~\cite{HABA2002,Oniga2016}. Furthermore, although we did not address the issue of renormalization in the master equation in this section, there is a discussion regarding which part of the derivation renormalization must take place~\cite{Fahn2025}. Finally, for a non-Markovian master equation for gravitational decoherence that is valid for arbitrary temperatures of the graviton bath, we refer the reader to Ref.~\cite{Cho2025}.

\section{Graviton-induced spatial localization} \label{Sec:Kanno}

In the previous section, we saw that gravitational decoherence occurs in the energy basis. Nevertheless, we can analyze how gravitons may lead to spatial decoherence for a system initially in a superposition of position states, as long as they have different energies. This is precisely the decoherence analysis that was conducted by Kanno et al.~\cite{Kanno2021}.

The approach of Kanno et al.~\cite{Kanno2021} is closer in spirit to our own, which is why we will simply outline some steps towards their conclusions while leaving the details for Part~\ref{Part:Dec-and-ent} of this thesis. It differs from the approaches discussed in Section~\ref{Sec:Master-eq-grav-dec} in two main aspects. First, Kanno et al.~\cite{Kanno2021} describe the quantum system as a non-relativistic point particle from the beginning. From the action for a free particle in curved spacetime, the interaction with gravitons arises from the introduction of Fermi normal coordinates with respect to the geodesic of another test particle and the usual metric expansion around the Minkowski background. Now, the use of this coordinate system works as long as the incoming gravitational radiation has wavelengths that are not smaller than some characteristic separation length, which is why we need to introduce an energy cutoff $\Lambda$ (we will make this clearer in Chapter~\ref{chap:quantum-system}). Second, the decoherence rate is obtained directly from the influence functional without the necessity of deriving a master equation.

By considering a case in which the system can move only along two classically distinguishable paths $\xi^{(1)}(t)$ and $\xi^{(2)}(t)$, and the gravitons to be initially in the vacuum state, Kanno et al.~\cite{Kanno2021} obtained the decoherence function\footnote{The authors work in natural units in which $\hbar=c=1$, but $G$ is held explicit.}
\begin{equation}
    \Gamma(t)\sim\frac{m^2\Lambda^6\Xi^2}{M_{\rm P}^2}\int_0^t\dd t_1\dd t_2\,\Delta\xi(t_1)\Delta\xi(t_2)F_5[\Lambda(t_1-t_2)],
\end{equation}
where $m$ is the mass of the particle, $M_{\rm P}$ is the Planck mass, $\Xi=(\xi^{(1)}+\xi^{(2)})/2$ (assumed to be approximately time independent) and $\Delta\xi(t)=\xi^{(1)}(t)-\xi^{(2)}(t)$. The function $F$ comes from the evaluation of the noise kernel for the initial graviton state, being defined by
\begin{equation}
    F_n(x)\equiv\frac{1}{x^{n+1}}\int_0^x\dd y\,y^n\cos y.
\end{equation}

For the specific configuration of the superposition state, Kanno et al.~\cite{Kanno2021} (see also Ref.~\cite{Breuer2001}) take
\begin{equation}
    \Delta\xi(t')=\left\{
    \begin{array}{ll}
        2vt' &\textrm{for}\hspace{0.2cm}0<t'\leq t/2\\
        2v(t-t') &\textrm{for}\hspace{0.2cm}t/2<t'<t
    \end{array}
    \right.,
\end{equation}
for some constant velocity $v$. One then finds
\begin{equation}
    \Gamma(t)\sim\frac{m^2v^2}{M_{\rm P}^2}f_{\rm v}^{(I)}(\Lambda t),
\end{equation}
where\footnote{The subscript and the superscript will become clear in Chapter~\ref{chap:decoherence}.}
\begin{equation}
    f_{\rm v}^{(I)}(x)=1+\frac{2}{3x}\qty[\sin x-8\sin\qty(\frac{x}{2})]+\frac{1}{x^2}\qty[\frac{2}{3}\cos x-\frac{32}{3}\cos\qty(\frac{x}{2})+10].
\end{equation}
Note that $f_{\rm v}^{(I)}(x)\sim O(1)$, and thus $\Gamma(t)\ll1$ holds as long as $mv\ll M_{\rm P}$, for which case decoherence does not occur. The off-diagonal density matrix elements, whose time evolution goes as $e^{-\Gamma(t)}$, will decay only for systems with momentum greater than the Planck mass, of order $M_{\rm P}\sim10^{-8}$ kg. However, the most massive quantum systems to have ever been put into spatial superposition are molecules with a mass of the order of $m\sim10^{-22}$ kg~\cite{Gerlich2011,Fein2019,Pedalino2025}, which illustrates the immense challenge of observing such gravitational decoherence. Nevertheless, as pointed out by Kanno et al.~\cite{Kanno2021}, the decoherence rate can be enhanced by considering other configurations and also initial squeezed graviton states.

Kanno et al.~\cite{Kanno2021} considered the quantum system to be a point particle. In Part~\ref{Part:Dec-and-ent} of this thesis, we will show how the internal degrees of freedom of a quantum system actually enhance such a mechanism of gravitational decoherence due to the universal character of gravity. For this reason, we will discuss another kind of gravitational decoherence, which occurs due to the coupling between internal and external variables induced by gravitational time dilation.

\section{Decoherence due to gravitational time dilation} \label{Sec:Dec-grav-time-dilation}

Contrary to the works we discussed in the previous sections, gravitational decoherence, as analyzed by Pikovski et al.~\cite{Pikovski2015,Pikovski2017}, is not due to spacetime fluctuations and occurs even in static gravitational fields. The idea is to consider a \emph{composite} quantum system, by which we mean a point-like particle described by a single center-of-mass coordinate that contains dynamical internal degrees of freedom. For instance, a system with internal oscillatory motion (which can be used to track time) models what we understand as a clock. In general, a composite system is described by the state
\begin{equation}
    \ket{\Psi}=\ket{\psi_{\rm ext}}\otimes\ket{\psi_{\rm int}},
\end{equation}
where $\ket{\psi_{\rm ext}}$ describes the quantum state of the center-of-mass variable, while $\ket{\psi_{\rm int}}$ describes the internal degrees of freedom.

According to non-relativistic quantum mechanics, in the absence of interactions between the external and internal variables, the time evolution of the composite system is described by the Hamiltonian
\begin{equation}
    \hat{H}=\hat{H}_{\rm ext}\otimes \hat{I}_{\rm int}+\hat{I}_{\rm ext}\otimes \hat{H}_{\rm int},
\end{equation}
where $\hat{H}_{\rm ext}$ ($\hat{H}_{\rm int}$) dictates the free evolution of the external (internal) degrees of freedom, and $\hat{I}_{\rm ext}$ ($\hat{I}_{\rm int}$) is the identity operator in the external (internal) Hilbert space. The total time evolution is simply
\begin{align}
    \ket{\Psi(t)}&=e^{-i\hat{H}t}\ket{\Psi} \nonumber \\
    &=e^{-i\hat{H}_{\rm ext}t}\ket{\psi_{\rm ext}}\otimes e^{-i\hat{H}_{\rm int}t}\ket{\psi_{\rm int}}.
\end{align}
The final state is still a product state.

The situation is different in a relativistic scenario (both special and general). Let us now consider the composite system in a general background spacetime and choose coordinates $x^\mu=(t,\vb{x})$ to describe its worldline with respect to some laboratory frame. The system will evolve according to its proper time $\tau$. Now, as long as its constituents are enclosed in a spacetime region that is small enough such that one can neglect metric variations along the extension of the composite particle\footnote{This condition needs to be satisfied in order to have a precise definition of center-of-mass in the first place~\cite{Zych2019}.}, the total action will be given by
\begin{equation}
    S=\int L_{\rm rest}\dd\tau,
\end{equation}
where $L_{\rm rest}=-m+\mathscr{L}$, with $\mathscr{L}$ denoting the Lagrangian describing the internal DoFs, and
\begin{equation}
    \dd\tau=\dd t\sqrt{-g_{\mu\nu}\dv{x^\mu}{t}\dv{x^\nu}{t}}.
\end{equation}
Note that gravity couples with the total rest energy of the system, which is compatible with the equivalence principle (see Ref.~\cite{Zych2019} for the precise definition of gravitational mass for composite systems).

Pikovski et al.~\cite{Pikovski2015,Pikovski2017} proceed to consider the Newtonian limit,
\begin{equation}
    g_{\mu\nu}=\eta_{\mu\nu}-2\phi\delta_{\mu\nu},
\end{equation}
with $\phi$ being the static Newtonian potential, while also employing a $c^{-1}-$expansion. This allows us to avoid the full machinery of quantum field theory (in curved spacetime). Up to $O(c^{-2})$ we have
\begin{equation}
    \dd\tau\simeq\dd t\qty[1+\phi(x)-\frac{v^2}{2}],
\end{equation}
with $v$ denoting the velocity of the composite particle in the laboratory frame. The extra terms inside brackets are responsible for the phenomenon of time dilation. Within this limit, the total Hamiltonian describing a composite system of mass $m$ is found to be given by
\begin{equation}
    \hat{H}=\hat{H}_{\rm ext}\otimes \hat{I}_{\rm int}+\hat{I}_{\rm ext}\otimes \hat{H}_{\rm int}+\qty[\phi(\hat{X})-\frac{\Hat{P}^2}{2m^2}]\otimes\hat{H}_{\rm int}.
\end{equation}
For a free particle, for instance, we have
\begin{equation}
    \hat{H}_{\rm ext}=m+\frac{\hat{P}^2}{2m^2}+m\phi(\hat{X}).
\end{equation}
Note that, in this scenario, the time evolution induces a coupling between internal and external degrees of freedom.

If we are interested in the superposition of external variables, such as spatial superposition, then the center-of-mass coordinate is the system of interest while the internal degrees of freedom act as an environment. By integrating out the latter while considering a stationary center-of-mass (in order to neglect momentum terms) and within the Born-Markov approximation, Pikovski et al.~\cite{Pikovski2015,Pikovski2017} obtain the decoherence time for an initial superposition of two different heights $x_1$ and $x_2$ given by
\begin{equation}
    t_{\rm dec}\simeq\frac{\sqrt{2}\hbar}{\Delta E}\frac{c^2}{\Delta\phi},
\end{equation}
where $\Delta E=\sqrt{\expval{\hat{H}_{\rm int}^2}-\expval{\hat{H}_{\rm int}}^2}$, $\Delta\phi=\phi(x_2)-\phi(x_1)$, and we restored the universal constants. For $N$ degrees of freedom in thermal equilibrium at temperature $T_{\rm int}$, the three-dimensional Einstein solid model gives $\Delta E=\sqrt{N}k_BT_{\rm int}$, for instance~\cite{Pikovski2017}, and decoherence increases with the number of internal degrees of freedom and with internal temperature. As is the case with other models that receive the name of gravitational decoherence, the experimental realization of this time dilation induced effect faces the problem of controlling competing sources of decoherence (see Ref.~\cite{Carlesso2016} for a discussion).

The approach of Pikovski et al.~\cite{Pikovski2015,Pikovski2017} shows that gravity can induce decoherence in quantum systems even through indirect manifestations, such as coupling its center-of-mass coordinate with dynamical internal degrees of freedom, without needing to exhibit (classical or quantum) fluctuations itself. But we saw in previous sections that such fluctuations also lead to decoherence. Additionally, these can couple internal and external variables of a quantum composite system in the same way as the static Newtonian potential, and one may wonder how this contributes even further to the loss of quantum coherence in such systems. This will be the subject of Part~\ref{Part:Dec-and-ent} of this thesis, together with the consequences of gravitational decoherence on entropy production in quantum systems.
\let\oldthispagestyle\thispagestyle
\renewcommand{\thispagestyle}[1]{\oldthispagestyle{empty}}
\part{Graviton-induced decoherence and entropy production} \label{Part:Dec-and-ent}
\let\thispagestyle\oldthispagestyle
\chapter{Quantum system interacting with a graviton environment}
\label{chap:quantum-system}

Let us start by considering an open quantum system interacting with both a Newtonian gravitational potential and a bath of gravitons. To be specific, here we assume a non-relativistic quantum particle described by external and internal degrees of freedom (DoFs). This could represent a molecule for which the external DoFs are its center-of-mass coordinate, while the internal DoFs could be its vibrational modes, for instance. We will simply refer to it as a composite particle. Our main goal in this chapter is to obtain an expression for the reduced density matrix describing the external DoFs only. In order to do that, we start with the classical action describing the simultaneous interaction of a composite system with gravitational radiation and a static Newtonian potential in Section~\ref{Sec:Classical action}. Then, in Section~\ref{Sec:Gravitational-influence-functional}, we proceed to integrate over the gravitational variables, treating them as an environment. Finally, we integrate over the internal DoFs of the system in Section~\ref{Sec:External-DoFs-density-matrix} and obtain the desired reduced density matrix at time $t$. This will serve as the starting point for analyzing decoherence and entropy production in subsequent chapters.

\section{The classical action} \label{Sec:Classical action}

We could begin by writing down the Lagrangian of a single composite particle coupled with a classical gravitational wave. However, as discussed in Chapter~\ref{chap:quantum-grav}, a single particle is not enough to probe the effects of a gravitational field, since a particle at rest will remain at rest at all times in the TT frame. Let us start then by writing the classical action of a weak gravitational field coupled to a pair of free-falling massive particles. The total action takes the form
\begin{equation} \label{Total-action-matter+h}
    S=S_{\rm matter}+S_{\rm EH},
\end{equation}
where the first term describes the pair of freely falling test masses, while the second term is the Einstein-Hilbert action, which describes the dynamics of the metric field $g_{\mu\nu}$.

\subsection{The matter action}

By denoting the spacetime coordinates of each particle by $\zeta^\mu$ and $\xi^\mu$, the matter action can be written as~\cite{Zych2019}
\begin{equation} \label{Matter-action}
    S_{\rm matter}=-M\int \dd t\,\sqrt{-g_{\mu\nu}\dot{\zeta}^\mu\dot{\zeta}^\nu}+\int \dd t\,L_{\rm rest}\sqrt{-g_{\mu\nu}\dot{\xi}^\mu\dot{\xi}^\nu},
\end{equation}
where $M$ is the mass of the first particle (with coordinates $\zeta^\mu$), and $L_{\rm rest}$ denotes the rest Lagrangian of the second particle (with coordinates $\xi^\mu$). In writing down the action~\eqref{Matter-action}, we assumed that the first particle has no (dynamical) internal DoFs. For the second particle, the rest Lagrangian is of the form
\begin{equation} \label{L_rest}
    L_{\rm rest}(\varrho,\dot{\varrho}\,\bar{t})=-m+\mathscr{L}(\varrho,\dot{\varrho}\,\bar{t}).
\end{equation}
In this equation, $m$ is the mass of the second particle, while $\mathscr{L}(\varrho,\dot{\varrho}\,\bar{t})$ describes its internal degrees of freedom with coordinate $\varrho$ (relative to the center-of-mass coordinate) and generalized velocity $\dot{\varrho}=\dv*{\varrho}{t}$. We have also defined $\bar{t}=\dv*{t}{\tau}$, with $\tau$ being the particle's proper time.

Now, let us take the first particle to be at rest at the origin of our coordinate system, $\zeta^\mu(t)=t\,\delta_0^\mu$, such that the coordinate time $t$ is interpreted as its proper time, and let us assume that $M\gg L_{\rm rest}$. Under these assumptions, the first term in the action~\eqref{Matter-action} essentially has no dynamics. Since our interest relies on the composite particle, we will simply refer to it as the system from now on.

In the context we described above, it becomes appropriate to think of $(t,\xi^i)$ as the Fermi normal coordinates defined with respect to the worldline of the heavier particle (see Appendix~\ref{app:Differential}). In these coordinates, we can write the metric components as [Eqs.~\eqref{Metric-in-Fermi-Normal-Coordinates}]
\begin{subequations}
\begin{align}
    g_{00}(t,\xi^i)&=-1-R_{i0j0}(t,0)\xi^i\xi^j+O(\xi^3), \\
    g_{0i}(t,\xi^i)&=-\frac{2}{3}R_{0jik}(t,0)\xi^j\xi^k+O(\xi^3), \\
    g_{ij}(t,\xi^i)&=\delta_{ij}-\frac{1}{3}R_{ikjl}(t,0)\xi^k\xi^l+O(\xi^3),
\end{align}
\end{subequations}
where $R_{\mu\nu\rho\sigma}$ is the Riemann curvature tensor. Physically, the use of Fermi normal coordinates allows us to interpret the coordinates $\xi^i$ as not simply describing a single particle in an arbitrary coordinate system, but rather as the geodesic deviation between two test masses (Figure~\ref{Fig:geodesic-deviation}). Following the discussion at the end of Appendix~\ref{app:Differential}, we emphasize that this expansion of the metric components holds as long as $\xi\ll R_0$, with $R_0$ being the scale over which the metric changes appreciably.

\begin{figure}[!ht]
    \centering
    \includegraphics[width=0.35\linewidth]{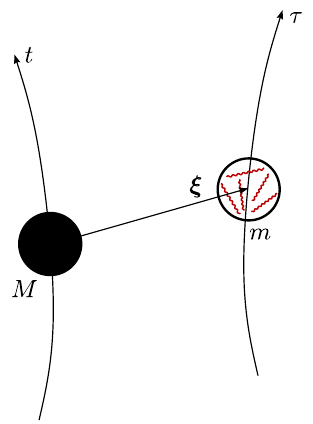}
    \caption{Two test masses $M$ and $m$, with $M\gg m$, and their geodesic deviation in Fermi normal coordinates, represented by the vector $\boldsymbol{\xi}$. The mass $m$ is also described by internal degrees of freedom, represented by the curly red lines.}
    \label{Fig:geodesic-deviation}
\end{figure}

In our parametrization, $\xi^0(t)=t$, thus resulting
\begin{align} \label{dot-tau}
    \sqrt{-g_{\mu\nu}(\xi)\dot{\xi}^\mu\dot{\xi}^\nu}&\simeq\sqrt{1+R_{i0j0}(t,0)\xi^i\xi^j-\delta_{ij}\dot{\xi}^i\dot{\xi}^j} \nonumber \\
    &\simeq1-\frac{1}{2}\delta_{ij}\dot{\xi}^i\dot{\xi}^j+\frac{1}{2}R_{i0j0}(t,0)\xi^i\xi^j.
\end{align}

Next, let us specify our metric field to describe small perturbations $h_{\mu\nu}$ around some background spacetime $\gamma_{\mu\nu}$ as in Eq.~\eqref{g=gamma+h}, namely $g_{\mu\nu}=\gamma_{\mu\nu}+h_{\mu\nu}$, with $\abs{h_{\mu\nu}}\ll\abs{\gamma_{\mu\nu}}$. Recall from Eq.~\eqref{Riemann-in-terms-of-background-Riemann} that the Riemann tensor associated with $g_{\mu\nu}$ is given by
\begin{equation}
    {R^\rho}_{\sigma\mu\nu}={R^\rho}_{\sigma\mu\nu}[\gamma]+2D_{[\mu}{C^\rho}_{\nu]\sigma}+2{C^\rho}_{\delta[\mu}{C^\delta}_{\nu]\sigma}.
\end{equation}
where $D_\mu$ is the covariant derivative compatible with $\gamma_{\mu\nu}$ and whose commutator defines ${R^\rho}\,_{\sigma\mu\nu}[\gamma]$. The tensor ${C^\rho}_{\mu\nu}$ reads, up to second order in the metric perturbation,
\begin{align}
    {C^\rho}_{\mu\nu}&=\frac{1}{2}g^{\rho\sigma}\qty(D_\mu g_{\nu\sigma}+D_\nu g_{\sigma\mu}-D_\sigma g_{\mu\nu}) \nonumber \\
    &=\frac{1}{2}\gamma^{\rho\sigma}\qty(D_\mu h_{\nu\sigma}+D_\nu h_{\sigma\mu}-D_\sigma h_{\mu\nu}),
\end{align}
which follows from metric compatibility. Then, an explicit calculation yields
\begin{align}
    R_{\rho\sigma\mu\nu}&=R_{\rho\sigma\mu\nu}[\gamma]+\frac{1}{2}\qty({R^\lambda}\,_{\sigma\mu\nu}[\gamma]h_{\lambda\rho}-{R^\lambda}\,_{\rho\mu\nu}[\gamma]h_{\sigma\lambda}) \nonumber \\
    &+\frac{1}{2}\qty(D_\mu D_\sigma h_{\rho\nu}-D_\mu D_\rho h_{\nu\sigma}-D_\nu D_\sigma h_{\rho\mu}+D_\nu D_\rho h_{\mu\sigma}),
\end{align}
where $R_{\rho\sigma\mu\nu}=g_{\rho\eta}{R^\eta}_{\sigma\mu\nu}$ as usual. The interaction between the system variables and the spacetime curvature then takes the form
\begin{align} \label{Rxixi}
    \frac{1}{2}R_{i0j0}(t,0)\xi^i\xi^j&=\frac{1}{2}R_{i0j0}[\gamma](t,0)\xi^i\xi^j+\frac{1}{4}\eval{\qty({R^\lambda}\,_{0i0}[\gamma]h_{\lambda j}-{R^\lambda}\,_{ij0}[\gamma]h_{0\lambda})}_{\xi=0}\xi^i\xi^j \nonumber \\
    &+\frac{1}{4}\eval{\qty(D_iD_0h_{j0}+D_0D_ih_{j0}-D_iD_jh_{00}-D_0D_0h_{ij})}_{\xi=0}\xi^i\xi^j.
\end{align}

\begin{figure}[!ht]
    \centering
    \includegraphics[width=0.45\linewidth]{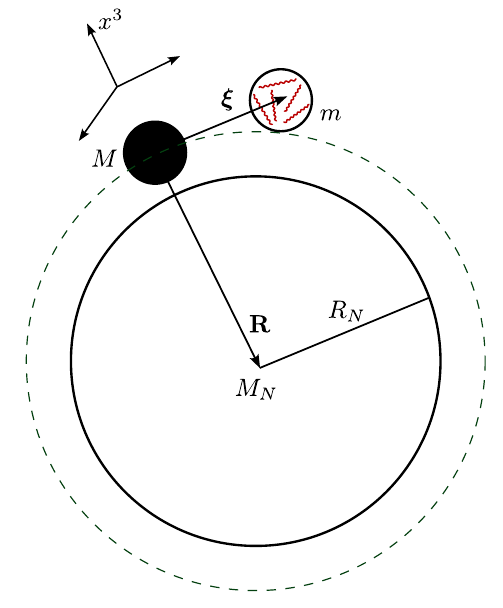}
    \caption{The two test masses $M$ and $m$ are in the vicinity of a much bigger and much more massive spherical mass $M_N$, with radius $R_N$, located at $\vb{R}\simeq R_N\vu{e}_3$ with respect to the mass $M$.}
    \label{Fig:Newtonian-source}
\end{figure}

Up to this point, everything works for a general background. Ultimately, we are interested in considering the background metric in the Newtonian limit, for which $\gamma_{\mu\nu}=\eta_{\mu\nu}-2\phi\delta_{\mu\nu}$, with $\phi(x)$ being the time-independent gravitational potential. In that case, as we saw in Chapter~\ref{chap:quantum-grav}, the metric perturbation $h_{\mu\nu}$ can be chosen to satisfy the TT gauge conditions\footnote{Since we will only work within the TT gauge, we will not use the superscript 'TT'. From this point on, the metric perturbation field $h_{ij}$ is understood to satisfy the TT gauge conditions, Eq.~\eqref{TT-gauge-curved}.}~\eqref{TT-gauge-curved}, and we find
\begin{equation}
    R_{i0j0}[\gamma](\xi)=\partial_j\partial_i\phi(\xi).
\end{equation}
Putting everything back together in Eq.~\eqref{Rxixi} yields
\begin{equation}
    \frac{1}{2}R_{i0j0}(t,0)\xi^i\xi^j=\eval{\qty[\frac{1}{2}\partial_i\partial_j\phi+\frac{1}{4}\qty(\partial^k\partial_i\phi)h_{kj}-\frac{1}{4}D_0D_0h_{ij}]}_{\xi=0}\xi^i\xi^j.
\end{equation}
The covariant derivatives in the last term differ from partial derivatives by the Christoffel symbols. Since those are evaluated along the geodesic, they vanish by Fermi's second condition [Eq.~\eqref{Fermi-second-condition}] and, finally, the interaction term can be written as
\begin{equation} \label{Rxixi-2}
    \frac{1}{2}R_{i0j0}(t,0)\xi^i\xi^j=-\qty[\frac{1}{2}\Phi_{ij}+\frac{1}{4}{\Phi^k}_ih_{kj}(t,0)+\frac{1}{4}\Ddot{h}_{ij}(t,0)]\xi^i\xi^j,
\end{equation}
where we introduced the \textbf{tidal tensor} $\Phi_{ij}=-\eval{\qty(\partial_i\partial_j\phi)}_{\xi=0}$~\cite{Hartle2003,Cho_2023}. Specifically, for two test masses close to a spherically symmetric (even if only approximately) Newtonian source, such as Earth, the much lighter one is under the influence of the gravitational potential given by
\begin{equation} \label{Newtonian-potential}
    \phi(\boldsymbol{\xi})=-\frac{M_N}{\abs{\boldsymbol{\xi}-\vb{R}}},
\end{equation}
with $M_N$ being the source's mass and $\vb{R}$ being the radius vector that points from the mass $M$ (the origin of our coordinate system) to the center of the Newtonian source, such that $\abs{\vb{R}}\simeq R_N$, which represents its radius (Figure~\ref{Fig:Newtonian-source}). From Eq.~\eqref{Newtonian-potential} one can easily show that the tidal tensor takes the form
\begin{align} \label{Tidal-tensor}
    \Phi_{ij}&=-\eval{\qty(\partial_i\partial_j\phi)}_{\xi=0}=\frac{M_N}{R^3}\qty(\frac{3R_iR_j}{R^2}-\delta_{ij}) \nonumber \\
    &=\frac{M_N}{R_N^3}(3\delta_{i3}\delta_{j3}-\delta_{ij}).
\end{align}

Note from Eq.~\eqref{Rxixi-2} that the validity of the metric expansion in Fermi normal coordinates holds as long as $\xi^2\ll R_N^3/M_N$, and also $\xi^2\ll\omega^{-2}$, with $\omega$ denoting the angular frequency of the gravitational waves. Hence, we need to introduce an energy cutoff for the incident gravitational radiation of the order $\Lambda\sim L_0^{-1}$, where $L_0$ is some typical geodesic separation, which we sometimes refer to as the "detector size".

At last, using Eqs.~\eqref{dot-tau} and~\eqref{Rxixi-2}, the matter action becomes
\begin{align} \label{Matter-action-2}
    S_{\rm matter}&=\int\dd t\,\qty[\frac{1}{2}m\delta_{ij}\dot{\xi}^i\dot{\xi}^j+\mathscr{L}(\varrho,\dot{\varrho}\Bar{t})-\frac{1}{2}\mathscr{L}(\varrho,\dot{\varrho}\Bar{t})\delta_{ij}\dot{\xi}^i\dot{\xi}^j] \nonumber \\
    &-\frac{1}{2}\int\dd t\,L_{\rm rest}(\varrho,\dot{\varrho}\Bar{t})\Phi_{ij}\xi^i\xi^j \nonumber \\
    &-\frac{1}{4}\int\dd t\,h_{ij}(t,0)\qty{L_{\rm rest}(\varrho,\dot{\varrho}\Bar{t}){\Phi^i}_k\xi^k\xi^j+\dv[2]{t}\qty[L_{\rm rest}(\varrho,\dot{\varrho}\Bar{t})\xi^i\xi^j]},
\end{align}
where we dropped non-dynamical terms and integrated by parts while dropping boundary contributions.

\subsection{The gravitational action}

Now, for the gravitational field, we saw in Chapter~\ref{chap:quantum-grav} that the Einstein-Hilbert action for the metric expansion around a Newtonian background, and with the perturbation in the TT gauge, is given by Eq.~\eqref{Graviton-action-position-space}, which we repeat here for convenience:
\begin{equation} \label{Graviton-action-position-space-again}
    S_{\rm EH}=\frac{1}{64\pi}\int\dd^4x\,\qty(h_{ij}\Box h^{ij}+2\phi\,h_{ij}\delta_{\mu\nu}\partial^\mu\partial^\nu h^{ij}).
\end{equation}
Note that the interaction between the gravitational radiation and the classical Newtonian potential is of order $O(\phi h^2)$, while in Eq.~\eqref{Matter-action-2} they simultaneously couple with the system through an interaction of order $O(\phi h)$, which is therefore the dominant one in perturbation theory. The interaction term in Eq.~\eqref{Graviton-action-position-space-again} leads to coupling between graviton modes. In fact, we showed in Appendix~\ref{app:Scattering} that this term is physically associated with graviton scattering, leading to a differential cross section that behaves as $\dv*{\sigma}{\Omega}\sim M_N^2\theta^{-4}$, with $\theta\ll1$ being the scattering angle. Since this angle is very small, we will neglect this less dominant interaction\footnote{This interaction term is also neglected in Ref.~\cite{Chawla2023} where the authors compute quantum gravity corrections to the fall of test masses, for instance.} and consider only the $O(\phi h)$ contribution in Eq.~\eqref{Matter-action-2}.

Hence, we consider the gravitons to be described by their usual flat-spacetime Lagrangian,
\begin{equation} \label{Eistein-Hilbert-action-TT-gauge}
    S_{\rm EH}=-\frac{1}{64\pi}\int \dd^4x\,\partial_\mu h_{ij}\partial^\mu h^{ij},
\end{equation}
from which we immediately obtain the wave equation $\Box h_{ij}=0$. The general solution takes the form
\begin{equation} \label{Fourier-transform}
    h_{ij}(t,\vb{x})=\int \dd^3k\,\sum_{s=+,\cross}\epsilon_{ij}^{s}(\vb{k})q_s(t,\vb{k})e^{i\vb{k}\cdot\vb{x}},
\end{equation}
with $\epsilon_{ij}^{s}$ denoting the polarization tensor that satisfies the transversality and traceless conditions, Eqs.~\eqref{Polarization-tensor-conditions}, as well as the normalization condition~\eqref{Polarization-tensor-normalization}. Additionally, the reality of $h_{ij}$ implies that
\begin{equation} \label{Implication-of-reality-of-h-ij}
    \epsilon_{ij}^{*s}(\vb{k})q_s^*(t,\vb{k})=\epsilon_{ij}^{s}(-\vb{k})q_s(t,-\vb{k}).
\end{equation}

By plugging Eq.~\eqref{Fourier-transform} into Eq.~\eqref{Eistein-Hilbert-action-TT-gauge}, we obtain
\begin{align} \label{Graviton-action-Fourier-space}
    S_{\rm EH}&=-\frac{1}{64\pi}\int\dd^4x\,\int\dd^3k\,\dd^3k'\,\sum_{s,s'}\epsilon_{ij}^s(\vb{k})\epsilon^{ij}_{s'}(\vb{k}') \nonumber \\
    &\hspace{1cm}\times\qty[-\dot{q}_s(t,\vb{k})\dot{q}_{s'}(t,\vb{k}')-\vb{k}\cdot\vb{k}'q_s(t,\vb{k})q_{s'}(t,\vb{k}')]e^{i\qty(\vb{k}+\vb{k}')\cdot\vb{x}} \nonumber \\
    &=-\frac{(2\pi)^3}{64\pi}\int\dd t\int\dd^3k\,\sum_{s,s'}\epsilon_{ij}^s(\vb{k})\epsilon^{ij}_{s'}(-\vb{k}) \nonumber \\
    &\hspace{1cm}\times\qty[-\dot{q}_s(t,\vb{k})\dot{q}_{s'}(t,-\vb{k})+\vb{k}^2q_s(t,\vb{k})q_{s'}(t,-\vb{k})] \nonumber \\
    &=\frac{\pi^2}{4}\int\dd t\int\dd^3k\,\sum_s\qty[\abs{\dot{q}_s(t,\vb{k})}^2-\vb{k}^2\abs{q_s(t,\vb{k})}^2] \nonumber \\
    &=\int\dd t\int\dd^3k\,\sum_s\frac{1}{2}\qty[m_{\rm g}\abs{\dot{q}_s(t,\vb{k})}^2-m_{\rm g}\vb{k}^2\abs{q_s(t,\vb{k})}^2],
\end{align}
where we defined $m_{\rm g}=\pi^2/2$ and used Eq.~\eqref{Implication-of-reality-of-h-ij}, as well as the normalization condition~\eqref{Polarization-tensor-normalization}, and the identity $\int\dd^3x\,e^{i\qty(\vb{k}+\vb{k}')\cdot\vb{x}}=(2\pi)^3\delta^3(\vb{k}+\vb{k}')$.

Let us note that the action~\eqref{Graviton-action-Fourier-space} has the form $S_{\rm EH}=\int\dd t\int\dd^3k\,\sum_sL_s(t,\vb{k})$, with the Lagrangian $L_s(t,\vb{k})$ for each mode describing a harmonic oscillator with mass $m_{\rm g}$ and frequency $\omega=\abs{\vb{k}}$. The associated Hamiltonian takes the usual form
\begin{equation} \label{Hamiltonian-individual-mode}
    H_s(t,\vb{k})=\frac{\abs{p_s(t,\vb{k})}^2}{2m_{\rm g}}+\frac{1}{2}m_{\rm g}\vb{k}^2\abs{q_s(t,\vb{k})}^2,
\end{equation}
with $p_s(t,\vb{k})=\pdv{L_s(t,\vb{k})}{\dot{q}_s(t,\vb{k})}=m_{\rm g}\dot{q}_s(t,\vb{k})$.

\subsection{The total action}

At last, by plugging the general solution~\eqref{Fourier-transform} into the matter action~\eqref{Matter-action-2}, the total action~\eqref{Total-action-matter+h} takes the form
\begin{equation} \label{Total-action}
    S=S_{\rm sys}+S_{\rm grav}+S_{\textrm{s}+\textrm{g}},
\end{equation}
where
\begin{subequations}
\begin{equation} \label{System-action}
    S_{\rm sys}=\int\dd t\,\qty[\frac{1}{2}m\delta_{ij}\dot{\xi}^i\dot{\xi}^j+\mathscr{L}(\varrho,\dot{\varrho}\Bar{t})-\frac{1}{2}\mathscr{L}(\varrho,\dot{\varrho}\Bar{t})\delta_{ij}\dot{\xi}^i\dot{\xi}^j-\frac{1}{2}L_{\rm rest}(\varrho,\dot{\varrho}\Bar{t})\Phi_{ij}\xi^i\xi^j],
\end{equation}
\begin{equation}
    S_{\rm grav}=\int\dd t\int\dd^3k\,\sum_s\frac{1}{2}\qty[m_{\rm g}\abs{\dot{q}_s(t,\vb{k})}^2-m_{\rm g}\vb{k}^2\abs{q_s(t,\vb{k})}^2],
\end{equation}
\begin{equation} \label{Interaction-action}
    S_{\textrm{s}+\textrm{g}}=\int\dd t\int\dd^3k\sum_sq_s(t,\vb{k})X^s(t,\vb{k}),
\end{equation}
\end{subequations}
and we have defined
\begin{equation} \label{System-Xs-variable}
    X^s(t,\vb{k})=-\frac{1}{4}\epsilon_{ij}^s(\vb{k})\qty{L_{\rm rest}(\varrho,\dot{\varrho}\Bar{t}){\Phi^i}_k\xi^k\xi^j+\dv[2]{t}\qty[L_{\rm rest}(\varrho,\dot{\varrho}\Bar{t})\xi^i\xi^j]}.
\end{equation}

The action~\eqref{Total-action} describes the interaction of a system, characterized by both external and internal DoFs, with classical gravitational radiation in a Newtonian background. This will be our starting point for describing the interaction of such a system with gravitons. The quantization of the metric perturbation field, which we discussed in Chapter~\ref{chap:quantum-grav}, can be achieved by promoting the field amplitudes to operators in Hilbert space in the usual way. Since our ultimate goal is to describe the system variables alone, the weak quantized gravitational field shall be treated as an environment; for this case, we will follow the Feynman-Vernon influence functional approach to open quantum systems, as described in Chapter~\ref{chap:dec-and-FV}. In fact, one can anticipate the resemblance to quantum Brownian motion from the total action~\eqref{Total-action}.

\section{Gravitational influence functional} \label{Sec:Gravitational-influence-functional}

Let us now quantize both the system and the gravitational degrees of freedom. Suppose that, at initial time $t=0$, the system and the gravitons are uncorrelated, and the total density matrix can be written as
\begin{equation}
    \hat{\rho}(0)=\hat{\rho}_{\rm sys}(0)\otimes\hat{\rho}_{\rm grav}(0),
\end{equation}
with $\hat{\rho}_{\rm sys}(t)$ ($\hat{\rho}_{\rm grav}(t)$) denoting the reduced density matrix used to compute local system (gravitational) observables. This is obtained by taking a partial trace of the total density matrix with respect to the gravitational (system) variables.

For $t>0$, the system interacts with the gravitational field, and the total state becomes entangled in general. Since we are not interested in the gravitons final state, we must integrate over the gravitational variables such that the system becomes effectively open, and the weak quantum gravitational field is viewed as an environment. This description is accomplished by using the Feynman-Vernon influence functional (Chapter~\ref{chap:dec-and-FV}), which is written as
\begin{equation}
    \mathcal{F}[X,X']=e^{iS_{\rm IF}[X,X',t]},
\end{equation}
where $S_{\rm IF}$ is the influence action and $X$ and $X'$ denote two different histories of the system variables. In terms of the influence functional, the time evolution of the system reduced density matrix is given by
\begin{equation}
    \rho_{\rm sys}(X,X',t)=\int \dd X_0\dd X_0'\,\mathcal{J}_{\rm sys}(X,X',t|X_0,X_0',0)\rho_{\rm sys}(X_0,X_0',0),
\end{equation}
where $\rho_{\rm sys}(X,X',t)=\mel{X}{\hat{\rho}_{\rm sys}(t)}{X'}$ and the evolution operator for the reduced density matrix reads
\begin{equation}
    \mathcal{J}_{\rm sys}(X,X',t|X_0,X_0',0)\equiv\int\displaylimits_{\substack{X(0)\,=\,X_0 \\ X'(0)\,=\,X_0'}}^{\substack{X(t)\,=\,X \\ X'(t)\,=\,X'}}\mathcal{D}X\mathcal{D}X'\,e^{i\qty(S_{\rm sys}[X]-S_{\rm sys}[X'])}\mathcal{F}[X,X'],
\end{equation}
with
\begin{align}
    \mathcal{F}[X,X']&=\int_{-\infty}^\infty \dd q\dd q_0\dd q_0'\,\rho_{\rm grav}(q_0,q_0',0) \nonumber \\
    &\times\int\displaylimits_{\substack{q(0)\,=\,q_0 \\ q'(0)\,=\,q_0'}}^{q(t)\,=\,q'(t)\,=\,q}\mathcal{D}q\mathcal{D}q'\,e^{i\qty(S_{\rm grav}[q]+S_{\textrm{s}+\textrm{g}}[X,q]-S_{\rm grav}[q']-S_{\textrm{s}+\textrm{g}}[X',q'])},
\end{align}
with $q$ and $q'$ denoting two different histories of the gravitational environment variables.

The influence action encodes all influence of the environment on the system. In our case, the system is coupled with the infinite graviton modes. Since each mode (and polarization) is independent of all the others, they can be treated separately in such a way that the total influence action is the sum of the action corresponding to each mode\footnote{Note that this would no longer be true if we had kept the scattering term in the graviton action.} (and polarization). This is because the total influence functional for a system coupled with statistically and dynamically independent environments is simply the product of each individual influence functional [Eq.~\eqref{Independent-environments}]. Furthermore, we note that the environment action (the Einstein-Hilbert action) is quadratic in the field amplitudes, and the coupling with the system variable $X^s(t,\vb{k})$ is linear. This is then a special case of the linear coupling model, described in full detail in Section~\ref{Subsec:linear-coupling-model}, for which the path integrals can be computed analytically. Translating the results of that section in terms of the variables described by Eq.~\eqref{Total-action} leads to the influence action
\begin{align} \label{Grav-influence-action-1}
    S_{\rm IF}[X,X']&=\frac{1}{2}\int\dd t\dd t'\sum_s\int\dd^3k\,\qty[X^s(t,\vb{k})-{X'}^s(t,\vb{k})]d_s(t,t',\vb{k}) \nonumber \\
    &\hspace{5cm}\times\qty[X^s(t',\vb{k})+{X'}^s(t',\vb{k})] \nonumber \\
    &+\frac{i}{2}\int\dd t\dd t'\sum_s\int\dd^3k\,\qty[X^s(t,\vb{k})-{X'}^s(t,\vb{k})]n_s(t,t',\vb{k}) \nonumber \\
    &\hspace{5cm}\times\qty[X^s(t',\vb{k})-{X'}^s(t',\vb{k})],
\end{align}
where
\begin{subequations} \label{Grav-Dissipation-and-noise-kernels}
\begin{equation} \label{Grav-Dissipation-kernel}
    d_s(t,t',\vb{k})=i\expval{\comm{\hat{q}_s(t,\vb{k})}{\hat{q}_s(t',\vb{k})}}_{\rm g}\theta(t-t')
\end{equation}
\begin{equation} \label{Grav-Noise-kernel}
    n_s(t,t',\vb{k})=\frac{1}{2}\expval{\acomm{\hat{q}_s(t,\vb{k})}{\hat{q}_s(t',\vb{k})}}_{\rm g}
\end{equation}
are the \emph{dissipation} and \emph{noise kernels}.
\end{subequations}
In Eqs.~\eqref{Grav-Dissipation-and-noise-kernels}, the $\hat{q}$'s stand for position operators in the Heisenberg picture, and $\theta(x)$ is the Heaviside step function. The expectation values with the subscript 'g' are computed with respect to the initial state of the gravitons.

Both terms in the influence action~\eqref{Grav-influence-action-1} have the generic form
\begin{align}
    I(t,t')&=\int\dd t\dd t'\sum_s\int\dd^3k\,\qty[X^s(t,\vb{k})-{X'}^s(t,\vb{k})]j_s(t,t',\vb{k})\nonumber \\
    &\hspace{5cm}\times\qty[X^s(t',\vb{k})\pm{X'}^s(t',\vb{k})],
\end{align}
with $j_s(t,t',\vb{k})$ denoting some kernel (either dissipation or noise) and $X^s(t,\vb{k})$ defined in Eq.~\eqref{System-Xs-variable}. After some algebra, integrating by parts on the time variables and dropping second order terms on the Newtonian potential, this can be rewritten as
\begin{equation}
    I(t,t')=\int\dd t\dd t'\,\qty[x_{ij}(t)-x'_{ij}(t)]J^{ijkl}(t,t')\qty[x_{kl}(t')\pm x'_{kl}(t')],
\end{equation}
where
\begin{equation} \label{x-ij(t)-definition}
    x_{ij}(t)=L_{\rm rest}(\varrho,\dot{\varrho}\Bar{t})\xi_i(t)\xi_j(t)
\end{equation}
and
\begin{align}
    J_{ijkl}(t,t')&=\frac{1}{16}\sum_s\int\dd^3k\,\left[ \epsilon_{ij}(\vb{k})\epsilon_{kl}(\vb{k})\dv[2]{t}\dv[2]{{t'}}j_s(t,t',\vb{k})\right. \nonumber \\
    &\left. +\epsilon_{ij}(\vb{k})\epsilon_{nl}(\vb{k}){\Phi^n}_k\dv[2]{t}j_s(t,t',\vb{k})+\epsilon_{nj}(\vb{k})\epsilon_{kl}(\vb{k}){\Phi^n}_i\dv[2]{{t'}}j_s(t,t',\vb{k})\right] .
\end{align}

Then, after putting everything together, we find the influence action to be given by
\begin{equation} \label{Grav-influence-action-2}
\begin{split}
    S_{\rm IF}[x,x']=\int \dd t\dd t'\left\{ \frac{1}{2}\qty[x_{ij}(t)-x'_{ij}(t)]D^{ijkl}_{\rm g}(t,t')\qty[x_{kl}(t')+x'_{kl}(t')] \right. \\
    \left. +\frac{i}{2}\qty[x_{ij}(t)-x'_{ij}(t)]N^{ijkl}_{\rm g}(t,t')\qty[x_{kl}(t')-x'_{kl}(t')]\right\} ,
\end{split}
\end{equation}
where we have defined the \emph{gravitational dissipation} and \emph{noise kernels},
\begin{subequations} \label{Gravitational-dissipation-and-noise-kernels}
\begin{equation} \label{Gravitational-dissipation-kernel}
\begin{split}
    D_{\rm g}^{ijkl}(t,t')&=\frac{1}{16}\sum_s\int\dd^3k\,\left[ \epsilon^{ij}(\vb{k})\epsilon^{kl}(\vb{k})\dv[2]{t}\dv[2]{{t'}}d_s(t,t',\vb{k})\right. \\
    &\hspace{0.3cm}\left. +\epsilon^{ij}(\vb{k})\epsilon^{nl}(\vb{k}){\Phi_n}^k\dv[2]{t}d_s(t,t',\vb{k})+\epsilon^{nj}(\vb{k})\epsilon^{kl}(\vb{k}){\Phi_n}^i\dv[2]{{t'}}d_s(t,t',\vb{k})\right] ,
\end{split}
\end{equation}
\begin{equation} \label{Gravitational-noise-kernel}
\begin{split}
    N_{\rm g}^{ijkl}(t,t')&=\frac{1}{16}\sum_s\int\dd^3k\,\left[ \epsilon^{ij}(\vb{k})\epsilon^{kl}(\vb{k})\dv[2]{t}\dv[2]{{t'}}n_s(t,t',\vb{k})\right. \\
    &\hspace{0.3cm}\left. +\epsilon^{ij}(\vb{k})\epsilon^{nl}(\vb{k}){\Phi_n}^k\dv[2]{t}n_s(t,t',\vb{k})+\epsilon^{nj}(\vb{k})\epsilon^{kl}(\vb{k}){\Phi_n}^i\dv[2]{{t'}}n_s(t,t',\vb{k})\right] ,
\end{split}
\end{equation}
with $d_s(t,t',\vb{k})$ and $n_s(t,t',\vb{k})$ defined in Eqs.~\eqref{Grav-Dissipation-and-noise-kernels}.
\end{subequations}

Now, in the same way we proceeded in Section~\ref{Sec:Langevin-equation}, it will be useful to express the noise kernel contribution in terms of a stochastic variable $\mathcal{N}_{ij}(t)$ using the Gaussian functional identity~\cite{Cho2022}
\begin{align}
    &e^{-\frac{1}{2}\int \dd t\dd t'\,\qty[x_{ij}(t)-x'_{ij}(t)]N^{ijkl}_{\rm g}(t,t')\qty[x_{kl}(t')-x'_{kl}(t')]} \nonumber \\
    &\hspace{1cm}=\mathcal{C}\int\mathcal{D}\mathcal{N}\,e^{-\frac{1}{2}\int \dd t\dd t'\,\mathcal{N}_{ij}(t)(N_g^{-1})^{ijkl}(t,t')\mathcal{N}_{kl}(t')}e^{i\int \dd t\,\mathcal{N}^{ij}(t)\qty[x_{ij}(t)-x'_{ij}(t)]},
\end{align}
where $\mathcal{C}$ is a normalization constant, and $\mathcal{D}\mathcal{N}$ denotes the path integral measure for the stochastic variable $\mathcal{N}_{ij}(t)$. Stochastic averages are then computed using a Gaussian probability density $\mathscr{P}[\mathcal{N}]$, for which we have
\begin{subequations}  \label{Stochastic-averages}
\begin{equation}
    \expval{\mathcal{N}^{ij}(t)}_{\rm sto}=\int\mathcal{D}\mathcal{N}\,\mathscr{P}[\mathcal{N}]\mathcal{N}^{ij}(t)=0,
\end{equation}
\begin{equation}
    \expval{\mathcal{N}^{ij}(t)\mathcal{N}^{kl}(t')}_{\rm sto}=\int\mathcal{D}\mathcal{N}\,\mathscr{P}[\mathcal{N}]\mathcal{N}^{ij}(t)\mathcal{N}^{kl}(t')=N_{\rm g}^{ijkl}(t,t').
\end{equation}
\end{subequations}

The gravitational influence functional then becomes
\begin{align} \label{Influence-functional-final}
    e^{iS_{\rm IF}[x,x']}&=\int\mathcal{D}\mathcal{N}\,\mathscr{P}[\mathcal{N}]\,e^{i\int \dd t\,\mathcal{N}^{ij}(t)\qty[x_{ij}(t)-x'_{ij}(t)]} \nonumber \\
    &\hspace{0.5cm}\times e^{\frac{i}{2}\int \dd t\dd t'\qty[x_{ij}(t)-x'_{ij}(t)]D^{ijkl}_{\rm g}(t,t')\qty[x_{kl}(t')+x'_{kl}(t')]}.
\end{align}
Let us now note that the term involving the dissipation kernel in Eq.~\eqref{Influence-functional-final} is of order $O(\xi^4)$ since $x_{ij}$ is already of order $O(\xi^2)$. Thus, the leading order contribution comes from the term involving the noise kernel, and we may approximate Eq.~\eqref{Influence-functional-final} as
\begin{equation} \label{Influence-functional-final-approximated}
    e^{iS_{\rm IF}[x,x']}\simeq\int\mathcal{D}\mathcal{N}\,\mathscr{P}[\mathcal{N}]\,e^{i\int \dd t\,\mathcal{N}^{ij}(t)\qty[x_{ij}(t)-x'_{ij}(t)]}.
\end{equation}
This means that we are only considering the influence of the gravitational field encoded in the noise kernel, which is responsible for describing decoherence, as we saw in Chapter~\ref{chap:dec-and-FV}, and whose explicit form depends on the initial state of the gravitons. In Appendix~\ref{app:noise-kernels}, we obtain explicit expressions for the gravitational noise kernel by considering four different possible initial states: vacuum, thermal, coherent, and squeezed states.

\section{The external DoFs density matrix} \label{Sec:External-DoFs-density-matrix}

The total density matrix of the particle (including external and internal degrees of freedom) at time $t$ is given by
\begin{align}
    \rho_{\rm sys}(\xi,\varrho,\xi',\varrho,t)=\int \dd\xi(0)\dd\xi'(0)\dd\varrho(0)\dd\varrho'(0)\,\rho_{\rm sys}(\xi(0),\varrho(0),\xi'(0),\varrho'(0),0) \nonumber \\
    \times\int\mathcal{D}\xi\mathcal{D}\xi'\mathcal{D}\varrho\mathcal{D}\varrho'\,e^{i\qty(S_{\rm sys}[\xi,\varrho]-S_{\rm sys}[\xi',\varrho'])}e^{iS_{\rm IF}[\xi,\varrho,\xi',\varrho']},
\end{align}
with $e^{iS_{\rm IF}}$ given in Eq.~\eqref{Influence-functional-final} in terms of the variable $x_{ij}(t)$, defined in Eq.~\eqref{x-ij(t)-definition}, and
\begin{align}
    S_{\rm sys}[\xi,\varrho]&=\frac{1}{2}m\int \dd t\,\qty(\delta_{ij}\dot{\xi}^i\dot{\xi}^j+\Phi_{ij}\xi^i\xi^j)+\int \dd t\,\mathscr{L}(\varrho,\dot{\varrho}\Bar{t}) \nonumber \\
    &-\frac{1}{2}\int \dd t\,\mathscr{L}(\varrho,\dot{\varrho}\Bar{t})\qty(\delta_{ij}\dot{\xi}^i\dot{\xi}^j+\Phi_{ij}\xi^i\xi^j).
\end{align}
The path integral over $\xi(t)$ is taken from $\xi(0)$ to $\xi$, and similarly for the others.

Now, within the approximation~\eqref{Influence-functional-final-approximated}, we find
\begin{align} \label{Total-density-matrix-of-system}
    \rho_{\rm sys}(\xi,\varrho,\xi',\varrho',t)=\int \dd\xi(0)\dd\xi'(0)\dd\varrho(0)\dd\varrho'(0)\,\rho_{\rm sys}(\xi(0),\varrho(0),\xi'(0),\varrho'(0),0) \nonumber \\
    \times\int\mathcal{D}\xi\mathcal{D}\xi'\mathcal{D}\varrho\mathcal{D}\varrho'\mathcal{D}\mathcal{N}\,\mathscr{P}[\mathcal{N}]\,e^{i\qty(S_{\rm eff}[\xi,\varrho,\mathcal{N}]-S_{\rm eff}[\xi',\varrho',\mathcal{N}])},
\end{align}
where
\begin{align} \label{Effective-system-action-1}
    S_{\rm eff}[\xi,\varrho,\mathcal{N}]&=\int\dd t\,\qty[\frac{1}{2}m\delta_{ij}\dot{\xi}^i\dot{\xi}^j+m\qty(\frac{1}{2}\Phi_{ij}-\mathcal{N}_{ij})\xi^i\xi^j]+\int\dd t\,\mathscr{L}(\varrho,\dot{\varrho}\Bar{t}) \nonumber \\
    &-\int\dd t\,\mathscr{L}(\varrho,\dot{\varrho}\Bar{t})\qty[\frac{1}{2}\delta_{ij}\dot{\xi}^i\dot{\xi}^j+\qty(\frac{1}{2}\Phi_{ij}-\mathcal{N}_{ij})\xi^i\xi^j],
\end{align}
with $\mathcal{N}_{ij}(t)$ being the Gaussian stochastic variable satisfying Eqs.~\eqref{Stochastic-averages}. Let us now note that the tidal tensor can be absorbed into the stochastic variable by making $\mathcal{N}_{ij}(t)\to\mathcal{N}_{ij}(t)-\frac{1}{2}\Phi_{ij}$ in Eq.~\eqref{Total-density-matrix-of-system}. Since this is a linear transformation, the new variable is still Gaussian with stochastic averages
\begin{subequations}  \label{New-Stochastic-averages}
\begin{equation}
    \expval{\mathcal{N}^{ij}(t)}_{\rm sto}=-\frac{1}{2}\Phi_{ij},
\end{equation}
\begin{equation}
    \expval{\mathcal{N}^{ij}(t)\mathcal{N}^{kl}(t')}_{\rm sto}=N_{\rm g}^{ijkl}(t,t')+O(\phi^2).
\end{equation}
\end{subequations}
The effective system action~\eqref{Effective-system-action-1} then becomes
\begin{align} \label{Effective-system-action-2}
    S_{\rm eff}[\xi,\varrho,\mathcal{N}]&=\int\dd t\,\qty(\frac{1}{2}m\delta_{ij}\dot{\xi}^i\dot{\xi}^j-m\mathcal{N}_{ij}\xi^i\xi^j)+\int\dd t\,\mathscr{L}(\varrho,\dot{\varrho}\Bar{t}) \nonumber \\
    &-\int\dd t\,\mathscr{L}(\varrho,\dot{\varrho}\Bar{t})\qty(\frac{1}{2}\delta_{ij}\dot{\xi}^i\dot{\xi}^j-\mathcal{N}_{ij}\xi^i\xi^j).
\end{align}

We now proceed by considering only the external degrees of freedom of our system, as we are interested in the effects of the noise coming from both the internal DoFs and the quantum fluctuations of the gravitational field. In order to do this, we have to compute the reduced density matrix of the relevant degrees of freedom while tracing out all the others. Let us assume that initially the external and internal DoFs of our system are also uncorrelated, thus implying that 
\begin{equation}
    \rho_{\rm sys}(\xi(0),\varrho(0),\xi'(0),\varrho'(0),0)=\rho_{\rm ext}(\xi(0),\xi'(0),0)\rho_{\rm int}(\varrho(0),\varrho'(0),0),
\end{equation}
where $\hat{\rho}_{\rm ext}$ ($\hat{\rho}_{\rm int}$) stands for the external (internal) DoFs density matrix. The time evolution couples the external and internal variables, and we are left with the total density matrix~\eqref{Total-density-matrix-of-system}. The reduced external DoFs density matrix is obtained by taking the partial trace
\begin{equation}
    \rho_{\rm ext}(\xi,\xi',t)=\int \dd\varrho\,\rho_{\rm sys}(\xi,\varrho,\xi',\varrho,t),
\end{equation}
resulting in
\begin{align}
    \rho_{\rm ext}(\xi,\xi',t)&=\int \dd\xi(0)\dd\xi'(0)\,\rho_{\rm ext}(\xi(0),\xi'(0),0) \nonumber \\
    &\hspace{0.5cm}\times\int\mathcal{D}\xi\mathcal{D}\xi'\mathcal{D}\mathcal{N}\,\mathscr{P}[\mathcal{N}]\,e^{i\qty(S_{\rm eff}^{(1)}[\xi,\mathcal{N}]-S_{\rm eff}^{(1)}[\xi',\mathcal{N}])}e^{S_{\rm IF}^{(\textrm{int})}[\xi,\xi',\mathcal{N}]},
\end{align}
where we have defined the new influence functional
\begin{align} \label{Internal-DoF-IF}
    e^{iS_{\rm IF}^{(\textrm{int})}[\xi,\xi',\mathcal{N}]}&=\int \dd\varrho\dd\varrho(0)\dd\varrho'(0)\,\rho_{\rm int}(\varrho(0),\varrho'(0),0) \nonumber \\
    &\hspace{1cm}\times\int\mathcal{D}\varrho\mathcal{D}\varrho'\,e^{i\qty(S_{\rm eff}^{(2)}[\xi,\varrho,\mathcal{N}]-S_{\rm eff}^{(2)}[\xi',\varrho',\mathcal{N}])},
\end{align}
with
\begin{subequations}
\begin{equation}
    S_{\rm eff}^{(1)}[\xi,\mathcal{N}]=\int\dd t\,\qty(\frac{1}{2}m\delta_{ij}\dot{\xi}^i\dot{\xi}^j-m\mathcal{N}_{ij}\xi^i\xi^j),
\end{equation}
and
\begin{equation} \label{S-eff-(2)}
    S_{\rm eff}^{(2)}[\xi,\varrho,\mathcal{N}]=\int\dd t\,\mathscr{L}(\varrho,\dot{\varrho}\Bar{t})-\int\dd t\,\mathscr{L}(\varrho,\dot{\varrho}\Bar{t})\qty(\frac{1}{2}\delta_{ij}\dot{\xi}^i\dot{\xi}^j-\mathcal{N}_{ij}\xi^i\xi^j).
\end{equation}
\end{subequations}

Thus, we essentially have a similar problem to the one treated in Section~\ref{Sec:Gravitational-influence-functional}, namely a system interacting with a quantum environment. Therefore, we shall compute the Feynman-Vernon influence functional once again. It is worth remarking that the total influence functional (gravitons plus internal DoFs) is not simply the product of the individual functionals, since the gravitational field couples with all variables describing the system. When considering the system of interest to be the external degrees of freedom, we effectively end up with two environments that interact with the system and with each other. In such a case, the additive property of the influence action for multiple environments does not hold.

In order to proceed, let us assume that the Lagrangian describing the internal degrees of freedom is of the form
\begin{equation}
    \mathscr{L}(\varrho,\dot{\varrho}\,\bar{t})=\sum_\alpha\qty[\frac{1}{2}\mu_\alpha\qty(\dot{\varrho}_\alpha\,\bar{t})^2-\mathcal{V}(\varrho_\alpha)],
\end{equation}
with $\mu_{\alpha}$ representing the reduced masses of the system and $\mathcal{V}$ being a function of the coordinates. Since we want to keep terms only up to second order in the position and velocity coordinates, we may write
\begin{equation}
    \mathscr{L}(\varrho,\dot{\varrho}\,\bar{t})\simeq\sum_\alpha\qty(\frac{1}{2}\mu_\alpha\dot{\varrho}_\alpha^2-\vartheta_\alpha\varrho_\alpha-\frac{1}{2}\mu_\alpha\varpi_\alpha^2\varrho_\alpha^2),
\end{equation}
where $\vartheta_{\alpha}$ and $\varpi_{\alpha}$ are constants. Then, Eq.~\eqref{S-eff-(2)} becomes
\begin{equation}
    S_{\rm eff}^{(2)}=\sum_\alpha\left[ \int \dd t\qty(\frac{1}{2}\mu_\alpha\dot{\varrho}_\alpha^2-\frac{1}{2}\mu_\alpha\varpi_\alpha^2\varrho_\alpha^2) +\vartheta_\alpha\int\dd t\,Y(t)\varrho_\alpha(t)\right] ,
\end{equation}
with
\begin{equation}
    Y(t)=\frac{1}{2}\delta_{ij}\dot{\xi}^i\dot{\xi}^j-\mathcal{N}_{ij}\xi^i\xi^j-1.
\end{equation}

Note that we are essentially describing the internal degrees of freedom as a set of independent harmonic oscillators that couple linearly with the external ones. In this case, the internal degrees of freedom influence functional~\eqref{Internal-DoF-IF} is Gaussian, implying that we can write the influence action as
\begin{align}
    S^{(\textrm{int})}_{\rm IF}[Y,Y']&=\int \dd t\dd t'\left\{ \frac{1}{2}\qty[Y(t)-Y'(t)]D_{\rm int}(t,t')\qty[Y(t')+Y'(t')]\right. \nonumber \\
    &\hspace{1cm}\left. +\frac{i}{2}\qty[Y(t)-Y'(t)]N_{\rm int}(t,t')\qty[Y(t')-Y'(t')]\right\} ,
\end{align}
with
\begin{subequations}
    \begin{equation}
        D_{\rm int}(t,t')=i\sum_\alpha\vartheta_\alpha^2\expval{\comm{\hat{\varrho}_\alpha(t)}{\hat{\varrho}_\alpha(t')}}_{\rm int}\theta(t-t')
    \end{equation}
    and
    \begin{equation} \label{Internal-dofs-noise-kernel}
        N_{\rm int}(t,t')=\frac{1}{2}\sum_\alpha\vartheta_\alpha^2\expval{\acomm{\hat{\varrho}_\alpha(t)}{\hat{\varrho}_\alpha(t')}}_{\rm int}
    \end{equation}
\end{subequations}
being the internal DoFs dissipation and noise kernels. Now the $\hat{\varrho}(t)$'s are operators in the Heisenberg picture, and expectation values with the subscript 'int' are computed with respect to the initial state of the internal DoFs.

Similarly to what we did for the term containing the noise kernel for the gravitational influence functional, we can express the noise term in $S_{\rm IF}^{(\textrm{int})}$ in terms of a stochastic variable using the same Gaussian functional identity. Then, this term in the internal DoFs influence functional will lead to a Gaussian probability density and a linear term in the $Y$ variable. Therefore, just like in the gravitational case, the leading order contributions come from the noise term, and we may take
\begin{equation}
    e^{iS_{\rm IF}^{(\textrm{int})}[Y,Y']}\simeq e^{-\frac{1}{2}\int dtdt'\,\qty[Y(t)-Y'(t)]N_{\rm int}(t,t')\qty[Y(t')-Y'(t')]},
\end{equation}
resulting in
\begin{align} \label{rho(xi,xi',t)}
    &\rho_{\rm ext}(\xi,\xi',t)=\int \dd\xi(0)\dd\xi'(0)\,\rho_{\rm ext}(\xi(0),\xi'(0),0) \int\mathcal{D}\xi\mathcal{D}\xi'\,e^{\frac{i}{2}m\delta^{ij}\int \dd t\,\qty(\dot{\xi}_i\dot{\xi}_j-\dot{\xi}_i'\dot{\xi}_j')} \nonumber \\    &\hspace{0.5cm}\times\int\mathcal{D}\mathcal{N}\,\mathscr{P}[\mathcal{N}]\,e^{-im\int \dd t\,\mathcal{N}^{ij}\qty(\xi_i\xi_j-\xi_i'\xi_j')}e^{-\frac{1}{2}\int \dd t\dd t'\,\qty[Y(t)-Y'(t)]N_{\rm int}(t,t')\qty[Y(t')-Y'(t')]}.
\end{align}
The stochastic averages shown in Eqs.~\eqref{New-Stochastic-averages} can now be employed, provided we work in a perturbative regime (dropping higher order terms on $\xi$ and $\phi$). From this, we obtain the external degrees of freedom density matrix as
\begin{align} \label{External-DoFs-density-matrix}
    &\rho_{\rm ext}(\xi,\xi',t)=\int \dd\xi(0)\dd\xi'(0)\,\rho_{\rm ext}(\xi(0),\xi'(0),0) \int\mathcal{D}\xi\mathcal{D}\xi'\,e^{i\qty(S_{\rm ext}[\xi]-S_{\rm ext}[\xi'])} \nonumber \\
    &\times\exp\qty{-\frac{1}{4}\int\dd t\dd t'\,\delta^{ij}\qty[\frac{1}{2}\delta^{kl}y_{ij}(t)N_{\rm int}(t,t')y_{kl}(t')+\Phi^{kl}y_{ij}(t)N_{\rm int}(t,t')w_{kl}(t')]} \nonumber \\
    &\times\exp\qty{-\int\dd t\dd t'\,w_{ij}(t)\qty[\frac{m^2}{2}+N_{\rm int}(t,t')]N_{\rm g}^{ijkl}(t,t')w_{kl}(t')},
\end{align}
where
\begin{equation}
    S_{\rm ext}[\xi]=\frac{m}{2}\int\dd t\,\qty(\delta^{ij}\dot{\xi}_i\dot{\xi}_j+\Phi^{ij}\xi_i\xi_j),
\end{equation}
and we have defined
\begin{subequations}
\begin{equation}
    w_{ij}(t)=\xi_i(t)\xi_j(t)-\xi_i'(t)\xi_j'(t)
\end{equation}
and
\begin{equation}
    y_{ij}(t)=\dot{\xi}_i(t)\dot{\xi}_j(t)-\dot{\xi}_i'(t)\dot{\xi}_j'(t).
\end{equation}
\end{subequations}

The reduced density matrix~\eqref{External-DoFs-density-matrix} is the main result of this chapter. Note that its time evolution is not unitary due to the interaction with the environments. The non-unitarity sector of the time evolution is determined by the gravitational noise kernel $N_{\rm g}^{ijkl}(t,t')$ and the internal DoFs noise kernel $N_{\rm int}(t,t')$, which we also compute in Appendix~\ref{app:noise-kernels} by considering an internal thermal bath. In the next chapters, we explore the implications of this non-unitary time evolution.


\chapter{Graviton-induced decoherence of a composite particle}
\label{chap:decoherence}

Now that we have described the interaction of a composite particle with a bath of gravitons in a Newtonian background in Chapter~\ref{chap:quantum-system}, let us next obtain the time evolution of quantum superpositions of the center-of-mass variable. Our starting point will be the reduced density matrix~\eqref{External-DoFs-density-matrix} that was obtained by integrating over the gravitational and internal system degrees of freedom. Here, we will consider the system to be moving in a superposition of two classically distinguishable paths and study the decoherence arising from the interactions with both environments. In Section~\ref{Sec:The-decoherence-function}, we define the decoherence function and compute it for two possible configurations of the superposition state. We analyze the behavior of this function and compute the decoherence time in Section~\ref{Sec:Vacuum-state} by considering the gravitons to be initially in the vacuum state. We extend the analysis to other possible states in Section~\ref{Sec:Thermal-coherent-squeezed-states}. Lastly, in Section~\ref{Sec:Recoherence}, we discuss the possibility of gravitational recoherence for long times.

\section{The decoherence function} \label{Sec:The-decoherence-function}

Consider a special case in which the composite particle can move only along
two classically distinguishable paths $\xi^{(1)}(t)$ and $\xi^{(2)}(t)$. Then Eq.~\eqref{External-DoFs-density-matrix} can be written as
\begin{equation}
\begin{split}
    &\rho_{\rm ext}(\xi,\xi',t)=\int\dd\xi_0\dd\xi_0'\rho_{\rm ext}(\xi_0,\xi_0',0) \\
    &\hspace{0.3cm}\times\eval{\sum_{m,n=1}^2e^{i\qty{S_{\rm ext}\qty[\xi^{(m)}]-S_{\rm ext}\qty[\xi^{(n)}]}}e^{-\Gamma\qty[\xi^{(m)},\xi^{(n)},t]}}_{\xi^{(m)}(0)=\xi_0,\,\,\xi^{(n)}(0)=\xi_0'}^{\xi^{(m)}(t)=\xi,\,\,\xi^{(n)}(t)=\xi'},
\end{split}
\end{equation}
where we have introduced the functional
\allowdisplaybreaks
\begin{align}
    &\Gamma\qty[\xi^{(m)},\xi^{(n)},t]=\int\dd t\dd t'\,\left\{ \frac{1}{8}\delta^{ij}\delta^{kl}\qty[\dot{\xi}_i^{(m)}(t)\dot{\xi}_j^{(m)}(t)-\dot{\xi}_i^{(n)}(t)\dot{\xi}_j^{(n)}(t)]N_{\rm int}(t,t')\right. \nonumber \\
    &\hspace{5cm}\times\qty[\dot{\xi}_k^{(m)}(t')\dot{\xi}_l^{(m)}(t')-\dot{\xi}_k^{(n)}(t')\dot{\xi}_l^{(n)}(t')] \nonumber \\
    &\hspace{0.3cm}+\frac{1}{4}\delta^{ij}\Phi^{kl}\qty[\dot{\xi}_i^{(m)}(t)\dot{\xi}_j^{(m)}(t)-\dot{\xi}_i^{(n)}(t)\dot{\xi}_j^{(n)}(t)]N_{\rm int}(t,t') \nonumber \\
    &\hspace{5cm}\times\qty[\xi_k^{(m)}(t')\xi_l^{(m)}(t')-\xi_k^{(n)}(t')\xi_l^{(n)}(t')] \nonumber \\
    &\hspace{0.3cm}+\qty[\xi_i^{(m)}(t)\xi_j^{(m)}(t)-\xi_i^{(n)}(t)\xi_j^{(n)}(t)]\qty[\frac{m^2}{2}+N_{\rm int}(t,t')] \nonumber \\
    &\hspace{5cm}\left. \times N_{\rm g}^{ijkl}(t,t')\qty[\xi_k^{(m)}(t')\xi_l^{(m)}(t')-\xi_k^{(n)}(t')\xi_l^{(n)}(t')]\right\} .
\end{align}
This is a functional of paths $\xi^{(1)}(t)$ and $\xi^{(2)}(t)$, which satisfies
\begin{subequations} \label{Properties-decoherence-functional}
\begin{equation}
    \Gamma\qty[\xi^{(m)},\xi^{(n)},t]=\Gamma\qty[\xi^{(n)},\xi^{(m)},t],
\end{equation}
and
\begin{equation}
    \Gamma\qty[\xi^{(m)},\xi^{(m)},t]=0.
\end{equation}
\end{subequations}
Because of these properties, we only need to consider $\Gamma\qty[\xi^{(1)},\xi^{(2)},t]\equiv\Gamma(t)$. When $\Gamma(t)\geq0$, this function describes the decay of the off-diagonal density matrix elements, which is why we call it the \textbf{decoherence function}.

For simplicity, let us take $\xi_i^{(m)}(t)=\xi^{(m)}(t)\delta_{i3}$, namely unidimensional paths in the $x^3-$direction. Now, consider a scheme in which both paths start at the same point $\xi_0=\xi_0'$ and both end at another point $\xi=\xi'$. This is the typical scenario in which an initially localized quantum system undergoes a superposition of paths and is then recombined after time $t_f$ in order for its interference patterns to be analyzed.

Let us define the variables
\begin{subequations} \label{Configurations-of-paths}
\begin{equation}
    \Xi(t)\equiv\frac{1}{2}\qty[\xi^{(1)}(t)+\xi^{(2)}(t)],\hspace{0.5cm}\Delta\xi(t)\equiv\xi^{(1)}(t)-\xi^{(2)}(t),
\end{equation}
and
\begin{equation}
    V(t)\equiv\dv{t}\Xi(t),\hspace{0.5cm}\Delta v(t)\equiv\dv{t}\Delta\xi(t),
\end{equation}
\end{subequations}
such that we can write
\begin{align}
    \Gamma(t_f)&=\frac{1}{2}\int_0^{t_f}\dd t\dd t'\,V(t)\Delta v(t)N_{\rm int}(t,t')V(t')\Delta v(t') \nonumber \\
    &+\Phi_{zz}\int_0^{t_f}\dd t\dd t'\,V(t)\Delta v(t)N_{\rm int}(t,t')\Xi(t')\Delta\xi(t') \nonumber \\
    &+4\int_0^{t_f}\dd t\dd t'\,\Xi(t)\Delta\xi(t)\qty[\frac{m^2}{2}+N_{\rm int}(t,t')]N_{\rm g}(t,t')\Xi(t')\Delta\xi(t'),
\end{align}
with $N_{\rm g}(t,t')\equiv N_{\rm g}^{3333}(t,t')$.

In Appendix~\ref{app:noise-kernels}, the noise kernel $N_{\rm int}(t,t')$ was computed by considering the internal DoFs to represent an Ohmic bath described by the coupling constant $\eta$ in thermal equilibrium at high temperature $T_{\rm int}\gg\abs{t-t'}$. Explicitly, using Eq.~\eqref{Int-noise-high-T}, we have
\begin{align}
    \Gamma(t_f)&=\frac{1}{2}\eta\pi T_{\rm int}\int_0^{t_f}\dd t\,\qty[V(t)\Delta v(t)]^2+\eta\pi T_{\rm int}\Phi_{zz}\int_0^{t_f}\dd t\,V(t)\Delta v(t)\Xi(t)\Delta\xi(t) \nonumber \\
    &+2m^2\int_0^{t_f}\dd t\dd t'\,\Xi(t)\Delta\xi(t)N_{\rm g}(t,t')\Xi(t')\Delta\xi(t') \nonumber \\
    &+4\eta\pi T_{\rm int}\int_0^{t_f}\dd t\,\qty[\Xi(t)\Delta\xi(t)]^2N_{\rm g}(t),
\end{align}
where $N_{\rm g}(t)\equiv\lim_{t'\to t}N_{\rm g}(t,t')$.

The function $\Gamma(t)$ depends on the specific configurations of the superposition state through the variables~\eqref{Configurations-of-paths}, and also on the gravitational noise kernel $N_{\rm g}(t,t')$, which was computed in Appendix~\ref{app:noise-kernels}. In this work, we will consider the same configuration path as the one in Refs.~\cite{Kanno2021,Breuer2001} in order to explore the time evolution of the superposition state. However, we will also show how the decoherence function is modified for another configuration for illustrative purposes.

\subsection{Configuration 1}

For the first configuration, which we refer to as \emph{Configuration 1}, we choose
\begin{subequations} \label{Configuration1}
\begin{equation}
    \Xi(t)=\Xi=\textrm{constant in time}
\end{equation}
and
\begin{equation}
    \Delta\xi(t)=\left\{
    \begin{array}{ll}
        2vt &\textrm{for}\hspace{0.2cm}0<t\leq t_f/2\\
        2v(t_f-t) &\textrm{for}\hspace{0.2cm}t_f/2<t<t_f
    \end{array}
    \right.,
\end{equation}
for some constant velocity $v$. Note that this implies $V(t)=\dv*{\Xi}{t}=0$.
\end{subequations}
We then have
\begin{align} \label{Dec-rate-Conf-1-general}
    \Gamma_1(t_f)&=8m^2\Xi^2v^2\left[ \int_0^{t_f/2}\dd t\dd t'\,tt'N_{\rm g}(t,t')+\int_{t_f/2}^{t_f}\dd t\dd t'\,(t_f-t)(t_f-t')N_{\rm g}(t,t') \right. \nonumber \\
    &\hspace{2cm}\left. +2\int_0^{t_f/2}\dd t\int_{t_f/2}^{t_f}\dd t'\,t(t_f-t')N_{\rm g}(t,t')\right] \nonumber \\
    &+16\eta\pi T_{\rm int}\Xi^2v^2\qty[\int_0^{t_f/2}\dd t\,t^2N_{\rm g}(t)+\int_{t_f/2}^{t_f}\dd t\,(t_f-t)^2N_{\rm g}(t)].
\end{align}

The next step is to use the various expressions for the gravitational noise kernel to explicitly compute the function $\Gamma_1(t)$. These can be found in Appendix~\ref{app:noise-kernels} for gravitons initially in the vacuum, thermal, coherent, and squeezed states. Although long, this is a straightforward computation, so let us simply present the final results for the function $\Gamma_1(t)$.

First, let us recall from Eq.~\eqref{Tidal-tensor} that $\Phi_{zz}=2M_N/R_N^3$. Also, it will be interesting to return the universal constants $\hbar$, $c$, $G$, and $k_B$ so that we present the result in SI units rather than Planck units.

Let us introduce the index $A$, which can be either v, t, c, or s, representing vacuum, thermal, coherent, and squeezed states, respectively. Then, the results for the function $\Gamma_1(t)$ can be summarized in the following expression:
\begin{align} \label{Conf1-Dec-function}
    \Gamma_1^{(A)}(t)&=b_A(1-\delta_{A,\textrm{v}})\Gamma_1^{\rm (v)}(t) \nonumber \\
    &+\frac{8\Xi^2v^2m^2}{5\pi E_{\rm P}^2}K_{1,A}\left\{ \qty(\frac{\Lambda_A}{\hbar})^2\qty[f_A^{(I)}\qty(\frac{\Lambda_At}{\hbar})+\frac{\kappa_A}{E_{\rm rest}^2}f_A^{(II)}\qty(\frac{\Lambda_At}{\hbar})]\right. \nonumber \\
    &\hspace{2.5cm}\left. -\frac{GM_N}{R_N^3}\qty[f_A^{(III)}\qty(\frac{\Lambda_At}{\hbar})+\frac{\kappa_A}{E_{\rm rest}^2}f_A^{(IV)}\qty(\frac{\Lambda_At}{\hbar})]\right\} .
\end{align}
Here,
\begin{subequations} \label{Definitions-contants-Gamma(t)}
\begin{equation}
    b_A=\left\{ 
    \begin{array}{ll}
        1 & \textrm{for $A\neq\textrm{s}$} \\
        \cosh2r & \textrm{for $A=\textrm{s}$}
    \end{array}
    \right. ,
\end{equation}
\begin{equation}
    \Lambda_A=\left\{ 
    \begin{array}{ll}
        \Lambda & \textrm{for $A\neq\textrm{t}$} \\
        \pi k_BT_{\rm g} & \textrm{for $A=\textrm{t}$}
    \end{array}
    \right. ,
\end{equation}
and
\begin{equation}
    \kappa_A=\eta\pi k_BT_{\rm int}\Lambda_A.
\end{equation}
\end{subequations}
Also, $E_{\rm rest}=mc^2$ and $E_{\rm P}$ are the rest and Planck energies. The constant $\Lambda$ is the graviton energy cutoff, $T_{\rm g}$ is the graviton temperature for the initial thermal state, and $r$ is the real squeeze parameter for the initial squeezed state. The constants $K_{1,A}$ are
\begin{subequations}
\begin{align}
    &K_{1,\textrm{v}}=2, \\
    &K_{1,\textrm{t}}=\frac{4}{3}, \\
    &K_{1,\textrm{c}}=\frac{\alpha^2}{3}, \\
    &K_{1,\textrm{s}}=-\frac{2}{3}\sinh2r,
\end{align}
\end{subequations}
with $\alpha$ being the displacement parameter for the initial coherent state. For all states, the functions $f_A^{(I)}$ describe the contributions to decoherence that come from the gravitons alone, while the functions $f_A^{(II)}$ describe the contributions coming from the interplay between the gravitons and the internal DoFs of the system. The functions $f_A^{(III)}$ and $f_A^{(IV)}$, representing the contributions from gravitons and gravitons plus internal DoFs, respectively, also encode the contribution from the Newtonian potential. Note that, within this configuration, there are no contributions coming solely from the internal DoFs, and in the absence of quantum spacetime fluctuations, decoherence does not occur (for Configuration 1). The various functions $f_A$ are listed below:

\paragraph{Vacuum state}

\begin{subequations}
\begin{equation}
    f_{\rm v}^{(I)}(x)=1+\frac{2}{3x}\qty[\sin x-8\sin\qty(\frac{x}{2})]+\frac{1}{x^2}\qty[\frac{2}{3}\cos x-\frac{32}{3}\cos\qty(\frac{x}{2})+10],
\end{equation}
\begin{equation}
    f_{\rm v}^{(II)}(x)=\frac{1}{108}x^3,
\end{equation}
\begin{equation}
    f_{\rm v}^{(III)}(x)=8\gamma_E-\frac{4}{3}\ln4-\frac{32}{3}\textrm{Ci}\qty(\frac{x}{2})+\frac{8}{3}\textrm{Ci}(x)+8\ln\qty(\frac{x}{2}),
\end{equation}
\begin{equation}
    f_{\rm v}^{(IV)}(x)=\frac{1}{18}x^3.
\end{equation}
\end{subequations}

\paragraph{Thermal state}

\begin{subequations}
\begin{equation}
    f_{\rm t}^{(I)}(x)=\frac{1+16e^x+26e^{2x}+16e^{3x}+e^{4x}}{(e^{2x}-1)^2}-\frac{15}{x^2},
\end{equation}
\begin{equation}
    f_{\rm t}^{(II)}(x)=\frac{4}{189}x^3,
\end{equation}
\begin{equation}
    f_{\rm t}^{(III)}(x)=4\ln\qty[\frac{2(e^x-1)^3}{x^3(e^x+1)}]-4x,
\end{equation}
\begin{equation}
    f_{\rm t}^{(IV)}(x)=\frac{2}{45}x^3.
\end{equation}
\end{subequations}

\paragraph{Coherent state}

\begin{subequations}
\begin{align}
    f_{\rm c}^{(I)}(x)&=\frac{7}{2}+\frac{1}{6x}\qty{3\sin(2x)-16\qty[9\sin\qty(\frac{x}{2})-3\sin x+\sin\qty(\frac{3x}{2})]} \nonumber \\
    &+\frac{1}{36x^2}\qty[1495-1728\cos\qty(\frac{x}{2})+288\cos x-64\cos\qty(\frac{3x}{2})+9\cos(2x)],
\end{align}
\begin{align}
    f_{\rm c}^{(II)}(x)&=\frac{x^3}{36}+\frac{1}{8x^3}\left[ 49+24(x^2-2)\cos x+(2x^2-1)\cos(2x)\right. \nonumber \\
    &\left. -4x(12-2x^2+\cos x)\sin x\right] ,
\end{align}
\begin{equation}
    f_{\rm c}^{(III)}(x)=28\gamma_E+4\ln\qty(\frac{81x^7}{2^{17}})-48\textrm{Ci}\qty(\frac{x}{2})+32\textrm{Ci}(x)-16\textrm{Ci}\qty(\frac{3x}{2})+4\textrm{Ci}(2x),
\end{equation}
\begin{equation}
    f_{\rm c}^{(IV)}(x)=\frac{x^3}{6}+4\sin x+\frac{2}{x}(2\cos x+\cos^2x-3).
\end{equation}
\end{subequations}

\paragraph{Squeezed state}

\begin{subequations}
\begin{align}
    f_{\rm s}^{(I)}(x)&=\frac{1}{2}+\frac{1}{2x}\qty[\sin(2x)+12\sin x-16\sin\qty(\frac{x}{2})-\frac{16}{3}\sin\qty(\frac{3x}{2})] \nonumber \\
    &+\frac{1}{36x^2}\qty[415-576\cos\qty(\frac{x}{2})+216\cos x-64\cos\qty(\frac{3x}{2})+9\cos(2x)],
\end{align}
\begin{align}
    f_{\rm s}^{(II)}(x)&=\frac{1}{8x^3}\left[ 49+24(x^2-2)\cos x+(2x^2-1)\cos(2x)\right. \nonumber \\
    &\left. -4x(12-2x^2+\cos x)\sin x\right] ,
\end{align}
\begin{equation}
    f_{\rm s}^{(III)}(x)=4\gamma_E+4\ln\qty(\frac{81x}{2^{9}})-16\textrm{Ci}\qty(\frac{x}{2})+24\textrm{Ci}(x)-16\textrm{Ci}\qty(\frac{3x}{2})+4\textrm{Ci}(2x),
\end{equation}
\begin{equation}
    f_{\rm s}^{(IV)}(x)=4\sin x+\frac{2}{x}(2\cos x+\cos^2x-3).
\end{equation}
\end{subequations}
Here $\gamma_E\simeq0.577$ is the Euler-Mascheroni constant, and $\textrm{Ci}(z)=-\int_z^\infty\frac{\cos t}{t}\dd t$ is the cosine integral function.

\subsection{Configuration 2}

We can consider a different configuration for the superposition state; for instance, one that is described by linear paths with different constant velocities, $\xi^{(m)}=v_mt$. In that case, which we call \emph{Configuration 2}, we have
\begin{subequations}
\begin{equation}
    \Xi(t)=Vt,\hspace{0.5cm}\Delta\xi(t)=\Delta vt,
\end{equation}
and
\begin{equation}
    V=\frac{v_1+v_2}{2},\hspace{0.5cm}\Delta v=v_1-v_2.
\end{equation}
\end{subequations}

For Configuration 2, one finds
\begin{align} \label{Dec-rate-Conf-2-general}
    \Gamma_2(t_f)&=\frac{1}{4}\eta\pi T_{\rm int}\qty(v_1^2-v_2^2)^2\qty[\frac{t_f}{2}+\frac{\Phi_{zz}t_f^3}{3}+4\int_0^{t_f}\dd t\,t^4N_{\rm g}(t)] \nonumber \\
    &+\frac{m^2}{2}\qty(v_1^2-v_2^2)^2\int_0^{t_f}\dd t\dd t'\,(tt')^2N_{\rm g}(t,t').
\end{align}

Just as we did for Configuration 1, our next task is to plug in the expressions for the noise kernel into Eq.~\eqref{Dec-rate-Conf-2-general} in order to obtain explicit expressions for $\Gamma_2(t)$. Once again, we shall only summarize the results of this long but straightforward computation, which gives (after restoring the universal constants once more)
\begin{subequations}
\begin{equation}
    \Gamma_2^{(A)}(t)=\frac{\pi\eta k_BT_{\rm int}}{4\hbar}\frac{\qty(v_1^2-v_2^2)^2}{c^4}\qty(\frac{t}{2}+\frac{2}{3}\frac{GM_N}{R_N^3}t^3)+\Gamma_{2(g)}^{(A)}(t),
\end{equation}
with
\begin{equation}
\begin{split}
    \Gamma_{2(g)}^{(A)}(t)&=b_A(1-\delta_{A,\textrm{v}})\Gamma_{2(g)}^{\rm (v)}(t) \\
    &+\frac{m^2\qty(v_1^2-v_2^2)^2}{15\pi E_{\rm P}^2}K_{2,A}\left\{ g_A^{(I)}\qty(\frac{\Lambda_At}{\hbar})+\frac{\kappa_A}{E_{\rm rest}^2}g_A^{(II)}\qty(\frac{\Lambda_At}{\hbar})\right. \\
    &\hspace{2.5cm}\left. -\qty(\frac{\hbar}{\Lambda_A})^2\frac{GM_E}{R_E^3}\qty[g_A^{(III)}\qty(\frac{\Lambda_At}{\hbar})+\frac{\kappa_A}{E_{\rm rest}^2}g_A^{(IV)}\qty(\frac{\Lambda_At}{\hbar})]\right\} ,
\end{split}
\end{equation}
\end{subequations}
with $b_A$, $\Lambda_A$, and $\kappa_A$ defined in Eqs.~\eqref{Definitions-contants-Gamma(t)}.
The constants $K_{2,A}$ are
\begin{subequations}
\begin{align}
    &K_{2,\textrm{v}}=1, \\
    &K_{2,\textrm{t}}=12, \\
    &K_{2,\textrm{c}}=\alpha^2, \\
    &K_{2,\textrm{s}}=-\sinh2r.
\end{align}
\end{subequations}
The functions $g_A$ have the same interpretations as the functions $f_A$ with respect to the physical mechanism responsible for their contribution. However, as opposed to Configuration 1, the decoherence function for Configuration 2 exhibits contributions that come from the internal DoFs alone, in the same spirit as the gravitational decoherence mechanism proposed by Pikovski et al.~\cite{Pikovski2015,Pikovski2017} (Section~\ref{Sec:Dec-grav-time-dilation}). Since our interest will be in analyzing the decoherence resulting from quantum spacetime fluctuations, we will work with Configuration 1 for the remainder of this work. Nevertheless, for illustrative purposes (and as a reference for some possible future work), the various functions $g_A$ are listed below:

\paragraph{Vacuum state}

\begin{subequations}
\begin{equation}
    g_{\rm v}^{(I)}(x)=\frac{x^4}{4}+8\gamma_E-12-8\textrm{Ci}(x)+8\ln x+4x\sin x+12\cos x,
\end{equation}
\begin{equation}
    g_{\rm v}^{(II)}(x)=\frac{x^5}{15},
\end{equation}
\begin{equation}
    g_{\rm v}^{(III)}(x)=2x^4-8x^2+16\cos x+16x\sin x-16,
\end{equation}
\begin{equation}
    g_{\rm v}^{(IV)}(x)=\frac{2}{5}x^5.
\end{equation}
\end{subequations}

\paragraph{Thermal state}

\begin{subequations}
\begin{equation}
    g_{\rm t}^{(I)}(x)=1-\frac{2x}{3}+\frac{x^4}{90}+\frac{2}{3}\ln\qty(\frac{e^{2x}-1}{2x})-\frac{x}{3}\qty[\frac{\sinh(2x)+x}{\sinh^2x}],
\end{equation}
\begin{equation}
    g_{\rm t}^{(II)}(x)=\frac{8}{945}x^5,
\end{equation}
\begin{equation}
    g_{\rm t}^{(III)}(x)=\frac{x^4}{9}+\frac{4}{9}x^3+\frac{2}{3}x^2-\frac{4}{3}x^2\ln(1-e^{2x})-\frac{4}{3}x\textrm{Li}_2(e^{2x})+\frac{2}{3}\textrm{Li}_3(e^{2x})-\frac{2}{3}\zeta(3),
\end{equation}
\begin{equation}
    g_{\rm t}^{(IV)}(x)=\frac{4}{225}x^5.
\end{equation}
\end{subequations}

\paragraph{Coherent state}

\begin{subequations}
\begin{align}
    g_{\rm c}^{(I)}(x)&=\frac{x^4}{8}+2\gamma_E-\frac{59}{16}+\frac{1}{16}(59-22x^2)\cos(2x)-2\textrm{Ci}(2x)-2\ln\qty(\frac{x}{2})+4\ln x \nonumber \\
    &+\frac{x}{8}(27-2x^2)\sin(2x),
\end{align}
\begin{equation}
    g_{\rm c}^{(II)}(x)=\frac{x^5}{30}+\frac{1}{8x}\qty[15+\qty(-15+18x^2-2x^4)\cos(2x)]+(x^2-3)\sin(2x),
\end{equation}
\begin{equation}
    g_{\rm c}^{(III)}(x)=x^4-\frac{7}{2}x^2-4+\qty(4-\frac{9}{2}x^2)\cos(2x)-x(x^2-8)\sin(2x),
\end{equation}
\begin{equation}
    g_{\rm c}^{(IV)}(x)=\frac{3}{2}x+\frac{x^5}{5}+x\qty(\frac{9}{2}-x^2)\cos(2x)+3(x^2-1)\sin(2x).
\end{equation}
\end{subequations}

\paragraph{Squeezed state}

\begin{subequations}
\begin{align}
    g_{\rm s}^{(I)}(x)&=\frac{37}{8}-4\gamma_E-12\cos x+\frac{1}{8}(59-22x^2)\cos(2x)+8\textrm{Ci}(x)-4\textrm{Ci}(2x) \nonumber \\
    &-4\ln\qty(\frac{x}{2})-\frac{x}{2}\qty[8+(2x^2-27)\cos x]\sin x,
\end{align}
\begin{equation}
    g_{\rm s}^{(II)}(x)=\frac{1}{4x}\qty[15+\qty(-15+18x^2-2x^4)\cos(2x)]+2(x^2-3)\sin(2x),
\end{equation}
\begin{equation}
    g_{\rm s}^{(III)}(x)=x^2+8-16\cos x+(8-9x^2)\cos(2x)-4x\qty[4+(x^2-8)\cos x]\sin x,
\end{equation}
\begin{equation}
    g_{\rm s}^{(IV)}(x)=3x+x(9-2x^2)\cos(2x)+6(x^2-1)\sin(2x).
\end{equation}
\end{subequations}
Here $\zeta(z)$ is the Riemann zeta function, and $\textrm{Li}_n(z)$ is the polylogarithm function.

\section{Decoherence for initial vacuum state} \label{Sec:Vacuum-state}

\begin{figure}[!ht]
    \centering
    \includegraphics[width=0.8\linewidth]{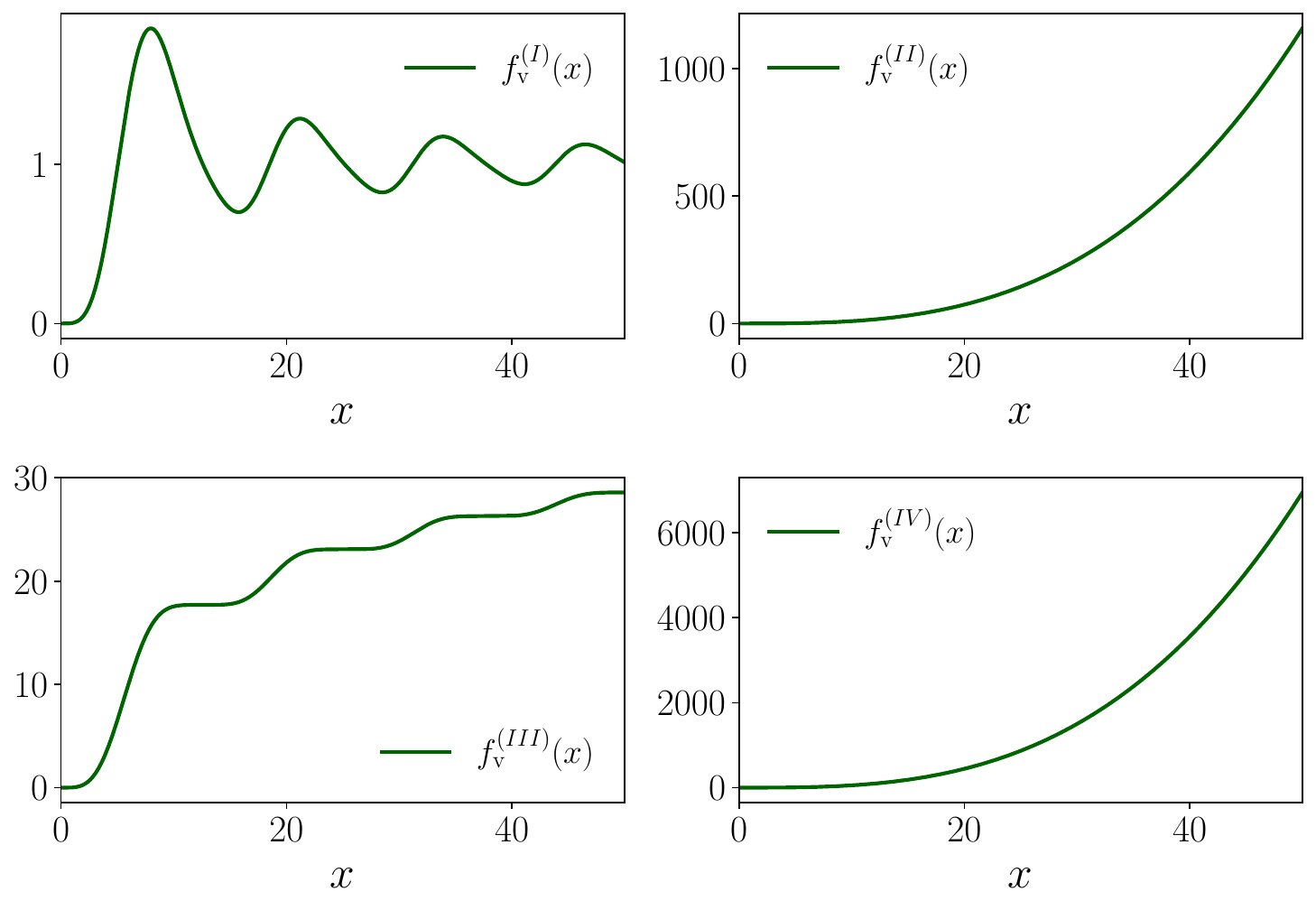}
    \caption{Different contributions for the decoherence function considering the gravitons to be initially in the \textbf{vacuum state}.}
    \label{Fig:vacuum_state}
\end{figure}

For gravitons initially in the vacuum state, the decoherence function (for Configuration 1) reads
\begin{align} \label{Vacuum-decoherence-function-Conf-1}
    &\Gamma_1^{(\textrm{v})}(t)=\frac{16\Xi^2v^2m^2}{5\pi E_{\rm P}^2}\left\{ \qty(\frac{\Lambda}{\hbar})^2\qty[f_{\rm v}^{(I)}\qty(\frac{\Lambda t}{\hbar})+\frac{\kappa}{E_{\rm rest}^2}f_{\rm v}^{(II)}\qty(\frac{\Lambda t}{\hbar})]\right. \nonumber \\
    &\hspace{2.5cm}\left. -\frac{GM_N}{R_N^3}\qty[f_{\rm v}^{(III)}\qty(\frac{\Lambda t}{\hbar})+\frac{\kappa}{E_{\rm rest}^2}f_{\rm v}^{(IV)}\qty(\frac{\Lambda t}{\hbar})]\right\} ,
\end{align}
with $\kappa\equiv\kappa_{\rm v}$. The explicit expressions for the functions $f_{\rm v}(x)$ were listed in Section~\ref{Sec:The-decoherence-function}, while some of their relevant properties are shown in Table~\ref{tab:vacuum_state} (see also Figure~\ref{Fig:vacuum_state}).

\begin{table*}[!ht]
\centering 

\begin{tabular}{llcc}
\toprule 

\multicolumn{4}{c}{\textbf{Vacuum state}} \\
\cmidrule(lr){1-4} 

\textbf{Contribution} & \textbf{Function} & \textbf{Behavior for $x \ll 1$} & \textbf{Behavior for $x \gg 1$} \\
\midrule 

G     & $f_{\rm v}^{(I)}(x)$   & $x^4/288$ & $1$ \\
G+I   & $f_{\rm v}^{(II)}(x)$  & $x^3/108$ & $x^3/108$ \\
G+N   & $f_{\rm v}^{(III)}(x)$ & $x^4/48$  & $8\gamma_E-\frac{32}{3}\ln2+8\ln x$ \\
G+N+I & $f_{\rm v}^{(IV)}(x)$  & $x^3/18$  & $x^3/18$ \\

\bottomrule 
\end{tabular}

\caption{Different contributions to the decoherence function considering the gravitons to be initially in the \textbf{vacuum state}. In the "Contribution" column, "G" means gravitons, "I" means internal DoFs, and "N" stands for Newtonian potential.}
\label{tab:vacuum_state}

\end{table*}

From Eq.~\eqref{Vacuum-decoherence-function-Conf-1}, we see that the contribution from the interplay between the gravitons and the internal degrees of freedom of the system is scaled by the ratio
\begin{equation} \label{Ratio-int}
    \mathscr{R}=\frac{\kappa}{E_{\rm rest}^2}=\frac{\eta\pi k_BT_{\rm int}\Lambda}{m^2c^4}.
\end{equation}
This ratio depends on the dimensionless coupling between the external and internal DoFs, $\eta$, as well as the internal temperature, which can be of order $T_{\rm int}\sim10^4$ K for complex molecules~\cite{Hornberger2012}. The graviton cutoff $\Lambda$ is related to the typical size of the geodesic deviation, $L_0$, as pointed out in Chapter~\ref{chap:quantum-system}. We take $\Lambda=\hbar c/L_0$ and refer to $L_0$ as the "detector size". Finally, the ratio also depends on the mass of the composite particle, $m$. As we discussed in Chapter~\ref{chap:gravdec}, quantum spatial superposition has been observed for complex molecules with masses up to $m\sim10^{-22}$ kg~\cite{Gerlich2011,Fein2019,Pedalino2025}. We can rewrite the ratio~\eqref{Ratio-int} as
\begin{equation} \label{Ratio-int-2}
    \mathscr{R}=\eta\frac{L(m,T_{\rm int})}{L_0},
\end{equation}
where $L(m,T_{\rm int})$ is a typical length scale determined by the mass of the system and its internal temperature. For instance, if we take $T_{\rm int}\sim10^4$ K and $m\sim10^{-22}$ kg, we find $L(m,T_{\rm int})\sim L_{\rm P}$, the Planck length. Therefore, although the ratio $\mathscr{R}$ can be increased by increasing the coupling between external and internal DoFs, and/or decreasing the detector size, it is clear that, for typical systems, one can expect to have $\mathscr{R}\ll1$. Nevertheless, the relevance of such contributions needs to be analyzed by looking at the behaviors of the functions listed in Table~\ref{tab:vacuum_state}.

\subsection{Short-time limit}

Let us begin by considering short times, $t\ll\hbar/\Lambda=L_0/c$. In this limit, we can use the expansions shown in Table~\ref{tab:vacuum_state} and write the decoherence function as
\begin{equation}
    \Gamma_1^{(\textrm{v})}(t)=\frac{8\Xi^2v^2m^2}{15\pi E_{\rm P}^2}\qty[\frac{1}{6}\qty(\frac{\Lambda}{\hbar})^2-\frac{GM_N}{R_N^3}]\qty(\frac{\Lambda t}{\hbar})^3\qty(\frac{1}{8}\frac{\Lambda t}{\hbar}+\frac{1}{3}\frac{\kappa}{E_{\rm rest}^2}).
\end{equation}

First, let us note that there seems to be a competition between the squared frequencies $\qty(\Lambda/\hbar)^2$ and $GM_N/R_N^3$, and decoherence occurs only if the former is greater than the latter. Typically, this is the case, as one can see from Table~\ref{tab:table-comparing-squared-frequencies}, where we estimate some values of both squared frequencies for different detector sizes and different sources of Newtonian gravitational potential. However, it seems to be possible, at least in principle, to have a situation in which the reversed scenario holds by increasing the detector size as well as the density of the Newtonian source. Of course, one must be careful not to violate any of our assumptions in doing so, such as the Newtonian approximation, which prevents us from increasing the density of the source to an arbitrarily large value. For instance, we show the result of the tidal squared frequency for neutron stars for informational purposes, but let us not forget that we are working in a perturbative regime and dropping higher order terms in the tidal tensor. We will return to this discussion at the end of the chapter.

\begin{table*}[!ht]
\centering 

\begin{minipage}[t]{0.4\textwidth} 
    \centering
    \begin{tabular}{cc}
    \toprule
    \multicolumn{2}{c}{\textbf{Estimating $\qty(\Lambda/\hbar)^2$}} \\
    \cmidrule(r){1-2}
    $L_0$ [m] & $\qty(\Lambda/\hbar)^2$ [$\textrm{s}^{-2}$] \\
    \midrule
    $10^{-6}$ & $9.0\times10^{28}$ \\
    $10^{3}$ & $9.0\times10^{10}$ \\
    $10^{9}$ & $9.0\times10^{-2}$ \\
    \bottomrule
    \end{tabular}
\end{minipage}
%
\hspace{0.05\textwidth} 
%
\begin{minipage}[t]{0.4\textwidth} 
    \centering
    \begin{tabular}{cc}
    \toprule
    \multicolumn{2}{c}{\textbf{Estimating $GM_N/R_N^3$}} \\
    \cmidrule(l){1-2}
    Source & $GM_N/R_N^3$ [$\textrm{s}^{-2}$] \\
    \midrule
    Sun & $3.9\times10^{-7}$ \\
    Earth & $1.5\times10^{-6}$ \\
    Neutron star & $1.7\times10^{8}$ \\
    \bottomrule
    \end{tabular}
\end{minipage}

\caption{Typical squared frequencies. On the left we estimate the magnitude of $\qty(\Lambda/\hbar)^2$ for some detector sizes $L_0$. On the right we estimate the magnitude of $GM_N/R_N^3$ for some sources of the Newtonian potential.}
\label{tab:table-comparing-squared-frequencies}

\end{table*}

Let us rewrite the decoherence function as
\begin{equation} \label{dec-func-short-time}
    \Gamma_1^{(\textrm{v})}(t)=\frac{4}{45\pi}\delta\Omega\qty(\frac{\Xi}{L_0})^2\qty(\frac{v}{c})^2\qty(\frac{m}{M_{\rm P}})^2\qty(\frac{\Lambda t}{\hbar})^3\qty(\frac{1}{8}\frac{\Lambda t}{\hbar}+\frac{1}{3}\frac{\kappa}{E_{\rm rest}^2}),
\end{equation}
where $M_{\textrm{P}}$ is the Planck mass, and we define
\begin{equation} \label{delta-Omega}
    \delta\Omega\equiv1-6\qty(\frac{\hbar}{\Lambda})^2\frac{GM_N}{R_N^3}.
\end{equation}
Note that $\delta\Omega\leq1$ and typically one has $\delta\Omega\simeq1$.

The decoherence function depends on many parameters, with the ratio $\Xi/L_0$ among them. Recall that, while $L_0$ represents a typical geodesic deviation with respect to the much more massive mass $M$ at the center of our reference frame, $\Xi$ basically denotes the size of the superposition. For simplicity, we can consider from now on that one of the paths is close enough to the reference mass $M$ such that we can take $\Xi\sim L_0$.

Now, let us note from Eq.~\eqref{dec-func-short-time} that, as long as $t\gg(8/3)\eta t(m,T_{\rm int})$, with $t(m,T_{\rm int})=L(m,T_{\rm int})/c$ denoting some typical time scale determined by the properties of the system, the contribution from the interplay between the gravitons and the internal DoFs becomes negligible. For instance, for the typical values of the mass and internal temperature we discussed before, $t(m,T_{\rm int})\sim t_{\rm P}$, the Planck time.

In summary, in the regime in which
\begin{equation}
    \frac{8}{3}\eta\frac{L(m,T_{\rm int})}{c}\ll t\ll\frac{L_0}{c},
\end{equation}
only the interaction with the graviton bath dominates, and the decoherence function becomes
\begin{equation} \label{Final-dec-func-short-time}
    \Gamma_1^{(\textrm{v})}(t)=\frac{1}{90\pi}\delta\Omega\qty(\frac{v}{c})^2\qty(\frac{m}{M_{\rm P}})^2\qty(\frac{\Lambda t}{\hbar})^4.
\end{equation}
We can compute the decoherence time $t_{\rm dec}$, defined by $\Gamma(t_{\rm dec})=1$. One finds\footnote{A word on notation: let $t_{\rm dec}$ denote decoherence times computed in the short-time limit, while $\tau_{\rm dec}$ denotes decoherence times computed in the long-time limit.}
\begin{equation} \label{Short-time-dec-time-vacuum}
    t^{(\textrm{v})}_{\rm dec}=\frac{\hbar}{\Lambda}\sqrt{\sqrt{\frac{90\pi}{\delta\Omega}}\frac{c}{v}\frac{M_{\rm P}}{m}}.
\end{equation}

Consistency demands that $t^{(\textrm{v})}_{\rm dec}\ll\hbar/\Lambda$, and, as a consequence, graviton-induced decoherence occurs for systems that satisfy
\begin{equation} \label{Short-time-decoherence-condition}
    mv\gg\sqrt{\frac{90\pi}{\delta\Omega}}M_{\rm P}c,
\end{equation}
which agrees with the conclusions obtained by Kanno et al.~\cite{Kanno2021} (Section~\ref{Sec:Kanno}). For $\delta\Omega\simeq1$ this means that we must have $mv\gg110\,\textrm{kg}\cdot\textrm{m}/\textrm{s}$.

Eq.~\eqref{Short-time-decoherence-condition} shows that the observation of graviton-induced decoherence requires the preparation of spatial quantum superpositions of macroscopic masses. Even if we were to extrapolate our results to the ultra-relativistic limit, $v\simeq c$, the decoherence condition would require masses satisfying $m\gg3.7\times10^{-7}$ kg.

In conclusion, the short-time limit is dominated solely by the graviton bath, and decoherence generally does not occur for microscopic, typical experimentally accessible masses. One can hope that the scenario improves for long times. However, if we consider only the pure graviton contribution, one can see from Table~\ref{tab:vacuum_state} that $f_{\rm v}^{(I)}$ tends to a constant value as $t\to\infty$ and the decoherence function saturates at
\begin{equation} \label{Gamma_sat}
    \Gamma^{(\textrm{v})}_{1,\rm sat}=\frac{16}{5\pi}\delta\Omega\qty(\frac{v}{c})^2\qty(\frac{m}{M_{\rm P}})^2,
\end{equation}
increasing no further. Therefore, the graviton bath alone cannot decohere spatial superpositions of a microscopic mass. The situation changes in the long-time limit when we consider the graviton interplay with the other contributions (internal DoFs and the Newtonian potential), since the other functions exhibit no such behavior as $t\to\infty$. Let us then return our attention to Eq.~\eqref{Vacuum-decoherence-function-Conf-1} and consider the long-time limit.

\subsection{Long-time limit}

We can see from Table~\ref{tab:vacuum_state} that, for $x\gg1$, the functions $f_{\rm v}^{(II)}(x)$ and $f_{\rm v}^{(IV)}(x)$, which include the internal DoFs contributions, scale as $x^3$, while $f_{\rm v}^{(I)}(x)$ remains constant and $f_{\rm v}^{(III)}(x)$ grows in a much slower logarithmic rate. This means that, regardless of how small the ratio $\mathscr{R}=\kappa/E_{\rm rest}^2$ is, there will always be a time, no matter how long, when the G+I contributions will dominate. Since these contributions will continue to increase without bound, decoherence will eventually occur.

In the long-time limit, using definition~\eqref{delta-Omega} and taking $\Xi\sim L_0$, Eq.~\eqref{Vacuum-decoherence-function-Conf-1} becomes
\begin{equation} \label{Final-dec-func-long-time}
    \Gamma_1^{(\textrm{v})}(t)=\frac{4}{135}\delta\Omega\qty(\frac{v}{c})^2\qty(\frac{\eta k_BT_{\rm int}\Lambda}{E_{\rm P}^2})\qty(\frac{\Lambda t}{\hbar})^3.
\end{equation}

Interestingly, we see that the decoherence function~\eqref{Final-dec-func-long-time} does not seem to contain any explicit dependence on the mass of the composite particle. This does not mean, however, that this function does not depend at all on the mass, since the long-time limit was established according to the ratio $\kappa/E_{\rm rest}^2$, which \emph{does} depend on it. Also, there could be some explicit dependence on $m$ through the dimensionless coupling constant $\eta$.

From Eq.~\eqref{Final-dec-func-long-time}, we can compute the decoherence time, for which $\Gamma(\tau_{\rm dec})=1$. One finds
\begin{align} \label{Long-time-dec-time-vacuum}
    \tau_{\rm dec}^{(\textrm{v})}&=\qty[\frac{135}{4}\frac{1}{\delta\Omega}\qty(\frac{c}{v})^2\frac{\hbar^4c^5}{Gk_B}\frac{1}{\eta T_{\rm int}\Lambda^4}]^{1/3} \nonumber \\
    &=\qty[\frac{135}{4}\frac{1}{\delta\Omega}\qty(\frac{c}{v})^2\frac{cL_0^4}{Gk_B\eta T_{\rm int}}]^{1/3}
\end{align}
This is still a typically long time. For instance, consider a strong coupling scenario and take $\eta\sim1$. Also, consider $T_{\rm int}\sim10^4$ K, a typical molecular speed of the order $v\sim10^{-6}c$, $L_0\sim10^{-9}$ m, and $\delta\Omega\sim1$. Then one obtains $\tau_{\rm dec}^{(\textrm{v})}\sim10^5$ s (about $1.2$ days). Nevertheless, this shows that decoherence eventually does happen when we consider the interplay between the gravitons and the internal DoFs of the system, as opposed to the case where only the graviton bath contributes directly.

This analysis was performed by considering gravitons initially in the vacuum state. However, other possible states can significantly decrease the decoherence time, as we will see now.

\section{Decoherence for initial thermal, coherent, and squeezed states} \label{Sec:Thermal-coherent-squeezed-states}

\begin{figure}[!ht]
    \centering
    \includegraphics[width=0.8\linewidth]{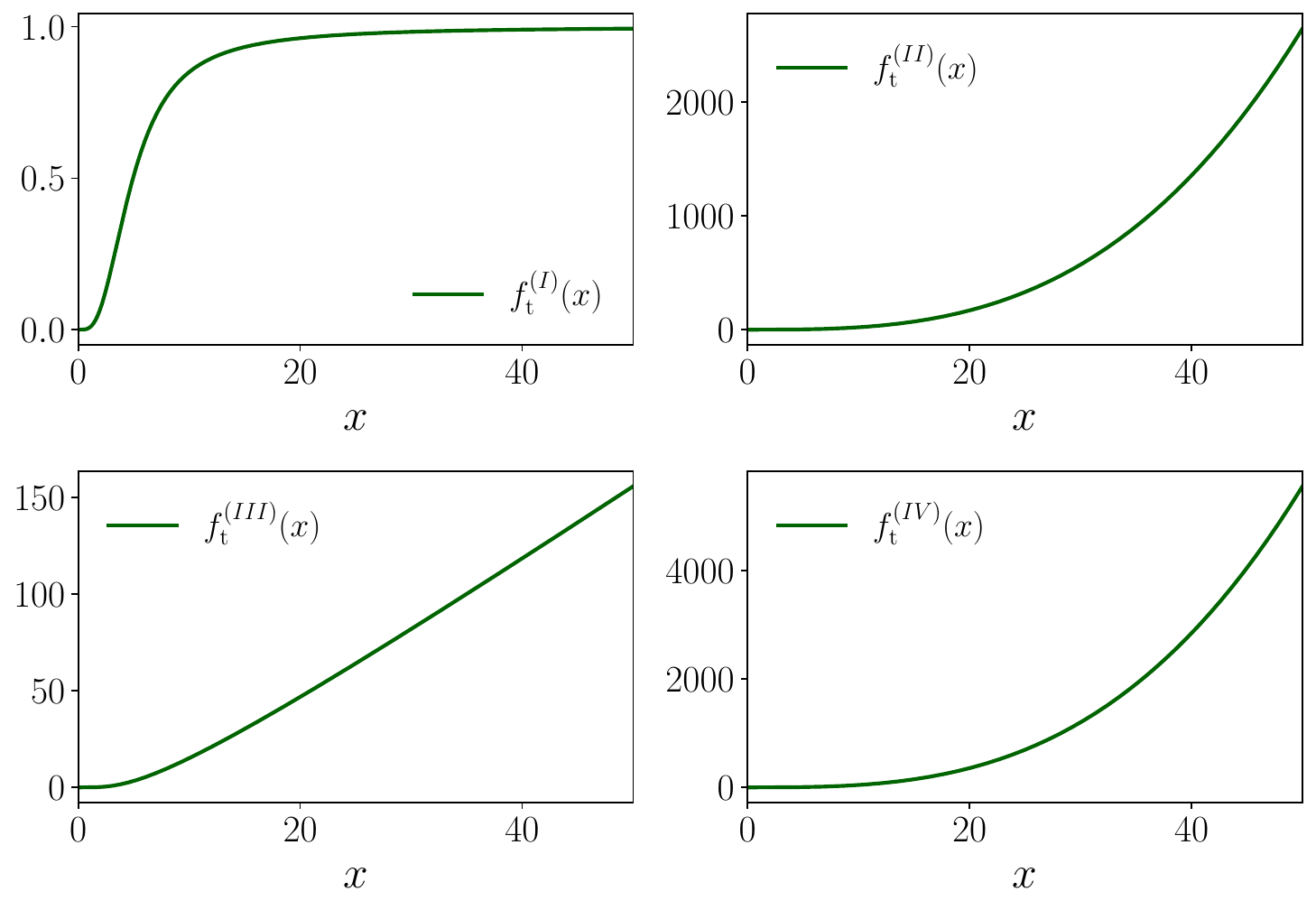}
    \caption{Different contributions for the decoherence function considering the gravitons to be initially in the \textbf{thermal state}.}
    \label{Fig:thermal_state}
\end{figure}

\begin{figure}[!ht]
    \centering
    \includegraphics[width=0.8\linewidth]{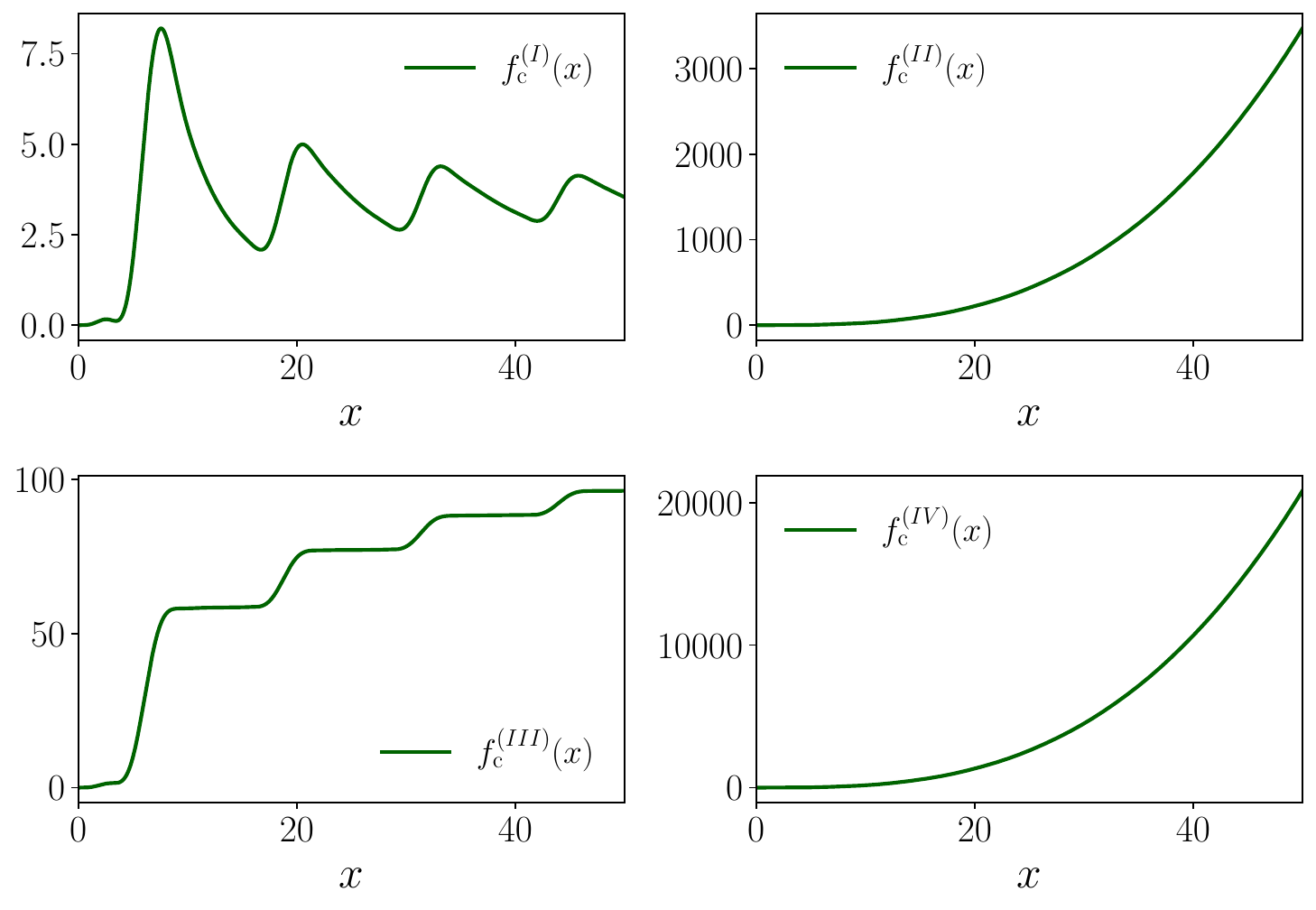}
    \caption{Different contributions for the decoherence function considering the gravitons to be initially in the \textbf{coherent state}.}
    \label{Fig:coherent_state}
\end{figure}

\begin{figure}[!ht]
    \centering
    \includegraphics[width=0.8\linewidth]{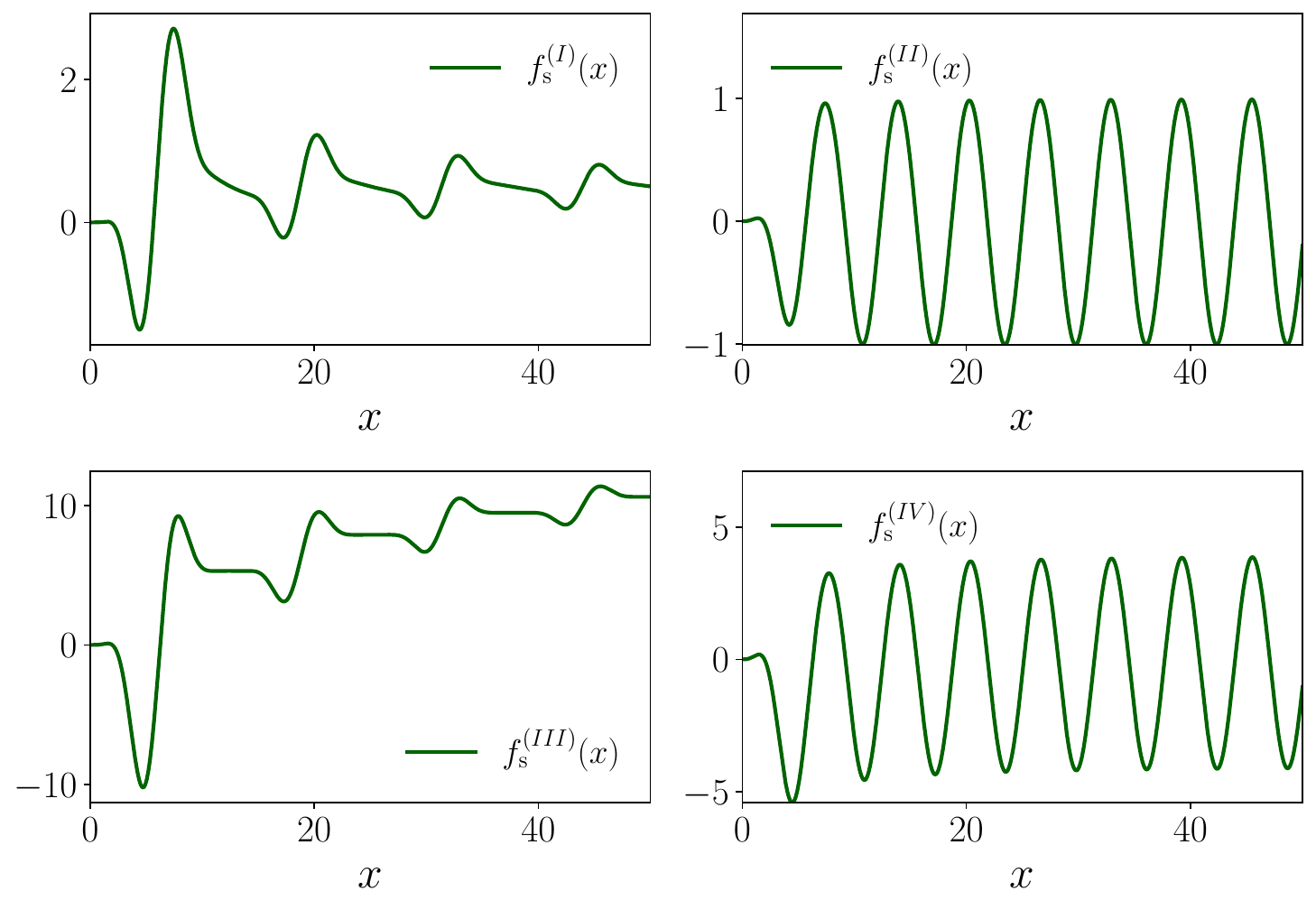}
    \caption{Different contributions for the decoherence function considering the gravitons to be initially in the \textbf{squeezed state}.}
    \label{Fig:squeezed_state}
\end{figure}

We can now repeat the same analysis for the other initial states for which the decoherence function is given by Eq.~\eqref{Conf1-Dec-function}. The functions $f_A$ were all listed in Section~\ref{Sec:The-decoherence-function}, and some of their relevant properties are shown in Tables~\ref{tab:thermal_state},~\ref{tab:coherent_state}, and~\ref{tab:squeezed_state} (see also Figures~\ref{Fig:thermal_state},~\ref{Fig:coherent_state}, and~\ref{Fig:squeezed_state}). The procedure is basically the same, so we will simply list the results.

\begin{table*}[!ht]
\centering 

\begin{tabular}{lcc}
\toprule 

\multicolumn{3}{c}{\textbf{Thermal state}} \\
\cmidrule(lr){1-3} 

\textbf{Function} & \textbf{Behavior for $x\ll1$} & \textbf{Behavior for $x\gg1$} \\
\midrule 

$f_{\rm t}^{(I)}(x)$            & $x^4/126$          & $1$      \\
$f_{\rm t}^{(II)}(x)$           & $4x^3/189$         & $4x^3/189$ \\
$f_{\rm t}^{(III)}(x)$          & $x^4/60$           & $4\ln2+4x-12\ln x$ \\
$f_{\rm t}^{(IV)}(x)$           & $2x^3/45$          & $2x^3/45$ \\

\bottomrule 
\end{tabular}

\caption{Different contributions for the decoherence function considering the gravitons to be initially in a \textbf{thermal state}.}
\label{tab:thermal_state}

\end{table*}

\begin{table*}[!ht]
\centering 

\begin{tabular}{lcc}
\toprule 

\multicolumn{3}{c}{\textbf{Coherent state}} \\
\cmidrule(lr){1-3} 

\textbf{Function} & \textbf{Behavior for $x\ll1$} & \textbf{Behavior for $x\gg1$} \\
\midrule 

$f_{\rm c}^{(I)}(x)$            & $x^4/48$          & $7/2$      \\
$f_{\rm c}^{(II)}(x)$           & $x^3/18$          & $x^3/36+\sin x$ \\
$f_{\rm c}^{(III)}(x)$          & $x^4/8$           & $28\gamma_E+16\ln3-68\ln2+28\ln x$ \\
$f_{\rm c}^{(IV)}(x)$           & $x^3/3$           & $x^3/6+4\sin x$ \\

\bottomrule 
\end{tabular}

\caption{Different contributions for the decoherence function considering the gravitons to be initially in a \textbf{coherent state}.}
\label{tab:coherent_state}

\end{table*}

\begin{table*}[!ht]
\centering 

\begin{tabular}{lcc}
\toprule 

\multicolumn{3}{c}{\textbf{Squeezed state}} \\
\cmidrule(lr){1-3} 

\textbf{Function} & \textbf{Behavior for $x\ll1$} & \textbf{Behavior for $x\gg1$} \\
\midrule 

$f_{\rm s}^{(I)}(x)$            & $x^4/96$          & $1/2$      \\
$f_{\rm s}^{(II)}(x)$           & $x^3/36$          & $\sin x$ \\
$f_{\rm s}^{(III)}(x)$          & $x^4/16$          & $4\gamma_E+4\ln\qty(81/2^9)+4\ln x$ \\
$f_{\rm s}^{(IV)}(x)$           & $x^3/6$           & $4\sin x$ \\

\bottomrule 
\end{tabular}

\caption{Different contributions for the decoherence function considering the gravitons to be initially in a \textbf{squeezed state}.}
\label{tab:squeezed_state}

\end{table*}

Just as we did for the case of the initial vacuum state, we begin by considering the short-time limit. For $t\ll\hbar/\Lambda_A$, we can use the expansions for the functions $f_A(x)$ around $x=0$ that are shown in Tables~\ref{tab:thermal_state},~\ref{tab:coherent_state}, and~\ref{tab:squeezed_state}. Once again, the purely graviton bath contribution dominates in this regime, and one finds the decoherence times to be given by
\begin{subequations}
\begin{equation} \label{Short-time-dec-time-thermal}
    t_{\rm dec}^{(\textrm{t})}=\qty[1+\frac{32}{21}\delta\Omega_{\rm t}\frac{(\pi k_B T_{\rm g})^6}{\Lambda^6}]^{-1/4}t_{\rm dec}^{(\textrm{v})},
\end{equation}
\begin{equation} \label{Short-time-dec-time-coherent}
    t_{\rm dec}^{(\textrm{c})}=\qty(1+\alpha^2)^{-1/4}t_{\rm dec}^{(\textrm{v})},
\end{equation}
\begin{equation} \label{Short-time-dec-time-squeezed}
    t_{\rm dec}^{(\textrm{s})}=e^{-r/2}t_{\rm dec}^{(\textrm{v})},
\end{equation}
\end{subequations}
where
\begin{equation}
    \delta\Omega_{\rm t}\equiv1+6\qty(\qty(\frac{\hbar}{\Lambda})^2-\frac{7}{20}\qty(\frac{\hbar}{\pi k_BT_{\rm g}})^2)\frac{GM_N}{R_N^3},
\end{equation}
and with $t_{\rm dec}^{(\textrm{v})}$ given by Eq.~\eqref{Short-time-dec-time-vacuum}. Note that there is a decrease in decoherence time when considering such states, with $t_{\rm dec}$ even exhibiting an exponential decay for the squeezed state.

Next, we turn to the long-time limit. Similar to the case of the initial vacuum state, one can see from Tables~\ref{tab:thermal_state} and~\ref{tab:coherent_state} that, for the thermal and coherent states, the contributions involving the internal DoFs keep increasing for $\Lambda_A t/\hbar\gg1$, while the others either saturate or exhibit a slower logarithmic behavior. Then, using the expressions from the tables, in the long-time limit, we find the decoherence times to be given by
\begin{subequations}
\begin{equation} \label{Long-time-dec-time-thermal}
    \tau_{\rm dec}^{(\textrm{t})}=\qty[1+\frac{32}{21}\delta\Omega_{\rm t}\frac{(\pi k_B T_{\rm g})^6}{\Lambda^6}]^{-1/3}\tau_{\rm dec}^{(\textrm{v})},
\end{equation}
\begin{equation} \label{Long-time-dec-time-coherent}
    \tau_{\rm dec}^{(\textrm{c})}=\qty(1+\frac{\alpha^2}{2})^{-1/3}\tau_{\rm dec}^{(\textrm{v})},
\end{equation}
\end{subequations}
with $\tau_{\rm dec}^{(\textrm{v})}$ given by Eq.~\eqref{Long-time-dec-time-vacuum}.

However, the situation is quite different for the initial squeezed state. The internal DoFs contributions behave as sine functions, and therefore they do not keep increasing. Due to the typical small values of the ratio $\mathscr{R}=\kappa/E_{\rm rest}^2$, this means that the contributions coming from the graviton bath alone also dominate in the long-time limit\footnote{It is important to remark that we are referring only to the additional contribution to the decoherence function for the squeezed state. One should keep in mind that, for all states, this is added to the vacuum term, for which the G+I contributions dominate for long times, as shown in Section~\ref{Sec:Vacuum-state}.}. Now, $f_{\rm s}^{(I)}(x)$ saturates at a constant value. Furthermore, although there is a logarithmic increase in $f_{\rm s}^{(III)}(x)$, this is still much smaller than the vacuum behavior, which goes as $\sim x^3$ as we saw in Section~\ref{Sec:Vacuum-state}. Therefore, the long-time limit simply reads
\begin{equation}
    \Gamma_1^{(\textrm{s})}(t)=(\cosh2r)\Gamma_1^{(\textrm{v})}(t),
\end{equation}
with $\Gamma_1^{(\textrm{v})}(t)$ given by Eq.~\eqref{Final-dec-func-long-time}. At last, the decoherence time is
\begin{equation} \label{Long-time-dec-time-squeezed}
    \tau_{\rm dec}^{(\textrm{s})}=\qty(\cosh2r)^{-1/3}\tau_{\rm dec}^{(\textrm{v})}.
\end{equation}
We see that, also in the long-time limit, there is an exponential decay of the decoherence time with respect to the vacuum state. In Ref.~\cite{Grishchuk1990}, the authors argue that relic gravitons should now be in squeezed states and also estimate a squeeze parameter of order $r\sim10^2$, for which $\qty(\cosh2r)^{-1/3}\sim10^{-29}$, a significant reduction in the graviton-induced decoherence time.

\section{The possibility of gravitational recoherence} \label{Sec:Recoherence}

Before we close this chapter, let us address one concern raised in Section~\ref{Sec:Vacuum-state}. We saw in Eq.~\eqref{Gamma_sat} that when we consider only the G contribution to the decoherence function for the vacuum state, it saturates and changes no further. The situation is a little different when we include the Newtonian potential, since the G+N contribution does not saturate to a fixed value but keeps increasing with time. So, let us now consider a particle with no internal DoFs (or simply frozen ones). Then, the decoherence function for the vacuum state reduces to
\begin{equation}
    \Gamma_1^{(\textrm{v})}(t)=\frac{16\Xi^2v^2m^2}{5\pi E_{\rm P}^2}\left\{ \qty(\frac{\Lambda}{\hbar})^2f_{\rm v}^{(I)}\qty(\frac{\Lambda t}{\hbar})-\frac{GM_N}{R_N^3}f_{\rm v}^{(III)}\qty(\frac{\Lambda t}{\hbar})\right\} .
\end{equation}

We already saw that, in the short-time limit, this reduces to Eq.~\eqref{Final-dec-func-short-time}. But now there are no internal DoFs to dominate in the long-time regime, and one may be concerned about the fact that the G contribution remains constant while the G+N contribution, which has a negative sign, keeps increasing with time. From Table~\ref{tab:vacuum_state} we see that $f_{\rm v}^{(I)}(x)\sim1$ and $f_{\rm v}^{(III)}(x)\sim8\ln x$ for $x\gg1$, and the decoherence function becomes
\begin{equation}
    \Gamma_1^{(\textrm{v})}(t)\simeq\frac{16}{5\pi}\qty(\frac{v}{c})^2\qty(\frac{m}{M_{\rm P}})^2\qty[1-8\qty(\frac{\hbar}{\Lambda})^2\frac{GM_N}{R_N^3}\ln\qty(\frac{\Lambda t}{\hbar})].
\end{equation}
This means that, when there are no internal DoFs to guarantee gravitational decoherence, not only does the function stop increasing with time, but it also starts to decrease (although very slowly). In light of this behavior, one may raise the concern that the "decoherence" function could ever become negative. In principle, this can happen, and we can compute the time it takes for that to occur. A quick analysis shows that $\Gamma_1^{(\textrm{v})}(t)<0$ for
\begin{equation}
    t\gtrsim\frac{\hbar}{\Lambda}\exp\qty[\frac{1}{8}\qty(\frac{\Lambda}{\hbar})^2\frac{R_N^3}{GM_N}].
\end{equation}
Now, even for a superposition size on the order of kilometers, $L_0\sim10^3$ m, we find $\qty(\frac{\Lambda}{\hbar})^2\frac{R_N^3}{GM_N}\sim10^{16}$ near Earth. Take the exponential of that number, and we are clearly talking about an infinitely long time before such recoherence can happen.

For the coherent and squeezed state, the situation is quite similar since the G+N contributions in those cases also grow as $\ln(\Lambda t/\hbar)$ for $t\gg\hbar/\Lambda$. For the thermal state, however, Table~\ref{tab:thermal_state} shows that $f_{\rm t}^{(III)}(x)\sim4x$, which grows faster than the logarithm. The thermal decoherence function, in the absence of internal DoFs and in the long-time limit, reads
\begin{equation}
    \Gamma_1^{(\textrm{t})}(t)\simeq\frac{16}{5\pi}\qty(\frac{v}{c})^2\qty(\frac{m}{M_{\rm P}})^2\qty[1+\qty(\frac{\pi k_B T_{\rm g}}{\Lambda})^2\qty(1-\frac{4\hbar}{\pi k_B T_{\rm g}}\frac{GM_N}{R_N^3}t)].
\end{equation}
In that case, the condition $\Gamma_1^{(\textrm{t})}(t)<0$ is achieved by
\begin{equation}
    t\gtrsim\frac{\pi k_BT_{\rm g}}{4\hbar}\frac{R_N^3}{GM_N}\qty[1+\qty(\frac{\Lambda}{\pi k_B T_{\rm g}})^2].
\end{equation}
For $T_{\rm g}\sim1$ K, we have $\frac{\pi k_BT_{\rm g}}{4\hbar}\frac{R_N^3}{GM_N}\sim10^{17}$ s (near Earth), which is about the age of the universe.

In conclusion, the presence of the Newtonian potential seems to slightly slow down the decoherence mechanism, which can eventually lead to the recoherence of the system. Although possible in principle, this is not expected to happen for typical parameters. Even if it were so, we saw in the previous sections that decoherence will inevitably prevail when we include the interaction with the system's own internal structure.

Nevertheless, we should mention that recoherence mechanisms can appear in non-Markovian models for open quantum systems, such as when one describes quantum fields in cosmological backgrounds (see Ref.~\cite{Colas2023}). The connection between memory (non-Markovian) effects and recoherence is explored in Ref.~\cite{Kranas2025}.
\chapter{Graviton-induced entropy production}
\label{chap:entropy}

Let us consider once again an open system coupled with a bath of gravitons, only now we include an external agent acting upon it according to some force protocol $f(t)$ in a given time interval. Due to the coupling with the environment, whose final state is not of interest, the dynamics of the system evolves according to the combined action of the deterministic force $f(t)$ and a stochastic force term encompassing the influence of the environment, as discussed in Section~\ref{Sec:Langevin-equation} on more general grounds. Now, this coarse-grained description of the environment inevitably leads to dissipation. We then turn to the question of how much work is dissipated when we try to move a system through a bath of gravitons, which we attempt to answer by verifying how the fluctuation theorem applies to the problem at hand (see Appendix~\ref{app:fluctuation-theorem} for a brief review of the fluctuation theorem, or, more precisely, the Crooks fluctuation theorem and Jarzynski's equality).

The fluctuation theorem was established for a quantum system under the effects of relativistic time dilation in Ref.~\cite{Basso2023}, and it was later generalized to include the effects of \emph{classical} spacetime curvature~\cite{Basso2025} and to apply to a quantum field~\cite{Costa2025}. While these references treated spacetime as a classical entity, here we aim to apply the fluctuation theorem to a system under the influence of \emph{quantum} spacetime fluctuations.

One of the main conceptual difficulties faced in establishing a quantum fluctuation theorem is the definition of work. While there seems to be general agreement in defining work via the two-time measurement scheme~\cite{Talkner2007} for closed systems, the proper definition for open quantum systems is still under debate. Here, we follow the proposal of Ref.~\cite{Hu2012} to define work as it is done in classical systems by making sense of the notion of trajectories in quantum mechanics. In order to do so, we shall turn to the \emph{decoherent} (or \emph{consistent}) \emph{histories} formalism, which we discussed in Section~\ref{Sec:Decoherent-histories}. As we saw in that section, the solutions of the Langevin equation describe trajectories of the system, provided they are followed with an accuracy of an order $\sim N^{-1}$, with $N$ being the noise kernel of the environment. The idea is then to use these solutions to define the work done on the open system.

Since we are going to need the expressions for the solutions to the Langevin equation, we begin by discussing them in Section~\ref{Sec:Langevin-gravitons}. Then, in Section~\ref{Sec:Work-and-entropy}, we use these solutions to define work and establish the fluctuation theorem in order to quantify the entropy production.

\section{The Langevin equation with graviton noise} \label{Sec:Langevin-gravitons}

We begin by going back to Eq.~\eqref{rho(xi,xi',t)}, which describes the external degrees of freedom density matrix of a quantum system coupled with both gravitational and its own internal degrees of freedom. As we have discussed in Chapter~\ref{chap:quantum-system}, the term containing the internal DoFs noise kernel may also be rewritten in terms of a Gaussian stochastic variable, just as we have done for the gravitational noise kernel. Indeed, Eq.~\eqref{rho(xi,xi',t)} is equivalent to
\begin{align}
    &\rho_{\rm ext}(\xi,\xi',t)=\int \dd\xi(0)\dd\xi'(0)\,\rho_{\rm ext}(\xi(0),\xi'(0),0) \int\mathcal{D}\xi\mathcal{D}\xi'\,e^{\frac{i}{2}m_0\delta^{ij}\int \dd t\,\qty(\dot{\xi}_i\dot{\xi}_j-\dot{\xi}_i'\dot{\xi}_j')} \nonumber \\    
    &\hspace{1cm}\times\int\mathcal{D}\mathcal{N}\,\mathscr{P}[\mathcal{N}]\mathscr{P}[\mathcal{N}_{\rm int}]\,e^{-im\int \dd t\,\mathcal{N}^{ij}\qty(\xi_i\xi_j-\xi_i'\xi_j')}e^{-\frac{i}{2}\int\dd t\,\mathcal{N}_{\rm int}\qty(\dot{\xi}_i\dot{\xi}_j-\dot{\xi}_i'\dot{\xi}_j')} \nonumber \\
    &\hspace{1cm}\times e^{i\int\dd t\,\mathcal{N}_{\rm int}\mathcal{N}^{ij}\qty(\xi_i\xi_j-\xi_i'\xi_j')},
\end{align}
where $\mathscr{P}$ denotes the Gaussian probability density and $\mathcal{N}$ ($\mathcal{N}_{\rm int}$) is the stochastic variable describing the interaction with the gravitons (internal degrees of freedom).

The system is then described by a stochastic effective action given by
\begin{align}
    S_{\rm SEA}[\xi,\xi']&=S_{\rm p}[\xi]-S_{\rm p}[\xi']-\frac{1}{2}\int\dd t\,\mathcal{N}_{\rm int}\qty(\dot{\xi}_i\dot{\xi}_j-\dot{\xi}_i'\dot{\xi}_j') \nonumber \\
    &-\int\dd t\,\qty(m-\mathcal{N}_{\rm int})\mathcal{N}^{ij}\qty(\xi_i\xi_j-\xi_i'\xi_j'),
\end{align}
with
\begin{equation}
    S_{\rm p}[\xi]=\int \dd t\,\qty[\frac{1}{2}m\,\delta_{ij}\dot{\xi}^i\dot{\xi}^j-V(\xi)],
\end{equation}
and we included a potential function $V(\xi)$ to account for the possibility of an external agent acting on the system.

The equation of motion is obtained by setting $\delta S_{\rm SEA}=0$ as usual, resulting in
\begin{equation}
    \qty[m-\mathcal{N}_{\rm int}(t)]\Ddot{\xi}_i(t)+\pdv{V}{\xi^i}-\dot{\mathcal{N}}_{\rm int}(t)\dot{\xi}_i(t)+2\qty[m-\mathcal{N}_{\rm int}(t)]\mathcal{N}_{ij}(t)\xi^j(t)=0.
\end{equation}
Analytically solving this second-order differential equation for the geodesic deviation $\xi(t)$ is no easy task. Since we are going to need an explicit form for the solution, we will drop the internal DoFs contribution and analyze the effects only due to the gravitons for simplicity. Additionally, we choose to work on a perturbative regime for $\xi_i(t)$. Then, by taking $\mathcal{N}_{\rm int}\to0$ one is left with\footnote{We changed the sign of the noise term for convenience, which is allowed since this is simply a change of variables in the path integral.}
\begin{equation} \label{Langevin-like-equation}
    m\Ddot{\xi}_i(t)=f_i(t)+2m\mathcal{N}_{ij}(t)\xi_0^j(t),
\end{equation}
where $f_i$ is defined such that $V(\xi)=-f_i(t)\xi^i(t)$ and $\xi_0^i$ is the solution to $m\Ddot{\xi}_0^i=f^i$.

Eq.~\eqref{Langevin-like-equation} is the Langevin equation for the time evolution of a system which is affected by a stochastic force term that comes from the interaction with the bath of gravitons. The effects of such stochastic noise are also analyzed in Refs.~\cite{Parikh2020,Parikh_2021,Parikh2021,Kanno2021,Cho2022,Cho_2023,Chawla2023}.

The solution to the Langevin equation for the initial conditions $\xi^i(0)=\dot{\xi}^i(0)=0$ is given by
\begin{equation} \label{Langevin-like-solution}
    \xi_i(t)=\frac{1}{m}\int_0^t\dd t'\,\qty(t-t')f_i(t')+2\int_0^t\dd t'\,\qty(t-t')\mathcal{N}_{ij}(t')\xi_0^j(t').
\end{equation}
This can be seen as the classical limit for the geodesic separation between two test masses, where the stochastic behavior comes from the quantum fluctuations of the gravitational field. However, as we discussed in the beginning of the chapter, for paths with accuracy determined by the noise kernel, we can use the solution to the Langevin equation to provide a definition of work for the \emph{quantum} system in the same way that is done for classical systems, which in turn is heavily dependent on the notion of a trajectory in space. This will allow us to establish the fluctuation theorem for the quantum system and we shall see how dissipation arises when an external agent tries to move the system through a bath of gravitons.

\section{Dissipated work and entropy} \label{Sec:Work-and-entropy}

In order to discuss the validity of the fluctuation theorem, we proceed as in Ref.~\cite{Hu2012}. By fluctuation theorem we mean Jarzynski's equality, which is the integral form of the Crooks fluctuation theorem~\cite{Jarzynski1997,Crooks1999,Horowitz2007,Jarzynski2007,Jarzynski2008,Campisi2011} (see Appendix~\ref{app:fluctuation-theorem}). For closed quantum systems, the derivation of the theorem relies on the hypothesis of initial thermal state of the entire system. However, our analysis of a quantum particle interacting with a bath of gravitons is built under the assumption of initial product state, which is obviously not a thermal one. In order to circumvent this issue, we take the initial time to be $t_0=-\infty$. At this time instant, we assume that the total state of the system is described by a tensor product, $\hat{\rho}(-\infty)=\hat{\rho}_{\rm sys}(-\infty)\otimes\hat{\rho}_{\rm grav}(-\infty)$, where $\hat{\rho}_{\rm grav}$ is a thermal state of gravitons. We then allow the system to evolve according to the total action with $f^i(t)=f^i(0)$ for $t<0$ so that we obtain a total thermal state at $t=0$. Then, the driving force starts changing according to some arbitrary protocol until a given time instant $t=\tau$. The work performed on the system by the external agent in the interval $\qty[0,\tau]$ is then defined by
\begin{equation}
    W\equiv-\int_0^\tau \dd t\,\dot{f}_i(t)\xi^i(t),
\end{equation}
where $\xi^i(t)$ is the solution to the Langevin equation given in Eq.~\eqref{Langevin-like-solution}. By using the explicit forms of the solutions, we can write
\begin{equation} \label{Work-done-on-the-sytem}
    W=-\frac{1}{m}\int_0^\tau \dd t \dd t'\,\dot{f}^i(t)g(t-t')f_i(t')-2\int_0^\tau \dd t\dd t'\,\dot{f}^i(t)g(t-t')\mathcal{N}_{ij}(t')\xi^j_0(t'),
\end{equation}
where $g(t-t')=\qty(t-t')\theta(t-t')$.

We can see from Eq.~\eqref{Work-done-on-the-sytem} that the work performed on the system is linear in $\mathcal{N}_{ij}(t)$, which is a Gaussian random process. Therefore, $W$ itself must also be a Gaussian random variable with its statistics specified by the first two moments $\expval{W}$ and $\sigma_W^2=\expval{W^2}-\expval{W}^2$. We can then write the work probability density function
\begin{equation}
    \mathscr{P}(W)=\frac{1}{\sqrt{2\pi\sigma_W^2}}e^{-\qty(W-\expval{W})^2/2\sigma_W^2}.
\end{equation}
Throughout this chapter, the brackets denote the stochastic average with a Gaussian probability density.

Before we proceed, it is worth emphasizing that the average with $\mathscr{P}(W)$ is the same stochastic average with $\mathscr{P}(\mathcal{N})$ due to the linear dependence of $W$ on $\mathcal{N}_{ij}(t)$. In usual formulations of the classical fluctuation theorem, such stochastic force is not considered, and the probabilistic aspect of the work done on the system comes from the initial thermal state assumption. Each sampling from the initial state gives rise to a trajectory, and the average is performed over an ensemble of such realizations. For the quantum fluctuation theorem, the inherent quantum uncertainty of the initial state contributes a further probabilistic aspect. However, we chose to follow an equivalent initial state preparation method based on a product initial state for the total system. This choice replaces the system's dependence on the initial state with the properties of noise statistics and, consequently, one is left with only one probabilistic element instead of two~\cite{Hu2012}.

The moments of the work distribution can be computed as follows. First, since $\expval{\mathcal{N}^{ij}(t)}=\frac{1}{2}\Phi_{ij}$, with $\Phi_{ij}$ being the tidal tensor\footnote{Recall that we changed the sign of the stochastic noise variable.}, we have
\begin{equation} \label{First-moment-of-work}
    \expval{W}=-\frac{1}{m}\int_0^\tau \dd t\dd t'\,\dot{f}^i(t)g(t-t')f_i(t')-\Phi_{ij}\int_0^\tau \dd t\dd t'\,\dot{f}^i(t)g(t-t')\xi^j_0(t').
\end{equation}
Additionally, by using\footnote{We dropped the index g since, in this section, there is to be no confusion with the noise kernel coming from any other environment.} $\expval{\mathcal{N}^{ij}(t)\mathcal{N}^{kl}(t')}=N^{ijkl}(t,t')$, which is the noise kernel of gravitons, we also find
\begin{equation} \label{Second-moment-of-work}
    \sigma_W^2=\int_0^\tau \dd t\dd t'\,\dot{f}^i(t)\sigma_{ij}(t,t')\dot{f}^j(t'),
\end{equation}
where
\begin{equation} \label{sigma_ij}
    \sigma_{ij}(t,t')=4\int_0^\tau \dd t_1\dd t _2\,g(t-t_1)g(t'-t_2)N_{ijkl}(t_1,t_2)\xi^k_0(t_1)\xi^l_0(t_2).
\end{equation}
Note that, while $\expval{W}$ is independent of the noise kernel of gravitons, the second moment $\sigma_W^2$ is not.

With the expressions for the first and second moments of work, the probability distribution is completely specified. One can now compute
\begin{equation} \label{Average-of-exponentiated-work}
    \expval{e^{-\beta W}}=\int \mathcal{D}W\,\mathscr{P}(W)e^{-\beta W},
\end{equation}
where $\beta$ is the inverse temperature of the bath of gravitons. Since Jarzynski's equality holds for the entire system, as argued in the beginning of this section, we must have
\begin{equation} \label{Jarzynski}
    \expval{e^{-\beta W}}=e^{-\beta\Delta F},
\end{equation}
where $\Delta F=F(\tau)-F(0)$ is the free energy difference between time instants $t=0$ and $t=\tau$, calculated quantum mechanically for the total system.

A direct calculation turns Eq.~\eqref{Average-of-exponentiated-work} into
\begin{equation}
    \expval{e^{-\beta W}}=e^{-\beta\qty(\expval{W}-\beta\sigma_W^2/2)},
\end{equation}
which, upon comparison with Eq.~\eqref{Jarzynski}, allows us to identify the free energy difference of the total system as
\begin{equation}
    \Delta F=\expval{W}-\beta\sigma_W^2/2.
\end{equation}
At last, we identify the dissipated work $W_{\rm diss}$ with
\begin{align} \label{Dissipated-work}
    W_{\rm diss}&=\expval{W}-\Delta F=\beta\sigma_W^2/2 \nonumber \\
    &=2\beta\int_0^\tau\dd t_1\cdots\dd t_4\,\dot{f}^i(t_1)\dot{f}^j(t_2)g(t_1-t_3)g(t_2-t_4)N_{ijkl}(t_3,t_4)\xi_0^k(t_3)\xi_0^l(t_4).
\end{align}
For movement along the $z-$direction, we computed the noise kernel $N(t,t')=N_{3333}(t,t')$ in Appendix~\ref{app:noise-kernels} (naturally, in this case we are interested in the initial thermal state for the gravitons).

Finally, we conclude that there is an entropy production in the system given by
\begin{equation} \label{Entropy}
    \expval{\Sigma}=\beta W_{\rm diss}=\frac{\beta^2\sigma_W^2}{2}.
\end{equation}
Note that when there are no gravitons, or when $f=0$, we have $\sigma_W^2=0$ and, consequently, there is no production of entropy. Additionally, we remark that the thermodynamical interpretation of this entropy is tied to the interpretation of the parameter $\beta$ as a proper temperature, as discussed in Section~\ref{Sec:Master-eq-grav-dec}. And, lastly, we point out that, since Eq.~\eqref{Entropy} arises due to the fluctuating spacetime, one expects this entropy production to be a universal and unavoidable feature of driven (classical and quantum) systems.
\cleardoublepage
\phantomsection
\chapter*{Conclusions and future work}
\addcontentsline{toc}{chapter}{Conclusions and future work}
\markboth{Conclusions and future work}{Conclusions and future work}

The quantum fluctuations of spacetime, however weak, cannot be avoided since spacetime is the background in which matter propagates and interacts. Here, we quantified the consequences of such realization for the decoherence of spatial superpositions and entropy production. We began by considering a composite quantum system in a spacetime described by a classical static Newtonian potential and the quantum gravitational radiation degrees of freedom. Such a spacetime also induces interaction between the center-of-mass variable and the internal degrees of freedom of the composite system. Since we were interested in studying spatial superpositions of the center-of-mass variable, we integrated out both the gravitational radiation degrees of freedom and the internal variables of the system itself. The description of this open quantum system coupled with two mutually interacting environments was accomplished within the influence functional approach, which allowed us to obtain the decoherence function for the center-of-mass variable in a superposition of two classically distinguishable paths.

We found that, although the interaction with gravitons alone cannot decohere spatial superpositions of microscopic systems (in agreement with previous results in the literature~\cite{Kanno2021}), the interplay between the gravitons and the system's internal structure will inevitably lead to decoherence, even if it happens for typically long times. Such decoherence times were found to depend on the initial state of the gravitons. In fact, for gravitons initially in thermal, coherent, and squeezed states, the decoherence time exhibits a decrease with respect to the value found by considering an initial vacuum state. This decrease can be significant, especially for an initial squeezed state, which is expected to describe relic gravitons~\cite{Grishchuk1990}.

Apart from the enhancement of gravitational decoherence arising from a suitable choice of the initial graviton state, we emphasize the role of the interplay between gravity and the internal degrees of freedom of the system in leading to decoherence, even in situations where it was not expected to occur at all. One can think of the system's internal structure acting as an environment that, due to the universal aspect of gravity, works as a mediator and amplifier of the effects of the quantum spacetime fluctuations. This opens the possibility of considering more general scenarios in which a system is in simultaneous interaction with both a gravitational and a non-gravitational environment. Since these will inevitably interact with each other, we are left to wonder what the effects of such interplay on the system are. Whether non-gravitational environments can be used to mediate and amplify gravitational decoherence, as the internal structure of the system can, is a subject of future investigation.

In this work, we also analyzed the entropy production coming from the driving of a quantum system through a bath of gravitons. This was done via the decoherent histories approach to quantum mechanics, which allowed us to give meaning to the concept of trajectory in space and to provide a suitable definition of work done on the open quantum system in complete analogy with classical mechanics. With this definition at hand, the fluctuation theorem was established in the form of Jarzynski's equality, from which the entropy production was estimated.  We remark on the universal character of this entropy production, which arises from the quantum degrees of freedom of the universal gravitational radiation.

Although we considered the entropy produced when the system is driven by an external agent through the graviton bath, it would be interesting to investigate the case in which the driving agent is spacetime itself. For instance, one could analyze the work done by the stochastic force that comes from the gravitational fluctuations. However, such a work distribution is not expected to be Gaussian, which would render the analysis more complex.

In general, it would be interesting to see how these results apply to relativistic particles and even quantum fields. For instance, one could study how graviton-induced decoherence affects wave-packet dispersion in neutrino propagation (and, in the spirit of the previous discussion, how other non-gravitational environments, like matter effects, could amplify such decoherence and modify the neutrino oscillation probability formula). Finally, it would also be interesting to investigate how different spacetimes could affect our results. Here, we have expanded the gravitational field around a metric in the Newtonian limit, but the same can be done for any other classical solution to Einstein's equation, such as the Schwarzschild solution or the Robertson-Walker expanding universe. Whether such curved backgrounds shall have a significant impact on the results of graviton-induced decoherence and entropy production remains to be seen.

\let\oldthispagestyle\thispagestyle
\renewcommand{\thispagestyle}[1]{\oldthispagestyle{empty}}
\part{Appendices} \label{Part:App}
\let\thispagestyle\oldthispagestyle
\appendix 


\chapter{Differential geometry}
\label{app:Differential}

The main goal of this appendix is to provide a brief review of the mathematics behind the general theory of relativity, namely differential geometry. This is by no means an extensive or complete presentation of the subject, with many mathematical concepts and demonstrations left out. Our approach is heavily based on mathematical physics textbooks such as \cite{Nakahara2003,Pires2015}, as well as some very good books on general relativity \cite{Carroll,Wald1984,Weinberg2013,Stewart1993}. For a more mathematically rigorous approach the reader is referred to Ref. \cite{Petersen2006}.

\section{Differentiable manifolds}

We begin with what is probably the most fundamental concept in differential geometry, the one of a differentiable manifold (sometimes we just write manifold, leaving the word "differentiable" implicit). The basic intuitive idea of a manifold is that of a space that may or may not be curved while looking flat in small enough regions. The entire manifold is then constructed by smoothly sewing together those regions. Now, in order to give this idea a precise mathematical meaning, we are going to need some preliminary definitions.

The first basic definition we are going to need is that of a topological space. A \textbf{topological space} $(X,\mathcal{T})$ consists of a set $X$ together with a collection $\mathcal{T}$ of subsets of $X$ satisfying the following three properties:
\begin{enumerate}
    \item If $O_\alpha\in\mathcal{T}$ for all $\alpha$, then
    \begin{equation}
        \bigcup_{\alpha}O_\alpha\in\mathcal{T}.
    \end{equation}
    \item If $O_1,\dots,O_n\in\mathcal{T}$ ($n$ is a finite number), then
    \begin{equation}
        \bigcap_{i=1}^nO_i\in\mathcal{T}.
    \end{equation}
    \item The entire set $X$ and the empty set $\emptyset$ are in $\mathcal{T}$.
\end{enumerate}
The collection $\mathcal{T}$ is referred to as a \textbf{topology} on $X$, and subsets of $X$ which are listed in the collection $\mathcal{T}$ are called \textbf{open sets}.

Another indispensable idea is the notion of a metric space. A \textbf{metric} $d:X\times X\to\mathbb{R}$ is a function that satisfies the conditions:
\begin{itemize}
    \item $d(x,y)=d(y,x)$,
    \item $d(x,y)\geq0$, where the equality holds if and only if $x=y$, and
    \item $d(x,y)+d(y,z)\geq d(x,z)$,
\end{itemize}
for all $x,y,z\in X$. If $X$ is endowed with a metric $d$, $X$ is made into a topological space whose open sets are given by open discs centered at each point $x$,
\begin{equation}
    O_\epsilon(x)=\qty{y\in X|d(x,y)<\epsilon},
\end{equation}
and all their possible unions. The topology $\mathcal{T}$ thus defined is called the \textbf{metric topology} determined by $d$. The topological space $(X,\mathcal{T})$ is called a \textbf{metric space}.

If $(X,\mathcal{T})$ and $(Y,\mathcal{J})$ are topological spaces, a map $f:X\rightarrow Y$ is said to be \textbf{continuous} if the inverse image, $f^{-1}(O)=\qty{x\in X|f(x)\in O}$, of every open set $O$ in $Y$ is an open set in $X$. If $(X,\mathcal{T})$ and $(Y,\mathcal{J})$ are topological spaces, a map $f:X\rightarrow Y$ is a \textbf{homeomorphism} if it is continuous and has an inverse $f^{-1}:Y\rightarrow X$ which is also continuous. If there is a homeomorphism between $X$ and $Y$, $X$ is said to be \textbf{homeomorphic} to $Y$ and vice versa.

\begin{figure}[!ht]
    \centering
    \includegraphics[width=0.6\linewidth]{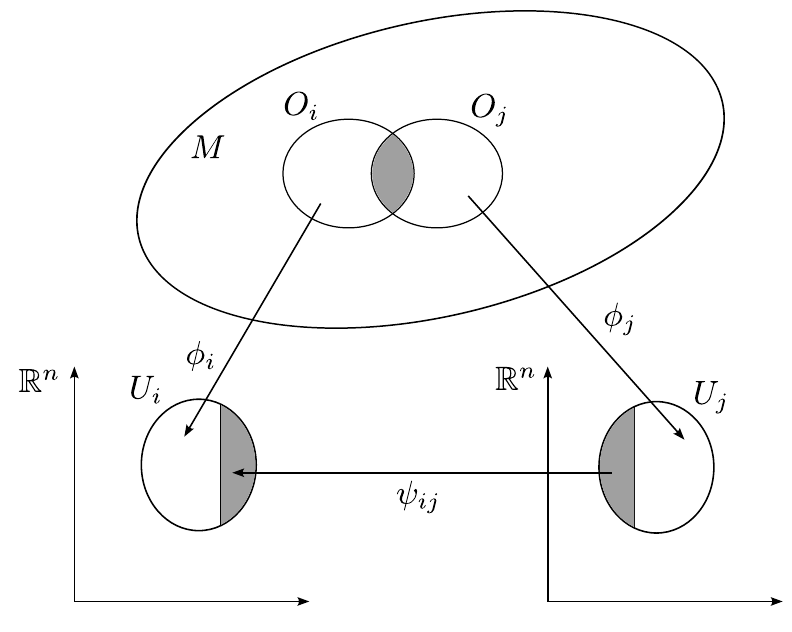}
    \caption{Homeomorphisms $\phi_i$ and $\phi_j$ map open sets $O_i$ and $O_j$ on the manifold $M$ into open sets $U_i$ and $U_j$ on Euclidean spaces.}
    \label{Manifold}
\end{figure}

We are now ready to give a precise definition of differentiable manifolds. A set $M$ is an $n-$dimensional \textbf{differentiable manifold} if
\begin{itemize}
    \item $M$, together with a collection of its subsets, is a topological space;
    \item $M$ is provided with a family of pairs $\qty{(O_i,\phi_i)}$ such that $\qty{O_i}$ is a family of open sets which covers $M$, that is, $\bigcup_iO_i=M$ and $\phi_i$ is a homeomorphism from $O_i$ onto an open subset $U_i$ of $\mathbb{R}^n$ (Figure \ref{Manifold});
    \item given $O_i$ and $O_j$ such that $O_i\cap O_j\neq\emptyset$ for $i\neq j$, the map $\psi_{ij}=\phi_i\circ\phi_j^{-1}:\phi_j(O_i\cap O_j)\mapsto\phi_i(O_i\cap O_j)$ is infinitely differentiable.
\end{itemize}

The pair $(O_i,\phi_i)$ is called a \textbf{chart} or a \textbf{coordinate system}, while the whole set $\qty{(O_i,\phi_i)}$ constitutes an \textbf{atlas}. The subset $O_i$ is called the \textbf{coordinate neighborhood} and $\phi_i$ the \textbf{coordinate function}. Given a point $p\in M$, the homeomorphism $\phi_i$ is represented by the set of coordinates $\qty{x^{\mu}(p)}=\qty{x^0(p),\dots,x^{n-1}(p)}$.

\begin{figure}[!ht]
    \centering
    \includegraphics[width=0.6\linewidth]{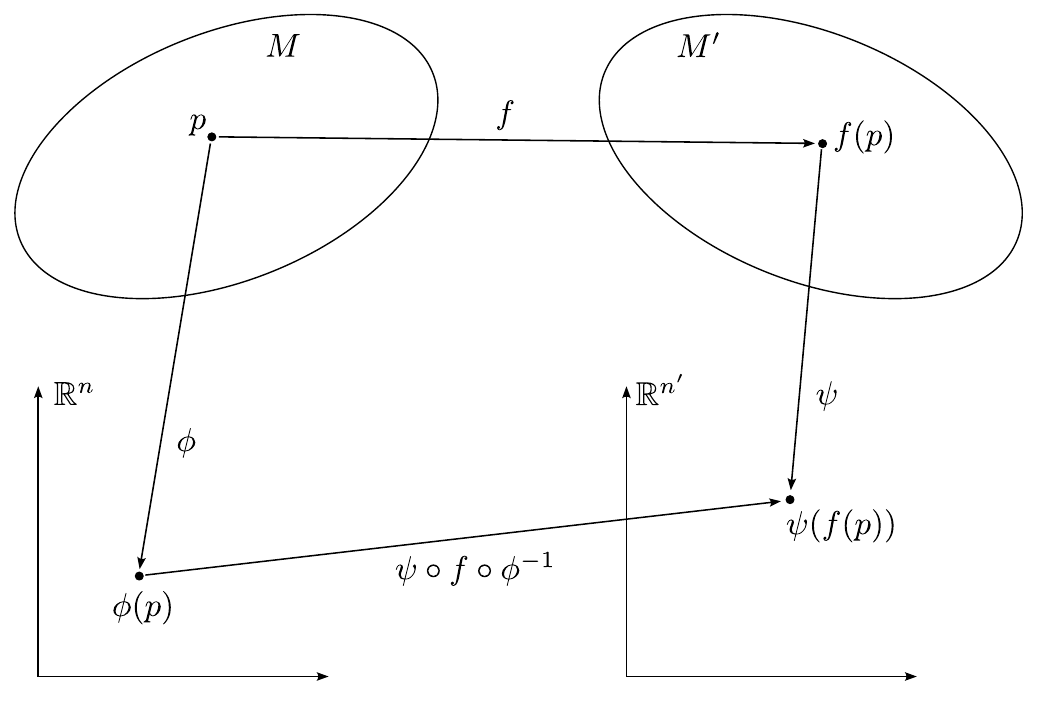}
    \caption{The map $f:M\to M'$ has a coordinate representation given by $\psi\circ f\circ\phi^{-1}:\mathbb{R}^n\to\mathbb{R}^{n'}$.}
    \label{Maps-on-Manifolds}
\end{figure}

A transformation that leaves invariant the differential structure described above is called a diffeomorphism. To be more precise, let $f:M\to M'$ be a homeomorphism and $\phi$ and $\psi$ coordinate functions, where the dimensions of $M$ and $M'$ are $n$ and $n'$, respectively (Figure \ref{Maps-on-Manifolds}). If $\psi\circ f\circ\phi^{-1}$ is invertible, and both $y=\psi\circ f\circ\phi^{-1}(x)$ and $x=\phi\circ f^{-1}\circ\psi^{-1}(y)$ are $C^{\infty}$ (infinitely differentiable), the map $f$ is said to be a \textbf{diffeomorphism} and $M$ and $M'$ are said to be \textbf{diffeomorphic}, which we denote as $M\equiv M'$. It follows that if $M\equiv M'$ then $n=n'$. In fact, two diffeomorphic manifolds are regarded as being the same manifold.

\begin{figure}[!ht]
    \centering
    \includegraphics[width=0.5\linewidth]{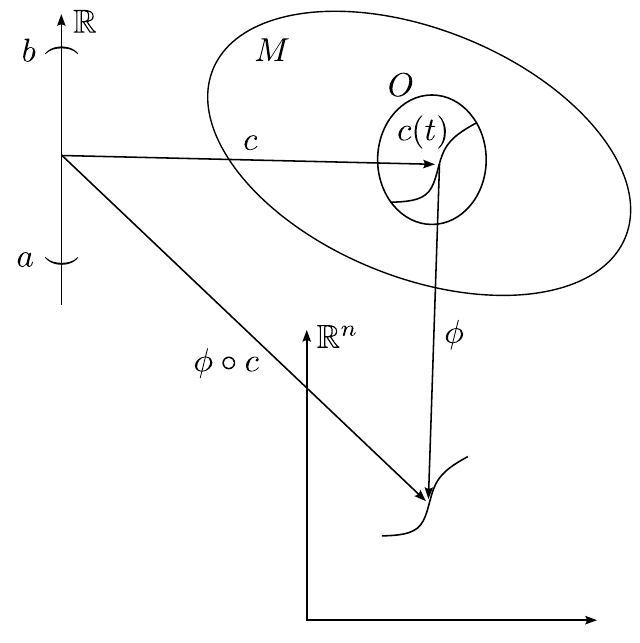}
    \caption{A curve $c$ in $M$, parametrized by $t$, and its coordinate representation $\phi\circ c$.}
    \label{Curves-on-manifolds}
\end{figure}

Since in physics the notion of a vector, and moreover a tensor, is a very useful one, it will be interesting to define vector spaces on a manifold. In order to do so, we need to define a very special class of maps called \textbf{curves}. An open curve on an $n-$dimensional manifold $M$ is the map $c:(a,b)\to M$, where $(a,b)$ is an open interval in $\mathbb{R}$ such that $a<0<b$ (Figure \ref{Curves-on-manifolds}). On a coordinate system $(O,\phi)$, a curve $c(t)$ has the coordinate representation $x=\phi\circ c:\mathbb{R}\to\mathbb{R}^n$.

At last, we define a \textbf{function} on $M$ as a smooth map from $M$ to $\mathbb{R}$. On a coordinate system $(O,\phi)$, we can build the coordinate representation of $f$ as $f\circ\phi^{-1}:\mathbb{R}^n\to\mathbb{R}$, which is a real-valued function of $n$ variables. The set of smooth functions on $M$ is denoted by $\mathcal{F}(M)$.

We are now ready to introduce the concept of tensors in a manifold.

\section{Tensors}

\begin{figure}[!ht]
    \centering
    \includegraphics[width=0.6\linewidth]{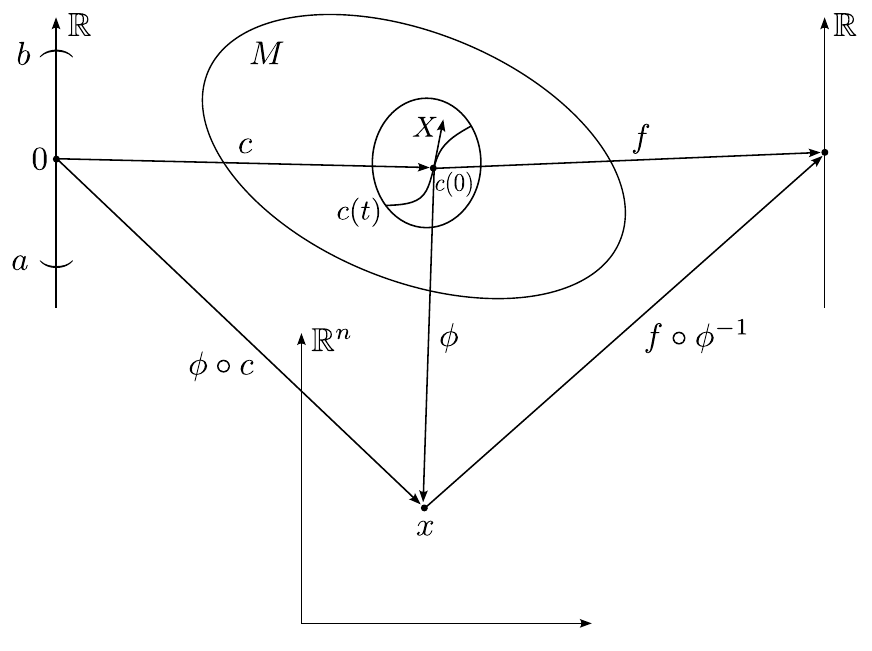}
    \caption{A curve $c$ along with a function $f$ and a local coordinate system $\phi$ define the tangent vector $X$ at a point of the manifold $M$ in the direction determined by $c(t)$.}
    \label{Vector-on-Manifolds}
\end{figure}

Let us begin with the more usual kind of tensor, namely a vector. In order to give a precise definition, let us consider a curve $c:(a,b)\to M$ and a function $f:M\to\mathbb{R}$, where $(a,b)$ is an open interval in $\mathbb{R}$ containing the point $t=0$, with $t$ being an arbitrary parametrization of the curve (Figure \ref{Vector-on-Manifolds}). By choosing a local coordinate function $\phi$ and using the chain rule, the rate of change of the function $f$ at $t=0$ along the curve $c$ is given by
\begin{align}
    \eval{\dv{f(c(t))}{t}}_{t=0}&=\eval{\dv{t}(f\circ c)}_{t=0}=\eval{\dv{t}\qty[(f\circ\phi^{-1})\circ(\phi\circ c)]}_{t=0} \nonumber \\
    &=\eval{\pdv{(f\circ\phi^{-1})}{x^\mu}\dv{(\phi\circ c)^\mu}{t}}_{t=0}=\eval{\dv{x^\mu(c(t))}{t}}_{t=0}\partial_\mu f,
\end{align}
with
\begin{equation}
    \partial_\mu f=\pdv{f}{x^\mu}\equiv\pdv{(f\circ\phi^{-1})}{x^\mu}.
\end{equation}
This means that the rate $\eval{(\dv*{f}{t})}_{t=0}$ is obtained by the application of the differential operator
\begin{equation} \label{X=X^mupartial_mu}
    X=X^\mu\qty(\pdv{x^\mu})\,\,\,\,\,\,\textrm{with}\,\,\,\,\,\,X^\mu=\eval{\dv{x^\mu(c(t))}{t}}_{t=0}
\end{equation}
to the map $f$, that is
\begin{equation} \label{X(f)}
    \eval{\dv{f(c(t))}{t}}_{t=0}=X^\mu\qty(\pdv{f}{x^\mu})\equiv X(f).
\end{equation}
We then define $X$ as the \textbf{tangent vector} to $M$ at the point $c(0)$ along the direction given by the curve $c(t)$, as shown in Figure \ref{Vector-on-Manifolds}.

In summary, each curve through $p$ defines an operator on $\mathcal{F}$, namely the tangent vector, which maps $f\mapsto\dv*{f}{t}$ at $p$. Basically, we are defining vectors at each point of $M$ as directional derivatives on that point. Therefore, a tangent vector $X$ at a point $p\in M$ is a map $X:\mathcal{F}\to\mathbb{R}$ which is linear and obeys the Leibniz rule,
\begin{itemize}
    \item $X(af+bg)=aX(f)+bX(g)$, $\forall\,f,g\in\mathcal{F}$, $\forall\,a,b\in\mathbb{R}$;
    \item $X(fg)=g(p)X(f)+f(p)X(g)$, $\forall\,f,g\in\mathcal{F}$.
\end{itemize}

If $X_1=\dv*{t_1}$ and $X_2=\dv*{t_2}$ represent derivatives along two curves $x^\mu(t_1)$ and $x^\nu(t_2)$ through $p$, one can define a new operator $Y=aX_1+bX_2$, $a,b\in\mathbb{R}$, which is manifestly linear and can be straightforwardly shown to satisfy the Leibniz rule. Thus, the set $T_pM$ of all the tangent vectors of $p\in M$ form a vector space, called the \textbf{tangent space} of $M$ at $p$, whose dimension is the one of $M$. From Eq. \eqref{X=X^mupartial_mu} it is clear that $\vu{e}_\mu\equiv\partial_\mu$ is a basis vector. The basis $\qty{\vu{e}_\mu}$ is called the \textbf{coordinate basis}.

We can now see how the vectors components change when we change coordinates. Let $p\in O_i\cap O_j$ and $\phi_i(p)$ and $\phi_j(p)$ be the two coordinate functions represented by the set of coordinates $x^\mu$ and $x^{\mu'}$, respectively. Since the basis vectors in the coordinate system $x^\mu$ are $\vu{e}_\mu=\partial_\mu$, the basis vectors in the new coordinate system $x^{\mu'}$ are given by
\begin{equation}
    \partial_{\mu'}=\pdv{x^\mu}{x^{\mu'}}\partial_\mu.
\end{equation}
The vector $X=X^\mu\partial_\mu$ itself must be unchanged by a change of basis, so that
\begin{equation}
    X^\mu\partial_\mu=X^{\mu'}\partial_{\mu'}=X^{\mu'}\pdv{x^\mu}{x^{\mu'}}\partial_\mu,
\end{equation}
and hence, since the matrix $\pdv*{x^{\mu'}}{x^\mu}$ is the inverse of the matrix $\pdv*{x^{\mu}}{x^{\mu'}}$,
\begin{equation} \label{Vector-transformation-law}
    X^{\mu'}=\pdv{x^{\mu'}}{x^\mu}X^\mu.
\end{equation}
The rule \eqref{Vector-transformation-law} for transforming components is what we call the \emph{vector transformation law}.

If a vector is smoothly defined on each point of the manifold, we refer to it as a \textbf{vector field}. In other words, $X$ is a vector field if $X[f]\in\mathcal{F}(M)$ for any $f\in\mathcal{F}(M)$.

Since $T_pM$ is a vector space, there exists a dual vector space to $T_pM$ whose elements are linear functions from $T_pM$ to $\mathbb{R}$. The dual space is called the \textbf{cotangent space} at $p$, denoted by $T_p^*M$. An element $\omega:T_pM\to\mathbb{R}$ of $T_p^*M$ is called a \textbf{dual vector} or a \textbf{cotangent vector}. The differential $\dd f$ of a function $f\in\mathcal{F}(M)$ is an example of a dual vector, with its action on $V\in T_pM$ being defined by
\begin{equation}
    \langle \dd f,V\rangle\equiv V(f)=V^\mu\partial_\mu f\in\mathbb{R}.
\end{equation}

In terms of the coordinate $x=\phi(p)$, $\dd f$ is expressed as $\dd f=\qty(\pdv*{f}{x^\mu})\dd x^\mu$, where $\phi$ is a coordinate function. One can naturally regard $\qty{\dd x^\mu}$ as a basis of $T_p^*M$, which is, in fact, a dual basis, since
\begin{equation}
    \left\langle \dd x^\nu,\pdv{x^\mu}\right\rangle =\pdv{x^\nu}{x^\mu}=\delta^\nu_\mu.
\end{equation}
An arbitrary dual vector $\omega$ is then written as
\begin{equation}
    \omega=\omega_\mu \dd x^\mu,
\end{equation}
where the $\omega_\mu$ are the components of $\omega$.

The notion of a dual vector allows us to define the inner product. For instance, take a vector $V=V^\mu\partial_\mu$ and a dual vector $\omega=\omega_\mu \dd x^\mu$. The \textbf{inner-product} $\langle\,, \rangle:T_p^*M\times T_pM\to\mathbb{R}$ is obtained by
\begin{equation} \label{Inner-product}
    \langle\omega,V\rangle=\omega_\mu V^\nu\left\langle \dd x^\mu,\pdv{x^\nu}\right\rangle=\omega_\mu V^\nu\delta^\mu_\nu=\omega_\mu V^\mu\equiv\omega(V).
\end{equation}

Next, let $p\in O_i\cap O_j$ and $\phi_i(p)$ and $\phi_j(p)$ be the two coordinate functions represented by the set of coordinates $x^\mu$ and $x^{\mu'}$, respectively. We thus have
\begin{equation}
    \omega=\omega_\mu \dd x^\mu=\omega_{\mu'}\dd x^{\mu'}.
\end{equation}
From the fact that $\dd x^{\mu'}=\qty(\pdv*{x^{\mu'}}{x^\mu})\dd x^\mu$ we can write down the transformation law for the components of the dual vector $\omega$ as
\begin{equation}
    \omega_{\mu'}=\pdv{x^{\mu}}{x^{\mu'}}\omega_\mu.
\end{equation}

At last, a \textbf{tensor} of type $(q,r)$ is a multilinear object which maps $q$ elements of $T_p^*M$ and $r$ elements of $T_pM$ to a real number,
\begin{equation}
    T:\qty[\times^qT_p^*M]\qty[\times^rT_pM]\to\mathbb{R},
\end{equation}
where the symbol $\times^qT_p^*M$ means the Cartesian product of the space $T_p^*M$ $q$ times, with a similar definition for the vector space $T_pM$. The set of all tensors of type $(q,r)$ at $p\in M$ is denoted by $T^q_{r,p}(M)$. The components of the tensor $T\in T^q_{r,p}(M)$ in a coordinate basis can be obtained by acting the tensor on basis dual vectors and vectors,
\begin{equation}
    {T^{\mu_1\dots\mu_q}}_{\nu_1\dots\nu_r}=T(\dd x^{\mu_1},\dots,\dd x^{\mu_q},\partial_{\nu_1},\dots,\partial_{\nu_r}).
\end{equation}
This is equivalent to the expansion
\begin{equation}
    T={T^{\mu_1\dots\mu_q}}_{\nu_1\dots\nu_r}\partial_{\mu_1}\otimes\dots\otimes\partial_{\mu_q}\otimes \dd x^{\nu_1}\otimes\dots\otimes \dd x^{\nu_r},
\end{equation}
where $\otimes$ denotes the tensor product. Now let $V_i=V_i^\mu\partial_\mu$ ($1\leq i\leq r$) and $\omega_i=\omega_{i\mu}\dd x^\mu$ ($1\leq i\leq q$). The action of $T$ on them results in the number
\begin{equation}
    T(\omega_1,\dots,\omega_q;V_1,\dots,V_r)={T^{\mu_1\dots\mu_q}}_{\nu_1\dots\nu_r}\omega_{1\mu_1}\dots\omega_{q\mu_q}V_1^{\nu_1}\dots V_r^{\nu_r}.
\end{equation}

If $p\in O_i\cap O_j$ and $\phi_i(p)$ and $\phi_j(p)$ are the two coordinate functions represented by the set of coordinates $x^\mu$ and $x^{\mu'}$, respectively, the components of the tensor $T$ change as
\begin{equation} \label{General-transformation-law-for-tensors}
    {T^{\mu_1'\dots\mu_q'}}_{\nu_1'\dots\nu_r'}=\pdv{x^{\mu_1'}}{x^{\mu_1}}\dots\pdv{x^{\mu_q'}}{x^{\mu_q}}\pdv{x^{\nu_1}}{x^{\nu_1'}}\dots\pdv{x^{\nu_r}}{x^{\nu_r'}}\,{T^{\mu_1\dots\mu_q}}_{\nu_1\dots\nu_r},
\end{equation}
which is the general transformation law for tensors.

We define a \textbf{tensor field} of type $(q,r)$ by the smooth assignment of an element of $T^q_{r,p}(M)$ to each point $p\in M$. The set of all tensor fields of type $(q,r)$ on $M$ is denoted by $T_r^q(M)$. For example, $T_1^0(M)$ is the set of the dual vector fields.

Since the tensors themselves are objects that are independent of the coordinate system, it is useful (although not strictly mandatory) to express physics laws in terms of tensor equations. But the laws of physics usually involve derivatives, and, unfortunately, the partial derivative of a tensor is not, in general, a new tensor, as we can see by considering, for example, the partial derivative of a dual vector, $\partial_\mu\omega_\nu$, and changing to a new coordinate system:
\begin{align}
    \pdv{x^{\mu'}}\omega_{\nu'}&=\pdv{x^\mu}{x^{\mu'}}\pdv{x^\mu}\qty(\pdv{x^\nu}{x^{\nu'}}\omega_\nu) \nonumber \\
    &=\pdv{x^\mu}{x^{\mu'}}\pdv{x^\nu}{x^{\nu'}}\qty(\pdv{x^\mu}\omega_\nu)+\omega_\nu\pdv{x^\mu}{x^{\mu'}}\pdv{x^\mu}\qty(\pdv{x^\nu}{x^{\nu'}}).
\end{align}
The second term should not be there if $\partial_\mu\omega_\nu$ were to transform as a $(0,2)$ tensor. Therefore we need to introduce a coordinate independent derivative operator and, to do so, we first need to see how tensor fields are carried along by maps between manifolds.

\section{Maps between manifolds}

\begin{figure}[!ht]
    \centering
    \includegraphics[width=0.6\linewidth]{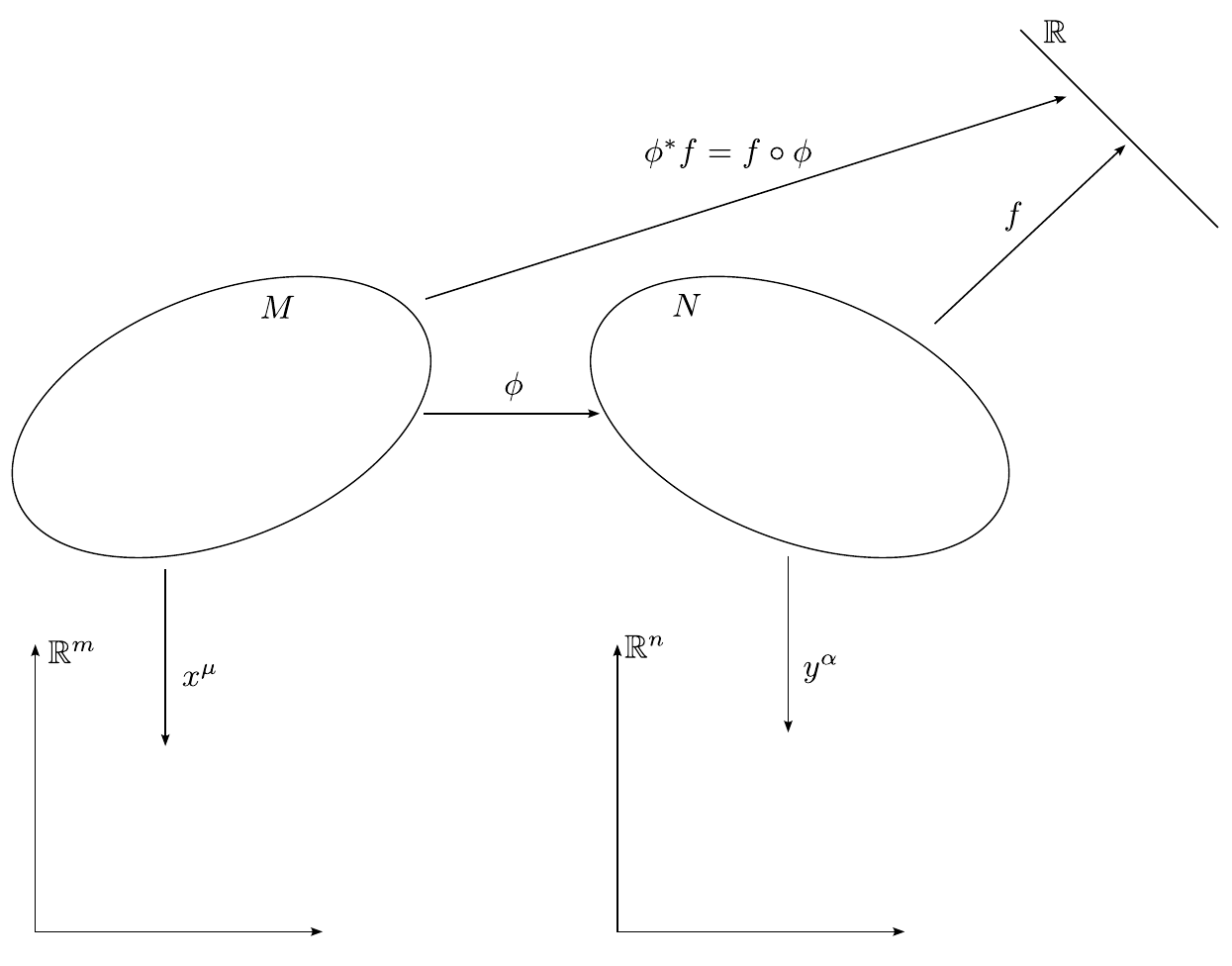}
    \caption{Pullback of a function $f$ by a map $\phi:M\to N$.}
    \label{pullback}
\end{figure}

Consider two manifolds $M$ and $N$, with dimensions $m$ and $n$, and coordinate systems $x^\mu$ and $y^\alpha$, respectively. Consider the map $\phi:M\to N$ and a function $f:N\to\mathbb{R}$. We define the \textbf{pullback} of $f$ by $\phi$, denoted $\phi^*f$, by
\begin{equation}
    \phi^*f=\qty(f\circ\phi):M\to\mathbb{R}.
\end{equation}
We think of $\phi^*$ as "pulling back" the function $f$ from $N$ to $M$ (Figure \ref{pullback}).

If $V(p)$ is a vector at a point $p$ on $M$, we define the \textbf{pushforward} vector $\phi_*V$ at the point $\phi(p)$ on $N$ by giving its action on functions on $N$,
\begin{equation}
    \qty(\phi_*V)(f)=V(\phi^*f).
\end{equation}
Basically, the action of $\phi_*V$ on the function $f$ is simply the action of $V$ on $\phi^*f$.

Since a basis for vectors on $M$ is given by the set of partial derivatives $\partial_\mu=\pdv*{x^\mu}$ and a basis on $N$ is given by the set of partial derivatives $\partial_\alpha=\pdv*{y^\alpha}$, we can relate the components of $V=V^\mu\partial_\mu$ to those of $(\phi_*V)=(\phi_*V)^\alpha\partial_\alpha$ by applying the pushed-forward vector to a test function and using the chain rule of partial derivatives,
\begin{equation}
    (\phi_*V)^\alpha\partial_\alpha f=V^\mu\partial_\mu(\phi^*f)=V^\mu\partial_\mu(f\circ\phi)=V^\mu\pdv{y^\alpha}{x^\mu}\partial_\alpha f.
\end{equation}
Thus the pushforward operation $\phi_*$ can be seen as a matrix operator, $(\phi_*V)^\alpha={\qty(\phi_*)^\alpha}_\mu V^\mu$, with the matrix being given by
\begin{equation} \label{Pushforward}
    {\qty(\phi_*)^\alpha}_\mu=\pdv{y^\alpha}{x^\mu}.
\end{equation}

The pullback $\phi^*\omega$ of a dual vector $\omega$ on $N$ is defined by its action on a vector $V$ on $M$: it is equal to the action of $\omega$ on the pushforward of $V$,
\begin{equation}
    (\phi^*\omega)(V)=\omega(\phi_*V).
\end{equation}
From this expression, we have
\begin{equation}
    (\phi^*\omega)(V)=\omega_\alpha(\phi_*V)^\alpha=\omega_\alpha{(\phi_*)^\alpha}_\mu V^\mu={(\phi_*)_\mu}^\alpha \omega_\alpha V^\mu.
\end{equation}
where ${(\phi_*)_\mu}^\alpha=\pdv*{y^\alpha}{x^\mu}$. It is the same matrix as the pushforward \eqref{Pushforward} but with a different index being contracted when the matrix acts to pull back dual vectors.

We can pull back tensors with an arbitrary number of lower indices. Since a $(0,r)$ tensor is a linear map from the direct product of $r$ vectors to $\mathbb{R}$, we define the pullback of such a tensor by the action of the original tensor on the pushed-forward vectors,
\begin{equation}
    (\phi^*T)(V^{(1)},\dots,V^{(r)})=T(\phi_*V^{(1)},\dots,\phi_*V^{(r)}),
\end{equation}
where $T_{\alpha_1\dots\alpha_r}$ is a $(0,r)$ tensor on $N$. We can similarly push forward any $(q,0)$ tensor $S^{\mu_1\dots\mu_q}$ on $M$ by acting it on pulled-back dual vectors,
\begin{equation}
    (\phi_*S)(\omega^{(1)},\dots,\omega^{(q)})=S(\phi^*\omega^{(1)},\dots,\phi^*\omega^{(q)}).
\end{equation}
For the matrix representation of the pullback of a $(0,r)$ tensor we have
\begin{equation} \label{Pullback-(0,l)}
    (\phi^*T)_{\mu_1\dots\mu_r}=\pdv{y^{\alpha_1}}{x^{\mu_1}}\dots\pdv{y^{\alpha_r}}{x^{\mu_r}}T_{\alpha_1\dots\alpha_r},
\end{equation}
while for the pushforward of a $(q,0)$ tensor,
\begin{equation} \label{Pushforward-(k,0)}
    (\phi_*S)^{\alpha_1\dots\alpha_q}=\pdv{y^{\alpha_1}}{x^{\mu_1}}\dots\pdv{y^{\alpha_q}}{x^{\mu_q}}S^{\mu_1\dots\mu_q}.
\end{equation}

In general, tensors with both upper and lower indices can be neither pushed forward nor pulled back since the matrix $\pdv*{y^\alpha}{x^\mu}$ might not be invertible. If it is invertible, however, then $\phi$ defines a diffeomorphism between $M$ and $N$, allowing us to define the pushforward and pullback of arbitrary tensors. Specifically, for a $(q,r)$ tensor field ${T^{\mu_1\dots\mu_q}}_{\nu_1\dots\nu_r}$ on $M$, we define the pushforward by
\begin{align}
    &(\phi_*T)(\omega^{(1)},\dots,\omega^{(q)},V^{(1)},\dots,V^{(r)}) \nonumber \\
    &\hspace{2cm}=T(\phi^*\omega^{(1)},\dots,\phi^*\omega^{(q)},[\phi^{-1}]_*V^{(1)},\dots,[\phi^{-1}]_*V^{(r)}).
\end{align}
In components, this becomes
\begin{equation} \label{Pushforward-tensor-in-a-diffeomorphism}
    {(\phi_*T)^{\alpha_1\dots\alpha_q}}_{\beta_1\dots\beta_r}=\pdv{y^{\alpha_1}}{x^{\mu_1}}\dots\pdv{y^{\alpha_q}}{x^{\mu_q}}\pdv{x^{\nu_1}}{y^{\beta_1}}\dots\pdv{x^{\nu_r}}{y^{\beta_r}}\,{T^{\mu_1\dots\mu_q}}_{\nu_1\dots\nu_r}.
\end{equation}
Note the resemblance to the tensor transformation law between coordinate systems, Eq. \eqref{General-transformation-law-for-tensors}. It is for this reason that we can think of a diffeomorphism as an "active" coordinate transformation.

A diffeomorphism also provides a way of comparing tensors at different points on a manifold, which we can use to define a way of computing derivatives of tensors.

\section{Lie derivatives}

Given a diffeomorphism $\phi:M\to M$ and a tensor field ${T^{\mu_1\dots\mu_q}}_{\nu_1\dots\nu_r}(x)$ we can define the difference between the value of the tensor at some point $p$ and its value at $\phi(p)$ pulled back to $p$, $\phi^*[{T^{\mu_1\dots\mu_q}}_{\nu_1\dots\nu_r}(\phi(p))]$. Therefore we can define some kind of derivative operator on tensor fields, which would require a one-parameter family of diffeomorphisms, $\phi_t$. This family can be thought of as a smooth map $\mathbb{R}\times M\to M$, such that for each $t\in\mathbb{R}$ we have a diffeomorphism $\phi_t$, satisfying
\begin{equation}
    \phi_s\circ\phi_t=\phi_{s+t}.
\end{equation}
Note that this condition implies that $\phi_0$ is the identity map.

With the entire family $\phi_t$, a point $p$ describes a curve in $M$ such that these curves for all points cover the entire manifold. We can then define a vector field $V^\mu(x)$ to be the set of tangent vectors to each of these curves at every point, evaluated at $t=0$. This entire line of thought can be reversed, however, and it will be useful to think of the one-parameter family of diffeomorphisms as being defined by the vector field itself. Explicitly, given an arbitrary $V^\mu(x)$, we define its \textbf{integral curves} to be those curves $x^\mu(t)$ that solve
\begin{equation}
    \dv{x^\mu}{t}=V^\mu.
\end{equation}
The vector field $V^\mu(x)$ is referred to as the \textbf{generator} of the diffeomorphism.

For a given a vector field $V^\mu(x)$, then, we have a family of diffeomorphisms parametrized by $t$, and we can ask how fast a tensor changes along the integral curves. For each $t$ we can define this change as the difference between the pullback of the tensor to $p$ and its original value at $p$,
\begin{equation}
    \Delta_t{T^{\mu_1\dots\mu_q}}_{\nu_1\dots\nu_r}(p)=\phi_t^*[{T^{\mu_1\dots\mu_q}}_{\nu_1\dots\nu_r}(\phi_t(p))]-{T^{\mu_1\dots\mu_q}}_{\nu_1\dots\nu_r}(p).
\end{equation}
Note that both terms on the right-hand side are tensors at $p$. We then define the \textbf{Lie derivative} of the tensor along the vector field as
\begin{equation} \label{Lie-derivative}
    \mathcal{L}_V{T^{\mu_1\dots\mu_q}}_{\nu_1\dots\nu_r}=\lim_{t\to0}\qty(\frac{\Delta_t{T^{\mu_1\dots\mu_q}}_{\nu_1\dots\nu_r}}{t}).
\end{equation}
The Lie derivative is a map from $(q,r)$ tensor fields to $(q,r)$ tensor fields, which is manifestly independent of coordinates. It is linear,
\begin{subequations}
\begin{equation}
    \mathcal{L}_V(aT+bS)=a\mathcal{L}_VT+b\mathcal{L}_VS,
\end{equation}
and obeys the Leibniz rule,
\begin{equation}
    \mathcal{L}_V(T\otimes S)=(\mathcal{L}_VT)\otimes S+T\otimes (\mathcal{L}_VS),
\end{equation}
where $S$ and $T$ are tensors and $a$ and $b$ are constants. Also, the Lie derivative reduces to the ordinary directional derivative when acting on functions,
\begin{equation}
    \mathcal{L}_Vf=\dv{f}{t}=\dv{x^\mu}{t}\partial_\mu f=V^\mu\partial_\mu f=V(f).
\end{equation}
\end{subequations}

The Lie derivative of a vector field $U(x)$ with respect to $V(x)$ can be shown to be
\begin{equation} \label{Lie-Bracket}
    \mathcal{L}_VU^\mu=\comm{V}{U}^\mu=V^\nu\partial_\nu U^\mu-U^\nu\partial_\nu V^\mu.
\end{equation}
The commutator is called the \textbf{Lie bracket}. Note that $\mathcal{L}_VU=-\mathcal{L}_UV$.

Finally, in terms of components in a given coordinate system, the Lie derivative of an arbitrary tensor field is given by
\begin{align} \label{Lie-derivative-tensors-partial}
    \mathcal{L}_V{T^{\mu_1\mu_2\dots\mu_q}}_{\nu_1\nu_2\dots\nu_r}&=V^\sigma\partial_\sigma {T^{\mu_1\mu_2\dots\mu_q}}_{\nu_1\nu_2\dots\nu_r} \nonumber \\
    &-(\partial_\lambda V^{\mu_1}){T^{\lambda\mu_2\dots\mu_q}}_{\nu_1\nu_2\dots\nu_r}-(\partial_\lambda V^{\mu_2}){T^{\mu_1\lambda\dots\mu_q}}_{\nu_1\nu_2\dots\nu_r}-\dots \nonumber \\
    &+(\partial_{\nu_1}V^\lambda){T^{\mu_1\mu_2\dots\mu_q}}_{\lambda\nu_2\dots\nu_r}+(\partial_{\nu_2}V^\lambda){T^{\mu_1\mu_2\dots\mu_q}}_{\nu_1\lambda\dots\nu_r}+\dots\,.
\end{align}
Although not manifestly tensorial, this expression can be shown to be covariant (see Section \ref{Sec:Covariant-derivatives}).

The Lie derivative is a successful result to our quest of obtaining a coordinate independent derivative operator. However, it can be argued that this operator lacks some generality since it depends on the specification of the vector field. It will then be useful to define another derivative operator, namely the covariant derivative. But before we do that, let us pause a moment to discuss an extremely important tensor in differential geometry, the metric tensor.

\section{The metric tensor}

The metric tensor is the mathematical object that provides the notion of the length of a path in a manifold, which we also call the line element
\begin{equation}
    \dd s^2=g_{\mu\nu}\dd x^\mu\otimes \dd x^\nu=g_{\mu\nu}\dd x^\mu \dd x^\nu.
\end{equation}
We may also use the metric to denote the inner product of two vectors $V^\mu$ and $W^\nu$ as
\begin{equation} \label{Inner-product-of-vectors}
    \dd s^2(V,W)=g_{\mu\nu}V^\mu W^\nu=V_\nu W^\nu.
\end{equation}
We refer to two vectors whose inner product vanishes as \textbf{orthogonal}, and the inner product of a vector with itself is called its \textbf{norm}.

The metric is a symmetric tensor, by which we mean that $g_{\mu\nu}=g_{\nu\mu}$. By imposing that the determinant of the metric does not vanish, $\det(g_{\mu\nu})\equiv g\neq0$, we can properly define the inverse of the metric, $g^{\mu\nu}$, such that
\begin{equation}
    g^{\mu\nu}g_{\nu\sigma}=g_{\lambda\sigma}g^{\lambda\mu}=\delta^\mu_\sigma.
\end{equation}
The symmetry of $g_{\mu\nu}$ implies that $g^{\mu\nu}$ is also symmetric. We can use the inverse of the metric to denote the inner product of two dual vectors $\omega_\mu$ and $\zeta_\nu$ as
\begin{equation} \label{Inner-product-of-dual-vectors}
    \dd s^2(\omega,\zeta)=g^{\mu\nu}\omega_\mu\zeta_\nu=\omega^\nu\zeta_\nu.
\end{equation}
Eqs. \eqref{Inner-product-of-vectors} and \eqref{Inner-product-of-dual-vectors} illustrate what we usually say that the metric is used to raise and lower indices.

The metric can be characterized by putting $g_{\mu\nu}$ into its \textbf{canonical form}. In this form the metric components become
\begin{equation} \label{Canonical-form}
    g_{\mu\nu}=\textrm{diag}\,\qty(-1,-1,\dots,-1,+1,+1,\dots,+1,0,\dots,0).
\end{equation}
The \textbf{signature} of the metric refer to the number of both positive and negative eigenvalues; we speak of "a metric with signature minus-plus-plus-plus" for Minkowski space, for example, for which
\begin{equation}
    g_{\mu\nu}=\eta_{\mu\nu}=\textrm{diag}\,\qty(-1,+1,+1,+1)
\end{equation}
for the entire manifold (in Cartesian coordinates). If any of the eigenvalues are zero, the metric is degenerate, and the inverse metric will not exist; if the metric is continuous and nondegenerate, its signature will be the same at every point. If all of the signs are positive, the metric is called \textbf{Euclidean} or \textbf{Riemannian} (or just positive definite), while if there is a single minus it is called \textbf{Lorentzian} or \textbf{pseudo-Riemannian}, and any metric with some $+1$'s and some $-1$'s is called indefinite.

For a Lorentzian metric, the norm of a vector is not positive definite. We say $V^\mu$ is \textbf{timelike} if $g_{\mu\nu}V^\mu V^\nu<0$, \textbf{lightlike} or \textbf{null} if $g_{\mu\nu}V^\mu V^\nu=0$ and \textbf{spacelike} if $g_{\mu\nu}V^\mu V^\nu>0$.

At a single point $p$ on the manifold, one may start with any set of components for $g_{\mu\nu}$, diagonalize the matrix and then rescale the basis vectors such that the metric components satisfy \eqref{Canonical-form}. Minkowski spacetime is just a four dimensional manifold with a Lorentzian metric such that this can be done globally for a single coordinate system, namely the Cartesian one. We say it is a flat manifold (or flat spacetime in a more physical context). This is not true for a general metric, however. At a single point one can always choose coordinates in which the metric tensor looks flat, but this has to be done in different coordinate systems for different points, in general. We say that such spacetime is curved. It will then be interesting to learn how we can extract the curvature information of the metric tensor. We start in the next section with the notion of covariant derivatives, which tells us how to parallel transport vectors along curves in a coordinate independent way that accounts for the curvature of the manifold.

\section{Covariant derivatives} \label{Sec:Covariant-derivatives}

We define the \textbf{covariant derivative} $\nabla$ as a map from $(q,r)$ tensor fields to $(q,r+1)$ tensor fields which has the following properties:
\begin{enumerate}
    \item (linearity) $\nabla(T+S)=\nabla T+\nabla S$;
    \item (Leibniz rule) $\nabla(T\otimes S)=(\nabla T)\otimes S+T\otimes(\nabla S)$;
    \item (commutativity with contractions) $\nabla_\mu({T^\lambda}_{\lambda\rho})={{\qty(\nabla T)_\mu}^\lambda}_{\lambda\rho}$;
    \item (reduction to the partial derivatives on scalars) $\nabla_\mu\phi=\partial_\mu\phi$.
\end{enumerate}
Here $\phi$ is a scalar field and $T$ and $S$ are arbitrary rank tensor fields. For the covariant derivative of a vector $V^\mu$, the general expression that can be easily shown to satisfy all requirements above is
\begin{equation} \label{Covariant-derivative-of-vector}
    \nabla_\mu V^\nu=\partial_\mu V^\nu+\Gamma_{\mu\lambda}^\nu V^\lambda,
\end{equation}
where the set of $n$ matrices ${(\Gamma_\mu)^\nu}_\lambda$ (one $n\times n$ matrix for each $\mu$, where $n$ is the dimension of the manifold) are known as the \textbf{connection coefficients}. The second term on the right-hand side of Eq. \eqref{Covariant-derivative-of-vector} can be thought as being a correction to the partial derivative of a vector due to the curvature of the manifold. We can determine the transformation properties of $\Gamma_{\mu\lambda}^\nu$ by demanding that the left-hand side of Eq. \eqref{Covariant-derivative-of-vector} be a $(1,1)$ tensor. In doing so, one finds that, in a new coordinate system,
\begin{equation} \label{Connection-transformation-law}
    \Gamma_{\mu'\lambda'}^{\nu'}=\pdv{x^\mu}{x^{\mu'}}\pdv{x^{\nu'}}{x^\nu}\pdv{x^\lambda}{x^{\lambda'}}\Gamma_{\mu\lambda}^\nu-\pdv{x^\mu}{x^{\mu'}}\pdv{x^\lambda}{x^{\lambda'}}\pdv[2]{x^{\nu'}}{x^\mu}{x^\lambda}.
\end{equation}
This is clearly not the tensor transformation law, so the connection coefficients are not really tensors.

By a similar reasoning to that used for vectors, the covariant derivative of a dual vector $\omega_\mu$ can also be expressed as a partial derivative plus some linear transformation. In order to find out what this linear transformation is, we apply the required properties of the covariant derivative of the scalar $\omega_\mu V^\mu$, which then tell us that the covariant derivative of $\omega_\mu$ must be of the form
\begin{equation}
    \nabla_\mu\omega_\nu=\partial_\mu\omega_\nu-\Gamma_{\mu\nu}^\lambda\omega_\lambda.
\end{equation}

The generalization to the covariant derivative of a tensor of arbitrary rank is
\begin{equation}
\begin{split}
    \nabla_\sigma {T^{\mu_1\mu_2\dots\mu_q}}_{\nu_1\nu_2\dots\nu_r}&=\partial_\sigma {T^{\mu_1\mu_2\dots\mu_q}}_{\nu_1\nu_2\dots\nu_r} \\
    &\,\,\,\,\,\,\,\,+\Gamma_{\sigma\lambda}^{\mu_1}{T^{\lambda\mu_2\dots\mu_q}}_{\nu_1\nu_2\dots\nu_r}+\Gamma_{\sigma\lambda}^{\mu_2}{T^{\mu_1\lambda\dots\mu_q}}_{\nu_1\nu_2\dots\nu_r}+\dots \\
    &\,\,\,\,\,\,\,\,-\Gamma_{\sigma\nu_1}^\lambda{T^{\mu_1\mu_2\dots\mu_q}}_{\lambda\nu_2\dots\nu_r}-\Gamma_{\sigma\nu_2}^\lambda{T^{\mu_1\mu_2\dots\mu_q}}_{\nu_1\lambda\dots\nu_r}-\dots\,.
\end{split}
\end{equation}

In order to define a covariant derivative we need to put a connection on our manifold, which is specified in some coordinate system by a set of coefficients $\Gamma_{\mu\nu}^\lambda$. This specification is generally not unique. In general relativity, however, we use a special kind of connection coefficients that emerge when we require two more properties:
\begin{itemize}
    \item (torsion-free) $\Gamma_{\mu\nu}^\lambda=\Gamma_{\nu\mu}^\lambda$;
    \item (metric compatibility) $\nabla_\rho g_{\mu\nu}=0$.
\end{itemize}

Given a metric $g_{\mu\nu}$, there exists a unique derivative operator compatible with it \cite{Wald1984}, for which
\begin{equation} \label{Christoffel-connection}
    \Gamma_{\mu\nu}^\sigma=\frac{1}{2}g^{\sigma\rho}\qty(\partial_\mu g_{\nu\rho}+\partial_\nu g_{\rho\mu}-\partial_\rho g_{\mu\nu}).
\end{equation}
The connection given by Eq. \eqref{Christoffel-connection} is known as the \textbf{Christoffel connection}, and the associated connection coefficients are called \textbf{Christoffel symbols}.

More generally, let $\nabla^{(1)}_\mu$ and $\nabla^{(2)}_\mu$ be two distinct covariant derivatives that are compatible with the distinct metric tensors $g^{(1)}_{\mu\nu}$ and $g^{(2)}_{\mu\nu}$, respectively. Then, the action of $\nabla^{(1)}_\mu$ on an arbitrary tensor field in terms of $\nabla^{(2)}_\mu$ and a tensor field ${C^\sigma}_{\mu\nu}$ can be written as~\cite{Wald1984}
\begin{equation}
\begin{split}
    \nabla^{(1)}_\sigma {T^{\mu_1\mu_2\dots\mu_q}}_{\nu_1\nu_2\dots\nu_r}&=\nabla^{(2)}_\sigma {T^{\mu_1\mu_2\dots\mu_q}}_{\nu_1\nu_2\dots\nu_r} \\
    &\,\,\,\,\,\,\,\,+{C^{\mu_1}}_{\sigma\lambda}{T^{\lambda\mu_2\dots\mu_q}}_{\nu_1\nu_2\dots\nu_r}+{C^{\mu_2}}_{\sigma\lambda}{T^{\mu_1\lambda\dots\mu_q}}_{\nu_1\nu_2\dots\nu_r}+\dots \\
    &\,\,\,\,\,\,\,\,-{C^\lambda}_{\sigma\nu_1}{T^{\mu_1\mu_2\dots\mu_q}}_{\lambda\nu_2\dots\nu_r}-{C^\lambda}_{\sigma\nu_2}{T^{\mu_1\mu_2\dots\mu_q}}_{\nu_1\lambda\dots\nu_r}-\dots\,.
\end{split}
\end{equation}
Metric compatibility implies that
\begin{equation}
    {C^\sigma}_{\mu\nu}=\frac{1}{2}g^{\sigma\rho}_{(1)}\qty(\nabla^{(2)}_\mu g^{(1)}_{\nu\rho}+\nabla^{(2)}_\nu g^{(1)}_{\rho\mu}-\nabla^{(2)}_\rho g^{(1)}_{\mu\nu}).
\end{equation}

We can use the covariant derivative in order to obtain a manifestly covariant expression for the action of the Lie derivative. First, note that the Lie Bracket of vector fields $V^\mu(x)$ and $U^\mu(x)$ can be written as
\begin{align}
    \mathcal{L}_VU^\mu&=\comm{V}{U}^\mu=V^\nu\partial_\nu U^\mu-U^\nu\partial_\nu V^\mu \nonumber \\
    &=V^\nu\nabla_\nu U^\mu-U^\nu\nabla_\nu V^\mu.
\end{align}
This holds true for any torsion-free covariant derivative since the commutator involves an antisymmetrization which cancels the nontensorial piece of the partial derivative transformation law. We can then write the action of the Lie derivative on tensors, Eq. \eqref{Lie-derivative-tensors-partial}, in a manifestly covariant form,
\begin{align}
    \mathcal{L}_V{T^{\mu_1\mu_2\dots\mu_q}}_{\nu_1\nu_2\dots\nu_r}&=V^\sigma\nabla_\sigma {T^{\mu_1\mu_2\dots\mu_q}}_{\nu_1\nu_2\dots\nu_r} \nonumber \\
    &-(\nabla_\lambda V^{\mu_1}){T^{\lambda\mu_2\dots\mu_q}}_{\nu_1\nu_2\dots\nu_r}-(\nabla_\lambda V^{\mu_2}){T^{\mu_1\lambda\dots\mu_q}}_{\nu_1\nu_2\dots\nu_r}-\dots \nonumber \\
    &+(\nabla_{\nu_1}V^\lambda){T^{\mu_1\mu_2\dots\mu_q}}_{\lambda\nu_2\dots\nu_r}+(\nabla_{\nu_2}V^\lambda){T^{\mu_1\mu_2\dots\mu_q}}_{\nu_1\lambda\dots\nu_r}+\dots\,.
\end{align}
The Lie derivative of the metric tensor, for instance, reads
\begin{align} \label{Lie-derivative-of-the-metric}
    \mathcal{L}_Vg_{\mu\nu}&=V^\sigma\nabla_\sigma g_{\mu\nu}+(\nabla_\mu V^\lambda)g_{\lambda\nu}+(\nabla_\nu V^\lambda)g_{\mu\lambda} \nonumber \\
    &=\nabla_\mu V_\nu+\nabla_\nu V_\mu.
\end{align}
This expression is particularly useful for studying gauge invariance of linearized gravity.

With the proper definition of a derivative operator at hand, we are now ready to introduce the idea of parallel transport of tensors, which shall lead to the ultimate notion of curvature in a manifold.

\section{Parallel transport and geodesics}

Let us start with the simplest non trivial example of a tensor field, namely the vector field. We may be tempted to compare and compute operations between vectors that are defined at different points, as we usually do in basic "flat-space physics". However, vectors that are defined at different points belong to different vector spaces. This is not a problem if spacetime is flat because we can simply transport vectors to the same point, so that they belong to the same tangent space, and then compare one with the other there. This is what we call a parallel transport, which basically means that we are moving the vector while keeping it "constant". Parallel transport can also be defined for curved manifolds, but the problem is that, in general, the end result of a tensor being parallel transported along a path will depend on our choice of the curve itself.

Given a curve $x^\mu(\lambda)$, the requirement of constancy of a tensor ${T^{\mu_1\dots\mu_q}}_{\nu_1\dots\nu_r}$ along this curve in flat space reads
\begin{equation}
    \dv{\lambda}\,{T^{\mu_1\dots\mu_q}}_{\nu_1\dots\nu_r}=\dv{x^\mu}{\lambda}\pdv{x^\mu}\,{T^{\mu_1\dots\mu_q}}_{\nu_1\dots\nu_r}=0.
\end{equation}
In a generic manifold we need to replace this partial derivative by a covariant one. We then define the \textbf{directional covariant derivative} to be
\begin{equation}
    \frac{\rm D}{\textrm{d}\lambda}=\dv{x^\mu}{\lambda}\nabla_\mu.
\end{equation}
This is a map, defined only along the path, from $(q,r)$ tensors to $(q,r)$ tensors. The \textbf{parallel transport} of the tensor $T$ along the path $x^\mu(\lambda)$ is defined via
\begin{equation} \label{Equation-of-parallel-transport}
    {\qty(\frac{\rm D}{\textrm{d}\lambda}T)^{\mu_1\dots\mu_q}}_{\nu_1\dots\nu_r}\equiv\dv{x^\sigma}{\lambda}\nabla_\sigma{T^{\mu_1\dots\mu_q}}_{\nu_1\dots\nu_r}=0.
\end{equation}
This is a tensor equation known as the \textbf{equation of parallel transport}. For a vector it takes the form
\begin{equation} \label{Parallel-transport-vector}
    \dv{\lambda}V^\mu+\Gamma_{\sigma\rho}^\mu\dv{x^\sigma}{\lambda}V^\rho=0.
\end{equation}
The idea of parallel transport gives us a way to interpret the covariant derivative of tensors: it quantifies the instantaneous rate of change of a tensor field in comparison to what the tensor would be if it were parallel transported \cite{Carroll}.

Parallel transport depends on the connection, since different connections define different covariant derivatives that lead to different ways on how to transport a tensor. Here we stick with metric-compatible connections, for which the metric is always parallel transported,
\begin{equation}
    \frac{\rm D}{\textrm{d}\lambda}g_{\mu\nu}=\dv{x^\sigma}{\lambda}\nabla_\sigma g_{\mu\nu}=0.
\end{equation}
It follows that the inner product of two parallel-transported vectors is preserved and parallel transport with respect to a metric-compatible connection preserves the norm of vectors.

A path $x^\mu(\lambda)$ that parallel-transports its own tangent vector is called a \textbf{geodesic}. Such parallel transport condition reads
\begin{equation} \label{Geodesic-equation}
    \dv[2]{x^\mu}{\lambda}+\Gamma_{\sigma\rho}^\mu\dv{x^\sigma}{\lambda}\dv{x^\rho}{\lambda}=0.
\end{equation}
This is the \textbf{geodesic equation}. It reproduces the usual notion of straight lines if the connection coefficients are the Christoffel symbols in Euclidean space. In particular, for any given space, if the connections in Eq. \eqref{Geodesic-equation} are the Christoffel connections, the geodesic is also the shortest distance between two points, by which we mean that it extremizes the length
\begin{equation}
    l=\int \sqrt{g_{\mu\nu}\dv{x^\mu}{\lambda}\dv{x^\nu}{\lambda}}\,\dd\lambda
\end{equation}

Since a curve is said to be timelike or null or spacelike at a point if its tangent vector is timelike or null or spacelike at that point, it follows from metric compatibility that geodesics preserve their own signature. For null curves the length is zero, while for timelike curves we change the sign in the square root above and use the term \textbf{proper time},
\begin{equation}
    \tau=\int \sqrt{-g_{\mu\nu}\dv{x^\mu}{\lambda}\dv{x^\nu}{\lambda}}\,\dd\lambda.
\end{equation}
Sometimes it is useful to specify the curve parameter to be the proper time itself. Then, any transformation of the form $\tau\to\lambda=\alpha\tau+\beta$, for constants $\alpha$ and $\beta$, leaves the geodesic equation invariant. Any parameter $\lambda$ related to the proper time as specified above is called an \textbf{affine parameter}. 

\section{The Riemann curvature tensor}

\begin{figure}[!ht]
    \centering
    \includegraphics[width=0.3\linewidth]{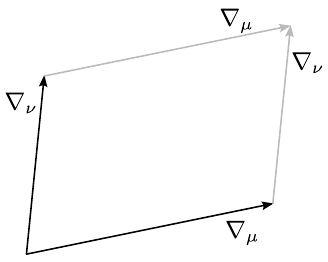}
    \caption{The commutator of two covariant derivatives.}
    \label{Commutator-of-covariant-derivatives}
\end{figure}

We are now ready to introduce our measure of curvature. Let us recall that the covariant derivative of a tensor in a certain direction measures how much the tensor changes relative to what it would have been if it had been parallel transported along that direction. The commutator of two covariant derivatives then measures the difference between parallel transporting the tensor first one way and then the other, versus the opposite ordering (Figure \ref{Commutator-of-covariant-derivatives}). This difference, which is expected to be non-existent in flat spaces, is thus related to the curvature of the manifold.

For a vector field $V^\rho$, we can compute
\begin{equation}
    \comm{\nabla_\mu}{\nabla_\nu}V^\rho=\nabla_\mu\qty(\nabla_\nu V^\rho)-\nabla_\nu\qty(\nabla_\mu V^\rho).
\end{equation}
A sequence of algebraic manipulations then leads to
\begin{equation}
    \comm{\nabla_\mu}{\nabla_\nu}V^\rho={R^\rho}_{\sigma\mu\nu}V^\sigma,
\end{equation}
where we have defined the \textbf{Riemann tensor} as
\begin{equation}
    {R^\rho}_{\sigma\mu\nu}=\partial_\mu\Gamma_{\nu\sigma}^\rho-\partial_\nu\Gamma_{\mu\sigma}^\rho+\Gamma_{\mu\lambda}^\rho\Gamma_{\nu\sigma}^\lambda-\Gamma_{\nu\lambda}^\rho\Gamma_{\mu\sigma}^\lambda.
\end{equation}
This can also be shown to be related to the change experienced by a vector when parallel transported around a closed loop \cite{Wald1984}.

In general, the action of $\comm{\nabla_\rho}{\nabla_\sigma}$ can be computed on a tensor of arbitrary rank as
\begin{align}
    \comm{\nabla_\rho}{\nabla_\sigma}{X^{\mu_1\dots\mu_q}}_{\nu_1\dots\nu_r}&={R^{\mu_1}}_{\lambda\rho\sigma}{X^{\lambda\mu_2\dots\mu_q}}_{\nu_1\dots\nu_r}+{R^{\mu_2}}_{\lambda\rho\sigma}{X^{\mu_1\lambda\dots\mu_q}}_{\nu_1\dots\nu_r}+\dots \nonumber \\
    &-{R^\lambda}_{\nu_1\rho\sigma}{X^{\mu_1\dots\mu_q}}_{\lambda\nu_2\dots\nu_r}-{R^\lambda}_{\nu_2\rho\sigma}{X^{\mu_1\dots\mu_q}}_{\nu_1\lambda\dots\nu_r}-\dots\,.
\end{align}

The Riemann tensor satisfies the following properties:
\begin{enumerate}
    \item $R_{\rho\sigma\mu\nu}=-R_{\sigma\rho\mu\nu}$,
    \item $R_{\rho\sigma\mu\nu}=-R_{\rho\sigma\nu\mu}$,
    \item $R_{\rho\sigma\mu\nu}=R_{\mu\nu\rho\sigma}$,
    \item $R_{\rho\sigma\mu\nu}+R_{\rho\mu\nu\sigma}+R_{\rho\nu\sigma\mu}=0$,
\end{enumerate}
where $R_{\rho\sigma\mu\nu}=g_{\rho\lambda}{R^\lambda}_{\sigma\mu\nu}$. In addition, it also obeys a very important differential identity known as \textbf{Bianchi identity}:
\begin{equation} \label{Bianchi-identity}
    \nabla_\lambda R_{\rho\sigma\mu\nu}+\nabla_\rho R_{\sigma\lambda\mu\nu}+\nabla_\sigma R_{\lambda\rho\mu\nu}=0.
\end{equation}

There are some other tensors that are useful in general relativity which are obtained from the Riemann tensor. For instance, we can take a contraction to form the \textbf{Ricci tensor}:
\begin{equation} \label{Ricci-tensor}
    R_{\mu\nu}={R^\lambda}_{\mu\lambda\nu},
\end{equation}
which is symmetric. The trace of the Ricci tensor is what we call the \textbf{Ricci scalar} or the \textbf{curvature scalar}:
\begin{equation} \label{Curvature-scalar}
    R={R^\mu}_\mu=g^{\mu\nu}R_{\mu\nu}.
\end{equation}

Now, by contracting twice on the Bianchi identity \eqref{Bianchi-identity}, we can write
\begin{align}
    0&=g^{\nu\sigma}g^{\mu\lambda}\qty(\nabla_\lambda R_{\rho\sigma\mu\nu}+\nabla_\rho R_{\sigma\lambda\mu\nu}+\nabla_\sigma R_{\lambda\rho\mu\nu}) \nonumber \\
    &=\nabla^\mu R_{\rho\mu}-\nabla_\rho R+\nabla^\nu R_{\rho\nu},
\end{align}
or
\begin{equation} \label{Twice-contracted-Bianchi-identity}
    \nabla^\mu R_{\rho\mu}=\frac{1}{2}\nabla_\rho R.
\end{equation}
This motivates us to define the \textbf{Einstein tensor} as
\begin{equation} \label{Einstein-tensor}
    G_{\mu\nu}=R_{\mu\nu}-\frac{1}{2}Rg_{\mu\nu},
\end{equation}
so that the twice-contracted Bianchi identity \eqref{Twice-contracted-Bianchi-identity} becomes equivalent to
\begin{equation}
    \nabla^\mu G_{\mu\nu}=0.
\end{equation}
This means that the Einstein tensor, fundamentally related to the curvature of the manifold, is conserved.

\section{Geodesic deviation}

Let us now address another important manifestation of curvature, namely the violation of Euclid's fifth postulate, which basically states that initially parallel lines remain parallel through the entire space. This is not true in a curved space and, since the generalization of a straight line is a geodesic, this violation is quantified by the geodesic deviation equation.

\begin{figure}[!ht]
    \centering
    \includegraphics[width=0.5\linewidth]{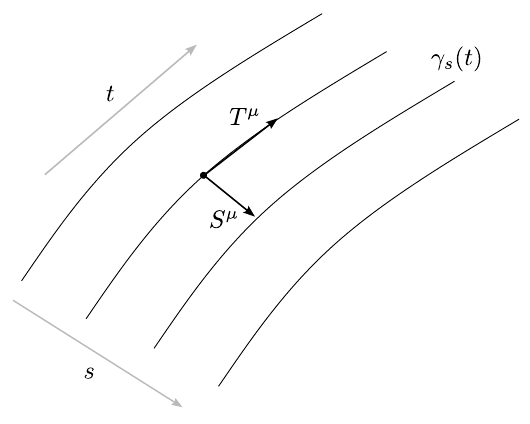}
    \caption{A set of geodesics $\gamma_s(t)$ with tangent vectors $T^\mu$ and deviation vectors $S^\mu$.}
    \label{Set-of-geodesics}
\end{figure}

Consider a one-parameter family of geodesics, $\gamma_s(t)$, that is, for each $s\in\mathbb{R}$, $\gamma_s$ is a geodesic parametrized by the parameter $t$. The collection of these curves defines a smooth two-dimensional surface on which the coordinates may be chosen to be $s$ and $t$, provided we have chosen a family of geodesics that do not cross (Figure \ref{Set-of-geodesics}). The entire surface is the set of points $x^\mu(s,t)\in M$. Let us then consider two vector fields: the tangent vectors to the geodesics,
\begin{equation}
    T^\mu=\pdv{x^\mu}{t},
\end{equation}
and the deviation vectors
\begin{equation}
    S^\mu=\pdv{x^\mu}{s}.
\end{equation}

We define the relative velocity of geodesics by
\begin{equation}
    V^\mu=T^\rho\nabla_\rho S^\mu,
\end{equation}
and the relative acceleration of geodesics by
\begin{equation}
    A^\mu=T^\rho\nabla_\rho V^\mu.
\end{equation}
Since $S$ and $T$ are basis vectors adapted to a coordinate system, their commutator vanishes \cite{Carroll}, which means that
\begin{equation}
    S^\rho\nabla_\rho T^\mu=T^\rho\nabla_\rho S^\mu.
\end{equation}
A direct computation then leads to
\begin{equation}
    A^\mu=\frac{\textrm{D}^2}{\textrm{d}t^2}S^\mu={R^\mu}_{\nu\rho\sigma}T^\nu T^\rho S^\sigma,
\end{equation}
which is known as the \textbf{geodesic deviation equation}. It shows that the relative acceleration between two neighboring geodesics is proportional to the curvature.

\section{Locally inertial coordinates}

Having discussed the curvature in a differentiable manifold, let us now show how to construct a set of coordinates in a small region that looks locally flat. In general relativity this is linked to the equivalence principle which, loosely speaking, concerns the impossibility of detecting the spacetime curvature by means of local experiments \cite{Carroll}. Here we show how to construct such reference frame at a point in the manifold. In the next section we show how we can do the same along an entire curve for small enough regions.

For a given point $p\in M$ we can define a locally flat set of coordinates for a small region that contains $p$, which we call \textbf{locally inertial coordinates}. In order to construct such a coordinate system, let $k^\mu$ be a tangent vector at $p$ for which $\lambda(p)=0$,
\begin{equation} \label{Initial-condition}
    k^\mu=\eval{\dv{x^\mu}{\lambda}}_{\lambda=0}.
\end{equation}
This tangent vector then defines a geodesic passing through $p$. Since we are specifying the initial conditions, $x^\mu(\lambda=0)$ and $k^\mu(\lambda=0)$, and the geodesic equation is a second-order differential equation, this geodesic is unique. Specifically, there will be a unique point in $M$ for which $\lambda=1$. We then define the \textbf{exponential map} at $p$, $\textrm{exp}_p:T_pM\to M$, as
\begin{equation}
    \exp_p(k)=x^\nu(\lambda=1),
\end{equation}
where $x^\nu(\lambda)$ solves the geodesic equation subject to the initial condition \eqref{Initial-condition}.

Since different geodesics emanating from a single point may eventually cross, the exponential map is not necessarily one-to-one. However, it is possible to show that there always exists a sufficiently small neighborhood of $p$ on which the exponential map is one-to-one \cite{Bishop2001}. This fact allows us to use the exponential map to construct locally inertial coordinates.

We have already argued how, at any given point $p\in M$, we can find basis vectors $\qty{\vu{e}_{\mu}}$ for $T_pM$ such that the components of the metric are those of the canonical form,
\begin{equation} \label{g=eta-at-p}
    g_{\mu\nu}=\eta_{\mu\nu},
\end{equation}
at $p$. Our next task is to find a coordinate system $x^{\mu}$ for which the basis vectors $\qty{\vu{e}_{\mu}}$ stand for a coordinate basis, $\vu{e}_{\mu}=\partial_{\mu}$, and such that $\partial_{\sigma}g_{\mu\nu}=0$. This is accomplished by considering a point $q\in M$ sufficiently close to $p$ and defining the coordinates $x^{\mu}(q)$ to be the components of the tangent vector $k=k^{\mu}\vu{e}_{\mu}$ that gets mapped to $q$ by the exponential map $\exp_p$. These coordinates are known as \textbf{Riemann normal coordinates} at $p$.

Now a parametrized set of vectors of the form $\lambda k^{\mu}$, for some fixed vector $k^{\mu}$, gets mapped to points in a geodesic by the exponential map. Therefore, in Riemann normal coordinates, any geodesic through $p$ may be expressed as
\begin{equation}
    x^{\mu}(\lambda)=\lambda k^{\mu},
\end{equation}
for which we have
\begin{equation}
    \dv[2]{x^{\mu}}{\lambda}=0
\end{equation}
along any geodesic through $p$ in this coordinate system. The geodesic equation \eqref{Geodesic-equation} then implies
\begin{equation}
    \Gamma_{\rho\sigma}^{\mu}(p)=0,
\end{equation}
from which metric compatibility yields
\begin{equation}
    \partial_{\sigma}g_{\mu\nu}=0
\end{equation}
at $p$. This result, together with \eqref{g=eta-at-p}, constitute the desired properties of locally inertial coordinates.

\section{Fermi normal coordinates}

Now that we have seen how to construct locally inertial coordinates in a small neighborhood of a single point, let us show that it is possible to introduce coordinates near any curve in a (pseudo-)Riemannian manifold in such a way that the metric is rectangular along the entire curve. These are called \textbf{Fermi normal coordinates}, and here we will be interested in the case where the curve is a geodesic $\gamma$. These coordinates will be constructed in order to satisfy the so-called Fermi conditions,
\begin{subequations} \label{Fermi-conditions}
\begin{equation} \label{Fermi-condition-1}
    \eval{g_{\mu\nu}}_\gamma=\eta_{\mu\nu},
\end{equation}
\begin{equation} \label{Fermi-condition-2}
    \eval{\Gamma_{\mu\nu}^\sigma}_\gamma=0,
\end{equation}
\end{subequations}
along the given geodesic. First we will show how such construction is possible, and then we will proceed to write the components of the metric tensor in a neighborhood of the geodesic in Fermi normal coordinates. In this section we follow the development of Ref. \cite{Manasse1963} (see also refs. \cite{GutierrezPineres2022,Misner,Zee2013}).

\subsection{Constructing Fermi normal coordinates}

We begin by choosing an arbitrary point $p_0$ in the spacetime manifold to be the origin of our coordinate system and an orthonormal set of vectors $\vu{e}_0,\dots,\vu{e}_3$ at $p_0$ to fix the coordinate axes. Now let $\gamma$ be the unique timelike geodesic which starts at $p_0$ with tangent vector $\vu{e}_0$ at that point. We denote the proper time along it by $t$ such that $p_0=\gamma(t=0)$. The orthonormal set at the origin is denoted by $\vu{e}_0(0),\dots,\vu{e}_3(0)$. Now, we know that the tangent vector to a geodesic at any two points on it is related by parallel transport along $\gamma$. Therefore, the vector $\vu{e}_0(t)$ will be tangent to the geodesic at any point $p=\gamma(t)$, and similarly the vectors $\vu{e}_i(0)$ are parallel transported to get $\vu{e}_{i}(t)$ at $p$. We assume for simplicity that $\vu{e}_0$ is a timelike vector and that all the $\vu{e}_i$'s are spacelike.

Now consider the point $q$, which is not necessarily a point of the geodesic curve $\gamma$, and whose Fermi normal coordinates are denoted by $x^\mu$. We construct, at the point $p=\gamma(x^0=t)$, the unit vector
\begin{equation} \label{v(x0)}
    \vb{v}(x^0)=v^i\vu{e}_i(x^0),\hspace{1cm}v^i=\frac{x^i}{s},
\end{equation}
with $s=\sqrt{x^jx_j}$. We can construct a family of spacelike geodesics $\beta(x^0,v^i)$ orthogonal to $\gamma$ at $p$ by varying $v^i$. Then, a unique geodesic intersects $\gamma$ orthogonally at $p$, is tangent to the vector $\vb{v}$, and passes through the point $q$ at a proper distance $s$,
\begin{equation} \label{Q=beta(x0,vi,s)}
    q=\beta(x^0,v^i,s).
\end{equation}
If the point $q$ is represented by its coordinate values $x^{\mu'}(q)$ in some coordinate system in which the metric components $g_{\mu'\nu'}(x^{\sigma'})$ are known, Eq. \eqref{Q=beta(x0,vi,s)} can be written in such coordinates as
\begin{equation}
    x^{\mu'}=x^{\mu'}(x^0,v^i,s).
\end{equation}

Since the geodesic equation is unchanged by a re-scaling of the affine parameter $\lambda$, it follows that
\begin{equation}
    x^{\mu'}(x^0,v^i,s\lambda)=x^{\mu'}(x^0,sv^i,\lambda).
\end{equation}
In particular, and recalling that $x^i=sv^i$, we find
\begin{equation}
    x^{\mu'}(x^0,v^i,s)=x^{\mu'}(x^0,sv^i,1)=x^{\mu'}(x^0,x^i,1)=x^{\mu'}(x^\mu),
\end{equation}
which gives the transformation law between the arbitrary coordinates $x^{\mu'}$ and Fermi normal coordinates $x^\mu$.

In the $x^{\mu'}$ coordinates the tangent vector to the geodesics $\beta(x^0,v^i,s)$ is
\begin{equation}
    v^{\mu'}=\dv{x^{\mu'}}{s},
\end{equation}
which calculated along $\gamma$ yields
\begin{equation} \label{v|gamma}
    \eval{v^{\mu'}}_\gamma=\eval{\dv{x^{\mu'}}{s}}_\gamma=\eval{\pdv{x^{\mu'}}{x^i}}_{s=0}\dv{x^i}{s}=\eval{\pdv{x^{\mu'}}{x^i}}_{s=0}v^i.
\end{equation}
A direct comparison between Eqs. \eqref{v(x0)} and \eqref{v|gamma} then gives
\begin{equation}
    \qty(\vu{e}_i)^{\mu'}=\eval{\pdv{x^{\mu'}}{x^i}}_{\gamma}.
\end{equation}
Additionally, since the tangent vector to $\gamma$ at $p=\gamma(x^0)$ is $\partial_0$, we find that, along $\gamma$,
\begin{equation} \label{e-along-gamma}
    \qty(\vu{e}_\mu)^{\mu'}=\eval{\pdv{x^{\mu'}}{x^\mu}}_\gamma,
\end{equation}
or $\vu{e}_\mu(t)=\eval{\partial_\mu}_\gamma$.

The relation between the metric components in the Fermi normal coordinates $x^\mu$ and the components of the metric in the arbitrary coordinates $x^{\mu'}$ is given by
\begin{equation}
    g_{\mu\nu}=\pdv{x^{\mu'}}{x^\mu}\pdv{x^{\nu'}}{x^\nu}g_{\mu'\nu'}.
\end{equation}
Evaluating it on the geodesic $\gamma$ and using Eq. \eqref{e-along-gamma}, we find
\begin{align}
    \eval{g_{\mu\nu}}_\gamma&=\eval{\pdv{x^{\mu'}}{x^\mu}}_\gamma\eval{\pdv{x^{\nu'}}{x^\nu}}_\gamma g_{\mu'\nu'}=\qty(\vu{e}_\mu)^{\mu'}\qty(\vu{e}_\nu)^{\nu'}g_{\mu'\nu'}=\qty(\vu{e}_\mu)^{\mu'}\qty(\vu{e}_\nu)_{\mu'} \nonumber \\
    &=\eta_{\mu\nu},
\end{align}
since the vectors $\vu{e}_\mu$ are orthogonal and we chose $\vu{e}_0$ ($\vu{e}_i$) to be timelike (spacelike). Therefore Fermi's first condition \eqref{Fermi-condition-1} is satisfied along the geodesic $\gamma$.

Next, recall that the curve $x^0=t$, $x^i=v^is$ satisfies the geodesic equation
\begin{equation}
    \dv[2]{x^\mu}{s}+\Gamma_{\sigma\rho}^\mu\dv{x^\sigma}{s}\dv{x^\rho}{s}=0,
\end{equation}
which implies that
\begin{equation} \label{Gamma-ij-vi-vj}
    \Gamma_{ij}^\mu v^iv^j=0,
\end{equation}
where $\Gamma_{ij}^\mu$ depends on $t$ and $x^i=v^is$. But along $\gamma$, $s=0$ and the dependence on $v^i$ vanishes. Then, in order to keep Eq. \eqref{Gamma-ij-vi-vj} satisfied, we must have
\begin{equation} \label{Gamma-ij-mu}
    \eval{\Gamma_{ij}^\mu}_\gamma=0.
\end{equation}
Now for the other symbols, we recall that the vector $\vu{e}_0(t)$, with components in Fermi normal coordinates given by $\qty(\vu{e}_0)^\mu=\delta_0^\mu$ according to Eq. \eqref{e-along-gamma}, must satisfy the equation of parallel transport along $\gamma$,
\begin{equation}
    \dv{t}\qty(\vu{e}_0)^\mu+\Gamma_{\sigma\rho}^\mu\dv{x^\sigma}{t}\qty(\vu{e}_0)^\rho=0\implies\Gamma_{\sigma\rho}^\mu\dv{x^\sigma}{t}\delta_0^\rho=0,
\end{equation}
which then implies
\begin{equation} \label{Gamma-sigma0-mu}
    \eval{\Gamma_{\sigma0}^\mu}_\gamma=\eval{\Gamma_{0\sigma}^\mu}_\gamma=0.
\end{equation}
Together, Eqs. \eqref{Gamma-ij-mu} and \eqref{Gamma-sigma0-mu} result in Fermi's second condition,
\begin{equation} \label{Fermi-second-condition}
    \eval{\Gamma_{\mu\nu}^\sigma}_\gamma=0.
\end{equation}

Therefore, given a geodesic, it is possible to introduce Fermi normal coordinates $x^\mu=(t,x^i)$ near it so that the Christoffel symbols vanish along it, with $t$ being the geodesic proper time. Also, along $\gamma$, $x^i=0$.

\subsection{Metric components} \label{Sec:Metric-components}

We are now interested in obtaining an expression for the components of the metric tensor in a neighborhood of the geodesic $\gamma$, written as a power series expansion up to second order in $x^i$.

Since $\Gamma_{\mu\nu}^\sigma=0$ holds for all $x^0$ at $x^i=0$, we may differentiate it with respect to $x^0$ to obtain
\begin{equation} \label{Temporal-derivative-Christoffel-symbols}
    \eval{\partial_0\Gamma_{\mu\nu}^\sigma}_\gamma=0.
\end{equation}
From the definition of the Christoffel symbols, this implies
\begin{equation} \label{partial-0partial-sigma-g-munu=0}
    \eval{\partial_0\partial_\sigma g_{\mu\nu}}_\gamma=0.
\end{equation}
It follows from Fermi's conditions \eqref{Fermi-conditions} that the zeroth order term will be the Minkowski metric while the linear terms vanish. Then, using Eq. \eqref{partial-0partial-sigma-g-munu=0} we can write
\begin{equation} \label{Metric-expansion}
    g_{\mu\nu}=\eta_{\mu\nu}+\frac{1}{2}\eval{\partial_i\partial_jg_{\mu\nu}}_\gamma x^ix^j+O(x^3).
\end{equation}

The Riemann curvature tensor reads
\begin{align}
    \eval{R_{\rho\sigma\mu\nu}}_\gamma&=\eval{g_{{\rho}{\lambda}}\qty(\partial_{{\mu}}\Gamma_{{\nu}{\sigma}}^{{\lambda}}-\partial_{{\nu}}\Gamma_{{\mu}{\sigma}}^{{\lambda}})}_\gamma \nonumber \\
    &=\eval{\frac{1}{2}\qty(\partial_{{\mu}}\partial_{{\sigma}}g_{{\rho}{\nu}}-\partial_{{\mu}}\partial_{{\rho}}g_{{\nu}{\sigma}}-\partial_{{\nu}}\partial_{{\sigma}}g_{{\rho}{\mu}}+\partial_{{\nu}}\partial_{{\rho}}g_{{\mu}{\sigma}})}_\gamma.
\end{align}
From this equation we immediately find
\begin{equation} \label{del-i-del-j-g-00}
    \eval{R_{i0j0}}_\gamma=-\frac{1}{2}\eval{\partial_i\partial_jg_{00}}_\gamma.
\end{equation}
Also, it follows from Eq. \eqref{Temporal-derivative-Christoffel-symbols} that
\begin{equation} \label{del-mu-Gamma-sigma-0-rho}
    \eval{{R^\rho}_{\sigma\mu0}}_\gamma=\eval{\partial_\mu\Gamma_{\sigma0}^\rho}_\gamma.
\end{equation}

In order to obtain the other second derivatives of the metric, we will need to use the geodesic deviation equation,
\begin{equation}
    \frac{\textrm{D}^2}{\textrm{d}s^2}n^\mu={R^\mu}_{\nu\rho\sigma}\dv{x^\nu}{s}\dv{x^\rho}{s}n^\sigma,
\end{equation}
where we are now denoting $n^\mu$ as the deviation vector. This equation can be rewritten as
\begin{equation} \label{Geodesic-deviation-equation}
    \dv[2]{s}n^\mu+2\Gamma_{\nu\sigma}^\mu\dv{x^\nu}{s}\dv{s}n^\sigma+\qty(\partial_\nu\Gamma_{\rho\sigma}^\mu-\Gamma_{\lambda\sigma}^\mu\Gamma_{\nu\rho}^\lambda+\Gamma_{\nu\lambda}^\mu\Gamma_{\rho\sigma}^\lambda-{R^\mu}_{\nu\rho\sigma})\dv{x^\nu}{s}\dv{x^\rho}{s}n^\sigma=0.
\end{equation}

Recall that the family of geodesics $\beta(x^0,v^i,s)$ is described in Fermi normal coordinates by $x^0=t$, $x^i=v^is$, and thus the deviation vectors are
\begin{equation}
    n_{(i)}^\mu=\pdv{x^\mu}{v^i}=s\delta_i^\mu,
\end{equation}
such that $\dv{s}n_{(i)}^\mu=\delta_i^\mu$ and $\dv[2]{s}n_{(i)}^\mu=0$. For that case, the geodesic deviation equation \eqref{Geodesic-deviation-equation} yields
\begin{equation}
    2\Gamma_{ji}^\mu v^j+s\qty(\partial_j\Gamma_{ki}^\mu-\Gamma_{\lambda i}^\mu\Gamma_{jk}^\lambda+\Gamma_{j\lambda}^\mu\Gamma_{ki}^\lambda-{R^\mu}_{jki})v^jv^k=0.
\end{equation}
In the first term we can expand the Christoffel symbols as
\begin{equation}
    \Gamma_{ji}^\mu=\eval{\Gamma_{ji}^\mu}_\gamma+s\eval{\partial_k\Gamma_{ji}^\mu}_\gamma v^k+O(s^2),
\end{equation}
such that dividing the equation by $s$ and then taking the limit $s\to0$ yields
\begin{equation}
    \eval{\qty(\partial_j\Gamma_{ki}^\mu+2\partial_k\Gamma_{ji}^\mu-{R^\mu}_{jki})}_\gamma v^jv^k=0,
\end{equation}
or
\begin{subequations}
\begin{equation} \label{Christoffel-derivatives-Riemann-tensor-a}
    \eval{\qty(\partial_k\Gamma_{ij}^\mu+\partial_j\Gamma_{ik}^\mu)}_\gamma=\eval{\frac{1}{3}\qty({R^\mu}_{jki}+{R^\mu}_{ikj})}_\gamma.
\end{equation}
We may also write the expressions obtained from the one above by cyclic permutations,
\begin{equation} \label{Christoffel-derivatives-Riemann-tensor-b}
    \eval{\qty(\partial_i\Gamma_{jk}^\mu+\partial_k\Gamma_{ji}^\mu)}_\gamma=\eval{\frac{1}{3}\qty({R^\mu}_{kij}+{R^\mu}_{jik})}_\gamma.
\end{equation}
and
\begin{equation} \label{Christoffel-derivatives-Riemann-tensor-c}
    \eval{\qty(\partial_j\Gamma_{ki}^\mu+\partial_i\Gamma_{kj}^\mu)}_\gamma=\eval{\frac{1}{3}\qty({R^\mu}_{ijk}+{R^\mu}_{kji})}_\gamma.
\end{equation}
\end{subequations}

Adding Eq. \eqref{Christoffel-derivatives-Riemann-tensor-b} to Eq. \eqref{Christoffel-derivatives-Riemann-tensor-a} and subtracting Eq. \eqref{Christoffel-derivatives-Riemann-tensor-c} from it finally yields
\begin{equation} \label{del-k-Gamma-ij-mu}
    \eval{\partial_k\Gamma_{ij}^\mu}_\gamma=\eval{\frac{1}{3}\qty({R^\mu}_{jki}+{R^\mu}_{ikj})}_\gamma.
\end{equation}

Now, for the second derivatives of the metric, we write the metric compatibility condition $\nabla_lg_{\mu\nu}=0$ as
\begin{equation}
    \partial_lg_{\mu\nu}-\Gamma_{l\mu}^\lambda g_{\lambda\nu}-\Gamma_{l\nu}^\lambda g_{\mu\lambda}=0,
\end{equation}
differentiate it with respect to $x^k$ and compute it along $\gamma$ to obtain
\begin{equation}
    \eval{\partial_k\partial_lg_{\mu\nu}}_\gamma=\eval{\partial_k\Gamma_{l\mu}^\lambda}_\gamma\eta_{\lambda\nu}+\eval{\partial_k\Gamma_{l\nu}^\lambda}_\gamma\eta_{\mu\lambda}.
\end{equation}
Then, using Eqs. \eqref{del-mu-Gamma-sigma-0-rho} and \eqref{del-k-Gamma-ij-mu} gives
\begin{align} \label{del-k-del-l-g-0i}
    \eval{\partial_k\partial_lg_{0i}}_\gamma&=\eval{\partial_k\Gamma_{l0}^\lambda}_\gamma\eta_{\lambda i}+\eval{\partial_k\Gamma_{li}^\lambda}_\gamma\eta_{0\lambda} \nonumber \\
    &=\eval{R_{ilk0}}_\gamma+\eval{\frac{1}{3}\qty(R_{0ikl}+R_{0lki})}_\gamma \nonumber \\
    &=\eval{\frac{1}{3}\qty(3R_{0kli}-R_{0kli}-R_{0lik}+R_{0lki})}_\gamma \nonumber \\
    &=\eval{\frac{2}{3}\qty(R_{0kli}+R_{0lki})}_\gamma,
\end{align}
where we used the properties of the Riemann tensor. Similarly we use Eq. \eqref{del-k-Gamma-ij-mu} to obtain
\begin{align} \label{del-k-del-l-g-ij}
    \eval{\partial_k\partial_lg_{ij}}_\gamma&=\eval{\partial_k\Gamma_{li}^\lambda}_\gamma\eta_{\lambda j}+\eval{\partial_k\Gamma_{lj}^\lambda}_\gamma\eta_{i\lambda} \nonumber \\
    &=-\eval{\frac{1}{3}\qty(R_{ikjl}+R_{iljk})}_\gamma.
\end{align}

Using the results from Eqs. \eqref{del-i-del-j-g-00}, \eqref{del-k-del-l-g-0i} and \eqref{del-k-del-l-g-ij} in the metric expansion \eqref{Metric-expansion}, we finally obtain
\begin{subequations} \label{Metric-in-Fermi-Normal-Coordinates}
\begin{align}
    g_{00}&=-1-\eval{R_{i0j0}}_\gamma x^ix^j+O(x^3), \\
    g_{0i}&=-\eval{\frac{2}{3}R_{0jik}}_\gamma x^jx^k+O(x^3), \\
    g_{ij}&=\delta_{ij}-\eval{\frac{1}{3}R_{ikjl}}_\gamma x^kx^l+O(x^3).
\end{align}
\end{subequations}
These are the desired expressions for the metric components written in Fermi normal coordinates $x^\mu$ up to second order.

Now, before we end this section, let us quickly discuss the concept of the "small enough region" over which this metric expansion is valid. Eqs. \eqref{Metric-in-Fermi-Normal-Coordinates} make it clear that the expansion is only valid for small values of $x^i$, but the natural question that arises is: small compared to \emph{what}?

To answer this, let us note that the metric tensor is dimensionless and, since the Riemann tensor involves second derivatives of $g_{\mu\nu}$, it has the dimension of one over length-squared. We can then introduce the concept of a typical variation scale of the metric, denoted $R_0$, such that $R_{\rho\sigma\mu\nu}\sim O(1/R_0^2)$. One can make this definition precise as, for instance, defining it in terms of the full contraction of the Riemann tensor, $R_0^{-4}\propto R^{\rho\sigma\mu\nu}R_{\rho\sigma\mu\nu}$, but such a precise definition is not necessary for our purposes. The important point is that $R_0$ measures the scale over which the metric changes appreciably. In this sense, the metric expansion \eqref{Metric-in-Fermi-Normal-Coordinates} is really an expansion over powers of the ratio $x/R_0$. This means that Fermi normal coordinates remain valid as long as $x\ll R_0$, i.e. it is the curvature itself that determines how "small enough" such region is. For stronger curvatures the smaller $R_0$ is, and consequently the smaller are the values of $x^i$ for which Eqs. \eqref{Metric-in-Fermi-Normal-Coordinates} remain valid.
\chapter{Graviton scattering by a Newtonian potential}
\label{app:Scattering}

In this appendix we explicitly compute the differential cross section for a graviton being scattered by a Newtonian source.

\section[The S-matrix]{The $S-$matrix}

In Chapter \ref{chap:quantum-grav} we showed that the action describing the interaction between gravitational radiation and a classical Newtonian potential is given by Eq. \eqref{Graviton-action-position-space}. Here we shall only work with the metric perturbation field in the TT gauge, Eqs. \eqref{TT-gauge-curved}, and so we will denote it by $h_{\mu\nu}$ instead of the explicit notation $h^{\rm TT}_{\mu\nu}$ since there is no room for confusion. Also, we will rescale the fields $h_{\mu\nu}\to\kappa_{\rm g}h_{\mu\nu}$, with $\kappa_{\rm g}=\sqrt{32\pi}$, so that they are canonically normalized. Hence, the action reads
\begin{equation}
    S_{\rm EH}=\frac{1}{2}\int\dd^4x\,\qty(h_{ij}\Box h^{ij}+2\phi\,h_{ij}\delta_{\mu\nu}\partial^\mu\partial^\nu h^{ij}),
\end{equation}
with $\phi(\vb{x})$ denoting the time-independent gravitational potential generated by a Newtonian source of mass $M_N$. Our goal now is to obtain the probability for a graviton to be scattered by an angle $\theta$.

As discussed in Chapter \ref{chap:quantum-grav}, the (canonical) quantization of gravitational radiation is done by writing the field operator
\begin{equation}
    \Hat{h}_{\mu\nu}(x)=\int\frac{\dd^3p}{(2\pi)^3}\frac{1}{\sqrt{2\omega_p}}\sum_s\qty[\epsilon_{\mu\nu}^s(p)\Hat{a}_{p,s}e^{ipx}+{\epsilon_{\mu\nu}^s}^*(p)\Hat{a}^\dagger_{p,s}e^{-ipx}],
\end{equation}
where $p^\mu=\qty(\omega_p,\vb{p})$ and $\epsilon_{\mu\nu}^s$ denotes the polarization tensors. The annihilation and creation operators satisfy the commutation relations
\begin{equation}
    \comm{\hat{a}_{p,s}}{\hat{a}_{p',s'}^\dagger}=(2\pi)^3\delta_{ss'}\delta^3(\vb{p}-\vb{p}').
\end{equation}

Now, consider an initial single graviton state $\ket{i}$ at $t=-\infty$, with momentum $p_1$ and polarization $s_1$, sufficiently far from the source of the gravitational potential $\phi$. After the interaction, one ends up with a final single graviton state $\ket{f}$ at $t=+\infty$, with momentum $p_2$ and polarization $s_2$. Suppose that the potential $\phi$ drops to zero sufficiently fast at infinity, so that we can consider the interaction dynamics to happen in some finite time interval and the theory to be free at $t\to\pm\infty$. The transition probability has the form
\begin{equation}
    P_{\rm transition}=\frac{\abs{\braket{f}{i}}^2}{\braket{f}{f}\braket{i}{i}},
\end{equation}
where the overlap $\braket{f}{i}$ is a $S-$matrix element with
\begin{subequations}
\begin{align}
    \ket{i}&=\sqrt{2\omega_{p_1}}\Hat{a}_{p_1,s_1}^\dagger(-\infty)\ket{\Omega}, \\
    \ket{f}&=\sqrt{2\omega_{p_2}}\Hat{a}_{p_2,s_2}^\dagger(+\infty)\ket{\Omega},
\end{align}
\end{subequations}
and $\ket{\Omega}$ is the vacuum state of the interacting theory. Explicitly, we have
\begin{equation}
    \braket{f}{i}=\sqrt{2\omega_{p_1}}\sqrt{2\omega_{p_2}}\mel{\Omega}{\Hat{a}_{p_2,s_2}(+\infty)\Hat{a}_{p_1,s_1}^\dagger(-\infty)}{\Omega}.
\end{equation}
It is interesting to notice that the operators inside brackets are in time order, so we can write this expression in a more convenient way as
\begin{align} \label{Transition-amplitude}
    \braket{f}{i}&=\sqrt{2\omega_{p_1}}\sqrt{2\omega_{p_2}} \nonumber \\
    &\times\mel{\Omega}{\textrm{T}\qty{\qty[\hat{a}_{p_2,s_2}(+\infty)-\hat{a}_{p_2,s_2}(-\infty)]\qty[\hat{a}_{p_1,s_1}^\dagger(-\infty)-\hat{a}_{p_1,s_1}^\dagger(+\infty)]}}{\Omega},
\end{align}
where T stands for the usual time ordering operation. This is essentially the same expression we had since time ordering sends the unwanted $\Hat{a}_{p_2,s_2}(-\infty)$ to the left, where it annihilates $\ket{\Omega}$, and the unwanted $\Hat{a}_{p_1,s_1}^\dagger(+\infty)$ to the right, where it annihilates $\bra{\Omega}$. This new expression will prove to be useful since we can relate it to the field operators as follows.

Assuming that the field dies off at $\vb{x}=\pm\infty$, we can write
\begin{align}
    i\int\dd^4x\,e^{-ipx}\Box\Hat{h}_{\mu\nu}(x)&=i\int\dd^4x\,e^{-ipx}\qty(-\partial_t^2+\laplacian)\Hat{h}_{\mu\nu}(x) \nonumber \\
    &=-i\int\dd^4x\,e^{-ipx}\qty(\partial_t^2+\omega_p^2)\Hat{h}_{\mu\nu}(x),
\end{align}
where $\omega_p^2=\vb{p}^2$ for a massless particle such as the graviton. Now, note that
\begin{align}
    \partial_t\qty[e^{-ipx}\qty(i\partial_t+\omega_p)\Hat{h}_{\mu\nu}(x)]&=\qty[i\omega_pe^{-ipx}\qty(i\partial_t+\omega_p)+e^{-ipx}\qty(i\partial_t^2+\omega_p\partial_t)]\Hat{h}_{\mu\nu}(x) \nonumber \\
    &=ie^{-ipx}\qty(\partial_t^2+\omega_p^2)\Hat{h}_{\mu\nu}(x),
\end{align}
and, therefore,
\allowdisplaybreaks
\begin{align}
    &i\int\dd^4x\,e^{-ipx}\Box\Hat{h}_{\mu\nu}(x)=-\int\dd^4x\,\partial_t\qty[e^{-ipx}\qty(i\partial_t+\omega_p)\Hat{h}_{\mu\nu}(x)] \nonumber \\
    &=-\eval{e^{i\omega_pt}\int\dd^3x\,e^{-i\vb{p}\vdot\vb{x}}\qty(i\partial_t+\omega_p)\Hat{h}_{\mu\nu}(x)}_{t=-\infty}^{t=+\infty} \nonumber \\
    &=-e^{i\omega_pt}\int\dd^3x\,e^{-i\vb{p}\vdot\vb{x}}\int\frac{\dd^3k}{(2\pi)^3}\sum_r\left[ \qty(\frac{\omega_k+\omega_p}{\sqrt{2\omega_k}})\epsilon_{\mu\nu}^r(\vb{k})\Hat{a}_{k,r}(t)e^{ik\cdot x}\right. \nonumber \\
    &\hspace{1.5cm}\eval{\left. +\qty(\frac{-\omega_k+\omega_p}{\sqrt{2\omega_k}}){\epsilon_{\mu\nu}^r}^*(\vb{k})\Hat{a}^\dagger_{k,r}(t)e^{-ik\cdot x}\right] }_{t=-\infty}^{t=+\infty} \nonumber \\
    &=-\sqrt{2\omega_p}\sum_r\epsilon_{\mu\nu}^r(\vb{p})\qty[\Hat{a}_{p,r}(+\infty)-\Hat{a}_{p,r}(-\infty)].
\end{align}
Here we used the fact that $\eval{\partial_t\Hat{a}_{k,r}(t)}_{t=-\infty}^{t=+\infty}=0$ since the theory is assumed to be free at $t\to\pm\infty$. Lastly, if we contract both sides of the equation above with $\epsilon^{\mu\nu}_s(p)$ and use the polarization tensors orthonormality condition \eqref{Polarization-tensor-normalization}, we arrive at
\begin{equation} \label{Useful-identity-for-LSZ}
    \sqrt{2\omega_p}\qty[\Hat{a}_{p,s}(+\infty)-\Hat{a}_{p,s}(-\infty)]=-\frac{i}{2}\epsilon_{\mu\nu}^s(\vb{p})\int\dd^4x\,e^{-ipx}\Box\Hat{h}^{\mu\nu}(x).
\end{equation}

Using Eq. \eqref{Useful-identity-for-LSZ} in \eqref{Transition-amplitude} leads to
\begin{align} \label{LSZ}
    \braket{f}{i}=\qty[-\frac{i}{2}{\epsilon_{\mu\nu}^{s_1}}^*(\vb{p}_1)\int\dd^4x_1\,e^{ip_1x_1}\Box_1]\qty[-\frac{i}{2}{\epsilon_{\rho\sigma}^{s_2}}(\vb{p}_2)\int\dd^4x_2\,e^{-ip_2x_2}\Box_2] \nonumber \\
    \times\mel{\Omega}{\textrm{T}\qty{\Hat{h}^{\mu\nu}(x_1)\Hat{h}^{\rho\sigma}(x_2)}}{\Omega},
\end{align}
where $\Box_j$ denotes the d'Alembertian with respect to\footnote{Here the index $j$ is not labeling components of a three dimensional vector, but simply different integration variables.} $x_j$. This is the \textbf{LSZ reduction formula} for gravitons, which relates $S-$matrix elements (left-hand side) to the Green functions involving the quantum fields (right-hand side)~\cite{Schwartz2013}. Hence, in order to compute the transition probability, we are going to need the time-ordered two-point function.

\section{The time-ordered two-point function}

\subsection{The free graviton propagator}

We will begin with the simplest case, namely the time-ordered two-point function of the free theory, also called the \textbf{Feynman propagator} for gravitons. It is defined by
\begin{equation}
    D^{\mu\nu\rho\sigma}(x_1-x_2)=\mel{0}{\textrm{T}\qty{\Hat{h}_0^{\mu\nu}(x_1)\Hat{h}_0^{\rho\sigma}(x_2)}}{0},
\end{equation}
where $\Hat{h}_0^{\mu\nu}(x)$ is the free field operator and $\ket{0}$ represents the vacuum of the free theory. From the plane-wave expansion of the field operator we find
\begin{align}
    \mel{0}{\Hat{h}_0^{\mu\nu}(x_1)\Hat{h}_0^{\rho\sigma}(x_2)}{0}&=\int\frac{\dd^3p_1}{(2\pi)^3}\frac{\dd^3p_2}{(2\pi)^3}\frac{1}{\sqrt{2\omega_{p_1}}}\frac{1}{\sqrt{2\omega_{p_2}}} \nonumber \\
    &\times\sum_{s_1,s_2}\epsilon^{\mu\nu}_{s_1}(\vb{p}_1){\epsilon^{\rho\sigma}_{s_2}}^*(\vb{p}_2)\mel{0}{\Hat{a}_{p_1,s_1}\Hat{a}_{p_2,s_2}^\dagger}{0}e^{i(p_1x_1-p_2x_2)},
\end{align}
or, by using the commutation relations for the creation and annihilation operators,
\begin{equation}
    \mel{0}{\Hat{h}_0^{\mu\nu}(x_1)\Hat{h}_0^{\rho\sigma}(x_2)}{0}=\int\frac{\dd^3p}{(2\pi)^3}\frac{1}{2\omega_p}A^{\mu\nu\rho\sigma}(\vb{p})e^{ip(x_1-x_2)},
\end{equation}
where
\begin{equation}
    A^{\mu\nu\rho\sigma}(\vb{p})\equiv\sum_s\epsilon^{\mu\nu}_{s}(\vb{p}){\epsilon^{\rho\sigma}_{s}}^*(\vb{p}).
\end{equation}

For the propagator, we have
\begin{align}
    &D^{\mu\nu\rho\sigma}(x_1-x_2)=\mel{0}{\textrm{T}\qty{\Hat{h}_0^{\mu\nu}(x_1)\Hat{h}_0^{\rho\sigma}(x_2)}}{0} \nonumber \\
    &=\mel{0}{\Hat{h}_0^{\mu\nu}(x_1)\Hat{h}_0^{\rho\sigma}(x_2)}{0}\theta(t_1-t_2)+\mel{0}{\Hat{h}_0^{\rho\sigma}(x_2)\Hat{h}_0^{\mu\nu}(x_1)}{0}\theta(t_2-t_1) \nonumber \\
    &=\int\frac{\dd^3p}{(2\pi)^3}\frac{1}{2\omega_p}\qty[A^{\mu\nu\rho\sigma}(\vb{p})e^{-i\omega_p\tau}\theta(\tau)+A^{\rho\sigma\mu\nu}(-\vb{p})e^{i\omega_p\tau}\theta(-\tau)]e^{i\vb{p}\vdot(\vb{x}_1-\vb{x}_2)},
\end{align}
where $\tau\equiv t_1-t_2$ and $\theta(t)$ is the Heaviside step function. Now, $\epsilon_{\mu\nu}(-\vb{p})=\epsilon_{\mu\nu}^*(\vb{p})$ and thus
\begin{equation}
    A^{\rho\sigma\mu\nu}(-\vb{p})=A^{\mu\nu\rho\sigma}(\vb{p}).
\end{equation}
We can also use the result
\begin{equation}
    e^{-i\omega_p\tau}\theta(\tau)+e^{i\omega_p\tau}\theta(-\tau)=\lim_{\varepsilon\to0}\frac{-2\omega_p}{2\pi i}\int_{-\infty}^\infty\frac{\dd\omega}{\omega^2-\omega_p^2+i\varepsilon}e^{i\omega\tau}
\end{equation}
to finally arrive at
\begin{equation} \label{Graviton-propagator}
    D^{\mu\nu\rho\sigma}(x_1-x_2)=\int\frac{\dd^4p}{(2\pi)^4}A^{\mu\nu\rho\sigma}(\vb{p})\frac{-i}{p^2+i\varepsilon}e^{-ip(x_1-x_2)}.
\end{equation}

\subsection{Interacting theory graviton propagator}

In quantum field theory, we can relate vacuum expectation values of time-ordered products in the interacting theory to those in the free theory by \cite{Schwartz2013,Peskin2007}
\begin{equation} \label{Relating-Green-functions}
    \mel{\Omega}{\textrm{T}\qty{\Hat{h}^{\mu\nu}(x_1)\Hat{h}^{\rho\sigma}(x_2)}}{\Omega}=\frac{\mel{0}{\textrm{T}\qty{\Hat{h}_0^{\mu\nu}(x_1)\Hat{h}_0^{\rho\sigma}(x_2)e^{i\int\dd^4x\,\mathcal{L}_{\rm int}[\Hat{h}_0]}}}{0}}{\mel{0}{\textrm{T}\qty{e^{i\int\dd^4x\,\mathcal{L}_{\rm int}[\Hat{h}_0]}}}{0}},
\end{equation}
where the interaction is described by the Lagrangian density $\mathcal{L}_{\rm int}$. In our case,
\begin{align}
    \mathcal{L}_{\rm int}[h(x)]&=\delta_{\mu\nu}\phi(\vb{x})h_{ij}(x)\partial^\mu\partial^\nu h^{ij}(x) \nonumber \\
    &=\delta_{\mu\nu}\phi(\vb{x})\eta_{\alpha\gamma}\eta_{\beta\delta}h^{\alpha\beta}(x)\partial^\mu\partial^\nu h^{\gamma\delta}(x),
\end{align}
which we wrote in a more convenient way.

Now, let us denote the free graviton propagator by $D_{12}^{\mu\nu\rho\sigma}\equiv D^{\mu\nu\rho\sigma}(x_1-x_2)$. Also, let $\int_x\equiv\int\dd^4x$ and $\delta^4_{xy}\equiv\delta^4(x-y)$. Then we can write
\begin{align}
    &\mel{0}{\textrm{T}\qty{\Hat{h}_0^{\mu\nu}(x_1)\Hat{h}_0^{\rho\sigma}(x_2)e^{i\int\dd^4x\,\mathcal{L}_{\rm int}[\Hat{h}_0]}}}{0}=D_{12}^{\mu\nu\rho\sigma} \nonumber \\
    &-i\eta_{\alpha\gamma}\eta_{\beta\delta}\int_x\int_y\phi(\vb{x})\delta^4_{xy}\delta^{\mu_y\nu_y}\partial_{\mu_y}\partial_{\nu_y}\mel{0}{\textrm{T}\qty{\Hat{h}_0^{\mu\nu}(x_1)\Hat{h}_0^{\rho\sigma}(x_2)\Hat{h}_0^{\alpha\beta}(x)\Hat{h}_0^{\gamma\delta}(y)}}{0},
\end{align}
up to first order in $\phi$. We may then use Wick's theorem to rewrite the time-ordered four-point function so as to obtain~\cite{Schwartz2013,Srednicki2007,Peskin2007}
\begin{align} \label{Relating-Green-functions-Numerator}
    &\mel{0}{\textrm{T}\qty{\Hat{h}_0^{\mu\nu}(x_1)\Hat{h}_0^{\rho\sigma}(x_2)e^{i\int\dd^4x\,\mathcal{L}_{\rm int}[\Hat{h}_0]}}}{0}=D_{12}^{\mu\nu\rho\sigma} \nonumber \\
    &-i\eta_{\alpha\gamma}\eta_{\beta\delta}\int_x\int_y\phi(\vb{x})\delta^4_{xy}\delta^{\mu_y\nu_y}\partial_{\mu_y}\partial_{\nu_y}\qty(D_{12}^{\mu\nu\rho\sigma}D_{xy}^{\alpha\beta\gamma\delta}+D_{1x}^{\mu\nu\alpha\beta}D_{2y}^{\rho\sigma\gamma\delta}+D_{1y}^{\mu\nu\gamma\delta}D_{2x}^{\rho\sigma\alpha\beta}).
\end{align}
Similarly,
\begin{equation} \label{Relating-Green-functions-Denominator}
    \mel{0}{\textrm{T}\qty{e^{i\int\dd^4x\,\mathcal{L}_{\rm int}[\Hat{h}_0]}}}{0}=1-i\eta_{\alpha\gamma}\eta_{\beta\delta}\int_x\int_y\phi(\vb{x})\delta^4_{xy}\delta^{\mu_y\nu_y}\partial_{\mu_y}\partial_{\nu_y}D_{xy}^{\alpha\beta\gamma\delta}.
\end{equation}

The first term on the second line of Eq. \eqref{Relating-Green-functions-Numerator}, proportional to $D_{12}^{\mu\nu\rho\sigma}$, corresponds to a Feynman diagram that includes what are usually called bubbles. However, as we can see from Eq. \eqref{Relating-Green-functions-Denominator}, and from the fact that $\frac{1}{1+kx}=1-kx+O(k^2)$, the bubble diagram gets canceled out, leaving us with
\begin{align}
    &\mel{\Omega}{\textrm{T}\qty{\Hat{h}^{\mu\nu}(x_1)\Hat{h}^{\rho\sigma}(x_2)}}{\Omega}=D_{12}^{\mu\nu\rho\sigma} \nonumber \\ &-2i\eta_{\alpha\gamma}\eta_{\beta\delta}\int_x\int_y\phi(\vb{x})\delta^4_{xy}\delta^{\mu_y\nu_y}\partial_{\mu_y}\partial_{\nu_y}D_{1x}^{\mu\nu\alpha\beta}D_{2y}^{\rho\sigma\gamma\delta}.
\end{align}
Using the expression for the free graviton propagator, Eq. \eqref{Graviton-propagator}, we arrive at\footnote{We are now leaving the $i\varepsilon$ prescription implicit.}
\begin{align}
    &\mel{\Omega}{\textrm{T}\qty{\Hat{h}^{\mu\nu}(x_1)\Hat{h}^{\rho\sigma}(x_2)}}{\Omega}=\int\frac{\dd^4p}{(2\pi)^4}A^{\mu\nu\rho\sigma}(\vb{p})\frac{-i}{p^2}e^{-ip(x_1-x_2)} \nonumber \\ 
    &-2i\int\frac{\dd^4p}{(2\pi)^4}\frac{\dd^4p'}{(2\pi)^4}\Tilde{\phi}(-p-p')B^{\mu\nu\rho\sigma}(\vb{p},\vb{p}')\frac{\omega'^2+\vb{p}'^2}{p^2p'^2}e^{-ipx_1}e^{-ip'x_2},
\end{align}
where
\begin{equation}
    B^{\mu\nu\rho\sigma}(\vb{p},\vb{p}')\equiv\eta_{\alpha\gamma}\eta_{\beta\delta}A^{\mu\nu\alpha\beta}(\vb{p})A^{\rho\sigma\gamma\delta}(\vb{p}')
\end{equation}
and $\Tilde{\phi}(q)$ is the four-dimensional Fourier transform of the Newtonian potential.

\section{The differential cross section}

With the expression for the time-ordered two-point function at hand, let us now return to the $S-$matrix elements via the reduction formula, Eq. \eqref{LSZ}. Explicitly,
\begin{align}
    \braket{f}{i}&=\frac{i}{4}{\epsilon_{\mu\nu}^{s_1}}^*(\vb{p}_1){\epsilon_{\rho\sigma}^{s_2}}(\vb{p}_2)\int\dd^4x_1\dd^4x_2\,e^{ip_1x_1}e^{-ip_2x_2}\Box_1\Box_2 \nonumber \\
    &\times\qty(i\mel{\Omega}{\textrm{T}\qty{\Hat{h}^{\mu\nu}(x_1)\Hat{h}^{\rho\sigma}(x_2)}}{\Omega})\nonumber \\
    &=\frac{i}{4}{\epsilon_{\mu\nu}^{s_1}}^*(\vb{p}_1){\epsilon_{\rho\sigma}^{s_2}}(\vb{p}_2)p_1^2A^{\mu\nu\rho\sigma}(\vb{p}_1)(2\pi)^4\delta^4(p_1-p_2) \nonumber \\
    &+\frac{i}{4}{\epsilon_{\mu\nu}^{s_1}}^*(\vb{p}_1){\epsilon_{\rho\sigma}^{s_2}}(\vb{p}_2)2\Tilde{\phi}(p_2-p_1)B^{\mu\nu\rho\sigma}(\vb{p}_1,-\vb{p}_2)(\omega_2^2+\vb{p}_2^2).
\end{align}

Then, by using the normalization condition \eqref{Polarization-tensor-normalization}, direct calculations give
\begin{subequations}
\begin{equation}
    {\epsilon_{\mu\nu}^{s_1}}^*(\vb{p}){\epsilon_{\rho\sigma}^{s_2}}(\vb{p})A^{\mu\nu\rho\sigma}(\vb{p})=4\delta_{s_1,s_2},
\end{equation}
and
\begin{equation}
    {\epsilon_{\mu\nu}^{s_1}}^*(\vb{p}_1){\epsilon_{\rho\sigma}^{s_2}}(\vb{p}_2)B^{\mu\nu\rho\sigma}(\vb{p}_1,-\vb{p}_2)=4\epsilon_{s_1}^{\mu\nu}(\vb{p}_1)\epsilon_{\mu\nu}^{s_2}(\vb{p}_2).
\end{equation}
\end{subequations}
Since the graviton is massless, we have $p_1^2=p_2^2=0$, and hence
\begin{equation}
    \abs{\braket{f}{i}}^2=\abs{4\Tilde{\phi}(p_2-p_1)\epsilon_{s_1}^{\mu\nu}(\vb{p}_1)\epsilon_{\mu\nu}^{s_2}(\vb{p}_2)}^2\omega_2^4.
\end{equation}

Additionally, we know that
\begin{subequations}
\begin{align}
    \braket{i}{i}&=2\omega_{1}(2\pi)^3\delta^3(0)=2\omega_{1}V, \\
    \braket{f}{f}&=2\omega_{2}(2\pi)^3\delta^3(0)=2\omega_{2}V,
\end{align}
\end{subequations}
where we are denoting $V=(2\pi)^3\delta^3(0)$, a (formally infinite) regularization volume. Therefore, the transition probability becomes
\begin{equation}
    P_{\rm transition}=\frac{4}{V^2}\abs{\Tilde{\phi}(p_2-p_1)\epsilon_{s_1}^{\mu\nu}(\vb{p}_1)\epsilon_{\mu\nu}^{s_2}(\vb{p}_2)}^2\frac{\omega_2^3}{\omega_1}.
\end{equation}

Now take the Newtonian potential to be given by $\phi(r)=-M_N/r$. Then
\begin{align}
    \Tilde{\phi}(p_2-p_1)&=-M_N\int\dd^4x\,\frac{1}{r}e^{i(p_1-p_2)x} \nonumber \\
    &=-M_N\int\dd t\,e^{-it(\omega_1-\omega_2)}\int\dd^3x\,\frac{1}{r}e^{i(\vb{p}_1-\vb{p}_2)\vdot\vb{x}} \nonumber \\
    &=-4\pi M_N\frac{1}{(\vb{p}_1-\vb{p}_2)^2}(2\pi)\delta(\omega_1-\omega_2),
\end{align}
and
\begin{equation} \label{Transition-probability-final-result}
    P_{\rm transition}=\frac{128\pi^3TM_N^2}{V^2}\abs{\frac{1}{(\vb{p}_1-\vb{p}_2)^2}\epsilon_{s_1}^{\mu\nu}(\vb{p}_1)\epsilon_{\mu\nu}^{s_2}(\vb{p}_2)}^2\frac{\omega_2^3}{\omega_1}\delta(\omega_1-\omega_2),
\end{equation}
with $T=(2\pi)\delta(0)$ (formally infinite) being the total time for the scattering process.

The differential cross section is defined by
\begin{equation}
    \dd\sigma=\frac{1}{T}\frac{1}{\Phi}\dd P,
\end{equation}
where $\Phi=1/V$ is the incoming flux, $\dd P$ is the quantum mechanical differential probability,
\begin{equation}
    \dd P=P_{\rm transition}\dd\Pi,
\end{equation}
and $\dd\Pi$ is the region of final momenta at which we are looking. For one outgoing particle with momentum $\vb{p}_2$, we have
\begin{equation}
    \dd\Pi=\frac{V}{(2\pi)^3}\dd^3p_2,
\end{equation}
defined such that $\int\dd\Pi=1$. Since $\dd^3p_2=\abs{\vb{p}_2}^2\dd\abs{\vb{p}_2}\dd\Omega$, we write
\begin{equation} \label{Cross-section}
    \dv{\sigma}{\Omega}=\frac{V}{T}\int_0^\infty\dd\abs{\vb{p}_2}\,\frac{V}{(2\pi)^3}\abs{\vb{p}_2}^2P_{\rm transition}(\abs{\vb{p}_2}).
\end{equation}

By plugging Eq. \eqref{Transition-probability-final-result} into \eqref{Cross-section} we find
\begin{align} \label{Cross-section-intermediate-step}
    \dv{\sigma}{\Omega}&=16M_N^2\int_0^\infty\dd\omega_2\frac{1}{(\omega_1^2+\omega_2^2-2\omega_1\omega_2\cos\theta)^2}\frac{\omega_2^5}{\omega_1}\delta(\omega_1-\omega_2) \nonumber \\
    &\hspace{2cm}\times\frac{1}{8}\sum_{s_1,s_2}\abs{\epsilon_{s_1}^{\mu\nu}(\vb{p}_1)\epsilon_{\mu\nu}^{s_2}(\vb{p}_2)}^2,
\end{align}
where $\theta$ is the angle between $\vb{p}_1$ and $\vb{p}_2$ and we averaged over the polarizations in order to obtain the unpolarized differential cross section\footnote{Note that, due to our normalization convention, $\sum_{s_1,s_2}\abs{\epsilon_{s_1}^{\mu\nu}(\vb{p})\epsilon_{\mu\nu}^{s_2}(\vb{p})}^2=8$.}. We can compute the sum over polarizations as follows.

Letting $u$ and $w$ denote the spatial polarization unit vectors orthogonal to the direction of propagation, $\vb{p}/\abs{\vb{p}}$, the polarization tensors can be written as
\begin{subequations}
\begin{align}
    \epsilon_{ij}^{+}(\vb{p})&=u_iu_j-w_iw_j, \\
    \epsilon_{ij}^{\cross}(\vb{p})&=u_iw_j+w_iu_j.
\end{align}
\end{subequations}
Then, we explicitly find
\begin{subequations}
\begin{align}
    \epsilon^{ij}_{+}(\vb{p}_1)\epsilon_{ij}^{+}(\vb{p}_2)&=(u_1\vdot u_2)^2+(w_1\vdot w_2)^2-(u_1\vdot w_2)^2-(w_1\vdot u_2)^2, \\
    \epsilon^{ij}_{\cross}(\vb{p}_1)\epsilon_{ij}^{\cross}(\vb{p}_2)&=2(u_1\vdot u_2)(w_1\vdot w_2)+2(u_1\vdot w_2)(w_1\vdot u_2), \\
    \epsilon^{ij}_{\cross}(\vb{p}_1)\epsilon_{ij}^{+}(\vb{p}_2)&=2(u_1\vdot u_2)(w_1\vdot u_2)-2(u_1\vdot w_2)(w_1\vdot w_2), \\
    \epsilon^{ij}_{+}(\vb{p}_1)\epsilon_{ij}^{\cross}(\vb{p}_2)&=2(u_1\vdot u_2)(u_1\vdot w_2)-2(w_1\vdot u_2)(w_1\vdot w_2),
\end{align}
\end{subequations}
where $u_1$, $w_1$ ($u_2$, $w_2$) are the unit vectors orthogonal to $\vb{p}_1$ ($\vb{p}_2$).

\begin{figure}[!ht]
    \centering
    \includegraphics[width=0.5\linewidth]{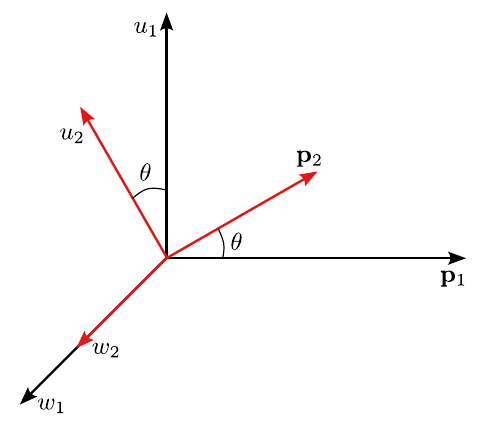}
    \caption{Vectors $\vb{p}_1$ and $\vb{p}_2$ and their corresponding spatial polarization orthogonal unit vectors.}
    \label{directions-pol}
\end{figure}

Choosing the systems $\qty(u_1,w_1,\frac{\vb{p}_1}{\abs{\vb{p}_1}})$ and $\qty(u_2,w_2,\frac{\vb{p}_2}{\abs{\vb{p}_2}})$ as the ones represented in Figure \ref{directions-pol}, so that the angle between $\vb{p}_1$ and $\vb{p}_2$ is $\theta$, we find
\begin{subequations}
\begin{align}
    u_1\vdot u_2&=\cos\theta, \\
    w_1\vdot w_2&=1, \\
    u_1\vdot w_2=w_1\vdot u_2&=0,
\end{align}
\end{subequations}
and thus, after some amount of trigonometric manipulations,
\begin{align}
    \sum_{s_1,s_2}\abs{\epsilon_{s_1}^{\mu\nu}(\vb{p}_1)\epsilon_{\mu\nu}^{s_2}(\vb{p}_2)}^2&=(1+\cos^2\theta)^2+4\cos^2\theta \nonumber \\
    &=8\qty(\cos^8\frac{\theta}{2}+\sin^8\frac{\theta}{2}).
\end{align}

Using this result in Eq. \eqref{Cross-section-intermediate-step} and computing the integral with the delta function finally leads to the differential cross section
\begin{equation}
    \dv{\sigma}{\Omega}=\frac{M_N^2}{\sin^4\frac{\theta}{2}}\qty(\cos^8\frac{\theta}{2}+\sin^8\frac{\theta}{2}).
\end{equation}
We can restore factors of $G$ and $c$ by noticing that the differential cross section has dimensions of area and letting $\dv*{\sigma}{\Omega}\to(1/L_{\rm P}^2)\dv*{\sigma}{\Omega}$ and $M_N\to M_N/M_{\rm P}$, with the Planck length and mass written in terms of the universal constants. The result is
\begin{equation}
    \dv{\sigma}{\Omega}=\frac{G^2M_N^2}{c^4\sin^4\frac{\theta}{2}}\qty(\cos^8\frac{\theta}{2}+\sin^8\frac{\theta}{2}).
\end{equation}
This differential cross section matches the one computed for scattering of \emph{classical} gravitational waves by a Newtonian potential~\cite{Westervelt_1971}.
\chapter{Computing the noise kernels}
\label{app:noise-kernels}

\section{Gravitational noise kernel} \label{SAppen:Gravitational-noise-kernel}

In this section, we compute the noise kernel encoding the influence of the quantized gravitational field, defined by Eq. \eqref{Gravitational-noise-kernel}, for different initial states. In what follows we shall consider an unpolarized and isotropic gravitational wave background such that the position operators are independent of the polarizations and of the direction of $\vb{k}$, that is, $\hat{q}_s(t,\vb{k})=\hat{q}_\omega(t)$, $\omega=\abs{\vb{k}}$. Within this assumption, we may write the gravitational noise kernel as
\begin{align}
    N_{\rm g}^{ijkl}(t,t')&=\frac{1}{32}\int_0^\infty\dd\omega\,\omega^2\left[ A_{\Omega}^{ijkl}\dv[2]{t}\dv[2]{{t'}}G_\omega(t,t')\right. \nonumber \\
    &\left. +A_{\Omega}^{ijnl}{\Phi_n}^k\dv[2]{t}G_\omega(t,t')+A_{\Omega}^{njkl}{\Phi_n}^i\dv[2]{{t'}}G_\omega(t,t')\right] ,
\end{align}
where
\begin{equation} \label{Hadamard-function}
    G_\omega(t,t')=\expval{\acomm{\hat{q}_\omega(t)}{\hat{q}_\omega(t')}}_{\rm g}
\end{equation}
is the Hadamard function, and the tensor structure is encoded in the angular integral given by
\begin{equation}
    A_{\Omega}^{ijkl}\equiv\int \dd \Omega\sum_s\epsilon_s^{ij}(\vb{k})\epsilon_s^{kl}(\vb{k}).
\end{equation}

Let us begin by explicitly evaluating the angular integral, since it does not depend on the initial graviton state. First, write the polarization tensors as in Eq. \eqref{Polarization-tensors},
\begin{equation}
\begin{split}
    \epsilon_{ij}^{+}&=\hat{\epsilon}_i^{(1)}\hat{\epsilon}_j^{(1)}-\hat{\epsilon}_i^{(2)}\hat{\epsilon}_j^{(2)} \\
    \epsilon_{ij}^{\cross}&=\hat{\epsilon}_i^{(1)}\hat{\epsilon}_j^{(2)}+\hat{\epsilon}_i^{(2)}\hat{\epsilon}_j^{(1)},
\end{split}
\end{equation}
where the spatial polarization unit vectors $\hat{\epsilon}^{(1)}$ and $\hat{\epsilon}^{(2)}$ are orthogonal to the direction of propagation $\hat{k}=\vb{k}/\abs{\vb{k}}$. They satisfy
\begin{equation}
    \hat{\epsilon}_i^{(1)}\hat{\epsilon}_j^{(1)}+\hat{\epsilon}_i^{(2)}\hat{\epsilon}_j^{(2)}=\delta_{ij}-\hat{k}_i\hat{k}_j\equiv P_{ij}.
\end{equation}
Then, a straightforward calculation yields
\begin{equation}
    \sum_s\epsilon_s^{ij}\epsilon_s^{kl}=P^{ik}P^{jl}+P^{il}P^{jk}-P^{ij}P^{kl}.
\end{equation}

We can use the following integration over solid angles in three spatial dimensions:
\begin{subequations}
\begin{equation}
    \int \dd\Omega\,\hat{k}_i\hat{k}_j=\frac{4\pi}{3}\delta_{ij},
\end{equation}
\begin{equation}
    \int \dd\Omega\,\hat{k}_i\hat{k}_j\hat{k}_k\hat{k}_l=\frac{4\pi}{15}\qty(\delta_{ij}\delta_{kl}+\delta_{ik}\delta_{jl}+\delta_{il}\delta_{jk}),
\end{equation}
\end{subequations}
to obtain
\begin{equation}
    \int \dd\Omega\,P_{ij}P_{kl}=\frac{8\pi}{5}\delta_{ij}\delta_{kl}+\frac{4\pi}{15}\qty(\delta_{ik}\delta_{jl}+\delta_{il}\delta_{jk}).
\end{equation}

At last, we find
\begin{equation} \label{Angular-integrals}
    A_{\Omega}^{ijkl}=\frac{8\pi}{15}\qty[3\qty(\delta^{ik}\delta^{jl}+\delta^{il}\delta^{jk})-2\delta^{ij}\delta^{kl}].
\end{equation}

In Chapter \ref{chap:decoherence} we argue that only the component $N_{\rm g}^{3333}(t,t')\equiv N_{\rm g}(t,t')$ will be of interest for the specific decoherence functions we work with, and so we shall focus on it from now on for simplicity. From Eq. \eqref{Angular-integrals} we have $A_{\Omega}^{3333}=32\pi/15$, and thus
\begin{equation} \label{Grav-noise-kernel-2}
    N_{\rm g}(t,t')=\frac{\pi}{15}\int_0^\infty\dd\omega\,\omega^2\qty{\dv[2]{t}\dv[2]{{t'}}G_\omega(t,t')+\Phi_{zz}\qty[\dv[2]{t}G_\omega(t,t')+\dv[2]{{t'}}G_\omega(t,t')]}.
\end{equation}
Here we lowered the index of the tidal tensor and dropped higher order terms in $\phi$ as usual.

Next, let us compute the Hadamard function. Since the free Hamiltonian for each mode of the gravitational field is that of a harmonic oscillator, Eq. \eqref{Hamiltonian-individual-mode}, we may write the position operators in the Heisenberg picture as
\begin{equation}
    \hat{q}_\omega(t)=\sqrt{\frac{1}{2m_{\rm g}\omega}}\qty(\hat{a}_\omega e^{-i\omega t}+\hat{a}_\omega^\dagger e^{i\omega t}),
\end{equation}
where the $\hat{a}$'s ($\hat{a}^\dagger$'s) are annihilation (creation) operators satisfying the usual commutation relations. A direct calculation then yields
\begin{equation} \label{G(t,t')}
    G_\omega(t,t')=\frac{2}{m_{\rm g}\omega^2}\langle\hat{H}_\omega\rangle_{\rm g}\cos\omega(t-t')+\frac{1}{m_{\rm g}\omega}\qty[\langle\hat{a}_\omega^2\rangle_{\rm g}\,e^{-i\omega(t+t')}+\langle\qty(\hat{a}^\dagger_\omega)^2\rangle_{\rm g}\,e^{i\omega(t+t')}],
\end{equation}
where $\hat{H}_\omega$ is the free Hamiltonian operator of the harmonic oscillator with frequency $\omega$. We may now compute the noise kernel for different initial states of the gravitational field.

\subsection{Vacuum state}

If the initial state is the vacuum state, we have $\langle \hat{a}_\omega^2\rangle_{\rm g}=\langle\qty(\hat{a}^\dagger_\omega)^2\rangle_{\rm g}=0$ and $\langle \hat{H}_\omega\rangle_{\rm g}=\omega/2$, and thus
\begin{equation} \label{Green-function-vacuum}
    G_\omega^{\rm (vac)}(t,t')=\frac{1}{m_{\rm g}\omega}\cos{\omega(t-t')}.
\end{equation}
Plugging this Hadamard function into Eq. \eqref{Grav-noise-kernel-2} leads to
\begin{align} \label{N-vac(t,t')}
    N_{\rm g}^{\rm (vac)}(t,t')&=\frac{2}{15\pi}\qty[\int_0^\Lambda\dd\omega\,\omega^5\cos{\omega(t-t')}-2\Phi_{zz}\int_0^\Lambda\dd\omega\,\omega^3\cos{\omega(t-t')}] \nonumber \\
    &=\frac{2\Lambda^4}{15\pi}\qty{\Lambda^2F_5\qty[\Lambda(t-t')]-2\Phi_{zz}F_3\qty[\Lambda(t-t')]},
\end{align}
where $\Lambda$ is the graviton energy cutoff (see Section \ref{Sec:Classical action}) and we used $m_{\rm g}=\pi^2/2$. Also, we defined
\begin{equation} \label{Fn(x)}
    F_n(x)\equiv\frac{1}{x^{n+1}}\int_0^x\dd y\,y^n\cos y.
\end{equation}
Explicitly,
\begin{subequations} \label{F5(x)eF3(x)}
\begin{equation} \label{F5(x)}
    F_5(x)=\frac{1}{x^6}\left[ \qty(5x^4-60x^2+120)\cos x+x\qty(x^4-20x^2+120)\sin x-120\right] ,
\end{equation}
and
\begin{equation} \label{F3(x)}
    F_3(x)=\frac{1}{x^4}\qty[\qty(3x^2-6)\cos x+\qty(x^3-6x)\sin x+6].
\end{equation}
\end{subequations}

Now, from Eq. \eqref{External-DoFs-density-matrix} we see that there are contributions to the external DoFs density matrix that are proportional to the product of the gravitational and the internal DoFs noise kernels. In Section \ref{SAppen:Internal-DoFs-noise-kernel} we will explicitly compute the latter, modeling it as a white noise. Therefore we will be interested in the coincidence limit $t'\to t$ for the gravitational noise kernel. From
\begin{equation} \label{Limit-Fn(x)-x-to-0}
    \lim_{x\to0}F_n(x)=\frac{1}{n+1},
\end{equation}
it follows that
\begin{equation} \label{N-vac(t)}
    N^{\rm (vac)}_{\rm g}(t)=\lim_{t'\to t}N^{\rm (vac)}_{\rm g}(t,t')=\frac{\Lambda^4}{15\pi}\qty(\frac{\Lambda^2}{3}-\Phi_{zz}).
\end{equation}

\subsection{Thermal state}

Let us now consider the gravitons to be initially in a thermal state with temperature $T_{\textrm{g}}=\beta_{\textrm{g}}^{-1}$ (which, as we discussed in Section \ref{Sec:Master-eq-grav-dec}, is a parameter that is not necessarily tied to a precise thermodynamical definition). In that case we have $\langle \hat{a}_\omega^2\rangle_{\rm g}=\langle\qty(\hat{a}^\dagger_\omega)^2\rangle_{\rm g}=0$ once again, and
\begin{equation}
    \langle \hat{H}_\omega\rangle_{\rm g}=\omega\qty(\frac{1}{2}+\frac{1}{e^{\omega\beta_{\rm g}}-1}).
\end{equation}
Plugging into Eq. \eqref{G(t,t')} leads to
\begin{equation} \label{Green-function-thermal}
    G_\omega^{\rm (th)}(t,t')=G_\omega^{\rm (vac)}(t,t')+\frac{2}{m_{\rm g}\omega}\frac{1}{e^{\omega/T_{\rm g}}-1}\cos{\omega(t-t')}.
\end{equation}
The noise kernel is found to be
\begin{align} \label{N-th(t,t')}
    N^{\rm (th)}_{\rm g}(t,t')&=N^{\rm (vac)}_{\rm g}(t,t') \nonumber \\
    &+\frac{4}{15\pi}\int_0^\infty\dd\omega\qty[\frac{\omega^5}{e^{\omega/T_{\rm g}}-1}\cos\omega(t-t')-2\Phi_{zz}\frac{\omega^3}{e^{\omega/T_{\rm g}}-1}\cos\omega(t-t')] \nonumber \\
    &=N^{\rm (vac)}_{\rm g}(t,t') \nonumber \\
    &+\frac{8(\pi T_{\rm g})^4}{5\pi}\qty{10(\pi T_{\rm g})^2F_1^{(\rm th)}\qty[\pi T_{\rm g}(t-t')]-\Phi_{zz}F_2^{(\rm th)}\qty[\pi T_{\rm g}(t-t')]},
\end{align}
with
\begin{subequations} \label{Fth1-e-2}
\begin{equation} \label{Fth1}
    F_1^{(\rm th)}(x)=\frac{1}{x^6}-\frac{1}{15\sinh^6x}\qty(2\cosh^4x+11\cosh^2x+2),
\end{equation}
and
\begin{equation} \label{Fth2}
    F_2^{(\rm th)}(x)=\frac{1}{3\sinh^4x}\qty(2\cosh^2x+1)-\frac{1}{x^4}.
\end{equation}
\end{subequations}
Also, using $\lim_{x\to0}F_1^{(\rm th)}(x)=2/945$ and $\lim_{x\to0}F_2^{(\rm th)}(x)=1/45$, we find
\begin{equation} \label{N-th(t)}
    N^{\rm (th)}_{\rm g}(t)=\lim_{t'\to t}N^{\rm (th)}_{\rm g}(t,t')=N^{\rm (vac)}_{\rm g}(t)+\frac{8(\pi T_{\rm g})^4}{45\pi}\qty(\frac{4}{21}\pi^2T_{\rm g}^2-\frac{1}{5}\Phi_{zz}).
\end{equation}

\subsection{Coherent state}

In quantum optics, the coherent states are those whose electric field variation approaches that of a classical electromagnetic wave. They are important because a single-mode laser usually generates a coherent state excitation~\cite{Loudon2000}. In complete analogy, one may say that a coherent graviton state is the quantum-mechanical state whose properties most closely resemble those of a classical gravitational wave.

For a quantum harmonic oscillator, the coherent state is defined by
\begin{equation}
    \ket{\alpha_\omega}=\Hat{\mathscr{D}}(\alpha_\omega)\ket{0},
\end{equation}
where $\Hat{\mathscr{D}}(\alpha_\omega)$ is called the \emph{coherent-state displacement operator}, defined as \cite{Loudon2000}
\begin{equation}
    \mathscr{D}(\alpha_\omega)=\exp(\alpha_\omega \hat{a}_\omega^\dagger-\alpha_\omega^*\hat{a}_\omega).
\end{equation}
Here we take the displacement parameter to be independent of $\omega$, $\alpha_\omega=\alpha$, for simplicity. The displacement operator satisfies a number of properties,
\begin{subequations}
\begin{align}
    \Hat{\mathscr{D}}^\dagger(\alpha)\Hat{\mathscr{D}}(\alpha)&=1, \\
    \Hat{\mathscr{D}}^\dagger(\alpha)\hat{a}_\omega^n\Hat{\mathscr{D}}(\alpha)&=(\hat{a}_\omega+\alpha)^n, \\
    \Hat{\mathscr{D}}^\dagger(\alpha)(\hat{a}_\omega^\dagger)^n\Hat{\mathscr{D}}(\alpha)&=(\hat{a}_\omega^\dagger+\alpha^*)^n,
\end{align}
\end{subequations}
for $n\in\mathbb{N}$, from which we find
\begin{subequations}
\begin{align}
    \langle\hat{a}_\omega^2\rangle_{\rm g}&=\alpha^2, \\
    \langle(\hat{a}_\omega^\dagger)^2\rangle_{\rm g}&=(\alpha^*)^2, \\
    \langle\hat{H}_\omega\rangle_{\rm g}&=\omega\qty(\abs{\alpha}^2+\frac{1}{2}),
\end{align}
\end{subequations}
where we also used the fact that the coherent state is an eigenstate of the annihilation operator, $\hat{a}_{\omega}\ket{\alpha}=\alpha\ket{\alpha}$.

For simplicity, let us further assume that the displacement parameter is real. In that case, we find
\begin{equation} \label{Green-function-coherent}
    G_\omega^{\rm (coh)}(t,t')=G_\omega^{\rm (vac)}(t,t')+\frac{\alpha^2}{m_{\rm g}\omega}\cos(\omega t)\cos(\omega t').
\end{equation}
For the noise kernel, one has
\begin{align} \label{N-coh(t,t')}
    N^{\rm (coh)}_{\rm g}(t,t')&=N^{\rm (vac)}_{\rm g}(t,t')+\frac{2\alpha^2}{15\pi}\left[ \int_0^\Lambda\dd\omega\,\omega^5\cos(\omega t)\cos(\omega t')\right. \nonumber \\
    &\hspace{4cm}\left. -2\Phi_{zz}\int_0^\Lambda\dd\omega\,\omega^3\cos(\omega t)\cos(\omega t')\right] \nonumber \\
    &=N^{\rm (vac)}_{\rm g}(t,t')+\frac{\alpha^2\Lambda^4}{15\pi}\left\{ \Lambda^2\qty[F_5\qty(\Lambda(t+t'))+F_5\qty(\Lambda(t-t'))]\right. \nonumber \\
    &\hspace{3.5cm}\left. -2\Phi_{zz}\qty[F_3\qty(\Lambda(t+t'))+F_3\qty(\Lambda(t-t'))]\right\} ,
\end{align}
where the functions $F_n(x)$ are defined in Eq. \eqref{Fn(x)}. By using Eq. \eqref{Limit-Fn(x)-x-to-0}, we also find
\begin{align} \label{N-coh(t)}
    N^{\rm (coh)}_{\rm g}(t)&=\lim_{t'\to t}N^{\rm (coh)}_{\rm g}(t,t')\nonumber \\
    &=N^{\rm (vac)}_{\rm g}(t)+\frac{\alpha^2\Lambda^4}{15\pi}\left\{ \Lambda^2\qty[F_5\qty(2\Lambda t)+\frac{1}{6}]-2\Phi_{zz}\qty[F_3\qty(2\Lambda t)+\frac{1}{4}]\right\} .
\end{align}

\subsection{Squeezed state}

Returning once again to the analogy with quantum optics, a squeezed state is the one for which the electric field has a quantum uncertainty smaller than the minimal uncertainty of a coherent state for some phases. For a weak quantized gravitational field, it has been argued that gravitons created from quantum fluctuations in the course of cosmological evolution should now be in strongly squeezed states~\cite{Grishchuk1990}.

For a quantum harmonic oscillator, the squeezed state is defined by
\begin{equation}
    \ket{\zeta_\omega}=\Hat{\mathscr{S}}(\zeta_\omega)\ket{0},
\end{equation}
where $\Hat{\mathscr{S}}(\zeta_\omega)$ is called the \emph{squeeze operator}, defined as \cite{Loudon2000}
\begin{equation}
    \Hat{\mathscr{S}}(\zeta_\omega)=\exp[\frac{1}{2}\zeta_\omega^*\hat{a}_\omega^2-\frac{1}{2}\zeta_\omega (\hat{a}_\omega^\dagger)^2],
\end{equation}
and $\zeta_\omega$ is the \emph{complex squeeze parameter}
\begin{equation}
    \zeta_\omega=r_\omega e^{i\varphi_\omega}.
\end{equation}

Here we take the squeeze parameter to be independent of $\omega$, $\zeta_\omega=\zeta$. The squeeze operator satisfies a number of properties,
\begin{subequations}
\begin{align}
    \Hat{\mathscr{S}}^\dagger(\zeta)\Hat{\mathscr{S}}(\zeta)&=1, \\
    \Hat{\mathscr{S}}^\dagger(\zeta)\hat{a}_\omega^n\Hat{\mathscr{S}}(\zeta)&=(\hat{a}\cosh r-\hat{a}^\dagger e^{i\varphi}\sinh r)^n, \\
    \Hat{\mathscr{S}}^\dagger(\zeta)(\hat{a}_\omega^\dagger)^n\Hat{\mathscr{S}}(\zeta)&=(\hat{a}^\dagger\cosh r-\hat{a}\,e^{-i\varphi}\sinh r)^n,
\end{align}
\end{subequations}
for $n\in\mathbb{N}$, from which we find
\begin{subequations}
\begin{align}
    \langle \hat{a}_\omega^2\rangle_{\rm g}&=-e^{i\varphi}\sinh r\cosh r, \\
    \langle(\hat{a}_\omega^\dagger)^2\rangle_{\rm g}&=-e^{-i\varphi}\sinh r\cosh r, \\
    \langle \hat{H}_\omega\rangle_{\rm g}&=\frac{\omega}{2}\cosh2r,
\end{align}
\end{subequations}
so that
\begin{equation} \label{Green-function-squeezed}
    G_\omega^{\rm (sq)}(t,t')=(\cosh{2r})G_\omega^{\rm (vac)}(t,t')-\frac{\sinh{2r}}{m_{\rm g}\omega}\cos[\omega(t+t')-\varphi].
\end{equation}

Assuming the squeeze parameter to be real for simplicity from now on ($\varphi=0$), the noise kernel becomes
\begin{align} \label{N-sq(t,t')}
    N^{\rm (sq)}_{\rm g}(t,t')&=(\cosh2r)N^{\rm (vac)}_{\rm g}(t,t') \nonumber \\
    &-\frac{2}{15\pi}\sinh2r\qty[\int_0^\Lambda\omega^5\cos\omega(t+t')-2\Phi_{zz}\int_0^\Lambda\omega^3\cos\omega(t+t')] \nonumber \\
    &=(\cosh2r)N^{\rm (vac)}_{\rm g}(t,t') \nonumber \\
    &-\frac{2\Lambda^4}{15\pi}\sinh2r\qty{\Lambda^2F_5\qty[\Lambda(t+t')]-2\Phi_{zz}F_3\qty[\Lambda(t+t')]}.
\end{align}
Also,
\begin{align} \label{N-sq(t)}
    N^{\rm (sq)}_{\rm g}(t)&=\lim_{t'\to t}N^{\rm (sq)}_{\rm g}(t,t') \nonumber \\
    &=(\cosh2r)N^{\rm (vac)}_{\rm g}(t)-\frac{2\Lambda^4}{15\pi}\sinh2r\qty[\Lambda^2F_5\qty(2\Lambda t)-2\Phi_{zz}F_3\qty(2\Lambda t)].
\end{align}

\section{Internal DoFs noise kernel} \label{SAppen:Internal-DoFs-noise-kernel}

In this section, we obtain an explicit expression for the internal degrees of freedom noise kernel, defined in Eq. \eqref{Internal-dofs-noise-kernel}. Recall that the free Lagrangian for each position operator $\hat{\varrho}_\alpha$ is that of a harmonic oscillator. Furthermore, we take the initial internal state to be a thermal one with temperature $T_{\rm int}=\beta_{\rm int}^{-1}$. In that case, the noise kernel is precisely the one we obtained in Eq. \eqref{Thermal-noise-kernel}, namely
\begin{equation}
    N_{\rm int}(t,t')=\frac{1}{2}\int_0^\infty\dd\varpi\,J(\varpi)\coth\qty(\frac{\varpi\beta_{\rm int}}{2})\cos\varpi(t-t'),
\end{equation}
where
\begin{equation}
    J(\varpi)=\sum_\alpha\frac{\vartheta_\alpha^2}{\mu_\alpha\varpi_\alpha}\delta(\varpi-\varpi_\alpha)
\end{equation}
is the spectral density.

If we consider the simple case of the internal DoFs being described by an Ohmic bath, we have
\begin{equation}
    J(\varpi)=\eta\varpi
\end{equation}
for some effective coupling constant $\eta$. For an Ohmic bath, the noise kernel reads
\begin{equation} \label{Int-noise}
    N_{\rm int}(t,t')=\frac{1}{2}\eta\int_0^\infty \dd\varpi\,\varpi\coth\qty(\frac{\varpi\beta}{2})\cos\varpi(t-t').
\end{equation}
At high temperatures, $T_{\rm int}=\beta_{\rm int}^{-1}\gg\abs{t-t'}$, the integral is dominated by low frequencies (and one does need to introduce a frequency cutoff in such case), and we find
\begin{equation} \label{Int-noise-high-T}
    N_{\rm int}(t,t')=\eta\pi T_{\rm int}\delta(t-t'),
\end{equation}
namely a white noise.

\chapter{The fluctuation theorem}
\label{app:fluctuation-theorem}

The term \emph{fluctuation theorems} stands for relations between probability distributions of fluctuating non-equilibrium quantities and some kind of equilibrium information of starting and ending points of a given thermodynamical process. In this appendix we present the fluctuation theorem following the approach due to Gavin E. Crooks \cite{Crooks1999}, which in turn leads to the Jarzynski equality \cite{Jarzynski1997}. We begin by presenting its classical version before the quantum one. The reader is referred to Ref. \cite{Campisi2011} for a review and a brief historical description of the field.

\section{Classical fluctuation theorem}

The fluctuation theorem follows from two basic assumptions: (i) the initial description of the system by a thermal state and (ii) the time reversal invariance of the equations of microscopic motion. Before presenting the theorem itself, we shall begin by making this last assumption more precise, which we call \emph{microreversibility}. We will assume Hamiltonian dynamics, although the results remain valid for a more general Markovian stochastic evolution \cite{Jarzynski2007,Klages2013}.

\subsection{Microreversibility of driven classical systems}

Consider a classical system described by a Hamiltonian
\begin{equation}
    H(\vb{z},f)=H_0(\vb{z})-f(t)X(\vb{z}),
\end{equation}
where $H_0(\vb{z})$ is referred to as the unperturbed Hamiltonian, $f(t)$ is a time-dependent external force and $X(\vb{z})$ is its conjugate coordinate. Also, we are denoting the points in phase space by $\vb{z}=\qty(\vb{q},\vb{p})$, where $\vb{q}$ represents the generalized coordinates $(q_1,\dots,q_n)$ of a system with $n$ degrees of freedom while $\vb{p}=(p_1,\dots,p_n)$ stands for their conjugated momenta. We assume that the force acts within a temporal interval $\qty[0,\tau]$ and that its instantaneous values $f(t)$ are specified by a given force protocol.

The solution of Hamilton's equations of motion assigns to each initial point $\vb{z}(0)$ in phase space a point $\vb{z}(t)$ at a later time $t\in[0,\tau]$, which is specified by the values of the force protocol in all times prior to $t$, that is
\begin{equation} \label{Forward-flow}
    \vb{z}(t)=\varphi_{t,0}[\vb{z}(0);f],
\end{equation}
where $\varphi$ represents the Hamiltonian flow in phase space \cite{Campisi2011}.

In order to write the microreversibility condition we need to describe the time reversed process, which is a trajectory that runs \emph{forward} in time such that the signs of momenta are reverted. We first assume that the unperturbed Hamiltonian $H_0$ is time reversal invariant. Additionally, we also assume the conjugate coordinate $X$ to transform under time reversal with a definite parity $\epsilon_X=\pm1$. Now, while the flow \eqref{Forward-flow} describes how the system evolves from $\vb{z}_0$ under the forward protocol $f(t)$, we define the reverse flow
\begin{equation}
    \varphi_{\tau-t}[\epsilon\vb{z}(\tau);\epsilon_X\Tilde{f}]
\end{equation}
as the one describing how the system evolves from the time-reversed final state $\epsilon\vb{z}(\tau)$, where $\epsilon\vb{z}=\epsilon\qty(\vb{q},\vb{p})=\qty(\vb{q},-\vb{p})$, under the reverse protocol $\Tilde{f}(t)=f(\tau-t)$.

The principle of microreversibility states that the trajectory followed in the forward process, when time-reversed, must coincide with the trajectory followed in the reverse process. Therefore, applying the time-reversal operation to the forward trajectory leads to \cite{Campisi2011,Stratonovich1994}
\begin{equation} \label{Microreversibility}
    \epsilon\varphi_{\tau-t,0}[\epsilon\vb{z}(\tau);\epsilon_X\Tilde{f}]=\varphi_{t,0}[\vb{z}(0);f].
\end{equation}

\subsection{Crooks fluctuation theorem}

When moving from $\vb{z}(0)$ at $t=0$ to $\vb{z}(\tau)$ at $t=\tau$, the work done on the system is given by\footnote{This definition is referred to as \emph{inclusive work} in the literature, as opposed to the \emph{exclusive work} defined by $W_0=H_0(\vb{z}(\tau))-H_0(\vb{z}(0))$. To see how the two approaches relate to each other the reader is referred to refs. \cite{Horowitz2007,Jarzynski2007}.}
\begin{equation} \label{Exclusive-work-1}
    W[\vb{z}(0);f]=H(\vb{z}(\tau),f(\tau))-H(\vb{z}(0),f(0)).
\end{equation}
Let us note that
\begin{align}
    \dv{t}H(\vb{z},f)&=\sum_i\qty(\pdv{H}{q_i}\dot{q}_i+\pdv{H}{p_i}\dot{p}_i)+\pdv{H}{f}\dot{f}=\dot{f}(t)\pdv{H(\vb{z},f)}{f(t)} \nonumber \\
    &=-\dot{f}(t)X(\vb{z}(t)),
\end{align}
where we used Hamilton's equation of motion \cite{Goldstein2001}
\begin{equation}
    \dot{q}_i=\pdv{H}{p_i},\hspace{1cm}\dot{p}_i=-\pdv{H}{q_i}.
\end{equation}
Eq. \eqref{Exclusive-work-1} can then be written as
\begin{equation} \label{Exclusive-work-2}
    W[\vb{z}(0);f]=\int_0^\tau\dd t\,\dot{f}(t)\pdv{H(\vb{z},f)}{f(t)}=-\int_0^\tau\dd t\,\dot{f}(t)X(\vb{z}(t)).
\end{equation}

At $t=0$ we assume the system to be in thermal equilibrium at inverse temperature $\beta$ described by a canonical distribution of the form
\begin{equation} \label{Classical-thermal-initial-ditribution}
    \rho(\vb{z}(0))=\frac{e^{-\beta H(\vb{z}(0),f(0))}}{Z_0},
\end{equation}
where
\begin{equation}
    Z_0=\int\dd\vb{z}(0)\,e^{-\beta H(\vb{z}(0),f(0))}
\end{equation}
is the classical partition function.

The initial conditions are randomly sampled from the distribution \eqref{Classical-thermal-initial-ditribution}, each of which leads to a trajectory in phase space from $t=0$ to $t=\tau$. Since the definition of work \eqref{Exclusive-work-2} depends on the trajectory, it acquires a probabilistic aspect when considering the ensemble of processes with initial preparation \eqref{Classical-thermal-initial-ditribution}. The complete information concerning the statistics is then contained in the work probability density function
\begin{equation} \label{Work-probability-density-function}
    P[W;f]=\int\dd\vb{z}(0)\rho(\vb{z}(0))\delta\qty[W-H(\vb{z}(\tau),f(\tau))+H(\vb{z}(0),f(0))].
\end{equation}

As usual, we define the characteristic function of work as the Fourier transform of the work probability density function,
\begin{subequations}
\begin{equation}
    \Phi[u;f]=\int\dd W\,e^{iuW}P[W;f],
\end{equation}
with
\begin{equation}
    P[W;f]=\frac{1}{2\pi}\int\dd u\,e^{-iuW}\Phi[u;f].
\end{equation}
\end{subequations}

By using Eqs. \eqref{Work-probability-density-function}, \eqref{Classical-thermal-initial-ditribution} and \eqref{Exclusive-work-1} the characteristic function becomes
\begin{align}
    \Phi[u;f]&=\int\dd\vb{z}(0)\,e^{iu\qty[H(\vb{z}(\tau),f(\tau))-H(\vb{z}(0),f(0))]}\frac{e^{-\beta H(\vb{z}(0),f(0))}}{Z_0} \nonumber \\
    &=\frac{1}{Z_0}\int\dd\vb{z}(0)\,e^{iuW[\vb{z}(0);f]}e^{-\beta H(\vb{z}(\tau),f(\tau))}e^{\beta W[\vb{z}(0);f]} \nonumber \\
    &=\frac{Z_\tau}{Z_0}\int\dd\vb{z}(0)\,e^{iuW\qty[\varphi_{\tau,0}[\vb{z}(0);f]]}e^{\beta W\qty[\varphi_{\tau,0}[\vb{z}(0);f]]}\rho(\vb{z}(\tau)),
\end{align}
where $Z_\tau=\int\dd\vb{z}(\tau)\,e^{-\beta H(\vb{z}(\tau),f(\tau))}$. In the expression above, $W\qty[\varphi_{\tau,0}[\vb{z}(0);f]]$ denotes the work done on the system from $t=0$ to $t=\tau$ in the forward process. By using the microreversibility condition \eqref{Microreversibility} and changing variables $\vb{z}(0)\to\vb{z}(\tau)$ (for which the Jacobian is equal to unity since time evolution in classical mechanics is a canonical transformation), we find
\begin{align}
    Z_0\Phi[u;f]&=Z_\tau\int\dd\vb{z}(\tau)\,e^{iuW\qty[\epsilon\varphi_{\tau-t,0}[\epsilon\vb{z}(\tau);\epsilon_X\Tilde{f}]]}e^{\beta W\qty[\epsilon\varphi_{\tau-t,0}[\epsilon\vb{z}(\tau);\epsilon_X\Tilde{f}]]}\rho(\vb{z}(\tau)) \nonumber \\
    &=Z_\tau\int\dd\vb{z}(\tau)\,e^{i(u-i\beta)\qty[H(\vb{z}(0),f(0))-H(\vb{z}(\tau),f(\tau))]}\rho(\vb{z}(\tau)),
\end{align}
where $W\qty[\epsilon\varphi_{\tau-t,0}[\epsilon\vb{z}(\tau);\epsilon_X\Tilde{f}]]=H(\vb{z}(0),f(0))-H(\vb{z}(\tau),f(\tau))$ is the work done on the reverse process starting at $t=\tau$ with initial thermal distribution $\rho(\vb{z}(\tau))$. By introducing a delta function to recover the characteristic function of work, we find
\begin{equation} \label{Identity-for-characteristic-function-of-work}
    Z_0\Phi[u;f]=Z_\tau\Phi[-u+i\beta;\epsilon_X\Tilde{f}].
\end{equation}
Applying the inverse Fourier transform on both sides yields
\begin{equation}
    Z_0P[W;f]=Z_\tau e^{\beta W}P[-W;\epsilon_X\Tilde{f}],
\end{equation}
or
\begin{equation}
    \frac{P[W;f]}{P[-W;\epsilon_X\Tilde{f}]}=\frac{Z_\tau}{Z_0}e^{\beta W}.
\end{equation}

We can rewrite this by recalling that the Helmholtz free energy is obtained from the partition function as $F(t)=-\frac{1}{\beta}\ln Z_t$ \cite{Pathria2021} and thus
\begin{equation}
    \frac{Z_\tau}{Z_0}=\frac{e^{-\beta F(\tau)}}{e^{-\beta F(0)}}=e^{-\beta\Delta F},
\end{equation}
where $\Delta F=F(\tau)-F(0)$ is the free energy difference between initial and final points. Finally we may write
\begin{equation} \label{Crooks-classical-fluctuation-theorem}
    \frac{P[W;f]}{P[-W;\epsilon_X\Tilde{f}]}=e^{\beta(W-\Delta F)}.
\end{equation}
This is the Crooks fluctuation theorem, which establishes that the probability of observing the forward process is exponentially larger than the probability of observing the reverse one. The exponent $W_{\rm diss}=W-\Delta F$, called \emph{dissipated work}, quantifies such statistical irreversibility.

\subsection{Jarzynski equality}

We can write Crooks fluctuation theorem \eqref{Crooks-classical-fluctuation-theorem} as
\begin{equation}
    P[W;f]e^{-\beta W}=P[-W;\epsilon_X\Tilde{f}]e^{-\beta\Delta F}
\end{equation}
and integrate both sides over $W$. This leads to the Jarzynski equality
\begin{equation} \label{Classical-Jarzynski-equality}
    \expval{e^{-\beta W}}=e^{-\beta\Delta F},
\end{equation}
where the brackets denote average with respect to the work probability density function.

The Jarzynski equality \eqref{Classical-Jarzynski-equality} states that the equilibrium free energy difference may be inferred by measurements of non-equilibrium work in many realizations of the system trajectory according to the force protocol $f(t)$. This result has many practical applications in the fields of physics, biology, chemistry and so on \cite{Jarzynski2008,Campisi2011}. It also allows us to describe the statistical nature of the second law of thermodynamics by using Jensen's inequality\footnote{Let $X$ be a real integrable random variable and let $g:\mathbb{R}\to\mathbb{R}$ be a convex function such that $Y=g(X)$ is also integrable. Then Jensen's inequality states that $E\qty[g(X)]\geq g\qty(E[X])$, where $E[X]$ denotes the expectation value of $X$.},
\begin{equation}
    e^{-\beta\Delta F}=\expval{e^{-\beta W}}\geq e^{-\beta\expval{W}}\implies e^{-\beta(\expval{W}-\Delta F)}\leq1.
\end{equation}
This implies that, on average, the entropy production is always positive
\begin{equation}
    \expval{\Sigma}=\beta(\expval{W}-\Delta F)\geq0.
\end{equation}
Also, as we can see from Eq. \eqref{Crooks-classical-fluctuation-theorem}, $\expval{\Sigma}=0$ holds only for reversible processes.

\section{Quantum fluctuation theorem}

On establishing the quantum version of the fluctuation theorem, as expressed by Crooks relation and Jarzynski equality, one is faced with an immediate question: how to define work done on quantum systems? Work describes a specific process instead of characterizing the state of the system. Thus it cannot be represented by a Hermitian operator whose eigenvalues can be determined in a single projective measurement \cite{Talkner2007,Campisi2011}. For open quantum systems the proper definition is still under debate, but consensus seems to have been reached on how to define work done on a closed quantum system according to the two-time measurements scheme. In this approach, we measure the energy of the system twice, first at $t=0$ and then at $t=\tau$, the end point of the force protocol. The difference of the outcomes of these two measurements is the work performed on the system in a particular realization. This inevitably yields a second probabilistic aspect to the work distribution, which comes from quantum uncertainty.

Before we make these ideas more precise, we follow the approach of the previous section and turn now to the question of microreversibility of driven \emph{quantum} systems.

\subsection{Quantum microreversibility}

Let $\hat{U}_f(t,0)$ be the unitary time evolution operator of the quantum system driven by the force protocol $f(t)$ from $t=0$ to $t=\tau$. In order to write the quantum microreversibility condition, we split the time interval in discrete steps and express the time evolution operator as a time ordered product,
\begin{equation}
    \hat{U}_{\Tilde{f}}(\tau-t,0)=\lim_{N\to\infty}e^{-i\hat{H}(\Tilde{f}(\tau-N\varepsilon))\varepsilon}\dots e^{-i\hat{H}(\Tilde{f}(\varepsilon))\varepsilon}e^{-i\hat{H}(\Tilde{f}(0))\varepsilon},
\end{equation}
where $\varepsilon=t/N$ denotes the time step and $\hat{H}(f(t))$ is the Hamiltonian operator. By recalling that $\Tilde{f}(t)=f(\tau-t)$ we may write
\begin{equation}
    \hat{U}_{\Tilde{f}}(\tau-t,0)=\lim_{N\to\infty}e^{-i\hat{H}(f(N\varepsilon))\varepsilon}\dots e^{-i\hat{H}(f(\tau-\varepsilon))\varepsilon}e^{-i\hat{H}(f(\tau))\varepsilon}.
\end{equation}

Next let us introduce the time reversal operator $\hat{\Theta}$ and assume that, at any time $t$, the Hamiltonian is invariant under time reversal,
\begin{equation}
    \comm{\hat{H}(f(t))}{\hat{\Theta}}=0.
\end{equation}
Then,
\begin{equation}
    \hat{\Theta}^\dagger \hat{U}_{\Tilde{f}}(\tau-t,0)\hat{\Theta}=\lim_{N\to\infty}\hat{\Theta}^\dagger e^{-i\hat{H}(f(N\varepsilon))\varepsilon}\hat{\Theta}\hat{\Theta}^\dagger\dots\hat{\Theta}^\dagger e^{-i\hat{H}(f(\tau))\varepsilon}\hat{\Theta},
\end{equation}
where we made $N-1$ insertions of the identity $1=\hat{\Theta}\hat{\Theta}^\dagger$. Since $\hat{\Theta}^\dagger e^{-i\hat{H}(f(t))\varepsilon}\hat{\Theta}=e^{i\hat{H}(f(t))\varepsilon}$ for real $\varepsilon$ \cite{Sakurai2020}, we have
\begin{align}
    \hat{\Theta}^\dagger \hat{U}_{\Tilde{f}}(\tau-t,0)\hat{\Theta}&=\lim_{N\to\infty}e^{i\hat{H}(f(N\varepsilon))\varepsilon}\dots e^{i\hat{H}(f(\tau))\varepsilon} \nonumber \\
    &=\lim_{N\to\infty}\qty[e^{-i\hat{H}(f(\tau))\varepsilon}\dots e^{-i\hat{H}(f(N\varepsilon))\varepsilon}]^\dagger=\hat{U}_f^\dagger(\tau,t)
\end{align}
and the microreversibility condition reads
\begin{equation} \label{Quantum-microreversibility}
    \hat{U}_f(t,\tau)=\hat{\Theta}^\dagger \hat{U}_{\Tilde{f}}(\tau-t,0)\hat{\Theta}.
\end{equation}

Now, note that Eq. \eqref{Quantum-microreversibility} can be rewritten as
\begin{equation}
    \hat{U}_f(t,0)=\hat{\Theta}^\dagger \hat{U}_{\Tilde{f}}(\tau-t,0)\hat{\Theta} \hat{U}_f(\tau,0),
\end{equation}
where we used $\hat{U}_f(t,\tau)=\hat{U}_f(t,0)\hat{U}_f(0,\tau)=\hat{U}_f(t,0)\hat{U}_f^\dagger(\tau,0)$. Next, let us apply this operator to the pure initial state $\ket{\psi_0}$,
\begin{equation}
    \ket{\psi_t}=\hat{U}_f(t,0)\ket{\psi_0}=\hat{\Theta}^\dagger \hat{U}_{\Tilde{f}}(\tau-t,0)\hat{\Theta} \hat{U}_f(\tau,0)\ket{\psi_0}=\hat{\Theta}^\dagger \hat{U}_{\Tilde{f}}(\tau-t,0)\hat{\Theta}\ket{\psi_\tau},
\end{equation}
where $\ket{\psi_\tau}=\hat{U}_f(\tau,0)\ket{\psi_0}$ is the final state. Multiplying by $\hat{\Theta}$ from the left leads to
\begin{equation}
    \hat{\Theta}\ket{\psi_t}=\hat{U}_{\Tilde{f}}(\tau-t,0)\hat{\Theta}\ket{\psi_\tau}.
\end{equation}
This allows us to interpret the quantum microreversibility in terms of states: under the evolution generated by the reverse protocol $\Tilde{f}$, the time reversed final state $\hat{\Theta}\ket{\psi_\tau}$ evolves from $0$ to $\tau-t$ to the state $\hat{\Theta}\ket{\psi_t}$ \cite{Campisi2011}.

\subsection{Work probability density function}

Consider a system whose time evolution is generated by the Hamiltonian operator $\hat{H}(f(t))$ initially described by the thermal density matrix
\begin{equation} \label{Initial-thermal-density-matrix}
    \hat{\rho}(0)=\frac{e^{-\beta \hat{H}(f(0))}}{Z_0},
\end{equation}
where $Z_0=\textrm{Tr}\,e^{-\beta \hat{H}(f(0))}$ is the partition function and Tr denotes trace over the Hilbert space of the system. The Hamiltonian eigenvalue equation can be written as
\begin{equation}
    \hat{H}(f(t))\ket{\psi_{n,\alpha}(t)}=E_n(t)\ket{\psi_{n,\alpha}(t)},
\end{equation}
where $n$ labels the quantum numbers that specify the energy eigenvalues while $\alpha$ labels all further quantum numbers.

Let us now describe the two-time measurement scheme from which we will define the work done on the system. At $t=0$ we measure the energy of the system, which yields the eigenvalue $E_n(0)$ with probability
\begin{equation}
    p_n(0)=\textrm{Tr}\qty[\hat{\Pi}_n^0\hat{\rho}(0)].
\end{equation}
Here, $\hat{\Pi}_n^0=\sum_\alpha\ket{\psi_{n,\alpha}(0)}\bra{{\psi_{n,\alpha}(0)}}$ is the projector onto the eigenspace spanned by the eigenvectors associated with the eigenvalue $E_n(0)$. Once the energy measurement is performed, with result $E_n(0)$, the system is described by the state
\begin{equation}
    \hat{\rho}_n=\frac{\hat{\Pi}_n^0\hat{\rho}(0)\hat{\Pi}_n^0}{p_n(0)}.
\end{equation}

Since we are considering the system to be closed, the time evolution is unitary,
\begin{equation}
    \hat{\rho}_n(t;f)=\hat{U}_f(t,0)\hat{\rho}_n\hat{U}_f^\dagger(t,0).
\end{equation}
At time $t=\tau$ a second energy measurement yields the eigenvalue $E_m(\tau)$ with probability
\begin{equation}
    p_{m|n}^\tau[f]=\textrm{Tr}\qty[\hat{\Pi}_m^\tau\hat{\rho}_n(\tau;f)].
\end{equation}

The work done on the system is then simply the difference between the energy eigenvalues, $w=E_m(\tau)-E_n(0)$. The probability distribution function of work reads
\begin{equation} \label{Quantum-Work-probability-density-function}
    P[w;f]=\sum_{m,n}\delta\qty[w-E_m(\tau)+E_n(0)]p_{m|n}^\tau[f]p_n^0.
\end{equation}

\subsection{Quantum versions of the Crooks fluctuation relation and Jarzynski equality}

From the probability distribution function of work \eqref{Quantum-Work-probability-density-function} we write the characteristic function as~\cite{Talkner2008}
\begin{align}
    \Phi[u;f]&=\int \dd w\,e^{iuw}P[w;f] \nonumber \\
    &=\sum_{m,n}e^{iuE_m(\tau)}e^{-iuE_n(0)}\textrm{Tr}\qty{\hat{\Pi}_m^\tau \hat{U}_f(\tau,0)\hat{\rho}_n\hat{U}_f^\dagger(\tau,0)}p_n^0 \nonumber \\
    &=\textrm{Tr}\qty{e^{iu\hat{H}(f(\tau))}\sum_m\hat{\Pi}_m^\tau \hat{U}_f(\tau,0)e^{-iu\hat{H}(f(0))}\sum_n\hat{\rho}_n p_n^0\hat{U}_f^\dagger(\tau,0)} \nonumber \\
    &=\textrm{Tr}\qty{\hat{U}_f^\dagger(\tau,0)e^{iu\hat{H}(f(\tau))}\hat{U}_f(\tau,0)e^{-iu\hat{H}(f(0))}\hat{\rho}(0)},
\end{align}
where we used the fact that $\sum_m\hat{\Pi}_m^\tau=1$ and $\sum_n\hat{\rho}_np_n^0=\hat{\rho}(0)$, which holds since the initial density matrix commutes with the Hamiltonian at $t=0$.

By using the explicit form of the initial density matrix, Eq. \eqref{Initial-thermal-density-matrix}, we can write
\begin{equation} \label{Z0Phi-quantum}
    Z_0\Phi[u;f]=\textrm{Tr}\qty{\hat{U}_f^\dagger(\tau,0)e^{iu\hat{H}(f(\tau))}\hat{U}_f(\tau,0)e^{-iu\hat{H}(f(0))}e^{-\beta \hat{H}(f(0))}}.
\end{equation}
Setting $t=0$ in the quantum microreversibility condition \eqref{Quantum-microreversibility} leads to
\begin{equation*}
    \hat{U}_f(0,\tau)=\hat{\Theta}^\dagger \hat{U}_{\Tilde{f}}(\tau,0)\hat{\Theta}\implies \hat{U}_f^\dagger(\tau,0)=\hat{\Theta}^\dagger \hat{U}_{\Tilde{f}}(\tau,0)\hat{\Theta},
\end{equation*}
\begin{equation}
    \implies \hat{U}_f(\tau,0)=\hat{\Theta}^\dagger \hat{U}_{\Tilde{f}}^\dagger(\tau,0)\hat{\Theta}.
\end{equation}
Using this expression back in Eq. \eqref{Z0Phi-quantum} results in
\begin{align}
    Z_0\Phi[u;f]&=\textrm{Tr}\qty{\hat{\Theta}^\dagger \hat{U}_{\Tilde{f}}(\tau,0)\hat{\Theta} e^{iu\hat{H}(f(\tau))}\hat{\Theta}^\dagger \hat{U}_{\Tilde{f}}^\dagger(\tau,0)\hat{\Theta} e^{-iu\hat{H}(f(0))}e^{-\beta \hat{H}(f(0))}} \nonumber \\
    &=\textrm{Tr}\qty{\hat{\Theta}^\dagger \hat{U}_{\Tilde{f}}(\tau,0)e^{-iu^*\hat{H}(f(\tau))}\hat{U}_{\Tilde{f}}^\dagger(\tau,0)e^{iu^*\hat{H}(f(0))}e^{-\beta \hat{H}(f(0))}\hat{\Theta}} \nonumber \\
    &=\textrm{Tr}\qty{e^{-\beta \hat{H}(f(0))}e^{-iu\hat{H}(f(0))}\hat{U}_{\Tilde{f}}(\tau,0)e^{iu\hat{H}(f(\tau))}\hat{U}_{\Tilde{f}}^\dagger(\tau,0)} \nonumber \\
    &=\textrm{Tr}\qty{\hat{U}_{\Tilde{f}}^\dagger(\tau,0)e^{i(-u+i\beta)\hat{H}(f(0))}\hat{U}_{\Tilde{f}}(\tau,0)e^{-i(-u+i\beta)\hat{H}(f(\tau))}e^{-\beta \hat{H}(f(\tau))}} \nonumber \\
    &=Z_\tau\Phi[-u+i\beta;\Tilde{f}],
\end{align}
where we used the anti-linearity\footnote{For any trace class operator $\hat{A}$ the anti-linearity of $\hat{\Theta}$ implies $\textrm{Tr}\qty(\hat{\Theta}^\dagger \hat{A}\hat{\Theta})=\textrm{Tr}\hat{A}^\dagger$.} of $\Theta$. This result is precisely the same identity we obtained for the classical case, Eq. \eqref{Identity-for-characteristic-function-of-work}. If we follow the same steps as before we are led to the quantum version of the Crooks fluctuation theorem,
\begin{equation}
    \frac{P[w;f]}{P[-w;\Tilde{f}]}=e^{\beta(w-\Delta F)},
\end{equation}
from which we also obtain the quantum Jarzynski equality
\begin{equation}
    \expval{e^{-\beta w}}=e^{-\beta\Delta F}.
\end{equation}

\medskip
\printbibliography[heading=bibintoc]

@Article{Kanno2021,
  author    = {Sugumi Kanno and Jiro Soda and Junsei Tokuda},
  title     = {Noise and decoherence induced by gravitons},
  journal   = {Physical Review D},
  year      = {2021},
  volume    = {103},
  number    = {4},
  doi       = {10.1103/physrevd.103.044017},
  publisher = {American Physical Society ({APS})},
}

@Article{Pikovski2017,
  author    = {Igor Pikovski and Magdalena Zych and Fabio Costa and {\v{C}}aslav Brukner},
  title     = {Time dilation in quantum systems and decoherence},
  journal   = {New Journal of Physics},
  year      = {2017},
  volume    = {19},
  number    = {2},
  pages     = {025011},
  doi       = {10.1088/1367-2630/aa5d92},
  publisher = {{IOP} Publishing},
}

@Article{Parikh2021,
  author    = {Maulik Parikh and Frank Wilczek and George Zahariade},
  title     = {Signatures of the quantization of gravity at gravitational wave detectors},
  journal   = {Physical Review D},
  year      = {2021},
  volume    = {104},
  number    = {4},
  doi       = {10.1103/physrevd.104.046021},
  publisher = {American Physical Society ({APS})},
}

@Article{Pikovski2015,
  author    = {Igor Pikovski and Magdalena Zych and Fabio Costa and {\v{C}}aslav Brukner},
  title     = {Universal decoherence due to gravitational time~dilation},
  journal   = {Nature Physics},
  year      = {2015},
  volume    = {11},
  number    = {8},
  pages     = {668--672},
  doi       = {10.1038/nphys3366},
  publisher = {Springer Science and Business Media {LLC}},
}

@Book{Feynman2010,
  author    = {R. P. Feynman and A. R. Hibbs},
  publisher = {Dover Publications},
  title     = {Quantum mechanics and path integrals},
  year      = {2010},
  isbn      = {9780486477220},
}

@Book{Calzetta2008,
  title     = {Nonequilibrium Quantum Field Theory},
  publisher = {Cambridge University Press},
  year      = {2008},
  author    = {Calzetta, Esteban A. and Hu, Bei-Lok B.},
  isbn      = {9780521641685},
}

@Article{Feynman1963,
  author    = {R.P Feynman and F.L Vernon},
  title     = {The theory of a general quantum system interacting with a linear dissipative system},
  journal   = {Annals of Physics},
  year      = {1963},
  volume    = {24},
  pages     = {118--173},
  doi       = {10.1016/0003-4916(63)90068-x},
  publisher = {Elsevier {BV}},
}

@Article{Manasse1963,
  author    = {F. K. Manasse and C. W. Misner},
  title     = {Fermi Normal Coordinates and Some Basic Concepts in Differential Geometry},
  journal   = {Journal of Mathematical Physics},
  year      = {1963},
  volume    = {4},
  number    = {6},
  pages     = {735--745},
  doi       = {10.1063/1.1724316},
  publisher = {{AIP} Publishing},
}

@Article{Cho2022,
  author    = {Hing-Tong Cho and Bei-Lok Hu},
  title     = {Quantum noise of gravitons and stochastic force on geodesic separation},
  journal   = {Physical Review D},
  year      = {2022},
  volume    = {105},
  number    = {8},
  pages     = {086004},
  doi       = {10.1103/physrevd.105.086004},
  publisher = {American Physical Society ({APS})},
}

@Book{Misner,
  author    = {Misner, Charles W. and Thorne, Kip S. and Wheeler, John Archibald},
  publisher = {Princeton University Press},
  title     = {Gravitation},
  year      = {2017},
  isbn      = {9780691177793},
}

@Book{Carroll,
  title     = {Spacetime and Geometry: An Introduction to General Relativity},
  publisher = {Cambridge University Press},
  year      = {2019},
  author    = {Carroll, Sean M.},
  isbn      = {9781108488396},
}

@Misc{GutierrezPineres2022,
  author    = {Gutiérrez-Piñeres, Antonio C.},
  title     = {Fermi-normal coordinates for the Newtonian approximation of gravity},
  year      = {2022},
  copyright = {Creative Commons Attribution 4.0 International},
  doi       = {10.48550/ARXIV.2211.12780},
  publisher = {arXiv},
}

@Book{Zee2013,
  title     = {Einstein Gravity in a Nutshell},
  publisher = {Princeton University Press},
  year      = {2013},
  author    = {Zee, A.},
  isbn      = {9780691145587},
}

@Article{Campisi2011,
  author    = {Michele Campisi and Peter Hänggi and Peter Talkner},
  title     = {Quantum fluctuation relations: Foundations and applications},
  journal   = {Reviews of Modern Physics},
  year      = {2011},
  volume    = {83},
  number    = {3},
  pages     = {771--791},
  doi       = {10.1103/revmodphys.83.771},
  publisher = {American Physical Society ({APS})},
}

@Article{Christodoulou2023,
  author    = {Marios Christodoulou and Andrea Di Biagio and Richard Howl and Carlo Rovelli},
  title     = {Gravity entanglement, quantum reference systems, degrees of freedom},
  journal   = {Classical and Quantum Gravity},
  year      = {2023},
  volume    = {40},
  number    = {4},
  pages     = {047001},
  doi       = {10.1088/1361-6382/acb0aa},
  publisher = {{IOP} Publishing},
}

@Article{Zych2019,
  author    = {Magdalena Zych and {\L}ukasz Rudnicki and Igor Pikovski},
  title     = {Gravitational mass of composite systems},
  journal   = {Physical Review D},
  year      = {2019},
  volume    = {99},
  number    = {10},
  pages     = {104029},
  doi       = {10.1103/physrevd.99.104029},
  publisher = {American Physical Society ({APS})},
}

@Book{Loudon2000,
  title     = {The quantum theory of light},
  publisher = {Oxford University Press},
  year      = {2000},
  author    = {Loudon, Rodney},
  isbn      = {0198501773},
}

@Article{Breuer2001,
  author    = {Heinz-Peter Breuer and Francesco Petruccione},
  title     = {Destruction of quantum coherence through emission of bremsstrahlung},
  journal   = {Physical Review A},
  year      = {2001},
  volume    = {63},
  number    = {3},
  pages     = {032102},
  doi       = {10.1103/physreva.63.032102},
  publisher = {American Physical Society ({APS})},
}

@Article{Schwinger_1961,
  author    = {Julian Schwinger},
  journal   = {Journal of Mathematical Physics},
  title     = {Brownian Motion of a Quantum Oscillator},
  year      = {1961},
  number    = {3},
  pages     = {407--432},
  volume    = {2},
  doi       = {10.1063/1.1703727},
  publisher = {{AIP} Publishing},
}

@Article{Hu_1992,
  author    = {B. L. Hu and Juan Pablo Paz and Yuhong Zhang},
  title     = {Quantum Brownian motion in a general environment: Exact master equation with nonlocal dissipation and colored noise},
  journal   = {Physical Review D},
  year      = {1992},
  volume    = {45},
  number    = {8},
  pages     = {2843--2861},
  doi       = {10.1103/physrevd.45.2843},
  publisher = {American Physical Society ({APS})},
}

@Article{Anastopoulos_2013,
  author    = {C Anastopoulos and B L Hu},
  title     = {A master equation for gravitational decoherence: probing the textures of spacetime},
  journal   = {Classical and Quantum Gravity},
  year      = {2013},
  volume    = {30},
  number    = {16},
  pages     = {165007},
  doi       = {10.1088/0264-9381/30/16/165007},
  publisher = {{IOP} Publishing},
}

@Article{Blencowe_2013,
  author    = {M. P. Blencowe},
  title     = {Effective Field Theory Approach to Gravitationally Induced Decoherence},
  journal   = {Physical Review Letters},
  year      = {2013},
  volume    = {111},
  number    = {2},
  doi       = {10.1103/physrevlett.111.021302},
  publisher = {American Physical Society ({APS})},
}

@Article{Bassi_2017,
  author    = {Angelo Bassi and Andr{\'{e}} Gro{\ss}ardt and Hendrik Ulbricht},
  title     = {Gravitational decoherence},
  journal   = {Classical and Quantum Gravity},
  year      = {2017},
  volume    = {34},
  number    = {19},
  pages     = {193002},
  doi       = {10.1088/1361-6382/aa864f},
  publisher = {{IOP} Publishing},
}

@Article{Parikh_2021,
  author    = {Maulik Parikh and Frank Wilczek and George Zahariade},
  title     = {Quantum Mechanics of Gravitational Waves},
  journal   = {Physical Review Letters},
  year      = {2021},
  volume    = {127},
  number    = {8},
  doi       = {10.1103/physrevlett.127.081602},
  publisher = {American Physical Society ({APS})},
}

@Article{Dyson_2013,
  author    = {Freeman Dyson},
  title     = {Is a graviton detectable?},
  journal   = {International Journal of Modern Physics A},
  year      = {2013},
  volume    = {28},
  number    = {25},
  pages     = {1330041},
  doi       = {10.1142/s0217751x1330041x},
  publisher = {World Scientific Pub Co Pte Lt},
}

@Article{Danielson_2022,
  author    = {Daine L. Danielson and Gautam Satishchandran and Robert M. Wald},
  title     = {Gravitationally mediated entanglement: Newtonian field versus gravitons},
  journal   = {Physical Review D},
  year      = {2022},
  volume    = {105},
  number    = {8},
  doi       = {10.1103/physrevd.105.086001},
  publisher = {American Physical Society ({APS})},
}

@Book{Kiefer2012,
  title     = {Quantum Gravity (International Series of Monographs on Physics)},
  publisher = {Oxford Science Publications},
  year      = {2012},
  author    = {Claus Kiefer},
}

@Article{Moreira_2023,
  author    = {Moreira, Thiago H and Céleri, Lucas C},
  title     = {Decoherence of a composite particle induced by a weak quantized gravitational field},
  journal   = {Classical and Quantum Gravity},
  year      = {2023},
  volume    = {41},
  number    = {1},
  pages     = {015006},
  month     = nov,
  issn      = {1361-6382},
  doi       = {10.1088/1361-6382/ad0a53},
  publisher = {IOP Publishing},
}

@Article{Hu2012,
  author    = {Subaşı, Y. and Hu, B. L.},
  title     = {Quantum and classical fluctuation theorems from a decoherent histories, open-system analysis},
  journal   = {Physical Review E},
  year      = {2012},
  volume    = {85},
  number    = {1},
  month     = jan,
  issn      = {1550-2376},
  doi       = {10.1103/physreve.85.011112},
  publisher = {American Physical Society (APS)},
}

@Book{Nakahara2003,
  title     = {Geometry, Topology and Physics},
  publisher = {Taylor \& Francis Group},
  year      = {2003},
  author    = {M. Nakahara},
  edition   = {Second edition},
}

@Book{Pires2015,
  title     = {Geometria diferencial para físicos},
  publisher = {Editora Livraria da Física},
  year      = {2015},
  author    = {Antônio S. T. Pires},
}

@Book{Stewart1993,
  title     = {Advanced General Relativity},
  publisher = {Cambridge University Press},
  year      = {1993},
  author    = {John Stewart},
}

@Book{Petersen2006,
  title     = {Riemannian Geometry},
  publisher = {Springer},
  year      = {2006},
  author    = {Peter Petersen},
}

@Book{Weinberg2013,
  title     = {Gravitation and Cosmology: Principles and Applications of the General Theory of Relativity},
  publisher = {Wiley},
  year      = {2013},
  author    = {Steven L. Weinberg},
}

@Book{Wald1984,
  title     = {General Relativity},
  publisher = {University of Chicago Press},
  year      = {1984},
  author    = {Robert M. Wald},
}

@Book{Bishop2001,
  title     = {Geometry of Manifolds},
  publisher = {American Mathematical Society},
  year      = {2001},
  author    = {Richard J. Crittenden and Richard L. Bishop},
}

@Book{Hartle2003,
  title     = {Gravity: An Introduction to {E}instein's General Relativity},
  publisher = {Addison-Wesley Professional},
  year      = {2003},
  author    = {James B. Hartle},
}

@Article{Isi_2023,
  author    = {Isi, Maximiliano},
  title     = {Parametrizing gravitational-wave polarizations},
  journal   = {Classical and Quantum Gravity},
  year      = {2023},
  volume    = {40},
  number    = {20},
  pages     = {203001},
  month     = sep,
  issn      = {1361-6382},
  doi       = {10.1088/1361-6382/acf28c},
  publisher = {IOP Publishing},
}

@Book{Jackson1998,
  title     = {Classical Electrodynamics},
  publisher = {Wiley},
  year      = {1998},
  author    = {John David Jackson},
  edition   = {3rd edition},
}

@Article{Cahillane2022,
  author    = {Cahillane, Craig and Mansell, Georgia},
  title     = {Review of the Advanced {L}{I}{G}{O} Gravitational Wave Observatories Leading to Observing Run Four},
  journal   = {Galaxies},
  year      = {2022},
  volume    = {10},
  number    = {1},
  pages     = {36},
  month     = feb,
  issn      = {2075-4434},
  doi       = {10.3390/galaxies10010036},
  publisher = {MDPI AG},
}

@Article{Bailes2021,
  author    = {Bailes, M. and others},
  title     = {Gravitational-wave physics and astronomy in the 2020s and 2030s},
  journal   = {Nature Reviews Physics},
  year      = {2021},
  volume    = {3},
  number    = {5},
  pages     = {344--366},
  month     = apr,
  issn      = {2522-5820},
  doi       = {10.1038/s42254-021-00303-8},
  publisher = {Springer Science and Business Media LLC},
}

@Article{Abbott_2016,
  author    = {B. P. Abbott and others},
  journal   = {Physical Review Letters},
  title     = {Observation of Gravitational Waves from a Binary Black Hole Merger},
  year      = {2016},
  issn      = {1079-7114},
  month     = feb,
  number    = {6},
  volume    = {116},
  doi       = {10.1103/physrevlett.116.061102},
  publisher = {American Physical Society (APS)},
}

@Book{Srednicki2007,
  title     = {Quantum Field Theory},
  publisher = {Cambridge University Press},
  year      = {2007},
  author    = {Mark Srednicki},
}

@Book{Schwartz2013,
  title     = {Quantum Field Theory and the Standard Model},
  publisher = {Cambridge University Press},
  year      = {2013},
  author    = {Matthew D. Schwartz},
}

@Book{Peskin2007,
  title     = {An Introduction to Quantum Field Theory},
  publisher = {Westview Press},
  year      = {2007},
  author    = {Michael E. Peskin and Daniel V. Schroeder},
}

@Book{Weinberg12005,
  title     = {The Quantum Theory of Fields: Foundations},
  publisher = {Cambridge University Press},
  year      = {2005},
  author    = {Steven Weinberg},
  volume    = {01},
}

@Book{Jeevanjee2015,
  title     = {An Introduction to Tensors and Group Theory for Physicists},
  publisher = {Birkhauser},
  year      = {2015},
  author    = {Nadir Jeevanjee},
}

@Book{Costa2012,
  title     = {Symmetries and Group Theory in Particle Physics: An Introduction to Space-Time and Internal Symmetries},
  publisher = {Springer},
  year      = {2012},
  author    = {Giovanni Costa and Gianluigi Fogli},
}

@Article{Fierz1939,
  author    = {Markus Fierz and Wolfgang Ernst Pauli},
  title     = {On relativistic wave equations for particles of arbitrary spin in an electromagnetic field},
  journal   = {Proceedings of the Royal Society of London. Series A. Mathematical and Physical Sciences},
  year      = {1939},
  volume    = {173},
  number    = {953},
  pages     = {211--232},
  month     = nov,
  issn      = {2053-9169},
  doi       = {10.1098/rspa.1939.0140},
  publisher = {The Royal Society},
}

@Article{Wigner1939,
  author    = {Eugene Wigner},
  title     = {On Unitary Representations of the Inhomogeneous {L}orentz Group},
  journal   = {The Annals of Mathematics},
  year      = {1939},
  volume    = {40},
  number    = {1},
  pages     = {149},
  month     = jan,
  issn      = {0003-486X},
  doi       = {10.2307/1968551},
  publisher = {JSTOR},
}

@Article{Carney2024,
  author    = {Carney, Daniel and Domcke, Valerie and Rodd, Nicholas L.},
  title     = {Graviton detection and the quantization of gravity},
  journal   = {Physical Review D},
  year      = {2024},
  volume    = {109},
  number    = {4},
  month     = feb,
  issn      = {2470-0029},
  doi       = {10.1103/physrevd.109.044009},
  publisher = {American Physical Society (APS)},
}

@Book{Cohen1,
  title     = {Quantum Mechanics, Volume 1: Basic Concepts, Tools, and Applications},
  publisher = {Wiley-VCH},
  year      = {2019},
  author    = {Claude Cohen-Tannoudji and Bernard Diu and Franck Laloë},
}

@Book{Breuer2002,
  title     = {The Theory of Open Quantum Systems},
  publisher = {Oxford University Press},
  year      = {2002},
  author    = {Heinz-Peter Breuer and Francesco Petruccione},
}

@Article{Grishchuk1990,
  author    = {Grishchuk, L. P. and Sidorov, Y. V.},
  title     = {Squeezed quantum states of relic gravitons and primordial density fluctuations},
  journal   = {Physical Review D},
  year      = {1990},
  volume    = {42},
  number    = {10},
  pages     = {3413--3421},
  issn      = {0556-2821},
  doi       = {10.1103/physrevd.42.3413},
  publisher = {American Physical Society (APS)},
}

@Book{Schlosshauer2008,
  title     = {Decoherence and the Quantum-To-Classical Transition},
  publisher = {Springer},
  year      = {2008},
  author    = {Maximilian A. Schlosshauer},
}

@Article{Bose2017,
  author    = {S. Bose and others},
  journal   = {Physical Review Letters},
  title     = {Spin Entanglement Witness for Quantum Gravity},
  year      = {2017},
  issn      = {1079-7114},
  number    = {24},
  volume    = {119},
  doi       = {10.1103/physrevlett.119.240401},
  publisher = {American Physical Society (APS)},
}

@Article{Marletto2017,
  author    = {Marletto, C. and Vedral, V.},
  title     = {Gravitationally Induced Entanglement between Two Massive Particles is Sufficient Evidence of Quantum Effects in Gravity},
  journal   = {Physical Review Letters},
  year      = {2017},
  volume    = {119},
  number    = {24},
  issn      = {1079-7114},
  doi       = {10.1103/physrevlett.119.240402},
  publisher = {American Physical Society (APS)},
}

@Article{Christodoulou2023b,
  author    = {Christodoulou, Marios and Di Biagio, Andrea and Aspelmeyer, Markus and Brukner, Časlav and Rovelli, Carlo and Howl, Richard},
  title     = {Locally Mediated Entanglement in Linearized Quantum Gravity},
  journal   = {Physical Review Letters},
  year      = {2023},
  volume    = {130},
  number    = {10},
  issn      = {1079-7114},
  doi       = {10.1103/physrevlett.130.100202},
  publisher = {American Physical Society (APS)},
}

@Article{Jarzynski1997,
  author    = {Jarzynski, C.},
  title     = {Nonequilibrium Equality for Free Energy Differences},
  journal   = {Physical Review Letters},
  year      = {1997},
  volume    = {78},
  number    = {14},
  pages     = {2690--2693},
  month     = apr,
  issn      = {1079-7114},
  doi       = {10.1103/physrevlett.78.2690},
  publisher = {American Physical Society (APS)},
}

@Article{Crooks1999,
  author    = {Crooks, Gavin E.},
  title     = {Entropy production fluctuation theorem and the nonequilibrium work relation for free energy differences},
  journal   = {Physical Review E},
  year      = {1999},
  volume    = {60},
  number    = {3},
  pages     = {2721--2726},
  month     = sep,
  issn      = {1095-3787},
  doi       = {10.1103/physreve.60.2721},
  publisher = {American Physical Society (APS)},
}

@Article{Jarzynski2007,
  author    = {Jarzynski, Christopher},
  title     = {Comparison of far-from-equilibrium work relations},
  journal   = {Comptes Rendus. Physique},
  year      = {2007},
  volume    = {8},
  number    = {5–6},
  pages     = {495--506},
  month     = jun,
  issn      = {1878-1535},
  doi       = {10.1016/j.crhy.2007.04.010},
  publisher = {Cellule MathDoc/Centre Mersenne},
}

@Article{Talkner2007,
  author    = {Talkner, Peter and Lutz, Eric and Hänggi, Peter},
  title     = {Fluctuation theorems: Work is not an observable},
  journal   = {Physical Review E},
  year      = {2007},
  volume    = {75},
  number    = {5},
  month     = may,
  issn      = {1550-2376},
  doi       = {10.1103/physreve.75.050102},
  publisher = {American Physical Society (APS)},
}

@Article{Jarzynski2008,
  author    = {Jarzynski, C.},
  title     = {Nonequilibrium work relations: foundations and applications},
  journal   = {The European Physical Journal B},
  year      = {2008},
  volume    = {64},
  number    = {3–4},
  pages     = {331--340},
  month     = jul,
  issn      = {1434-6036},
  doi       = {10.1140/epjb/e2008-00254-2},
  publisher = {Springer Science and Business Media LLC},
}

@Article{Griffiths1984,
  author    = {Griffiths, Robert B.},
  title     = {Consistent histories and the interpretation of quantum mechanics},
  journal   = {Journal of Statistical Physics},
  year      = {1984},
  volume    = {36},
  number    = {1–2},
  pages     = {219--272},
  month     = jul,
  issn      = {1572-9613},
  doi       = {10.1007/bf01015734},
  publisher = {Springer Science and Business Media LLC},
}

@Article{Omnes1990,
  author    = {Omnès, Roland},
  title     = {From hilbert space to common sense: A synthesis of recent progress in the interpretation of quantum mechanics},
  journal   = {Annals of Physics},
  year      = {1990},
  volume    = {201},
  number    = {2},
  pages     = {354--447},
  month     = aug,
  issn      = {0003-4916},
  doi       = {10.1016/0003-4916(90)90045-p},
  publisher = {Elsevier BV},
}

@Article{Omnes1992,
  author    = {Omnès, Roland},
  title     = {Consistent interpretations of quantum mechanics},
  journal   = {Reviews of Modern Physics},
  year      = {1992},
  volume    = {64},
  number    = {2},
  pages     = {339--382},
  month     = apr,
  issn      = {1539-0756},
  doi       = {10.1103/revmodphys.64.339},
  publisher = {American Physical Society (APS)},
}

@Book{Gell-Mann-Hartle1990,
  title     = {Complexity, Entropy, and the Physics of Information},
  publisher = {Addison-Wesley},
  year      = {1990},
  author    = {M. Gell-Mann and J. B. Hartle},
  editor    = {W. H. Zurek},
}

@Article{Dowker1992,
  author    = {Dowker, H. F. and Halliwell, J. J.},
  title     = {Quantum mechanics of history: The decoherence functional in quantum mechanics},
  journal   = {Physical Review D},
  year      = {1992},
  volume    = {46},
  number    = {4},
  pages     = {1580--1609},
  month     = aug,
  issn      = {0556-2821},
  doi       = {10.1103/physrevd.46.1580},
  publisher = {American Physical Society (APS)},
}

@Book{Klages2013,
  title     = {Nonequilibrium Statistical Physics of Small Systems: Fluctuation Relations and Beyond},
  publisher = {Wiley},
  year      = {2013},
  editor    = {Rainer Klages and Wolfram Just and Christopher Jarzynski},
  month     = feb,
  isbn      = {9783527658701},
  doi       = {10.1002/9783527658701},
}

@Book{Stratonovich1994,
  title     = {Nonlinear Nonequilibrium Thermodynamics {II}: Advanced Theory},
  publisher = {Springer},
  year      = {1994},
  author    = {Rouslan L. Stratonovich},
}

@Book{Goldstein2001,
  title     = {Classical Mechanics},
  publisher = {Pearson},
  year      = {2001},
  author    = {Herbert Goldstein and Charles Poole and John Safko},
}

@Article{Horowitz2007,
  author    = {Horowitz, Jordan and Jarzynski, Christopher},
  title     = {Comparison of work fluctuation relations},
  journal   = {Journal of Statistical Mechanics: Theory and Experiment},
  year      = {2007},
  volume    = {2007},
  number    = {11},
  pages     = {P11002–P11002},
  month     = nov,
  issn      = {1742-5468},
  doi       = {10.1088/1742-5468/2007/11/p11002},
  publisher = {IOP Publishing},
}

@Book{Sakurai2020,
  title     = {Modern Quantum Mechanics},
  publisher = {Cambridge University Press},
  year      = {2020},
  author    = {J. J. Sakurai and Jim Napolitano},
}

@Article{Talkner2008,
  author    = {Talkner, Peter and Hänggi, Peter and Morillo, Manuel},
  title     = {Microcanonical quantum fluctuation theorems},
  journal   = {Physical Review E},
  year      = {2008},
  volume    = {77},
  number    = {5},
  month     = may,
  issn      = {1550-2376},
  doi       = {10.1103/physreve.77.051131},
  publisher = {American Physical Society (APS)},
}

@Book{Polchinski2005,
  title     = {String Theory, Volume 1: An Introduction to the Bosonic String},
  publisher = {Cambridge University Press},
  year      = {2005},
  author    = {Joseph Polchinski},
}

@Book{Polchinski2005_2,
  title     = {String Theory, Volume 2: Superstring Theory and Beyond},
  publisher = {Cambridge University Press},
  year      = {2005},
  author    = {Joseph Polchinski},
}

@Book{Rovelli2007,
  title     = {Quantum Gravity},
  publisher = {Cambridge University Press},
  year      = {2007},
  author    = {Carlo Rovelli},
}

@Article{Basso2023,
  author    = {Basso, Marcos L W and Maziero, Jonas and Céleri, Lucas C},
  title     = {The irreversibility of relativistic time-dilation},
  journal   = {Classical and Quantum Gravity},
  year      = {2023},
  volume    = {40},
  number    = {19},
  pages     = {195001},
  month     = aug,
  issn      = {1361-6382},
  doi       = {10.1088/1361-6382/acf089},
  publisher = {IOP Publishing},
}

@Book{Pathria2021,
  title     = {Statistical Mechanics},
  publisher = {Academic Press},
  year      = {2021},
  author    = {R. K. Pathria and P. D. Beale},
}

@Article{Chawla2023,
  author    = {Chawla, Samarth and Parikh, Maulik},
  title     = {Quantum gravity corrections to the fall of an apple},
  journal   = {Physical Review D},
  year      = {2023},
  volume    = {107},
  number    = {6},
  month     = mar,
  issn      = {2470-0029},
  doi       = {10.1103/physrevd.107.066024},
  publisher = {American Physical Society (APS)},
}

@Article{Westervelt_1971,
  author    = {Westervelt, Peter J.},
  title     = {Scattering of Electromagnetic and Gravitational Waves by a Static Gravitational Field: Comparison Between the Classical (General-Relativistic) and Quantum Field-Theoretic Results},
  journal   = {Physical Review D},
  year      = {1971},
  volume    = {3},
  number    = {10},
  pages     = {2319--2324},
  issn      = {0556-2821},
  doi       = {10.1103/physrevd.3.2319},
  publisher = {American Physical Society (APS)},
}

@Article{Moreira2024,
  author    = {Moreira, Thiago H and Céleri, Lucas C},
  title     = {Entropy production due to spacetime fluctuations},
  journal   = {Classical and Quantum Gravity},
  year      = {2024},
  volume    = {42},
  number    = {2},
  pages     = {025022},
  month     = dec,
  issn      = {1361-6382},
  doi       = {10.1088/1361-6382/ada083},
  publisher = {IOP Publishing},
}

@Article{D_Esposito_2024,
  author    = {D’Esposito, Vittorio and Gubitosi, Giulia},
  title     = {Constraints on quantum spacetime-induced decoherence from neutrino oscillations},
  journal   = {Physical Review D},
  year      = {2024},
  volume    = {110},
  number    = {2},
  month     = jul,
  issn      = {2470-0029},
  doi       = {10.1103/physrevd.110.026004},
  publisher = {American Physical Society (APS)},
}

@Article{Cho_2023,
  author    = {Cho, Hing-Tong and Hu, Bei-Lok},
  title     = {Graviton noise on tidal forces and geodesic congruences},
  journal   = {Physical Review D},
  year      = {2023},
  volume    = {107},
  number    = {8},
  month     = apr,
  issn      = {2470-0029},
  doi       = {10.1103/physrevd.107.084005},
  publisher = {American Physical Society (APS)},
}

@Book{Armas2021,
  title     = {Conversations on Quantum Gravity},
  publisher = {Cambridge University Press},
  year      = {2021},
  editor    = {Jácome Armas},
}

@Article{Navas_2024,
  author    = {S. Navas and others},
  journal   = {Physical Review D},
  title     = {Review of Particle Physics},
  year      = {2024},
  issn      = {2470-0029},
  month     = aug,
  number    = {3},
  volume    = {110},
  doi       = {10.1103/physrevd.110.030001},
  publisher = {American Physical Society (APS)},
}

@Article{Hsiang_2024,
  author    = {Hsiang, Jen-Tsung and Cho, Hing-Tong and Hu, Bei-Lok},
  title     = {Graviton Physics: A Concise Tutorial on the Quantum Field Theory of Gravitons, Graviton Noise, and Gravitational Decoherence},
  journal   = {Universe},
  year      = {2024},
  volume    = {10},
  number    = {8},
  pages     = {306},
  month     = jul,
  issn      = {2218-1997},
  doi       = {10.3390/universe10080306},
  publisher = {MDPI AG},
}

@Book{Percacci2017,
  title     = {An Introduction to Covariant Quantum Gravity and Asymptotic Safety},
  publisher = {World Scientific Publishing Company},
  year      = {2017},
  author    = {Roberto Percacci},
}

@Article{Loll_2019,
  author    = {Loll, R},
  title     = {Quantum gravity from causal dynamical triangulations: a review},
  journal   = {Classical and Quantum Gravity},
  year      = {2019},
  volume    = {37},
  number    = {1},
  pages     = {013002},
  month     = dec,
  issn      = {1361-6382},
  doi       = {10.1088/1361-6382/ab57c7},
  publisher = {IOP Publishing},
}

@Article{Salvio_2018,
  author    = {Salvio, Alberto},
  title     = {Quadratic Gravity},
  journal   = {Frontiers in Physics},
  year      = {2018},
  volume    = {6},
  month     = aug,
  issn      = {2296-424X},
  doi       = {10.3389/fphy.2018.00077},
  publisher = {Frontiers Media SA},
}

@Article{Caldeira_1983,
  author    = {Caldeira, A.O. and Leggett, A.J.},
  title     = {Path integral approach to quantum Brownian motion},
  journal   = {Physica A: Statistical Mechanics and its Applications},
  year      = {1983},
  volume    = {121},
  number    = {3},
  pages     = {587--616},
  month     = sep,
  issn      = {0378-4371},
  doi       = {10.1016/0378-4371(83)90013-4},
  publisher = {Elsevier BV},
}

@Book{Birrell1984,
  author    = {N. D. Birrell and P. C. W. Davies},
  publisher = {Cambridge University Press},
  title     = {Quantum Fields in Curved Space},
  year      = {1984},
}

@Article{Ragusa_2003,
  author    = {Ragusa, S. and Céleri, L. C.},
  title     = {Letter: Solar Gravitational Deflection of a Graviton},
  journal   = {General Relativity and Gravitation},
  year      = {2003},
  volume    = {35},
  number    = {6},
  pages     = {1125--1126},
  month     = jun,
  issn      = {1572-9532},
  doi       = {10.1023/a:1024029220395},
  publisher = {Springer Science and Business Media LLC},
}

@Article{Joos_1985,
  author    = {Joos, E. and Zeh, H. D.},
  title     = {The emergence of classical properties through interaction with the environment},
  journal   = {Zeitschrift für Physik B Condensed Matter},
  year      = {1985},
  volume    = {59},
  number    = {2},
  pages     = {223--243},
  issn      = {1434-6036},
  doi       = {10.1007/bf01725541},
  publisher = {Springer Science and Business Media LLC},
}

@Article{Zurek_1982,
  author    = {Zurek, W. H.},
  title     = {Environment-induced superselection rules},
  journal   = {Physical Review D},
  year      = {1982},
  volume    = {26},
  number    = {8},
  pages     = {1862--1880},
  issn      = {0556-2821},
  doi       = {10.1103/physrevd.26.1862},
  publisher = {American Physical Society (APS)},
}

@Article{Zurek_2003,
  author    = {Zurek, Wojciech Hubert},
  title     = {Decoherence, einselection, and the quantum origins of the classical},
  journal   = {Reviews of Modern Physics},
  year      = {2003},
  volume    = {75},
  number    = {3},
  pages     = {715--775},
  issn      = {1539-0756},
  doi       = {10.1103/revmodphys.75.715},
  publisher = {American Physical Society (APS)},
}

@Article{Zurek_1991,
  author    = {Zurek, Wojciech H.},
  title     = {Decoherence and the Transition from Quantum to Classical},
  journal   = {Physics Today},
  year      = {1991},
  volume    = {44},
  number    = {10},
  pages     = {36--44},
  issn      = {1945-0699},
  doi       = {10.1063/1.881293},
  publisher = {AIP Publishing},
}

@Article{Schlosshauer_2005,
  author    = {Schlosshauer, Maximilian},
  title     = {Decoherence, the measurement problem, and interpretations of quantum mechanics},
  journal   = {Reviews of Modern Physics},
  year      = {2005},
  volume    = {76},
  number    = {4},
  pages     = {1267--1305},
  issn      = {1539-0756},
  doi       = {10.1103/revmodphys.76.1267},
  publisher = {American Physical Society (APS)},
}

@Article{K_bler_1973,
  author    = {Kübler, O and Zeh, H.D},
  title     = {Dynamics of quantum correlations},
  journal   = {Annals of Physics},
  year      = {1973},
  volume    = {76},
  number    = {2},
  pages     = {405--418},
  issn      = {0003-4916},
  doi       = {10.1016/0003-4916(73)90040-7},
  publisher = {Elsevier BV},
}

@Article{Zurek_1981,
  author    = {Zurek, W. H.},
  title     = {Pointer basis of quantum apparatus: Into what mixture does the wave packet collapse?},
  journal   = {Physical Review D},
  year      = {1981},
  volume    = {24},
  number    = {6},
  pages     = {1516--1525},
  issn      = {0556-2821},
  doi       = {10.1103/physrevd.24.1516},
  publisher = {American Physical Society (APS)},
}

@Book{Kubo1995,
  title     = {Statistical Physics {II}: Nonequilibrium Statistical Mechanics},
  publisher = {Springer Nature},
  year      = {1995},
  author    = {Ryogo Kubo and Morikazu Toda and Natsuki Hashitsume},
  edition   = {2nd},
}

@Book{LandauStat1,
  title     = {Statistical Physics {I}},
  publisher = {Butterworth-Heinemann},
  year      = {1980},
  author    = {L. D. Landau and E. M. Lifshitz and L. P. Pitaevskii},
  volume    = {5},
  series    = {Course of Theoretical Physics},
  edition   = {3rd},
}

@Book{Schutz2022,
  author    = {Schutz, Bernard F.},
  publisher = {Cambridge University Press},
  title     = {A First Course in General Relativity},
  year      = {2022},
  address   = {Cambridge},
  edition   = {3rd ed.},
  isbn      = {9781108586597},
  ppn_gvk   = {1913441318},
}

@Book{Maggiore2018,
  author    = {Maggiore, Michele},
  publisher = {Oxford University Press},
  title     = {Gravitational Waves},
  year      = {2018},
  address   = {Oxford},
  isbn      = {9780191074479},
  ppn_gvk   = {1809561213},
  subtitle  = {Volume 2: Astrophysics and Cosmology},
}

@Book{Maggiore2007,
  author    = {Maggiore, Michele},
  publisher = {Oxford University Press},
  title     = {Gravitational Waves},
  year      = {2007},
  address   = {Oxford},
  isbn      = {9780191524745},
  ppn_gvk   = {1809530725},
  subtitle  = {Volume 1: Theory and Experiments},
}

@Book{Griffiths2008,
  author    = {Griffiths, David J.},
  publisher = {Wiley-VCH},
  title     = {Introduction to Elementary Particles},
  year      = {2008},
  address   = {Weinheim},
  isbn      = {9783527618477},
  ppn_gvk   = {1672044588},
}

@Article{Penrose2014,
  author  = {Roger Penrose},
  journal = {Foundations of Physics},
  title   = {On the Gravitization of Quantum Mechanics 1: Quantum State Reduction},
  year    = {2014},
  pages   = {557–575},
  volume  = {44},
  doi     = {10.1007/s10701-013-9770-0},
}

@Article{Oppenheim2023,
  author    = {Oppenheim, Jonathan},
  journal   = {Physical Review X},
  title     = {A Postquantum Theory of Classical Gravity?},
  year      = {2023},
  issn      = {2160-3308},
  month     = dec,
  number    = {4},
  pages     = {041040},
  volume    = {13},
  doi       = {10.1103/physrevx.13.041040},
  publisher = {American Physical Society (APS)},
}

@Article{Eppley1977,
  author    = {Eppley, Kenneth and Hannah, Eric},
  journal   = {Foundations of Physics},
  title     = {The necessity of quantizing the gravitational field},
  year      = {1977},
  issn      = {1572-9516},
  month     = feb,
  number    = {1–2},
  pages     = {51--68},
  volume    = {7},
  doi       = {10.1007/bf00715241},
  publisher = {Springer Science and Business Media LLC},
}

@Article{Albers2008,
  author    = {Albers, Mark and Kiefer, Claus and Reginatto, Marcel},
  journal   = {Physical Review D},
  title     = {Measurement analysis and quantum gravity},
  year      = {2008},
  issn      = {1550-2368},
  month     = sep,
  number    = {6},
  pages     = {064051},
  volume    = {78},
  doi       = {10.1103/physrevd.78.064051},
  publisher = {American Physical Society (APS)},
}

@Article{Carney2019,
  author    = {Carney, Daniel and Stamp, Philip C E and Taylor, Jacob M},
  journal   = {Classical and Quantum Gravity},
  title     = {Tabletop experiments for quantum gravity: a user’s manual},
  year      = {2019},
  issn      = {1361-6382},
  month     = jan,
  number    = {3},
  pages     = {034001},
  volume    = {36},
  doi       = {10.1088/1361-6382/aaf9ca},
  publisher = {IOP Publishing},
}

@Article{Pikovski2012,
  author    = {Pikovski, Igor and Vanner, Michael R. and Aspelmeyer, Markus and Kim, M. S. and Brukner, Časlav},
  journal   = {Nature Physics},
  title     = {Probing {P}lanck-scale physics with quantum optics},
  year      = {2012},
  issn      = {1745-2481},
  month     = mar,
  number    = {5},
  pages     = {393--397},
  volume    = {8},
  doi       = {10.1038/nphys2262},
  publisher = {Springer Science and Business Media LLC},
}

@Article{Chevalier2020,
  author    = {Chevalier, Hadrien and Paige, A. J. and Kim, M. S.},
  journal   = {Physical Review A},
  title     = {Witnessing the nonclassical nature of gravity in the presence of unknown interactions},
  year      = {2020},
  issn      = {2469-9934},
  month     = aug,
  number    = {2},
  pages     = {022428},
  volume    = {102},
  doi       = {10.1103/physreva.102.022428},
  publisher = {American Physical Society (APS)},
}

@Article{Carney2021,
  author    = {Carney, Daniel and Müller, Holger and Taylor, Jacob M.},
  journal   = {PRX Quantum},
  title     = {Using an Atom Interferometer to Infer Gravitational Entanglement Generation},
  year      = {2021},
  issn      = {2691-3399},
  month     = aug,
  number    = {3},
  pages     = {030330},
  volume    = {2},
  doi       = {10.1103/prxquantum.2.030330},
  publisher = {American Physical Society (APS)},
}

@Article{Anastopoulos2020,
  author    = {Anastopoulos, C and Hu, B L},
  journal   = {Classical and Quantum Gravity},
  title     = {Quantum superposition of two gravitational cat states},
  year      = {2020},
  issn      = {1361-6382},
  month     = nov,
  number    = {23},
  pages     = {235012},
  volume    = {37},
  doi       = {10.1088/1361-6382/abbe6f},
  publisher = {IOP Publishing},
}

@Article{Pedernales2022,
  author    = {Pedernales, Julen S. and Streltsov, Kirill and Plenio, Martin B.},
  journal   = {Physical Review Letters},
  title     = {Enhancing Gravitational Interaction between Quantum Systems by a Massive Mediator},
  year      = {2022},
  issn      = {1079-7114},
  month     = mar,
  number    = {11},
  pages     = {110401},
  volume    = {128},
  doi       = {10.1103/physrevlett.128.110401},
  publisher = {American Physical Society (APS)},
}

@Article{Kaku2023,
  author    = {Kaku, Youka and Fujita, Tomohiro and Matsumura, Akira},
  journal   = {Physical Review D},
  title     = {Enhancement of quantum gravity signal in an optomechanical experiment},
  year      = {2023},
  issn      = {2470-0029},
  month     = nov,
  number    = {10},
  pages     = {106014},
  volume    = {108},
  doi       = {10.1103/physrevd.108.106014},
  publisher = {American Physical Society (APS)},
}

@Article{Bose2025,
  author    = {S. Bose and others},
  journal   = {Reviews of Modern Physics},
  title     = {Massive quantum systems as interfaces of quantum mechanics and gravity},
  year      = {2025},
  issn      = {1539-0756},
  month     = feb,
  number    = {1},
  pages     = {015003},
  volume    = {97},
  doi       = {10.1103/revmodphys.97.015003},
  publisher = {American Physical Society (APS)},
}

@Misc{Beyer2025,
  author    = {Beyer, Konstantin and Kim, M. S. and Pikovski, Igor},
  title     = {A One-sided Witness for the Quantumness of Gravitational Dynamics},
  year      = {2025},
  copyright = {arXiv.org perpetual, non-exclusive license},
  doi       = {10.48550/ARXIV.2507.15588},
  publisher = {arXiv},
}

@Article{Aziz2025,
  author    = {Aziz, Joseph and Howl, Richard},
  journal   = {Nature},
  title     = {Classical theories of gravity produce entanglement},
  year      = {2025},
  issn      = {1476-4687},
  month     = oct,
  number    = {8086},
  pages     = {813--817},
  volume    = {646},
  doi       = {10.1038/s41586-025-09595-7},
  publisher = {Springer Science and Business Media LLC},
}

@Misc{Marletto2025,
  author    = {Marletto, Chiara and Oppenheim, Jonathan and Vedral, Vlatko and Wilson, Elizabeth},
  title     = {Classical gravity cannot mediate entanglement},
  year      = {2025},
  copyright = {arXiv.org perpetual, non-exclusive license},
  doi       = {10.48550/ARXIV.2511.07348},
  publisher = {arXiv},
}

@Book{Reuter2019,
  author    = {Reuter, Martin},
  publisher = {Cambridge University Press},
  title     = {Quantum gravity and the functional renormalization group: The road towards asymptotic safety},
  year      = {2019},
  series    = {Cambridge monographs on mathematical physics},
  ppn_gvk   = {1885353197},
}

@Article{Basile2025,
  author    = {Basile, Ivano and Buoninfante, Luca and Di Filippo, Francesco and Knorr, Benjamin and Platania, Alessia and Tokareva, Anna},
  journal   = {SciPost Physics Lecture Notes},
  title     = {Lectures in quantum gravity},
  year      = {2025},
  issn      = {2590-1990},
  month     = aug,
  doi       = {10.21468/scipostphyslectnotes.98},
  publisher = {Stichting SciPost},
}

@Article{Ambjoern2012,
  author    = {Ambjørn, J. and Görlich, A. and Jurkiewicz, J. and Loll, R.},
  journal   = {Physics Reports},
  title     = {Nonperturbative quantum gravity},
  year      = {2012},
  issn      = {0370-1573},
  month     = oct,
  number    = {4–5},
  pages     = {127--210},
  volume    = {519},
  doi       = {10.1016/j.physrep.2012.03.007},
  publisher = {Elsevier BV},
}

@Article{Ashtekar2021,
  author    = {Ashtekar, Abhay and Bianchi, Eugenio},
  journal   = {Reports on Progress in Physics},
  title     = {A short review of loop quantum gravity},
  year      = {2021},
  issn      = {1361-6633},
  month     = mar,
  number    = {4},
  pages     = {042001},
  volume    = {84},
  doi       = {10.1088/1361-6633/abed91},
  publisher = {IOP Publishing},
}

@Article{Donoghue2022,
  author    = {John F. Donoghue and Gabriel Menezes},
  journal   = {Il Nuovo Cimento C},
  title     = {On quadratic gravity},
  year      = {2022},
  issn      = {03905551, 03905551},
  pages     = {1–11},
  volume    = {45},
  doi       = {10.1393/ncc/i2022-22026-7},
  publisher = {SIF},
}

@Article{Buoninfante2025,
  author    = {Buoninfante, Luca and Knorr, Benjamin and Kumar, K. Sravan and Platania, Alessia and Anselmi, Damiano and Basile, Ivano and Bjerrum-Bohr, N. Emil J. and Brandenberger, Robert and Carrillo Gonzalez, Mariana and Davis, Anne-Christine and Dittrich, Bianca and Di Vecchia, Paolo and Donoghue, John and Dowker, Fay and Dvali, Gia and Eichhorn, Astrid and Giddings, Steven B. and Gnecchi, Alessandra and Gubitosi, Giulia and Heisenberg, Lavinia and Kallosh, Renata and Koshelev, Alexey S. and Liberati, Stefano and Loll, Renate and Modesto, Leonardo and Moniz, Paulo and Oriti, Daniele and Papadoulaki, Olga and Pawlowski, Jan M. and Percacci, Roberto and Rachwał, Lesław and Sakellariadou, Mairi and Salvio, Alberto and Stelle, Kellogg and Surya, Sumati and Tseytlin, Arkady and Turok, Neil and van Riet, Thomas and Woodard, Richard},
  journal   = {SciPost Physics Community Reports},
  title     = {Visions in quantum gravity},
  year      = {2025},
  issn      = {3051-3197},
  month     = aug,
  doi       = {10.21468/scipostphyscommrep.11},
  publisher = {Stichting SciPost},
}

@Article{Donoghue1994,
  author    = {John F. Donoghue},
  journal   = {Physical Review Letters},
  title     = {Leading quantum correction to the Newtonian potential},
  year      = {1994},
  issn      = {0031-9007},
  number    = {19},
  pages     = {2996--2999},
  volume    = {72},
  doi       = {10.1103/physrevlett.72.2996},
  publisher = {American Physical Society (APS)},
}

@Book{Poisson2004,
  author    = {Poisson, Eric},
  publisher = {Cambridge University Press},
  title     = {A relativist's toolkit},
  year      = {2004},
  address   = {Cambridge},
  isbn      = {9780511606601},
  ppn_gvk   = {883419300},
  subtitle  = {The mathematics of black-hole mechanics},
}

@Book{Levin2016,
  author    = {Levin, Janna},
  publisher = {Alfred A. Knopf},
  title     = {Black hole blues and other songs from outer space},
  year      = {2016},
  address   = {New York},
  edition   = {First edition},
  isbn      = {030794848X},
  ppn_gvk   = {844050628},
}

@Article{Hulse1975,
  author    = {Hulse, R. A. and Taylor, J. H.},
  journal   = {The Astrophysical Journal},
  title     = {Discovery of a pulsar in a binary system},
  year      = {1975},
  issn      = {1538-4357},
  month     = jan,
  pages     = {L51},
  volume    = {195},
  doi       = {10.1086/181708},
  publisher = {American Astronomical Society},
}

@Book{Weinberg2013a,
  author    = {Steven Weinberg},
  publisher = {Cambridge University Press},
  title     = {The Quantum Theory of Fields: Supersymmetry},
  year      = {2013},
  isbn      = {9781139644198},
  volume    = {03},
  ppn_gvk   = {148664774X},
}

@Book{Wald1994,
  author    = {Wald, Robert M.},
  publisher = {University of Chicago Press},
  title     = {Quantum field theory in curved spacetime and black hole thermodynamics},
  year      = {1994},
  isbn      = {9780226870274},
  series    = {Chicago lectures in physics},
  ppn_gvk   = {152112421},
}

@Article{Tobar2024,
  author    = {Tobar, Germain and Manikandan, Sreenath K. and Beitel, Thomas and Pikovski, Igor},
  journal   = {Nature Communications},
  title     = {Detecting single gravitons with quantum sensing},
  year      = {2024},
  issn      = {2041-1723},
  month     = aug,
  number    = {1},
  volume    = {15},
  doi       = {10.1038/s41467-024-51420-8},
  publisher = {Springer Science and Business Media LLC},
}

@Article{Diosi1989,
  author    = {Diósi, L.},
  journal   = {Physical Review A},
  title     = {Models for universal reduction of macroscopic quantum fluctuations},
  year      = {1989},
  issn      = {0556-2791},
  month     = aug,
  number    = {3},
  pages     = {1165--1174},
  volume    = {40},
  doi       = {10.1103/physreva.40.1165},
  publisher = {American Physical Society (APS)},
}

@Article{Diosi2007,
  author    = {Diósi, Lajos},
  journal   = {Journal of Physics A: Mathematical and Theoretical},
  title     = {Notes on certain Newton gravity mechanisms of wavefunction localization and decoherence},
  year      = {2007},
  issn      = {1751-8121},
  month     = mar,
  number    = {12},
  pages     = {2989--2995},
  volume    = {40},
  doi       = {10.1088/1751-8113/40/12/s07},
  publisher = {IOP Publishing},
}

@Article{Diosi2014,
  author    = {Diósi, Lajos},
  journal   = {Journal of Physics: Conference Series},
  title     = {Newton force from wave function collapse: speculation and test},
  year      = {2014},
  issn      = {1742-6596},
  month     = apr,
  pages     = {012020},
  volume    = {504},
  doi       = {10.1088/1742-6596/504/1/012020},
  publisher = {IOP Publishing},
}

@Article{Karolyhazy1966,
  author    = {Karolyhazy, F.},
  journal   = {Il Nuovo Cimento A},
  title     = {Gravitation and quantum mechanics of macroscopic objects},
  year      = {1966},
  issn      = {1826-9869},
  month     = mar,
  number    = {2},
  pages     = {390--402},
  volume    = {42},
  doi       = {10.1007/bf02717926},
  publisher = {Springer Science and Business Media LLC},
}

@Misc{Adler2014,
  author    = {Adler, Stephen L.},
  title     = {Gravitation and the noise needed in objective reduction models},
  year      = {2014},
  copyright = {arXiv.org perpetual, non-exclusive license},
  doi       = {10.48550/ARXIV.1401.0353},
  publisher = {arXiv},
}

@Article{Frenkel1990,
  author    = {Frenkel, Andor},
  journal   = {Foundations of Physics},
  title     = {Spontaneous localizations of the wave function and classical behavior},
  year      = {1990},
  issn      = {1572-9516},
  month     = feb,
  number    = {2},
  pages     = {159--188},
  volume    = {20},
  doi       = {10.1007/bf00731645},
  publisher = {Springer Science and Business Media LLC},
}

@Article{Diosi1984,
  author    = {Diósi, L.},
  journal   = {Physics Letters A},
  title     = {Gravitation and quantum-mechanical localization of macro-objects},
  year      = {1984},
  issn      = {0375-9601},
  month     = oct,
  number    = {4–5},
  pages     = {199--202},
  volume    = {105},
  doi       = {10.1016/0375-9601(84)90397-9},
  publisher = {Elsevier BV},
}

@Article{Penrose1996,
  author    = {Penrose, Roger},
  journal   = {General Relativity and Gravitation},
  title     = {On Gravity’s role in Quantum State Reduction},
  year      = {1996},
  issn      = {1572-9532},
  month     = may,
  number    = {5},
  pages     = {581--600},
  volume    = {28},
  doi       = {10.1007/bf02105068},
  publisher = {Springer Science and Business Media LLC},
}

@Article{Linet1976,
  author    = {Linet, B. and Tourrenc, P.},
  journal   = {Canadian Journal of Physics},
  title     = {Changement de phase dans un champ de gravitation: Possibilité de détection interférentielle},
  year      = {1976},
  issn      = {1208-6045},
  month     = jun,
  number    = {11},
  pages     = {1129--1133},
  volume    = {54},
  doi       = {10.1139/p76-136},
  publisher = {Canadian Science Publishing},
}

@Article{Stodolsky1979,
  author    = {Stodolsky, Leo},
  journal   = {General Relativity and Gravitation},
  title     = {Matter and light wave interferometry in gravitational fields},
  year      = {1979},
  issn      = {1572-9532},
  month     = dec,
  number    = {6},
  pages     = {391--405},
  volume    = {11},
  doi       = {10.1007/bf00759302},
  publisher = {Springer Science and Business Media LLC},
}

@Article{Cai1989,
  author    = {Cai, Y Q and Papini, G},
  journal   = {Classical and Quantum Gravity},
  title     = {Particle interferometry in weak gravitational fields},
  year      = {1989},
  issn      = {1361-6382},
  month     = mar,
  number    = {3},
  pages     = {407--418},
  volume    = {6},
  doi       = {10.1088/0264-9381/6/3/017},
  publisher = {IOP Publishing},
}

@Article{Reynaud2004,
  author    = {Reynaud, Serge and Lamine, Brahim and Lambrecht, Astrid and Neto, Paulo Maia and Jaekel, Marc-Thierry},
  journal   = {General Relativity and Gravitation},
  title     = {{HYPER} and Gravitational Decoherence},
  year      = {2004},
  issn      = {1572-9532},
  month     = oct,
  number    = {10},
  pages     = {2271--2288},
  volume    = {36},
  doi       = {10.1023/b:gerg.0000046183.31629.02},
  publisher = {Springer Science and Business Media LLC},
}

@Article{Lamine2006,
  author    = {Lamine, Brahim and Hervé, Rémy and Lambrecht, Astrid and Reynaud, Serge},
  journal   = {Physical Review Letters},
  title     = {Ultimate Decoherence Border for Matter-Wave Interferometry},
  year      = {2006},
  issn      = {1079-7114},
  month     = feb,
  number    = {5},
  pages     = {050405},
  volume    = {96},
  doi       = {10.1103/physrevlett.96.050405},
  publisher = {American Physical Society (APS)},
}

@Article{Power2000,
  author    = {Power, W. L. and Percival, I. C.},
  journal   = {Proceedings of the Royal Society of London. Series A: Mathematical, Physical and Engineering Sciences},
  title     = {Decoherence of quantum wave packets due to interaction with conformal space–time fluctuations},
  year      = {2000},
  issn      = {1471-2946},
  month     = apr,
  number    = {1996},
  pages     = {955--968},
  volume    = {456},
  doi       = {10.1098/rspa.2000.0544},
  publisher = {The Royal Society},
}

@Article{Breuer2009,
  author    = {Breuer, Heinz-Peter and Göklü, Ertan and Lämmerzahl, Claus},
  journal   = {Classical and Quantum Gravity},
  title     = {Metric fluctuations and decoherence},
  year      = {2009},
  issn      = {1361-6382},
  month     = apr,
  number    = {10},
  pages     = {105012},
  volume    = {26},
  doi       = {10.1088/0264-9381/26/10/105012},
  publisher = {IOP Publishing},
}

@Article{Gerlich2011,
  author    = {Gerlich, Stefan and Eibenberger, Sandra and Tomandl, Mathias and Nimmrichter, Stefan and Hornberger, Klaus and Fagan, Paul J. and Tüxen, Jens and Mayor, Marcel and Arndt, Markus},
  journal   = {Nature Communications},
  title     = {Quantum interference of large organic molecules},
  year      = {2011},
  issn      = {2041-1723},
  month     = apr,
  number    = {1},
  volume    = {2},
  doi       = {10.1038/ncomms1263},
  publisher = {Springer Science and Business Media LLC},
}

@Article{Fein2019,
  author    = {Fein, Yaakov Y. and Geyer, Philipp and Zwick, Patrick and Kiałka, Filip and Pedalino, Sebastian and Mayor, Marcel and Gerlich, Stefan and Arndt, Markus},
  journal   = {Nature Physics},
  title     = {Quantum superposition of molecules beyond 25 kDa},
  year      = {2019},
  issn      = {1745-2481},
  month     = sep,
  number    = {12},
  pages     = {1242--1245},
  volume    = {15},
  doi       = {10.1038/s41567-019-0663-9},
  publisher = {Springer Science and Business Media LLC},
}

@Article{Aguiar2025,
  author    = {Gabriel H. S. Aguiar and George E. A. Matsas},
  journal   = {Physical Review D},
  title     = {Simple gravitational self-decoherence model},
  year      = {2025},
  issn      = {2470-0029},
  month     = aug,
  number    = {4},
  volume    = {112},
  doi       = {10.1103/thcr-22nb},
  publisher = {American Physical Society (APS)},
}

@Article{HABA2002,
  author    = {Haba, Z. and Kleinert, H.},
  journal   = {International Journal of Modern Physics A},
  title     = {QUANTUM-{L}IOUVILLE AND {L}ANGEVIN EQUATIONS FOR GRAVITATIONAL RADIATION DAMPING},
  year      = {2002},
  issn      = {1793-656X},
  month     = oct,
  number    = {26},
  pages     = {3729--3736},
  volume    = {17},
  doi       = {10.1142/s0217751x02009862},
  publisher = {World Scientific Pub Co Pte Lt},
}

@Article{Oniga2016,
  author    = {Oniga, Teodora and Wang, Charles H.-T.},
  journal   = {Physical Review D},
  title     = {Quantum gravitational decoherence of light and matter},
  year      = {2016},
  issn      = {2470-0029},
  month     = feb,
  number    = {4},
  pages     = {044027},
  volume    = {93},
  doi       = {10.1103/physrevd.93.044027},
  publisher = {American Physical Society (APS)},
}

@Article{Fahn2025,
  author    = {Fahn, Max Joseph and Giesel, Kristina},
  journal   = {Classical and Quantum Gravity},
  title     = {Gravitationally induced decoherence of a scalar field: investigating the one-particle sector and its interplay with renormalisation},
  year      = {2025},
  issn      = {1361-6382},
  month     = sep,
  number    = {17},
  pages     = {175019},
  volume    = {42},
  doi       = {10.1088/1361-6382/adfb1d},
  publisher = {IOP Publishing},
}

@Article{Cho2025,
  author    = {Cho, Hing-Tong and Hu, Bei-Lok},
  journal   = {Physical Review D},
  title     = {Non-{M}arkovian quantum master and {F}okker-{P}lanck equation for gravitational systems and gravitational decoherence},
  year      = {2025},
  issn      = {2470-0029},
  month     = aug,
  number    = {4},
  volume    = {112},
  doi       = {10.1103/b4n2-426h},
  publisher = {American Physical Society (APS)},
}

@Article{Lagouvardos2021,
  author    = {Lagouvardos, Michalis and Anastopoulos, Charis},
  journal   = {Classical and Quantum Gravity},
  title     = {Gravitational decoherence of photons},
  year      = {2021},
  issn      = {1361-6382},
  month     = may,
  number    = {11},
  pages     = {115012},
  volume    = {38},
  doi       = {10.1088/1361-6382/abf2f3},
  publisher = {IOP Publishing},
}

@Article{Pfister2016,
  author    = {Pfister, C. and Kaniewski, J. and Tomamichel, M. and Mantri, A. and Schmucker, R. and McMahon, N. and Milburn, G. and Wehner, S.},
  journal   = {Nature Communications},
  title     = {A universal test for gravitational decoherence},
  year      = {2016},
  issn      = {2041-1723},
  month     = oct,
  number    = {1},
  volume    = {7},
  doi       = {10.1038/ncomms13022},
  publisher = {Springer Science and Business Media LLC},
}

@Article{Arteaga2004,
  author    = {Arteaga, Daniel and Parentani, Renaud and Verdaguer, Enric},
  journal   = {Physical Review D},
  title     = {Propagation in a thermal graviton background},
  year      = {2004},
  issn      = {1550-2368},
  month     = aug,
  number    = {4},
  pages     = {044019},
  volume    = {70},
  doi       = {10.1103/physrevd.70.044019},
  publisher = {American Physical Society (APS)},
}

@Article{Carlesso2016,
  author    = {Carlesso, Matteo and Bassi, Angelo},
  journal   = {Physics Letters A},
  title     = {Decoherence due to gravitational time dilation: Analysis of competing decoherence effects},
  year      = {2016},
  issn      = {0375-9601},
  month     = jul,
  number    = {31–32},
  pages     = {2354--2358},
  volume    = {380},
  doi       = {10.1016/j.physleta.2016.05.034},
  publisher = {Elsevier BV},
}

@Article{Hornberger2012,
  author    = {Hornberger, Klaus and Gerlich, Stefan and Haslinger, Philipp and Nimmrichter, Stefan and Arndt, Markus},
  journal   = {Reviews of Modern Physics},
  title     = {Colloquium: Quantum interference of clusters and molecules},
  year      = {2012},
  issn      = {1539-0756},
  month     = feb,
  number    = {1},
  pages     = {157--173},
  volume    = {84},
  doi       = {10.1103/revmodphys.84.157},
  publisher = {American Physical Society (APS)},
}

@Article{Colas2023,
  author    = {Colas, Thomas and Grain, Julien and Vennin, Vincent},
  journal   = {Europhysics Letters},
  title     = {Quantum recoherence in the early universe},
  year      = {2023},
  issn      = {1286-4854},
  month     = jun,
  number    = {6},
  pages     = {69002},
  volume    = {142},
  doi       = {10.1209/0295-5075/acdd94},
  publisher = {IOP Publishing},
}

@Article{Kranas2025,
  author    = {Kranas, Dimitrios and Grain, Julien and Vennin, Vincent},
  journal   = {Physical Review D},
  title     = {Recoherence, adiabaticity, and Markovianity in Gaussian maps},
  year      = {2025},
  issn      = {2470-0029},
  month     = oct,
  number    = {7},
  volume    = {112},
  doi       = {10.1103/r8pq-kms4},
  publisher = {American Physical Society (APS)},
}

@Article{Basso2025,
  author    = {Marcos L. W. Basso and Jonas Maziero and Lucas C. Céleri},
  journal   = {Physical Review Letters},
  title     = {Quantum Detailed Fluctuation Theorem in Curved Spacetimes: The Observer Dependent Nature of Entropy Production},
  year      = {2025},
  issn      = {1079-7114},
  month     = feb,
  number    = {5},
  pages     = {050406},
  volume    = {134},
  doi       = {10.1103/physrevlett.134.050406},
  publisher = {American Physical Society (APS)},
}

@Article{Parikh2020,
  author    = {Parikh, Maulik and Wilczek, Frank and Zahariade, George},
  journal   = {International Journal of Modern Physics D},
  title     = {The noise of gravitons},
  year      = {2020},
  issn      = {1793-6594},
  month     = jul,
  number    = {14},
  pages     = {2042001},
  volume    = {29},
  doi       = {10.1142/s0218271820420018},
  publisher = {World Scientific Pub Co Pte Ltd},
}

@InBook{Keldysh2023,
  author    = {Keldysh, L. V.},
  pages     = {47--55},
  publisher = {WORLD SCIENTIFIC},
  title     = {Diagram technique for nonequilibrium processes},
  year      = {2023},
  isbn      = {9789811279461},
  month     = nov,
  booktitle = {Selected Papers of {L}eonid {V} {K}eldysh},
  doi       = {10.1142/9789811279461_0007},
}

@Article{Costa2025,
  author  = {Rafael L. S. Costa and Marcos L. W. Basso and Jonas Maziero and Lucas C. Céleri},
  journal = {Physical Review D},
  title   = {Work distribution of quantum fields in static curved spacetimes},
  year    = {2026},
  issn    = {2470-0029},
  month   = jan,
  pages   = {025010},
  volume  = {113},
  doi     = {10.1103/d26w-cj7z},
}

@Article{Pedalino2025,
  author  = {Sebastian Pedalino and Bruno E. Ramírez-Galindo and Richard Ferstl and Klaus Hornberger and Markus Arndt and Stefan Gerlich},
  journal = {Nature},
  title   = {Probing quantum mechanics with nanoparticle matter-wave interferometry},
  year    = {2026},
  issn    = {1476-4687},
  month   = jan,
  number  = {8098},
  pages   = {866--870},
  volume  = {649},
  doi     = {10.1038/s41586-025-09917-9},
}

@Article{Domi2024,
  author  = {Alba Domi and Thomas Eberl and Max Joseph Fahn and Kristina Giesel and Lukas Hennig and Ulrich Katz and Roman Kemper and Michael Kobler},
  journal = {Journal of Cosmology and Astroparticle Physics},
  title   = {Understanding gravitationally induced decoherence parameters in neutrino oscillations using a microscopic quantum mechanical model},
  year    = {2024},
  issn    = {1475-7516},
  month   = nov,
  number  = {11},
  pages   = {006},
  volume  = {2024},
  doi     = {10.1088/1475-7516/2024/11/006},
}

@Article{Moreira2026,
  author        = {Moreira, Thiago H. and Céleri, Lucas Chibebe},
  title         = {Gravitational decoherence and recoherence of a composite particle: the interplay between gravitons and a classical Newtonian potential},
  year          = {2026},
  month         = feb,
  archiveprefix = {arXiv},
  copyright     = {Creative Commons Attribution 4.0 International},
  doi           = {10.48550/ARXIV.2602.22517},
  eprint        = {2602.22517},
  primaryclass  = {quant-ph},
  publisher     = {arXiv},
}

@Article{Flanagan2005,
  author    = {Flanagan, Éanna É and Hughes, Scott A},
  journal   = {New Journal of Physics},
  title     = {The basics of gravitational wave theory},
  year      = {2005},
  issn      = {1367-2630},
  month     = sep,
  pages     = {204--204},
  volume    = {7},
  doi       = {10.1088/1367-2630/7/1/204},
  publisher = {IOP Publishing},
}

\end{document}